\documentclass[11pt,openbib]{article}

\usepackage{amsmath,amssymb,amsthm,cancel,hyperref,graphicx,xcolor}
\usepackage{mathrsfs,wasysym}
\usepackage{booktabs}
\usepackage{tikz}
\usetikzlibrary{calc}
\usepackage{placeins}
\usepackage{bm}

\usepackage[nottoc,numbib]{tocbibind}

\numberwithin{equation}{section}

\usepackage{fullpage}
\usepackage{slashed}
\usepackage{graphicx}

\usepackage{soul}

\DeclareRobustCommand{\Sec}[1]{Sec.~\ref{#1}}

\DeclareRobustCommand{\Fig}[1]{Fig.~\ref{#1}}

\DeclareRobustCommand{\Eq}[1]{Eq.~(\ref{#1})}
\DeclareRobustCommand{\Eqs}[2]{Eqs.~(\ref{#1}) and (\ref{#2})}
\DeclareRobustCommand{\InRef}[1]{Ref.~\cite{#1}}
\DeclareRobustCommand{\Refs}[1]{Refs.~\cite{#1}}

\newcommand{\be}{\begin{eqnarray}}
\newcommand{\ee}{\end{eqnarray}}

\allowdisplaybreaks

\usepackage{color}
\definecolor{darkblue}{rgb}{0,0,0.5}
\definecolor{darkgreen}{rgb}{0,0.5,0}

\title{QCD Masterclass Lectures on Jet Physics and Machine Learning}
\author{Andrew J.~Larkoski}
\date{%
{\it Email:} \url{larkoa@gmail.com}\\[2ex]
    \today
}

\begin{document}
\maketitle

\begin{abstract}
\noindent These lectures were presented at the 2024 QCD Masterclass in Saint-Jacut-de-la-Mer, France.  They introduce and review fundamental theorems and principles of machine learning within the context of collider particle physics, focused on application to jet identification and discrimination.  Numerous examples of binary discrimination in jet physics are studied in detail, including $H\to b\bar b$ identification in fixed-order perturbation theory, generic one- versus two-prong discrimination with parametric power counting techniques, and up versus down quark jet classification by assuming the central limit theorem, isospin conservation, and a convergent moment expansion of the single particle energy distribution.  Quark versus gluon jet discrimination is considered in multiple contexts, from using additive, infrared and collinear safe observables, to using hadronic multiplicity, and to including measurements of the jet charge.  While many of the results presented here are well known, some novel results are presented, the most prominent being a parametrized expression for the likelihood ratio of quark versus gluon discrimination for jets on which hadronic multiplicity and jet charge are simultaneously measured.  End-of-lecture exercises are also provided.
\end{abstract}

\clearpage

\tableofcontents

\section{Introduction and Motivation}

Machine learning (ML), or extremely high dimensional functional fitting and extrapolation through optimization of an objective function, is exploding as a research discipline in its own right, as well as through its application to particle physics.  Within particle physics, machine learning has been a relatively mature sub-discipline for nearly 10 years, shortly after the rise of graphical processing units (GPUs) used for extreme parallelization tasks in the early 2010s, to application of standard deep neural network architectures, like convolutional neural networks (CNNs) or recurrent neural networks (RNNs), to binary discrimination problems that are central to data interpretation in particle collision experiments.  Since this time in the mid-2010s, machine learning in particle physics has very literally increased exponentially, from a handful of papers each year, to now well over 1000.  As such, I will not attempt to thoroughly cite the particle physics machine learning literature, and instead point to reviews (some of which are now nearly historical documents) \cite{Larkoski:2017jix,Kogler:2018hem,Guest:2018yhq,Albertsson:2018maf,Radovic:2018dip,Carleo:2019ptp,Bourilkov:2019yoi,Schwartz:2021ftp,Karagiorgi:2021ngt,Boehnlein:2021eym,Shanahan:2022ifi,Plehn:2022ftl,Nachman:2022emq,DeZoort:2023vrm,Zhou:2023pti,Belis:2023mqs,Mondal:2024nsa} and to the HEP ML Living Review \cite{Feickert:2021ajf} which continually updates relevant references.  For a practical introduction to the methods and successes of ML applied to collider physics, there have been a number of community or public data analysis competitions, e.g., \Refs{oglhco,higgs-boson,flavours-of-physics,trackml-particle-identification,Aarrestad:2021oeb,Kasieczka:2021xcg}.  A number of yearly physics conferences are dedicated to particle physics analyses with machine learning, including BOOST \cite{boostconf}, ML4Jets \cite{ml4jetsconf}, and Hammer \& Nails \cite{handmconf}, and the large machine learning conferences now typically include a science workshop in which particle physics is well-represented, such as at ICML \cite{icmlconf} or NeurIPS \cite{nipsconf}.  This is already a very incomplete list that will assuredly grow in the future. 

\begin{figure}[t!]
\begin{center}
\includegraphics[width=12cm]{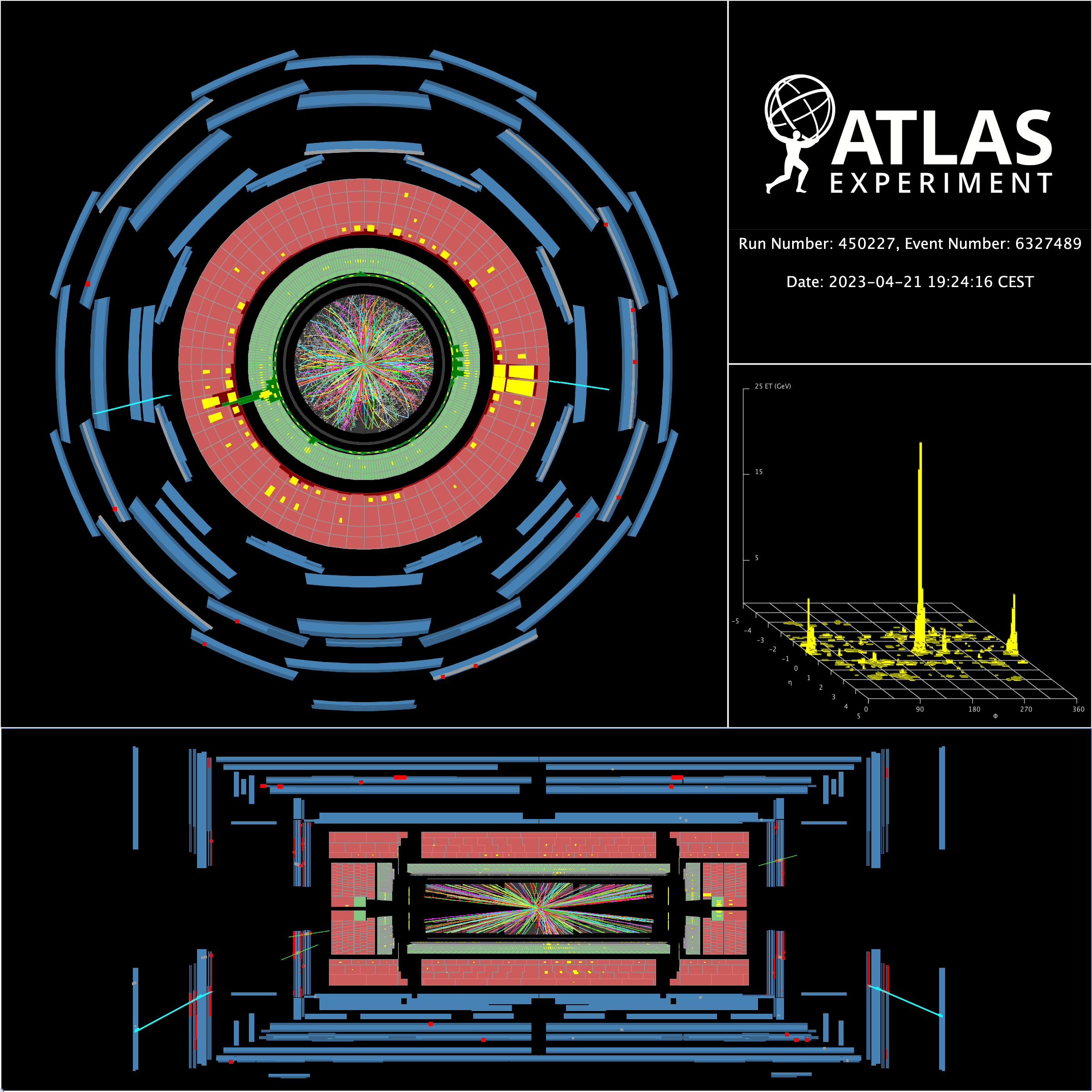}
\caption{\label{fig:jetdisplay}
Display of a di-jet event in the ATLAS experiment from early in Run 3 from April 2023.  Jets are most obviously identified in the inset on the center right, in the so-called ``lego plot'', which shows energy deposits in the calorimetry in the unrolled cylindrical detector.  Jets are the high-energy, collimated (tall, narrow) spikes in the lego plot.
}
\end{center}
\end{figure}

Within particle physics, machine learning was first explored and applied within jet substructure, the study of the internal structure of high-energy, collimated streams of particles ubiquitous in modern high-energy collision experiments like the Large Hadron Collider (LHC).  A display of a di-jet event from early Run 3 in the ATLAS experiment is shown in \Fig{fig:jetdisplay}.  Jets are a signature phenomena in QCD, whose observation validated the quark and parton models in deeply inelastic scattering experiments of the late 1960s \cite{PhysRevLett.23.930,PhysRevLett.23.935} and in $e^+e^-$ collisions of the mid-1970s \cite{Hanson:1975fe}, and in the late 1970s, led to the discovery of the gluon \cite{TASSO:1979zyf,Barber:1979yr,PLUTO:1979dxn,JADE:1979rke,Nilles:1980ys}, correspondingly validating the whole of QCD as the theory of the strong nuclear force.\footnote{The first reference I am aware of that uses the term ``jet'' in the QCD sense is Ref.~\cite{rochester1953new} from a study of cosmic ray air showers in 1953. As for the etymology of ``jet'', I do not know for sure, but a plausible source is from the Jet d'Eau, the spectacular water feature in Lake Geneva, just off the shore from downtown Geneva which itself dates from 1886.  If you have more information about the origin of the term ``jet'', please let me know.}  As experimental collision energies increased, the produced jets grew in complexity and representative phenomena, consisting of dozens of particles initiated by quarks, gluons, or a whole zoo of other particles that have been discovered in the decades since the 1970s.  The central goal of modern jet substructure is focused on establishing properties of the unmeasureable initiating particle from the properties, correlations, and dynamics of the actually experimentally observable particles.

These lectures then serve the dual goals of introducing jets, jet substructure, and techniques for analyzing them, as well as machine learning focusing on its application to the physics of jets.  To the latter goal, my review will be rather distinct from other approaches due almost exclusively to my idiosyncratic view on the utility of machine learning.  As I will describe through the sections that follow, we will, for the most part, treat the machine that analyzes jets for our particular task and its goals as a black box, merely controlling the input and interpreting the output.  We will learn to think like a machine, to sharpen and formalize our specific goal, guided by a collection of fundamental theorems of machine learning that we review.  To the former goal, we will introduce the physics of jets through a number of examples of binary discrimination problems, the problem of optimally separating signal from background in some mixed, unlabeled data sample.  We will find, through a textbook application of the scientific method, making simple, justifiable assumptions and hypotheses, and following the theory of QCD, that we can deeply and robustly understand a wide range of results and will make new predictions that we can test.

Nevertheless, these lectures do not exist in a vacuum, and there are numerous other extant review articles that introduce jets and jet substructure from a variety of viewpoints.  Several references on jets for theorists that I revisit myself are \Refs{Sterman:1995fz,Salam:2010nqg,Abdesselam:2010pt,Plehn:2011tg,Altheimer:2012mn,Shelton:2013an,Altheimer:2013yza,Adams:2015hiv,Cacciari:2015jwa,Marzani:2019hun}.  Though now over 30 years old, the ``QCD Pink Book'' \cite{Ellis:1996mzs} is an absolutely invaluable reference for the history of the topic, a compendium of fundamental results, and correct normalization of important quantities.  I have lectured on jets at several previous summer and winter schools, and several of those lectures are on arXiv  \cite{Larkoski:2017fip,Larkoski:2020jyz,Larkoski:2021aav}.  My philosophy for both teaching and presenting theoretical physics topics is expounded on in my textbooks on particle physics \cite{Larkoski:2019jnv} and quantum mechanics \cite{larkoski2022quantum}. Additionally, with the rise of machine learning and artificial intelligence, many theoretical physicists have made the transition from academia to industry and have further performed the service of translation of results into a form familiar to physicists.  A few reviews of modern machine learning by theoretical physicists for theoretical physicists are \Refs{kaplannotes,Roberts:2021fes,Douglas:2023olt}.

The outline of these lectures is as follows.  In \Sec{sec:mlfundy}, I will review what I consider the fundamental theorems of machine learning, specifically for the application of binary discrimination problems in particle physics.  These results are the universal approximation theorem, the central limit theorem, and the Neyman-Pearson lemma.  In \Sec{sec:qftfundy}, I will review fundamental principles from quantum field theory and QCD necessary to make theoretical predictions in jet physics.  These results include the master formula for differential cross sections (probability distributions), infrared and collinear safety, collinear and soft factorization, and some remarks about jet algorithms (though we won't worry much about specific algorithms in applications). With this foundation set, the following four sections study four different binary discrimination problems and stitch together the machine learning and quantum field theory results in various contexts.  In \Sec{sec:qvgdisc}, we introduce quark versus gluon jet discrimination through a simple picture of scale invariant emissions off of the initiating particle.  In \Sec{sec:hvsg}, we study binary discrimination within fixed-order perturbation theory with application to identification of $H\to b\bar b$ decays, which was in fact the problem that jump-started all of modern jet substructure.  In \Sec{sec:qvgagain}, we return to quark versus gluon discrimination, but this time focusing on particle multiplicity itself and in doing so take our first small steps away from the warmth and comfort of perturbation theory.  In \Sec{sec:uvsd} we forgo perturbation theory altogether, which will be necessary for discrimination of up-flavor quark jets from down-flavor jets, but, guided by the solid handrails of statistics, we will nevertheless be able to make many powerful predictions.  We summarize in \Sec{sec:summ}, and present some thoughts on the future of machine learning in particle physics.

\section{Some Fundamental Ideas from Machine Learning and Statistics} \label{sec:mlfundy}

\begin{figure}
\begin{center}
\includegraphics[height=3.5cm]{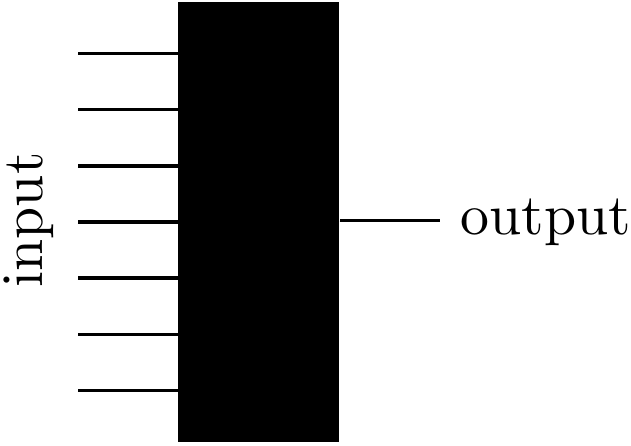}
\hspace{4cm} \includegraphics[height=3.5cm]{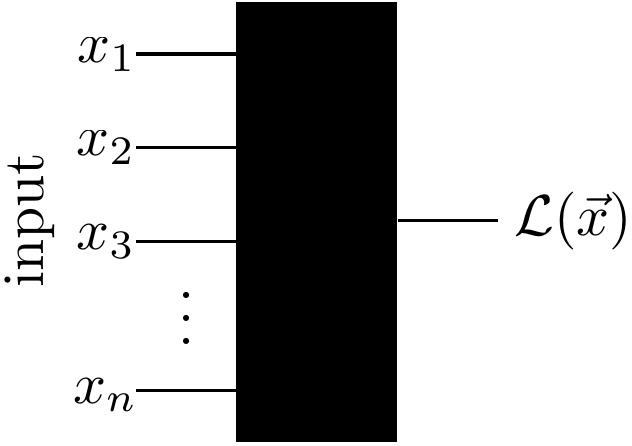}
\caption{\label{fig:mlblackbox} Our theorist's impression of a machine learning architecture.}
\end{center}
\end{figure}

We start this lecture in earnest by defining machine learning, especially in the context we will use throughout these notes.  At its most basic form, our picture of a ``machine'' is represented in \Fig{fig:mlblackbox}.  The machine, the black box, takes input on the left, analyzes it in some way, and returns some output on the right.  Here, we will not inquire as to {\it how} the machine is analyzing the data, or how it is trained, or how many parameters it has, but will rather use this incredibly simple picture to reframe our way of thinking about specific problems in physics to establish their solution.  A bit concretely, the more detailed picture on the right of \Fig{fig:mlblackbox} is what we will focus on specifically.  Here, at left, a large number $n$ of data points is input to the machine, each of which live in some enormous-dimensional space (hundreds or thousands of dimensions).  The machine analyzes these data and then outputs a function ${\cal L}(\vec x)$ which classifies the known data and extrapolates to unseen data elsewhere in the space.

We will be particularly interested in the situation when the input data is divided into two classes, which we call signal and background, respectively, and the function the machine learns is the optimal observable on the space to discriminate these two classes.  This is a very general problem in statistics called {\bf binary discrimination} and a machine's success in solving it depends on two fundamental results.  First, the optimal observable for discrimination is in general some arbitrary function of the input data on its high-dimensional domain.  Therefore, the machine must be able to approximate this optimal discriminant observable to arbitrary accuracy.  Second, the machine needs to know what its target is, it needs to know what exactly is this optimal discrimination observable.  Knowing its target, the machine can then optimize its internal parameters, or {\it learn}, to match the optimal discriminant as closely as possible.  Fortunately for us, both of these requirements have been solved, the first referred to as the universal approximation theorem and the second as the Neyman-Pearson lemma, and we will review both in detail below.  We will also review the central limit theorem and the three of these results form the basis for everything that follows here.

\subsection{Universal Approximation Theorem}

\begin{figure}
\begin{center}
\includegraphics[width=8cm]{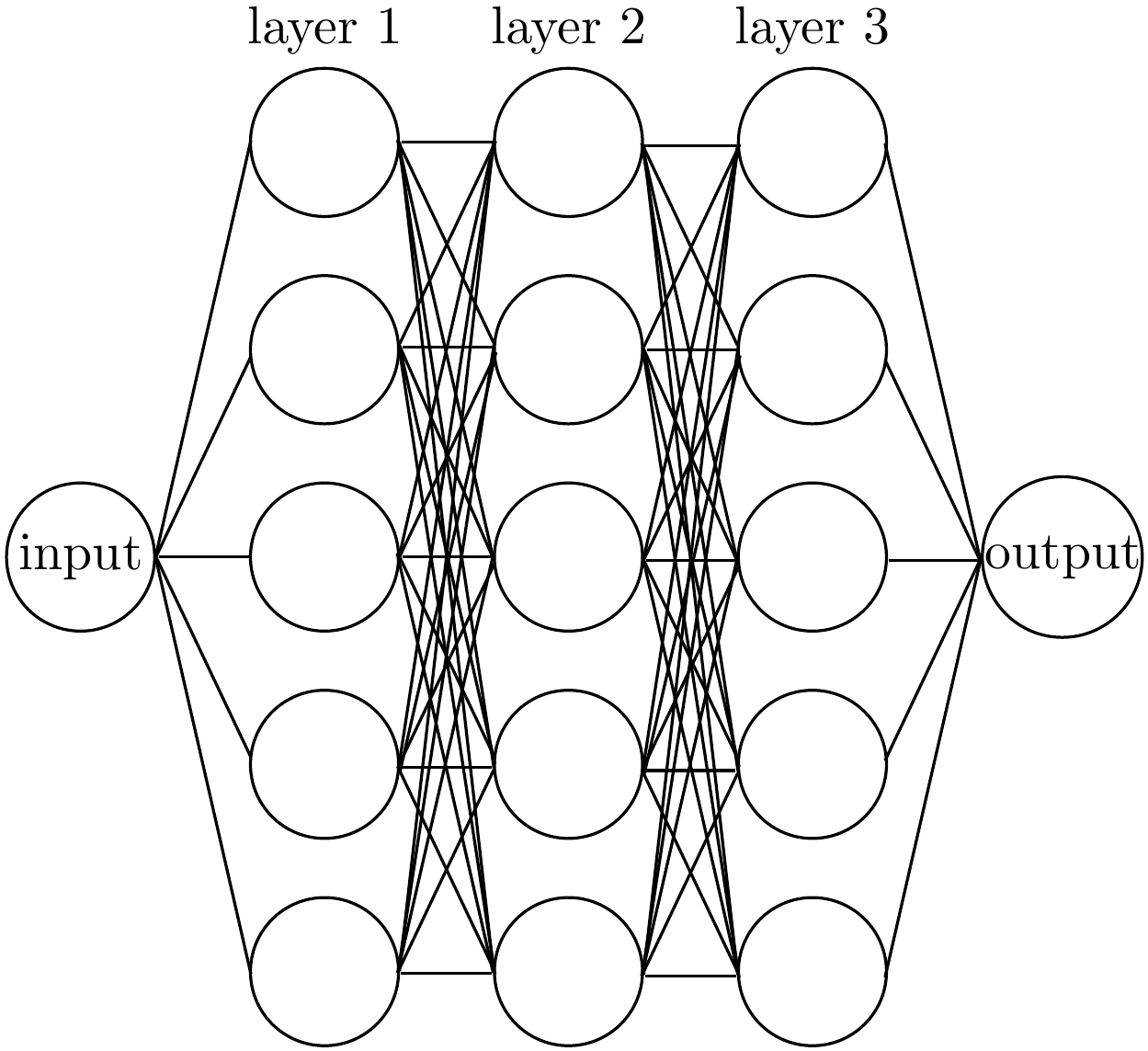}
\caption{\label{fig:mlp} A schematic drawing of a three-layer fully-connected multilayer perceptron architecture.  The empty circles represent individual neurons and the lines represent connections between them.  The input, at left, is distributed across the first layer of neurons, then passed to the next two ``hidden'' layers, and then linearly combined to form the output at right.}
\end{center}
\end{figure}

To understand the validity of our requirement that a machine can indeed approximate any function of the input to arbitrary accuracy, we need to peek beneath the black box curtain just a bit to understand its internal structure.  A modern, feed-forward, fully-connected deep neural network has an internal architecture as modeled by \Fig{fig:mlp}.  For historical reasons, this is also called a {\bf multilayer perceptron}, or MLP \cite{rosenblatt1958perceptron,ivakhnenko1966cybernetic,amari1967theory,linnainmaa1970representation}.  This will be our prototypical machine learning architecture and the only that we consider here, but should only be considered representative because there are now many architectures (CNN \cite{fukushima1980neocognitron}, RNN \cite{lenz1920beitrag,Ising:1925em,amari1972learning,little1974existence,hopfield1982neural}, generative adversarial networks \cite{goodfellow2014generative}, decoder-encoder transformers \cite{vaswani2017attention}, etc.) that are explored and used depending on the particular application (image recognition, language translation, prompt completion, etc.).\footnote{Very recently, there has been a proposal for a significantly different deep learning architecture from the MLP/activation function/universal approximation theorem paradigm \cite{liu2024kan}.  This network is based on the result of the Kolomogorov-Arnold theorem \cite{kolmogorov1956representation,arnold1957representation,lorentz1962metric,sprecher1965structure,braun2009constructive} which states that any $n$-dimensional multivariate function can be expressed as a linear combination of $2n+1$ continuous functions whose arguments are linear combinations of $n$ single-variable functions.  However, the simplicity, utility, or relevance of the Kolmorgorov-Arnold theorem as opposed to the universal approximation theorem to machine learning architecture is highly disputed \cite{girosi1989representation,kuurkova1991kolmogorov,lin1993realization,sprecher1996numerical,sprecher1997numerical,maiorov1999lower,koppen2002training,schmidt2021kolmogorov,kanreview} so much more work will need to be done to determine if it is indeed a game-changer.}

\subsubsection{Very Brief Description of Practical Machine Learning}

In \Fig{fig:mlp}, the input data are distributed across the first layer of neurons, then those data are processed and passed to the next layer, and continuing, until a linear combination is taken of the processed data to produce the output.  Layers after the first layer that directly interacts with the input are called ``hidden'' because they are hidden from the input. Each layer of the MLP processes the data from the previous layer by an affine transformation composed with a non-linear activation function $\sigma$.  That is, going from layer $\ell$ to layer $\ell+1$, the neurons act on the response from layer $\ell$ as:
\begin{align}
\vec x_{\ell+1} = \sigma\left(
\mathbb{W}_\ell\vec x_\ell + \vec b_\ell
\right)\,.
\end{align}
Here, $\mathbb{W}_\ell$ is called the weight matrix, $\vec b_\ell$ is a bias vector, and $\vec x_\ell$ is the vector of all neuron outputs at layer $\ell$.  For example, if layer $\ell$ has $n_\ell$ neurons, then $\vec x_\ell$ is $n_\ell$ dimensional.  The non-linear activation function $\sigma$ acts on all elements of its vector argument individually.

What makes machine learning fascinating and very distinct from traditional programming is that the weights $\mathbb{W}_\ell$ and the biases $\vec b_\ell$ are not preprogrammed; they are learned through minimization of an objective function ${\cal L}$ that compares input to output and is a function of all of the weights and biases.  With appropriate normalization, ${\cal L}  = 0$ means that the weights and biases have been learned to perfectly align the output with the input.  Learning is then accomplished through {\bf gradient descent} of the objective function.  Elements of the weights and biases are considered as parameters of the neural network and are initialized according to some expected probability distribution.  This is typically chosen to be Gaussian, and we will provide a physics interpretation of this in a bit.  Then, with initialized parameters set, the objective function is evaluated and the parameters are updated by flowing in the direction opposite to the gradient of the objective function, and then the objective function is evaluated again with the updated parameters.  Each complete pass through the dataset updating network parameters is called an {\bf epoch}, and a weight parameter/element ${\mathbb W}_{\ell, i j}$ is updated as
\begin{align}
{\mathbb W}_{\ell, i j}^{(m+1)} = {\mathbb W}_{\ell, i j}^{(m)} - \eta \nabla_{{\mathbb W}_{\ell, i j}}{\cal L}(\{{\mathbb{W}^{(m)},\vec b^{(m)}}\})\,.
\end{align}
Here, the superscript $(m)$ denotes the parameters at the $m$th epoch, $\nabla_{{\mathbb W}_{\ell, i j}}$ is the derivative with respect to the parameter of interest, and $\eta$ is called the {\bf learning rate}, a parameter that controls the rate of descent down the gradient.  If you pass through the network enough epochs, the value of the objective function will stop changing and you have therefore learned the appropriate function of the input data.

There are a few comments I want to make before making the statement of the universal approximation theorem.  First, this naive gradient descent is never used in modern applications because it is much, much too costly to evaluate if your data set and network parameters are large (e.g., billions or even trillions of elements).  Instead, methods like {\bf stochastic gradient descent} \cite{robbins1951stochastic,kiefer1952stochastic}, in which smaller {\bf batches} are randomly or stochastically selected during every epoch to train the network, are used.  Further, the gradient itself can be problematic to evaluate because the objective function on parameter space can be extremely non-smooth, so many methods for smoothing the gradient have been introduced like {\bf momentum} \cite{rumelhart1986learning} or {\bf Adam} \cite{kingma2014adam}.  Because of the recursive composition nature of the neurons on the network, the gradient can be evaluated through application of the Leibniz chain rule, which is very efficient if the derivative of the non-linear activation function $\sigma$ is simple.  A particularly efficient way to evaluate the chain rule is through {\bf backpropagation}, which simply means evaluating the derivatives of functions in composition backward through the network.  In terms of function composition, this means evaluating derivatives from the outside (the most recent evaluation) in (towards the earliest evaluation).  This ensures that derivatives of parameters early in the layers can be reused efficiently.

\subsubsection{Statement of the Universal Approximation Theorem}

With this setup for our neural network, we want to establish the conditions by which the output can indeed be an arbitrary function of the input.  The result is called the {\bf universal approximation theorem} whose original proofs made rather strong, non-physical assumptions on the network \cite{cybenko1989approximation,hornik1991approximation,leshno1993multilayer}.  Concretely, the original proofs assumed that the non-linear activation function was a sigmoid, with
\begin{align}
\sigma(x) = \frac{1}{1+e^{-x}}\,,
\end{align}
which was a popular activation function at the time (late 1980s, early 1990s).  Cybenko's proof assumed that the network consisted of a single layer, but was infinitely wide.  In this case, the statement of the universal approximation theorem is that such single-layer, infinitely-wide networks could approximate any function to arbitrary accuracy; or, that arbitrary linear combinations of sigmoids could approximate any function.

While we won't present a proof of it, let's be a bit more specific and state the universal approximation precisely (at least in the form of Cybenko's proof).  Let your data be represented by a vector $\vec x$.  We can then construct the function $G(\vec x)$ from a linear combination of sigmoids as
\begin{align}
G(\vec x) = \sum_{i=1}^N c_i\, \sigma\left(\mathbb{W}_i\vec x+\vec b_i\right)\,,
\end{align}
where $\mathbb{W}_i$ is a weight matrix, $\vec b_i$ is a bias vector, and $c_i$ is a real-valued coefficient.  The number of terms in the sum, $N$, is merely assumed to be finite, $N<\infty$.  Then, with appropriate choice of the weights, biases, and coefficients, the function $G(\vec x)$ can approximate any function $f(\vec x)$ to arbitrary accuracy.  Specifically, for any $\epsilon > 0$, there exist weights, biases, and coefficients such that
\begin{align}
\left|f(\vec x) - G(\vec x)\right| < \epsilon\,.
\end{align}
No relationship between the number of terms $N$ in the sum and the desired accuracy $\epsilon$ was established in the early proofs, and in Cybenko's paper, he specifically mentions that for any reasonable accuracy $\epsilon$, $N$ is likely enormous.  Indeed, for {\it arbitrary} accuracy, the number of terms in the sum must be infinite.

Today, sigmoids are basically never used anymore as activation functions and the machine learning community has explored a whole zoo of possible functions.  When using gradient descent for parameter/weight optimization, a problem with the sigmoid is that taking the gradient of many composed sigmoids uses the chain rule, that is, the gradient is sensitive to the derivatives at every level of the network.  Derivatives of the sigmoid are bounded on $\sigma'(x)\in[0,1]$, and away from its ``turn on'' at $x=0$, rapidly vanish.  This can lead to catastrophic loss of correlation between neurons even a few layers apart because just a few products of sigmoid derivatives can quickly get very small and is in fact exponential in the number of layers.  This problem with the sigmoid or other activation functions with small derivatives is called the {\bf vanishing gradient problem}.

There are many solutions to this problem but perhaps the easiest for these lectures is simply changing the non-linear function to something with a constant derivative almost anywhere.  The most popular activation function used now is the rectified linear unit, or {\bf ReLU}, which has the functional form \cite{4082265}
\begin{align}
\text{ReLU}(x) = x\, \Theta(x)\,.
\end{align}
This function is 0 when $x< 0$ and linear with slope 1 when $x>0$.  Thus, derivatives evaluated in backpropagation are simply 1 if the neuron is activated or ``fires'' (i.e., when $x>0$).  Because of its ubiquity, the universal approximation theorem has been extended to include ReLU and other activation functions, e.g., \Refs{yarotsky2017error,lu2017expressive,hanin2017approximating,kidger2020universal,park2020minimum,shen2022optimal}.

\subsubsection{Our Physicist's Interpretation of the Universal Approximation Theorem}

Unfortunately, the universal approximation theorem is of rather minimal relevance for practical machine learning.  At least as originally stated, it requires an infinitely wide network (clearly unphysical!) and in this limit, for any finite input dataset, we can always simply fix the parameters of the network to memorize the dataset.  So, indeed, it can output an arbitrary function of the input, but no active learning was taking place.  However, finite networks cannot simply memorize any input dataset, and so learning or training of parameters must take place and the rate of convergence of learning the objective function, residual error given network architecture, etc., are not in the purview of the universal approximation theorem. Further, the proofs are typically non-constructive: they tell you that a sufficiently large and powerful network can approximate any function, but not how to do it.

\begin{figure}
\begin{center}
\includegraphics[height=3.5cm]{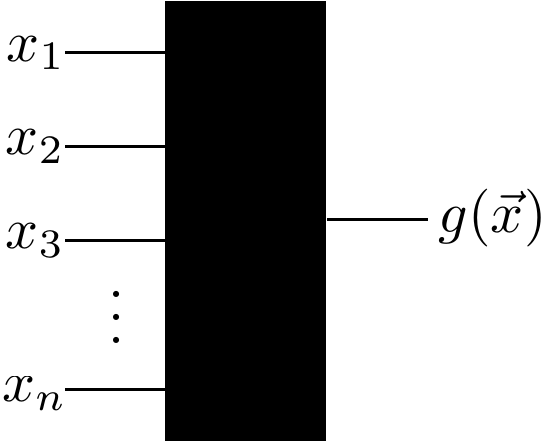}
\hspace{4cm} \includegraphics[height=3.5cm]{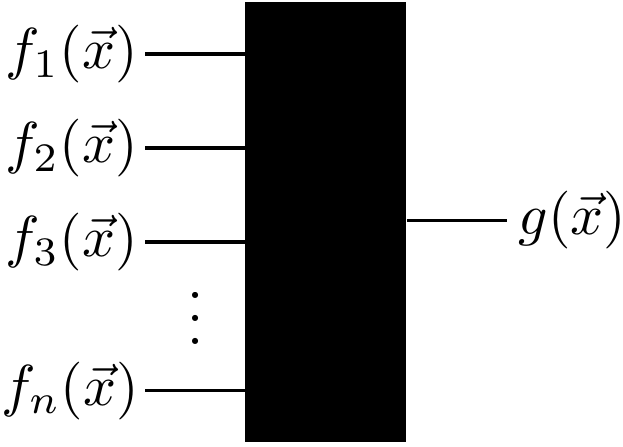}
\caption{\label{fig:uat_cons}The consequence of the universal approximation theorem: the output $g(\vec x)$ will always be learned by a sufficiently powerful machine as long as the input information spans the data space of interest.}
\end{center}
\end{figure}

So, what good is the universal approximation theorem for us?  Well, as physicists, the precise architecture and parameters of a practical network are not directly our focus; we want to learn (as humans!) the objective function for our specific task.  That is, given any possible machine we can determine optimal performance, given the laws of physics.  Additionally, with this understanding, the universal approximation theorem allows us to represent our data in whatever form is convenient for calculation or human interpretation more broadly.  Because a sufficiently powerful machine can learn {\it any} function of the input data, as long as the representation of the input data always completely spans its domain, a sufficiently powerful machine will always learn the same output or objective function.  This is represented in \Fig{fig:uat_cons}.  At left, the data are in the form of a vector $\vec x$ and the machine learns function $g(\vec x)$.  At right, we have transformed the initial data $\vec x$ by some set of functions $\{f_i(\vec x)\}_i$ and as long as those functions are non-degenerate, the machine will learn the exact same function $g(\vec x)$.

Thus, as physicists, we will always imagine working with a machine that has infinite and representative input data (the machine does no extrapolation) and is infinitely powerful (learns the objective function perfectly).  As such, we can choose to represent the input data in whatever form we want as long as it still completely spans the data domain in the sense that the original form of the data can be evaluated losslessly (i.e., the functions $\{f_i(\vec x)\}_i$ in \Fig{fig:uat_cons} are injective).  By considering what we can actually calculate and predict within quantum field theory and QCD, we will be lead to particular, preferred representations of the input data, but again, that is merely for convenience and our bias as theorists that we like to calculate.  That is, we want to use the theory of the physics we are studying to determine the absolute best possible performance that any such machine could produce.

With that, we can safely close the lid on the machine and for the rest of these lectures treat it as a black box.

\subsection{Neyman-Pearson Lemma and Binary Discrimination}

Now that our machine can output an arbitrary function of the input, we need to actually determine what output we want!  As mentioned earlier, throughout these lectures, we will focus on binary discrimination problems in which we would like to determine the observable (i.e., function of the inputs) that optimally separates these two classes.  This is a very general problem especially in particle physics where, given some data event from the LHC say, we would like to classify it as ``interesting'' or signal, or ``not interesting'' or background.  In a couple of lectures we will introduce perhaps the historically most important such classification problem in jet substructure, namely, distinguishing the dominant decay mode of the Higgs boson, $H\to b\bar b$, from the QCD background process of gluon splitting to bottom quarks, $g\to b\bar b$.

Fortunately for us, the optimal discrimination observable has been identified through the {\bf Neyman-Pearson lemma} as the {\bf likelihood ratio}, or the ratio of the probability distributions of signal and background.  In the explicit examples we consider in the following lectures, our goal will then be to first determine the signal and background probability distributions on an appropriate data space and then from those distributions, determine the likelihood ratio.  However, simply given the likelihood ratio, this only tells us it is the optimal discriminant given the input data representation, but the actual performance for binary discrimination will depend on the specific functional relationships of the signal and background distributions.  Here, we will also introduce the {\bf receiver operating characteristic} as a very useful representation of the discrimination power, once we have identified the likelihood ratio.

\subsubsection{Statement and Proof of the Neyman-Pearson Lemma}

We will start, however, with simply the statement of the Neyman-Pearson lemma \cite{Neyman:1933wgr} and provide its simple proof.  We assume that we measure some variables $x$ and that on the domain of $x$, signal events are distributed according to the distribution $p_s(x)$ and background events are distributed as $p_b(x)$.  We will often use this subscript notation to denote event labels, but a more explicit notation is through conditional probabilities, where
\begin{align}
&p_s(x) = p(x|s)\,, &p_b(x) = p(x|b)\,.
\end{align}
That is, $p(x|s)$ is the distribution of observable variables $x$ given that events are drawn from the signal.  Additionally, observables $x$ here may be vectors, or some other data representation, but we will simply write $x$ as the specific form of the data is not relevant.

Given this setup, we now define the constraints in the Neyman-Pearson lemma.  We would like to determine the region on $x$ space that maximizes the signal, for a fixed amount of background contamination.  The region can be specified by the requirement that ${\cal L} > \Lambda(x)$, where $\Lambda(x)$ is some function of the data $x$ whose form we will optimize for.  The background contamination or {\bf false positive rate} is simply the probability that the background events lie in the region of interest, which is
\begin{align}
\alpha = \int dx\, p_b(x)\, \Theta\left({\cal L} - \Lambda(x)\right)\,,
\end{align}
and by assumption, the rate $\alpha$ is fixed.  The {\bf true positive rate} that we want to maximize is equivalent to maximizing the difference
\begin{align}\label{eq:npopt}
\int dx\, \left(p_s(x) - p_b(x)\right)\, \Theta\left({\cal L} - \Lambda(x)\right)\,,
\end{align}
because the false positive rate is fixed.

This is all the setup we need.  This is now a simple constrained optimization problem that can be solved by Lagrange multipliers.  We introduce the Lagrange multiplier $\lambda$ that multiplies the constraint in the function we wish to maximize as
\begin{align}
\int dx\, \left(p_s(x) - \lambda p_b(x)\right)\, \Theta\left({\cal L} - \Lambda(x)\right)\,.
\end{align}
Next, we want to optimize the region $\Lambda(x)$, so we take a variational derivative with respect to $\Lambda(x)$ and set it to 0:
\begin{align}
\frac{\delta}{\delta \Lambda(x)}\int dx\, \left(p_s(x) - \lambda p_b(x)\right)\, \Theta\left({\cal L} - \Lambda(x)\right) = -\int dx\, \left(p_s(x) - \lambda p_b(x)\right)\, \delta\left({\cal L} - \Lambda(x)\right) = 0\,.
\end{align}
Now, we immediately see that this is trivially satisfied, the integrand itself vanishes, if we set the Lagrange multiplier $\lambda = {\cal L}$ and the region of interest defined by 
\begin{align}
\Lambda(x) = \frac{p_s(x)}{p_b(x)}\,,
\end{align}
which is the likelihood ratio.  Thus, the likelihood ratio is the optimal binary discriminant that maximizes signal given a fixed false positive rate.  Actually, we can weaken the statement of the Neyman-Pearson lemma a bit because the original optimization problem, \Eq{eq:npopt}, is unchanged if $\Lambda(x)$ is any monotonic function of the likelihood ratio.   So, the optimal discrimination observable is monotonic in the likelihood, by the Neyman-Pearson lemma.

\subsubsection{ROC Curves and Representations of Discrimination Power}

Now that we know the optimal discrimination observable is the likelihood ratio, we would like to quantify its discrimination power.  One approach, given the discussion above, would be to plot the true positive rate $\Sigma_s(\alpha)$ as a function of the false positive rate $\alpha$ defined as
\begin{align}
&\Sigma_s(\alpha) = \int dx\, p_s(x)\,\Theta\left({\cal L} - \frac{p_s(x)}{p_b(x)}\right)\,,
&\alpha = \int dx\, p_b(x)\,\Theta\left({\cal L} - \frac{p_s(x)}{p_b(x)}\right)\,.
\end{align}
By the Neyman-Pearson lemma, for a fixed $\alpha$, the likelihood ratio maximizes the true positive rate, $\Sigma_s(\alpha) $.  Note that the dependence of the true positive rate on the false positive rate $\alpha$ is implicit, through the value of ${\cal L}$, so let's work to make it explicit.  To do this, let's massage the false positive rate as
\begin{align}
\int dx\, p_b(x)\,\Theta\left({\cal L} - \frac{p_s(x)}{p_b(x)}\right) & = \int_0^{\cal L} d{\cal L}'\int dx\, p_b(x)\,\delta\left({\cal L}' - \frac{p_s(x)}{p_b(x)}\right)\\
&=\int_0^{\cal L}d{\cal L}'\, p_b({\cal L}') = \Sigma_b({\cal L})\,,\nonumber
\end{align}
where $\Sigma_b({\cal L})$ is the cumulative probability distribution of the likelihood on the background sample.  Because probability density $p_b(x)$ is everywhere non-negative, the cumulative distribution $ \Sigma_b({\cal L})$ is monotonic, and so its inverse is well-defined.  Then, the value of the likelihood ratio at fixed false positive rate $\alpha$ is
\begin{align}
{\cal L} = \Sigma_b^{-1}(\alpha)\,.
\end{align}

From similar manipulations of the true positive rate, it is also a cumulative probability distribution and we give the true positive rate as a function of false positive rate a name, where
\begin{align}\label{eq:roc1}
\text{ROC}(\alpha) \equiv \Sigma_s\left(\Sigma_b^{-1}(\alpha)\right)\,,
\end{align}
which is called the {\bf receiver operating characteristic curve} or ROC curve.  It ranges from $\text{ROC}(x)\in[0,1]$ for $x\in[0,1]$ and is bounded from above by the line $\text{ROC}(x) = x$.  In fact, if $\text{ROC}(x) = x$ then the true positive and false positive rates are equal; that is, the observable completely randomly selects events as signal or background.

Because the ROC curve is the minimum false positive rate for a given true positive rate, its integral is a useful measure of discrimination, as well.  The integral of the ROC curve, or the area under the curve (AUC) is then
\begin{align}\label{eq:acudef}
\text{AUC} &= \int_0^1 dx\, \Sigma_s\left(\Sigma_b^{-1}(x)\right) = \int d{\cal L}_b\, p_b({\cal L}_b)\, \Sigma_s({\cal L}_b)\\
&=\int d{\cal L}_s\, d{\cal L}_b\, p_s({\cal L}_s)\, p_b({\cal L}_b)\,\Theta({\cal L}_b - {\cal L}_s)\nonumber\,.
\end{align}
By the definition of the likelihood ratio, 
\begin{align}
{\cal L} =\frac{p_s(x)}{p_b(x)}\,,
\end{align}
the likelihood is large where the signal distribution is large, and small where the background distribution is large.  Thus, the AUC is a measure of the ``mis-ordering'' of events in the likelihood.  Perfect discrimination, for which all signal events lie at larger values of the likelihood than all background events, has an AUC of 0, while completely random discrimination, for which signal and background events are uniformly mixed in the likelihood, has an AUC of 1/2.  Therefore, a standard objective function for binary discrimination that is often used in machine learning is the AUC.

It's worth diving a little bit deeper into the relationship between the distributions of the likelihood ratio ${\cal L}$ on signal and background distributions.  First, from the definition of the likelihood, we have
\begin{align}
{\cal L} = \frac{p_s({\cal L})}{p_b({\cal L})}\,,
\end{align}
so, obviously, 
\begin{align}
p_s({\cal L}) = {\cal L}\, p_b({\cal L})\,.
\end{align}
Then, because both signal and background distributions of the likelihood ratio are normalized, this requires that the mean of the background distribution is unity:
\begin{align}
1 = \int_0^\infty d{\cal L}\, p_b({\cal L}) =  \int_0^\infty d{\cal L}\, p_s({\cal L}) =  \int_0^\infty d{\cal L}\,{\cal L}\, p_b({\cal L})\,.
\end{align}
The AUC can also be expressed exclusively in terms of the background distribution of the likelihood.  First, from \Eq{eq:acudef}, we can also write
\begin{align}
\text{AUC} = 1- \int_0^\infty d{\cal L}\, p_s({\cal L})\, \Sigma_b({\cal L}) = 1 - \int_0^\infty d{\cal L}\, {\cal L}\,p_b({\cal L})\,\Sigma_b({\cal L})\,.
\end{align}
Using integration-by-parts, this expression exclusively in terms of the background distribution can be expressed as
\begin{align}\label{eq:rocexp}
\text{AUC} = 1 -\frac{1}{2} \int_0^\infty d{\cal L}\left[
1 - \Sigma_b({\cal L})^2
\right]\,.
\end{align}
So, if you have an independent bound on the functional form of the background cumulative distribution of the likelihood, this can provide a useful estimate of the AUC.

For its intuitive definition, the ROC curve is rather unintuitively named, and one might wonder how such a seemingly useful statistical measure of discrimination power has such a clunky title.  The true positive versus false positive rate plot was developed in World War II for quantifying the efficacy (``operating characteristic'') of receivers (the signal processing units of a radar, which was at the time often a person) to identify whether a blip was a plane or not a plane.  The information contained in the ROC curve was formalized shortly thereafter, e.g., \Refs{woodward1953probability,1057460}, which was also shortly after Shannon's theory of information \cite{shannon1948mathematical}.  While the ROC is still widely used in radar applications today, its utility has extended far beyond, but the original name has stuck.

\subsubsection{Robust Lower Bound of the AUC}

With the expression for the AUC in \Eq{eq:rocexp}, we can rather simply determine a robust lower bound, as determined by the minimum ${\cal L}_{\min}\geq 0$ and maximum ${\cal L}_{\max}<\infty$ values of the likelihood ratio.  Assuming that the minimum and maximum are not 0 and infinite, respectively, the expression for the AUC can be expressed as an integral exclusively over the range $[{\cal L}_{\min},{\cal L}_{\max}]$ as
\begin{align}
\text{AUC} = 1+\frac{1}{2}\int_{{\cal L}_{\min}}^{{\cal L}_{\max}}d{\cal L}\,\Sigma_b({\cal L})^2
-\frac{{\cal L}_{\max}}{2}\,.
\end{align}
Additionally, the fact that the mean value of the likelihood of the background distribution is unity constrains the integral of the cumulative distribution, where
\begin{align}
\int_{{\cal L}_{\min}}^{{\cal L}_{\max}} d{\cal L}\, \Sigma_b({\cal L}) &=\left. {\cal L} \,\Sigma_b({\cal L})\right|_{{\cal L}_{\min}}^{{\cal L}_{\max}} - \int_{{\cal L}_{\min}}^{{\cal L}_{\max}} d{\cal L}\, {\cal L}\,p_b({\cal L})\\
&={\cal L}_{\max} - 1\,.\nonumber
\end{align}
That is, we would like to determine the cumulative distribution $\Sigma_b({\cal L})$ that minimizes the AUC with fixed integral on the domain ${\cal L}\in[{\cal L}_{\min},{\cal L}_{\max}]$.  As a constrained minimization problem, we can again use Lagrange multipliers to solve it.

Let's introduce the ``action'' as the sum of the value of the AUC and the Lagrange multiplier $\lambda$ times the integral of the cumulative distribution:
\begin{align}
S[\Sigma_b,\lambda] = 1+\frac{1}{2}\int_{{\cal L}_{\min}}^{{\cal L}_{\max}} d{\cal L}\left[ \Sigma_b({\cal L})^2+2 \lambda\,  \Sigma_b({\cal L})\right]-\frac{{\cal L}_{\max}}{2}\,.
\end{align}
Extremizing the variational derivative with respect to $\Sigma_b$ then constrains the Lagrange multiplier as
\begin{align}
\frac{\delta S[\Sigma_b,\lambda]}{\delta \Sigma_b} = \int_{{\cal L}_{\min}}^{{\cal L}_{\max}} d{\cal L}\left(
\Sigma_b({\cal L}) + \lambda
\right) = {\cal L}_{\max} - 1+\lambda\left(
{\cal L}_{\max}-{\cal L}_{\min}
\right)=0\,.
\end{align}
The integrand of this expression vanishes identically if
\begin{align}
\Sigma_b^{(\min)}({\cal L}) =- \lambda = \frac{{\cal L}_{\max}-1}{{\cal L}_{\max}-{\cal L}_{\min}}\,,
\end{align}
which corresponds to the cumulative distribution that minimizes the AUC with the constraints.  This constant cumulative distribution corresponds to the background probability distribution of
\begin{align}
p_b^{(\min)}({\cal L}) = \frac{{\cal L}_{\max}-1}{{\cal L}_{\max}-{\cal L}_{\min}}\,\delta\left(
{\cal L}-{\cal L}_{\min}
\right)+ \frac{1-{\cal L}_{\min}}{{\cal L}_{\max}-{\cal L}_{\min}}\,\delta\left(
{\cal L}-{\cal L}_{\max}
\right),
\end{align}
which is normalized and has unit mean.

\begin{figure}[t!]
\begin{center}
\includegraphics[width=6cm]{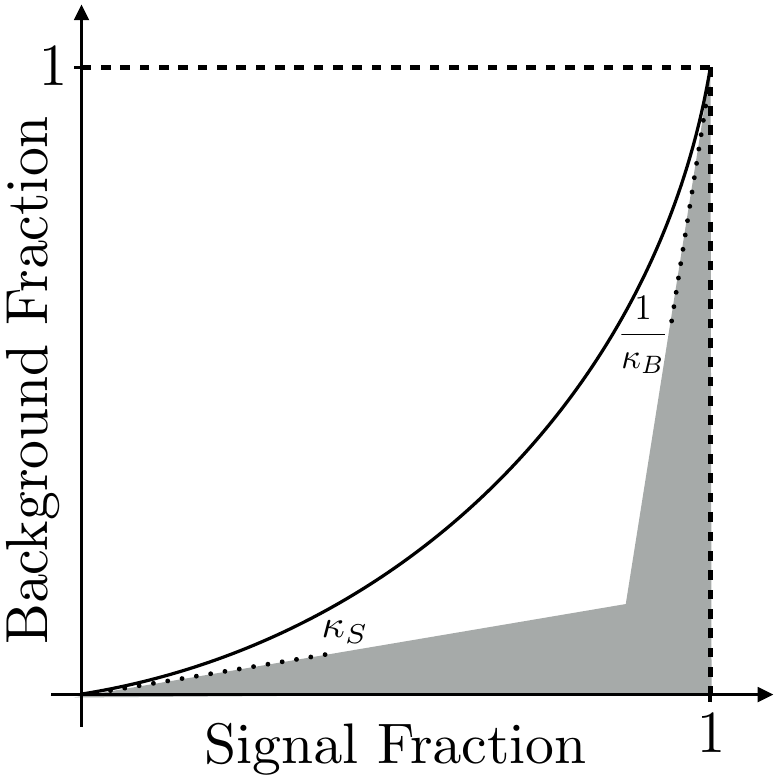}
\caption{\label{fig:redfacts}
Illustration of a ROC curve and a robust lower bound of the AUC as defined by the area of the gray quadrilateral defined by reducibility factors $\kappa_S = 1/{\cal L}_{\max}$ and $\kappa_B  = {\cal L}_{\min}$.  Figure from \InRef{Larkoski:2019nwj}.
}
\end{center}
\end{figure}

With this expression for the minimal background cumulative distribution, its AUC can be calculated that forms a lower bound for any AUC given that minimum and maximum likelihood:
\begin{align}\label{eq:aucmin}
\text{AUC} \geq \frac{1-2{\cal L}_{\min}+{\cal L}_{\max}{\cal L}_{\min}}{2({\cal L}_{\max}-{\cal L}_{\min})}\,.
\end{align}
This bound on the AUC was first derived (to my knowledge) in \InRef{Larkoski:2019nwj} graphically, from the plot of the ROC curve itself.  This is illustrated in \Fig{fig:redfacts}.  The slope of the ROC curve at its endpoints is set by the minimum and maximum values of the likelihood ratio, and from these slopes a minimal quadrilateral can be formed, as the derivative of the ROC curve is the value of the likelihood at a given background quantile $x$:
\begin{align}
\frac{d}{dx}\text{ROC}(x) = \frac{d}{dx}\Sigma_s\left(
\Sigma_b^{-1}(x)
\right) = \frac{p_s\left(
\Sigma_b^{-1}(x)
\right)}{p_b\left(
\Sigma_b^{-1}(x)
\right)} =\Sigma_b^{-1}(x)
\,.
\end{align}
By monotonicity of the likelihood, any ROC curve with those bounds on the likelihood must lie above this quadrilateral.  The area of this quadrilateral is precisely the expression of \Eq{eq:aucmin}.  These extremal values of the likelihood ratio are therefore useful measures of discrimination power in their own right, and have come to be called {\bf reducibility factors}, or maximal purity \cite{katz2019decontamination,Metodiev:2018ftz,Komiske:2018vkc}.  The minimum and maximum values of the likelihood ratio, by definition determine the maximal possible purity of signal or background samples that can be attained anywhere in the space.  

\subsubsection{Aside: Principle of Maximum Likelihood}

While I promised we wouldn't dive under the hood of our black box neural net again, we'll just take the tiniest of peeks into a useful objective function for binary discrimination.  In binary discrimination, events or instances of the data are labeled with a 0 or 1 to represent signal or background, respectively.  Then, using the non-label information of the data, the machine learns the correlation with labels and correspondingly outputs a probability that an event is signal, for instance.  To learn these probabilities, we want the machine to optimize an appropriate objective function that connects the labels with the probability distributions of the data. Such an objective function that accomplishes this is
\begin{align}
H(y,p) \equiv  - \sum_{i\in\text{data}} \left[y_i \log p_i + (1-y_i)\log(1-p_i)\right]\,,
\end{align}
which is called the {\bf cross entropy}, where $y_i$ is the event label (0 or 1) and $p_i$ is the probability output that the machine assigns to the data.  The cross entropy is non-negative (for the same reasons that Boltzmann or Shannon entropy is non-negative), and so we want to minimize it as an objective function.  Note that $H(y,p) = 0$ means that the machine perfectly identifies data labels ($y_i = 0$ means $p_i = 0$ and $y_i = 1$ means $p_i = 1$), along with the limiting relationship that $0\log 0 = 0$.  So minimization of the cross entropy means that the machine performs as best as possible label probability assignments.

The cross entropy is further monotonically related to the likelihood that the machine assigns event probabilities correctly, as it is the negative logarithm of the likelihood.  Thus, minimization of the cross entropy is equivalent to maximization of the likelihood, and this general feature of objective functions is called the {\bf principle of maximum likelihood} \cite{fisher1912absolute,10.1214/aoms/1177732360}.  The cross entropy is further monotonically related to the {\bf Kullback-Liebler divergence} \cite{kullback1951information}, which, for two probability distributions $q$ and $p$ on random variables $x$, is defined as
\begin{align}
D_\text{KL}(p|| q) = \int dx\, p(x)\, \log\frac{p(x)}{q(x)}\,.
\end{align}
For continuous distributions, the Kullback-Liebler divergence is often used as the objective function for optimization and its interpretation is that it is the information gained by using the distribution $p(x)$ than the current $q(x)$.  It is also the expected value of the logarithm of the likelihood ratio on the distribution $p(x)$.

So, what is this probability that the machine assigns to an event it analyzes?  We have discussed the conditional probability $p({\cal L}|s)$, which is the distribution of the likelihood given a signal event, for instance.  What we want instead is $p(s|{\cal L})$; that is, the probability that an event is signal, given the value of the likelihood.  These two probabilities are related by {\bf Bayes's theorem} \cite{bayes1763lii}, where
\begin{align}
p(s|{\cal L}) = \frac{p({\cal L}|s)p(s)}{p({\cal L})}\,.
\end{align}
The probability of the likelihood itself, $p({\cal L}) $, can be evaluated by summing over all possible event classes; namely,
\begin{align}
p({\cal L}) = p({\cal L}|s)p(s)+p({\cal L}|b)p(b)\,.
\end{align}
Now, $p(s)$ ($p(b)$), for example, is the probability that any single event in some mixed ensemble is signal (background).  Note that, because there are only two classes by fiat, $p(b) = 1-p(s)$.  Now, we can plug this into Bayes's theorem and find the result
\begin{align}
p(s|{\cal L}) = \frac{p({\cal L}|s)p(s)}{p({\cal L}|s)p(s)+p({\cal L}|b)p(b)} = \frac{{\cal L}\,p(s)}{{\cal L}\,p(s)+p(b)}\,.
\end{align}
On the right, we have used the definition of the likelihood where
\begin{align}
p({\cal L}|s) = {\cal L}\, p({\cal L}|b) \,.
\end{align}
We note that this probability $p(s|{\cal L})$ is itself monotonic in the likelihood, as well.

\subsection{Central Limit Theorem}\label{sec:clt}

With the universal approximation theorem and the Neyman-Pearson lemma, we now know what objective function to use for binary discrimination and that a sufficiently powerful machine will always be able to learn it.  In this section, we provide one more fundamental theorem of machine learning, namely, the {\bf central limit theorem}  \cite{de1733approximatio,de17381733,de1820theorie,lyapunov1901theoreme}, one of the earliest and most important results in all of probability and statistics.  The central limit theorem will appear again and again as the starting point or first approximation for limiting dynamics in jets in the examples we consider, and further, the central limit theorem provides a rather interesting physical interpretation for an infinitely wide neural network.

\subsubsection{Statement and Proof of the Central Limit Theorem}

The statement and proof of the central limit theorem is as follows.  Let the random variables $\{x_i\}_i$ each be independent and identically distributed (``i.i.d.'').  ``Identically distributed'' means that the value of each one of the $x_i$s is drawn from the same probability distribution $p(x_i)$ and ``independent'' means that the value of $x_i$ has no effect on the value of $x_j$ that is drawn from the distribution.  With this setup, we then consider the sum of $N$ of these i.i.d.~random variables and call it $X$:
\begin{align}
X = \sum_{i=1}^N x_i\,.
\end{align}
Our goal will be to determine the distribution of $X$ in the limit that the number of terms in the sum $N$ gets large, $N\to\infty$.

To prove the central limit theorem, it is useful to rescale and subtract the mean values, introducing a new variable $\tilde X$, where
\begin{align}
\tilde X = \sum_{i=1}^N \frac{x_i - \mu}{\sqrt{N}}\,.
\end{align}
Here, $\mu$ is the mean value of the $x$ random variables,
\begin{align}
\mu = \langle x\rangle = \int dx\, x\, p(x)\,.
\end{align}
Note that the mean of $\tilde X$ is 0:
\begin{align}
\langle \tilde X\rangle = \int \prod_{i=1}^N\left[
dx_i\, p(x_i)
\right]\sum_{j=1}^N \frac{x_j - \mu}{\sqrt{N}} = 0\,.
\end{align}
Its second moment is therefore also its variance, which we can calculate to be
\begin{align}
\langle \tilde X^2\rangle = \sigma_{\tilde X}^2 &= \int \prod_{i=1}^N\left[
dx_i\, p(x_i)
\right]\left(\sum_{j=1}^N \frac{x_j - \mu}{\sqrt{N}}\right)^2 \\
&= \int \prod_{i=1}^N\left[
dx_i\, p(x_i)
\right]\left(
\sum_{j=1}^N \frac{(x_j - \mu)^2}{N}+2\sum_{j<k=1}^N \frac{(x_j - \mu)(x_k - \mu)}{N}
\right)\nonumber\\
&=\sigma_x^2
\nonumber
\end{align}
On the second line, we have simply expanded out the square and noted that the cross term vanishes because $x_i$ and $x_j$ are independent, for $i\neq j$.  Therefore, the variance of $\tilde X$ is the same as the variance of $x$ itself, where
\begin{align}
\sigma_x^2 = \int dx\, (x-\mu)^2\,p(x)\,.
\end{align}

Now, we move to calculating the probability distribution of $\tilde X$, by integrating over all of the $x_i$ variables with the constraint to evaluate $\tilde X$ at its defined value:
\begin{align}
p(\tilde X) = \int \prod_{i=1}^N\left[dx_i\, p(x_i)\right]\, \delta\left(
\tilde X-\sum_{i=1}^N \frac{x_i-\mu}{\sqrt{N}}
\right)\,.
\end{align}
Now, we could work directly with this distribution and massage it in the large-$N$ limit, but note that because of the structure of the $\delta$-function's argument, this integral is effectively a high-dimensional convolution, or, an integral over a product of factors where the sum of arguments is fixed.  As a standard trick to deal with convolutions, we will Laplace transform this integral, as the Laplace transform of a convolution is simply a product of the Laplace transformed integrand factors.  In statistics, the Laplace transform of a probability distribution is its {\bf cumulant generating function}, by which repeated differentiation produces the distribution's cumulants.  In a more physics language, cumulants are a distribution's connected correlation functions, so the cumulant generating function is like evaluating the path integral with a field source, and then repeated variational differentiation with respect to the source produces the connected correlation functions.

Now, explicitly evaluating the Laplace transform, we find
\begin{align}
\int d\tilde X \, e^{t\tilde X}\, p(\tilde X) &=  \int d\tilde X \int \prod_{i=1}^N\left[dx_i\, p(x_i)\right]\, e^{t\tilde X}\,\delta\left(
\tilde X-\sum_{i=1}^N \frac{x_i-\mu}{\sqrt{N}}
\right)\\
&=\left[
\int dx\, p(x)\, e^{t\frac{x-\mu}{\sqrt{N}}}
\right]^N=\left(
1+\frac{t^2}{2}\frac{\sigma_x^2}{N}+{\cal O}(N^{-3/2})
\right)^N\nonumber\\
&=e^{\frac{t^2}{2}\sigma_x^2}\,,\text{ as }N\to\infty\nonumber
\end{align}
On the second line, we have expanded the Laplace transform of the distribution $p(x)$ in powers of the parameter $t$, and what is particularly interesting is that beyond order $t^2$, terms in the expansion are suppressed by inverse powers of $N$.  As $N\to\infty$, only the first two terms in powers of $t$ contribute, and produce an exponential function.  This exponential function on the third line is quadratic in $t$, and is therefore the form of a Gaussian function.  From years and years of Fourier and Laplace transforming, we know that the Fourier/Laplace transform of a Gaussian is a Gaussian (and inversely), and so we can immediately write down the probability distribution of $\tilde X$ in the large-$N$ limit:
\begin{align}
p(\tilde X) = \frac{1}{\sqrt{2\pi\sigma_x^2}}\, e^{-\frac{\tilde X^2}{2\sigma_x^2}}\,.
\end{align}
Changing variables from $\tilde X$ to the original $X$ (and remembering the Jacobian), we have
\begin{align}
p(X) = \frac{1}{\sqrt{2\pi N\sigma_x^2}}\, e^{-\frac{(X - N\mu)^2}{2N\sigma_x^2}}\,,
\end{align}
as $N\to\infty$.  This proves the central limit theorem, that the distribution of the sum of $N\to \infty$ i.i.d.~random variables is a Gaussian with mean $\langle X\rangle = N\mu$ and variance $\sigma_X^2 = N\sigma_x^2$.

\begin{figure}[t!]
\begin{center}
\includegraphics[width=8cm]{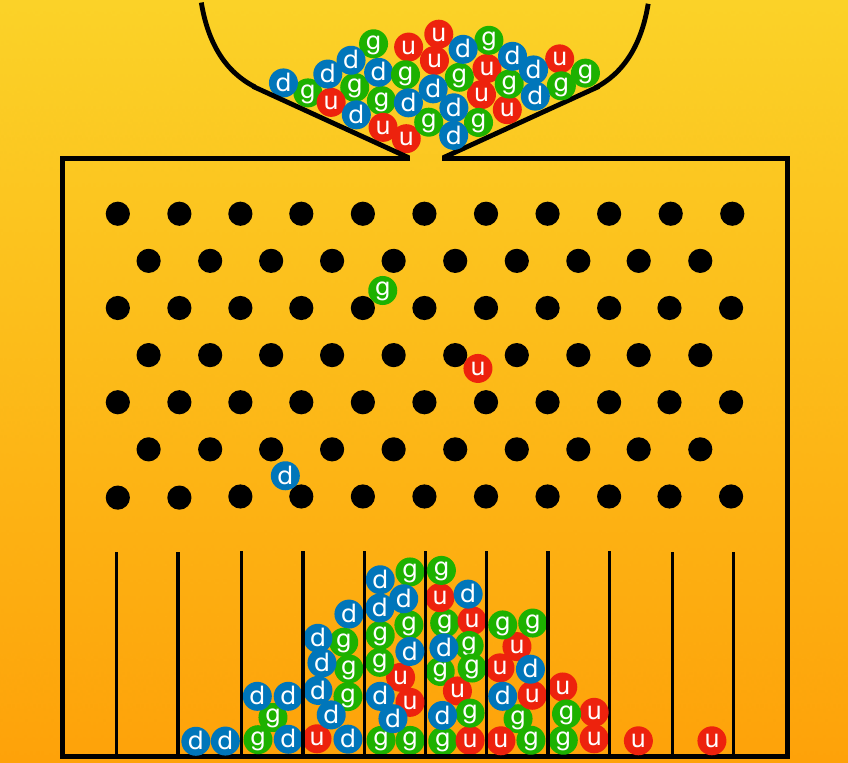}
\caption{\label{fig:plinko}
Illustration of a Galton board in which balls are dropped one-by-one from the hopper at top, and then strike the pegs as they fall, which randomly bump balls one space to the left or right.
}
\end{center}
\end{figure}

\Fig{fig:plinko} illustrates a Galton board \cite{galton1889natural} (used in games like pachinko or plinko) which is perhaps the most direct way to observe the central limit theorem in action.  Balls are dropped from the hopper at top one-by-one into the pegboard, and as they fall, they strike pegs which displaces the balls randomly one space left or right.  Then, as the balls continue their fall, they accumulate some displacement from their original horizontal position from the peg strikes.  For the Galton board of \Fig{fig:plinko}, the total horizontal displacement $X$ in terms of peg spacing is
\begin{align}
X = \sum_{i=1}^7 x_i\,,
\end{align}
where $x_i \in\{-1,1\}$ and there are 7 layers of pegs.  Even though 7 is not an especially large number, the resulting distribution of balls in the bins at the bottom closely follows the Gaussian distribution.  Actually, this Galton board represents the simplest version of a {\bf random walk}, where at every step you randomly choose to move forward or backward.  While on average, the displacement from the origin is 0, the standard deviation scales like $\sqrt{N}$ for $N$ total steps, and so in any given instance of a random walk, you will typically find yourself $\sqrt{N}$ steps from the origin.  In this particular figure of a Galton board, also note balls of different color and with letter labels; we will exploit this simple picture as a model for particle generation in \Sec{sec:uvsd}.

\subsubsection{An Infinitely Wide Neural Network as a Free Field Theory}

There is a rather interesting connection between neural networks and statistical field theory, which here we only sketch.  It has long been known that an infinitely wide neural network is a free field theory \cite{neal1996priors,williams1998computation,lee2017deep}.  For some intuition, here we will consider just a single-layer network, but with an infinite number of neurons. Further, we will initialize the weights of the neurons from a Gaussian distribution with 0 mean and variance $\sigma^2$.  From this starting point, we would like to determine the probability distribution of the response of this network, which we will denote as $p(\vec x)$, where $\vec x$ is a formally infinite dimensional vector with entries corresponding to the individual neurons.  Without loss of generality, we can write this probability as the exponential of some other function,
\begin{align}
p(\vec x) \propto e^{-S(\vec x)}\,.
\end{align}
In statistics, $S(\vec x)$ is the negative logarithm of the probability, but in physics we would call it the action.

The ``action'' can be directly constructed from the assumption of Gaussian initialization, where
\begin{align}
S(\vec x) = \frac{\vec x\cdot\vec x}{2\sigma^2}\,,
\end{align}
and so the probability distribution is 
\begin{align}
p(\vec x) = \frac{e^{-\frac{\vec x\cdot \vec x}{2\sigma^2}}}{Z}\,.
\end{align}
Here, $Z$ is the appropriate normalization factor, or, in physics language, the partition function,
\begin{align}
Z \equiv \lim_{N\to \infty}\int\prod_{i = 1}^N\left[
\frac{dx_i}{\sqrt{2\pi\sigma^2}}
\right]\, e^{-\frac{\vec x\cdot \vec x}{2\sigma^2}}\,.
\end{align}
Just like in statistical physics, this partition function encodes everything you could ever want to ask about the distribution of neurons, $p(\vec x)$.  For example, to evaluate correlation functions between neurons, we introduce a vector source $\vec J$ into the partition function,
\begin{align}
Z\left[\vec J\right] = \lim_{N\to \infty}\int\prod_{i = 1}^N\left[
\frac{dx_i}{\sqrt{2\pi\sigma^2}}
\right]\, e^{-\frac{\vec x\cdot \vec x}{2\sigma^2}+\vec J\cdot \vec x} = e^{\frac{\sigma^2}{2}\,\vec J\cdot\vec J}\,.
\end{align}
The two-point connected correlator, for instance, is the second derivative with respect to $\vec J$ and then setting $\vec J = 0$:
\begin{align}
\langle x_i x_j\rangle &=\left. \frac{\delta}{\delta J_i} \frac{\delta}{\delta J_j}\,Z\left[\vec J\right]\right|_{\vec J = 0}  = \delta_{ij}\,\sigma^2\,.
\end{align}

This interpretation of an infinitely-wide network as a free, Gaussian field theory additionally illustrates that such networks do no learning.  We can exactly evaluate the partition function, and can correspondingly evaluate any moment of the neuron response distribution we want, simply from initialization conditions.  To go beyond the free-field limit, we must consider finite-width networks and additionally incorporate multiple layers.  At finite width, effective four-point and higher interactions are generated, but they can be treated perturbatively, through an expansion about the infinite width limit.  Of course, interacting field theories in general do not have closed-form partition functions (at least for ``realistic'' theories that are non-integrable) and so honest learning or a systematic improvement in the neuron response must be considered in such cases.  The statistics of finite-width networks is fascinating in its own right, and I cannot attempt to do it more justice here, but point the interested reader to \InRef{Roberts:2021fes} for explicit details.

\section{Some Fundamental Ideas from Quantum Field Theory}\label{sec:qftfundy}

With our machine learning results established, we now move on to reviewing some fundamental results in quantum field theory and QCD, which will correspondingly provide the foundation for actually calculating likelihood ratios and probability distributions on signal and background event classes that we can define.  In the four binary discrimination examples that we study in the following sections, we will be guided by perturbation theory and so will need to define what to calculate and if it can be calculated at all.  Additionally, we will work in the soft and/or collinear limit, narrowly focusing on small angle regions of our detector and where the description of particle dynamics significantly simplifies and factorizes from the rest of the event.  Practically working in this collinear limit requires a jet algorithm to robustly define the region of interest, so we will review jet algorithms a bit here.  However, in the following sections, all of our results will hold for any reasonable jet algorithm definition, so we will not need to specify the algorithm in particular.

\subsection{Master Formula for Distributions from QFT}

Our goal for these lectures is to determine the likelihood ratio for binary discrimination problems.  As such, we need to calculate the probability distributions on signal and background events on which we have measured some collection of observables.  Therefore, we need a master formula for calculating probability distributions in QCD and quantum field theory more generally, which of course was worked out nearly a century ago by Dirac and is referred to as {\bf Fermi's Golden Rule} \cite{Dirac:1927dy,fermi1950nuclear}.  We will actually need to slightly modify Fermi's Golden Rule from its usual form, because we want not only the probability of {\it something} happening, but we also want the probability that something happened with a particular measured value for an observable.  In particular, Fermi's Golden Rule is typically stated as the rule for calculating scattering cross sections, while here we will need {\bf differential cross sections}.

Fermi's Golden Rule for the differential cross section of some observable ${\cal O}$ on a class of events is
\begin{align}
\frac{d\sigma}{d{\cal O}} = \int d\Phi\, |{\cal M}|^2\,\delta\left({\cal O} - \hat{\cal O}(\Phi)\right)\,.
\end{align}
This will be our master formula for making predictions in the sections that follow, so we want to ensure that we understand it deeply, in our bones.  There are three parts of the calculation to consider.  First, $d\Phi$ is {\bf differential phase space} and represents the volume measure of relativistic, on-shell particle four-momentum subject to global energy and/or momentum constraints.  Second, $|{\cal M}|^2$ is the squared {\bf S-matrix element} that represents the probability density on phase space for the initial, prepared, state to transform into the final, observed, state.  This is calculated with Feynman diagrams in perturbation theory.  Third, $\delta\left({\cal O} - \hat{\cal O}(\Phi)\right)$ is the measurement constraint, where ${\cal O}$ is the measured value and $ \hat{\cal O}(\Phi)$ is the functional form of the observable in phase space coordinates.  Finally, the differential cross section has dimensions of a cross section (an area) divided by the dimensions of the observable ${\cal O}$.  Without loss of generality, we can always work with dimensionless observables (which we do in these notes), and so to actually calculate the probability distribution of ${\cal O}$, we must normalize by the total cross section:
\begin{align}
p({\cal O}) = \frac{1}{\sigma}\, \frac{d\sigma}{d{\cal O}}\,,
\end{align}
where 
\begin{align}
\sigma = \int d\Phi\, |{\cal M}|^2\,.
\end{align}
We will discuss each of the three main parts of the differential cross section calculation.

\subsubsection{Differential Phase Space}

We prepare the initial state, like scattering protons at a fixed energy at the LHC, but the final state observed in our detectors is completely determined by the whims and whys of quantum mechanics.  As such, the particles that are produced must be allowed to have any energy or momentum consistent with the conservation laws of Nature.  Further, the S-matrix assumes that particles are measured on the {\bf celestial sphere}, points on the detector experiment that represent the direction of the particles formally an infinite distance and time away from the collision.  We can thus build up the phase space for a total of $N$ final state particles as follows.  We assume we can only measure particles' momentum four-vectors and so differential phase space is differential in all particles' momenta:
\begin{align}
d\Phi\supset \prod_{i=1}^N \left[
d^4p_i
\right]\,.
\end{align}
Next, every detected particle must be on-shell (detection is classical) so the square of each four vector must be the appropriate particle mass:
\begin{align}
d\Phi \supset \prod_{i=1}^N \left[
d^4p_i\, \delta(p_i^2 - m_i^2)
\right]\,.
\end{align}
Next, energies (the $p_0$ component) must be non-negative, which we can enforce with a step function:
\begin{align}
d\Phi \supset \prod_{i=1}^N \left[
d^4p_i\, \delta(p_i^2 - m_i^2)\,\Theta(p_{i,0})
\right]\,.
\end{align}
However, in the formulae that follow, we will not typically write the positive energy constraint explicitly, but just remember to include it in the bounds of integration.  Next, we sprinkle factors of $2\pi$ around to account for Fourier transform normalization from position space to momentum space:
\begin{align}
d\Phi \supset \prod_{i=1}^N \left[
\frac{d^4p_i}{(2\pi)^4}\, 2\pi\,\delta(p_i^2 - m_i^2)
\right]\,.
\end{align}

As written so far, particle momenta are still allowed to be anything, and are unconstrained by any conservation laws, so we need to fix this up.  However, once we actually write down the conservation law constraints, we have fixed ourselves to a particular frame.  Phase space is necessarily Lorentz-invariant (because the universe is), but the concrete form that it takes requires energies and momenta evaluated in a particular frame.  Here, we will just focus on two frames, the center-of-mass frame and, what we will study in great detail later, the {\bf infinite momentum frame} \cite{Fubini:1964boa,Weinberg:1966jm,Susskind:1967rg,Bardakci:1968zqb,Chang:1968bh,Kogut:1969xa}.

In the center-of-mass frame, the total energy is fixed and the total three-momentum is 0.  There are thus four additional constraints to impose:
\begin{align}\label{eq:cmps}
d\Phi _\text{CM} =  \prod_{i=1}^N \left[
\frac{d^4p_i}{(2\pi)^4}\, 2\pi\,\delta(p_i^2 - m_i^2)
\right] (2\pi)^4\, \delta\left(Q - \sum_{i=1}^N p_{i,0}\right)\delta^{(3)}\left(
\sum_{i=1}^N \vec p_i
\right)\,.
\end{align}
Here, the total energy of the event is $Q$.  It is illustrative to determine the dimensionality of this phase space.  First, there are $4N$ components of all $N$ particle four-vectors, but they are all on-shell, so there are only $3N$ independent components.  Energy-momentum conservation is an additional 4 constraints, and so the total dimensionality of $N$-body phase space is $3N-4$.  This result then gives you some appreciation for how challenging describing particle collision events is (at least naively).  For events in which dozens of particles are produced, phase space can easily be 100 dimensional!  Further, because of the non-linear relationship between energy and momentum, $N$-body phase space has a non-trivial topology, isomorphic to the product space of an $N-1$ dimensional simplex and a $2N-3$ dimensional sphere, $\Phi \simeq \Delta_{N-1}\times S^{2N-3}$ \cite{Cox:2018wce,Henning:2019mcv,Henning:2019enq,Larkoski:2020thc,Cai:2024xnt}.\footnote{This explicit form for differential phase space is not the most convenient for actually doing calculations, however.  A recursive formula that relates $N$ to $N-1$ body phase space through integration over the possible energy that particle $N$ can carry was first derived by Srivastava and Sudarshan \cite{Srivastava:1958ve}.  George Sudarshan made enormous contributions to theoretical physics, but have come to be largely ignored, such as with the $V-A$ theory \cite{Sudarshan:1958vf} (commonly credited to Feynman and Gell-Mann \cite{Feynman:1958ty}) and the formulation of quantum optical phase space \cite{Sudarshan:1963ts} (for which Roy Glauber won the 2005 Nobel Prize \cite{Glauber:1963tx}).}  It has been demonstrated that machine learning on topologically non-trivial manifolds can have obstructions, unless information about the topology is included \cite{olah_2014,korman2018autoencoding,moor2020topological,hajij2020topological,Batson:2021agz}, so this is a feature of studying particle collision events that may be important to consider.

Infinite momentum or collinear phase space is rather different.  Formally, we will expand phase space to lowest order in an expansion in relative angles between particles, thus considering the particles on phase space to effectively be collinear with one another.  Because we work only to lowest order in the expansion, there is no Lorentz boost that can be performed to go back to the center-of-mass frame, for example.  Mathematically, this means that we imagine particles to live exclusively on a tangent plane of the celestial sphere about a point fixed by the direction of the jet.  This collinear limit is also called the infinite momentum frame because if we boost an event by infinite momentum in some direction (i.e., the boost velocity is the speed of light), then all of the particles will be exactly collinear with one another.  Working with jets, which themselves are high-energy collimated sprays of particles, this phase space is the natural first approximation to work on, and will be the focus of the rest of these lectures.

To derive the collinear limit of phase space, we start with the full expression for $N$-body, massless phase space in the frame in which the particles have a total energy $E$ with net momentum along the $+\hat z$ axis:
\begin{align}
d\Phi =  \prod_{i=1}^N \left[
\frac{d^4p_i}{(2\pi)^4}\, 2\pi\,\delta(p_i^2)
\right] (2\pi)^4\, \delta\left(E - \sum_{i=1}^N p_{i,0}\right)\delta\left(
p_z- \sum_{i=1}^N  p_{i,z}
\right)\delta^{(2)}\left(
\sum_{i=1}^N \vec p_{i,\perp}
\right)\,.
\end{align}
In this expression, we have Lorentz-boosted the expression from the center-of-mass frame and so the total squared invariant mass is still $Q^2$:
\begin{align}
Q^2 = E^2 - p_z^2\,.
\end{align}
To proceed, we will change variables to convenient coordinates in which the collinear limit is directly manifest.  We define new momentum coordinates, called {\bf light-cone coordinates}, that are defined with respect to the boost axis.  We introduce the minus $-$, plus $+$, and perpendicular $\perp$ momenta as follows:
\begin{align}
p^- \equiv p_0 + p_3\,,  \qquad
p^+\equiv p_0 - p_3\,,  \qquad
\vec p_\perp \equiv (p_1, p_2)\,.
\end{align}
In light-cone coordinates, on-shell phase space is
\begin{align}
d\Phi &=  \prod_{i=1}^N \left[
\frac{dp_i^-\, dp_i^+\, d^2p_{i,\perp}}{2(2\pi)^3}\,\delta(p_i^- p_i^+ - p_{i,\perp}^2)
\right] \\
&\hspace{2cm}\times(2\pi)^4\, \delta\left(E - \sum_{i=1}^N \frac{p_{i}^-+p_i^+}{2}\right)\delta\left(
\sqrt{E^2-Q^2}- \sum_{i=1}^N  \frac{p_{i}^--p_i^+}{2}
\right)\delta^{(2)}\left(
\sum_{i=1}^N \vec p_{i,\perp}
\right)\,.\nonumber
\end{align}
where the new factor of 2 is the Jacobian.  We have also replaced the net momentum $p_z$ by its relationship to the total energy and invariant mass.

Next, we now work in the collinear or infinite momentum space limit and here we will see the convenience of the light-cone coordinates.  In terms of the angle $\theta$ of a particle from the boost axis (the $+\hat z$ direction), the light-cone momenta components are
\begin{align}
p^- = p_0(1+\cos\theta)\,, \qquad p^+ = p_0(1-\cos\theta)\,, \qquad \vec p_\perp = p_0\sin\theta(\cos\phi,\sin\phi)\,,
\end{align}
where $\phi$ is the azimuthal angle of the particle about the boost axis.  In the collinear limit, $\theta \to 0$, these variables reduce to
\begin{align}
p^- \to 2 p_0\,, \qquad p^+ \to p_0\,\frac{\theta^2}{2}\,, \qquad \vec p_\perp \to  p_0\,\theta(\cos\phi,\sin\phi)\,.
\end{align}
It is useful to track the scaling with angle from the boost axis through the parameter $\lambda \sim \theta \ll 1$, and then we can expand phase space to leading power in $\lambda$ everywhere to derive collinear or infinite momentum frame phase space.  Note that with this scaling, the light-cone components scale like
\begin{align}
p^- \sim \lambda^0\,, \qquad p^+ \sim \lambda^2 \,, \qquad \vec p_\perp \sim \lambda\,.
\end{align}
This power counting and expansion is generally very useful for expansion and evaluation of Feynman integrals in gauge theories and is referred to as the {\bf method of regions} \cite{Beneke:1997zp,Smirnov:1998vk,Smirnov:1999bza,Smirnov:2002pj}, by which one isolates different regions of an integral by asserting a power counting and uniformly expanding to leading power.  Power counting is also the first step of defining effective field theories, and this power counting in the collinear limit identifies and isolates the collinear modes of {\bf soft-collinear effective theory} \cite{Bauer:2000ew,Bauer:2000yr,Bauer:2001ct,Bauer:2001yt} (for reviews, see \Refs{scetnotes,Becher:2014oda}), or SCET.  We will apply power counting to deriving optimal discrimination observables from simple assumptions in \Sec{sec:d2sec}.

The first thing we will do is to simplify the momentum-conservation $\delta$-functions, expanding their arguments to leading power in the $\lambda\ll 1$ collinear limit.  Note that $p^+$ is parametrically smaller than $p^-$, and so the energy conservation $\delta$-function simplifies to
\begin{align}
\delta\left(E - \sum_{i=1}^N \frac{p_{i}^-+p_i^+}{2}\right) \to \delta\left(E - \sum_{i=1}^N \frac{p_{i}^-}{2}\right)\,.
\end{align}
Using this constraint, the second $\delta$-function, conservation of $p_z$, simplifies to
\begin{align}
\delta\left(
\sqrt{E^2-Q^2}- \sum_{i=1}^N  \frac{p_{i}^--p_i^+}{2}
\right) \to \delta\left(
\sqrt{E^2-Q^2}- E + \sum_{i=1}^N  \frac{p_i^+}{2}
\right)
\end{align}
In the strict collinear limit, where all particles are truly collinear, their total invariant mass $Q^2$ would be 0, regardless of the energy $E$.  Then, in the near-collinear limit, we take $Q \ll E$ and can expand the square-root factor as
\begin{align}
 \delta\left(
\sqrt{E^2-Q^2}- E + \sum_{i=1}^N  \frac{p_i^+}{2}
\right)\to   \delta\left(
\frac{Q^2}{2E} - \sum_{i=1}^N  \frac{p_i^+}{2}
\right)\,.
\end{align}
Finally, we note that the transverse momentum $\delta$-function
\begin{align}
\delta^{(2)}\left(
\sum_{i=1}^N \vec p_{i,\perp}
\right)
\end{align}
is homogeneous in the power counting and no expansion can be done.

At this point, differential collinear phase space can be expressed as
\begin{align}
d\Phi_\text{coll} &\to  \prod_{i=1}^N \left[
\frac{dp_i^-\, dp_i^+\, d^2p_{i,\perp}}{2(2\pi)^3}\,\delta(p_i^- p_i^+ - p_{i,\perp}^2)
\right] \\
&\hspace{2cm}\times(2\pi)^4\, \delta\left(E - \sum_{i=1}^N \frac{p_{i}^-}{2}\right)\delta\left(
\frac{Q^2}{2E}- \sum_{i=1}^N  \frac{p_i^+}{2}
\right)\delta^{(2)}\left(
\sum_{i=1}^N \vec p_{i,\perp}
\right)\,.\nonumber
\end{align}
We formally take the limit in which $\lambda \to 0$, and this correspondingly implies that $Q^2 \sim \lambda^2 \ll 1$.  Naively, from the original expression for total $p_z = \sqrt{E^2 - Q^2}$, this would imply that $Q < E$, for example.  However, taking our power counting and expansion seriously, to leading power, $Q^2$ loses all memory of relationship with total energy $E$, and is allowed to range over {\it all} values, $Q^2 \in [0,\infty)$.  Further, once this expansion is taken, with total energy effectively infinite compared to the invariant mass, the invariant mass no longer constrains the particles, and should be integrated over.  That is, to leading power in the collinear limit, there are no explicit angular scales.  Then, integrating over the invariant mass $Q^2$, collinear phase space becomes
\begin{align}
d\Phi_\text{coll} &\to  \int \frac{dQ^2}{2\pi}\, \prod_{i=1}^N \left[
\frac{dp_i^-\, dp_i^+\, d^2p_{i,\perp}}{2(2\pi)^3}\,\delta(p_i^- p_i^+ - p_{i,\perp}^2)
\right] \\
&\hspace{2cm}\times(2\pi)^4\, \delta\left(E - \sum_{i=1}^N \frac{p_{i}^-}{2}\right)\delta\left(
\frac{Q^2}{2E}- \sum_{i=1}^N  \frac{p_i^+}{2}
\right)\delta^{(2)}\left(
\sum_{i=1}^N \vec p_{i,\perp}
\right)\nonumber\\
&=2(2\pi)^3E\prod_{i=1}^N \left[
\frac{dp_i^-\, dp_i^+\, d^2p_{i,\perp}}{2(2\pi)^3}\,\delta(p_i^- p_i^+ - p_{i,\perp}^2)
\right]  \delta\left(E - \sum_{i=1}^N \frac{p_{i}^-}{2}\right)\delta^{(2)}\left(
\sum_{i=1}^N \vec p_{i,\perp}
\right)\nonumber\,.
\end{align}
We do the integral over the total squared invariant mass and divide by $2\pi$ because we effectively have to ``undo'' the on-shell $\delta$-function that would impose $Q^2 = E^2 - p_z^2$.

We now just have a couple more things to tidy up.  First, as all $p^+$ dependence in the momentum conservation $\delta$-functions has been removed, let's integrate over all $p^+$ components with the on-shell $\delta$-functions.  This produces
\begin{align}
d\Phi_\text{coll} &\to 2(2\pi)^3E\prod_{i=1}^N \left[
\frac{dp_i^-\, d^2p_{i,\perp}}{2p_i^-(2\pi)^3}
\right]  \delta\left(E - \sum_{i=1}^N \frac{p_{i}^-}{2}\right)\delta^{(2)}\left(
\sum_{i=1}^N \vec p_{i,\perp}
\right)\,.
\end{align}
We can also integrate over particle $N$'s transverse momentum with the appropriate $\delta$-function:
\begin{align}
d\Phi_\text{coll} &\to E\prod_{i=1}^{N-1} \left[
\frac{dp_i^-\, d^2p_{i,\perp}}{2p_i^-(2\pi)^3}
\right] \frac{dp_N^-}{p_N^-}\, \delta\left(E - \sum_{i=1}^N \frac{p_{i}^-}{2}\right)\,.
\end{align}
We can also introduce some convenient coordinates for the final, compact expression.  First, we will introduce the energy fractions
\begin{align}
z_i \equiv \frac{p_i^-}{2E}\,,
\end{align}
and correspondingly express the transverse momentum integration measure in polar coordinates:
\begin{align}
d^2p_{i,\perp} = p_{i,\perp}\, dp_{i,\perp}\, d\phi_i = z_i^2E^2\,\theta_i\,d\theta_i\, d\phi_i\,.
\end{align}
With these expressions, collinear phase space takes the nice form, where 
\begin{align}\label{eq:collpsN}
d\Phi_\text{coll} &= \frac{E^{2(N-1)}}{2^{N-1}(2\pi)^{3N-3}} \prod_{i=1}^{N-1} \left[
dz_i\, z_i \, \theta_i\,d\theta_i\, d\phi_i
\right] \frac{dz_N}{z_N}\, \delta\left(1 - \sum_{i=1}^N z_i\right)\,.
\end{align}
Recall that the polar angle $\theta_i$ is the angle of particle $i$ from the boost axis, the $+\hat z$ direction, and $\phi_i$ is its azimuthal angle about the boost axis.

\subsubsection{Two-Body Collinear Phase Space}

\begin{figure}
\begin{center}
\includegraphics[height=4cm]{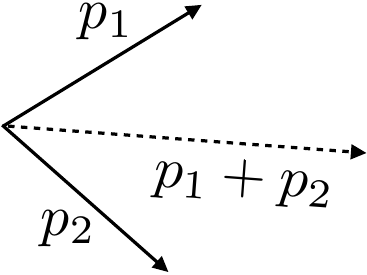}
\hspace{2cm}
\includegraphics[height=4cm]{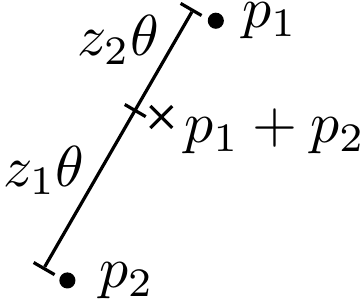}
\caption{\label{fig:2bodps} 
An illustration of a jet with two collinear particles.  Left: Representation of momentum vectors of particles 1 and 2 and their sum.  Right: Representation of the direction of momentum of particles 1 and 2 as points on the two-dimensional celestial sphere, and the direction of their sum.
}
\end{center}
\end{figure}

As an explicit example that we will revisit over and over in these lectures, let's evaluate this phase space for jets with two particles.  With $N = 2$, phase space simplifies to
\begin{align}
d\Phi^{(N=2)}_\text{coll} &= \frac{E^2}{2(2\pi)^3} \,
dz_1\, z_1 \, \theta_1\,d\theta_1\, d\phi_1\, \frac{dz_2}{z_2}\, \delta\left(1 -  z_1-z_2\right)\,.
\end{align}
We have drawn a representation of a jet with two particles in \Fig{fig:2bodps}, both as vectors and as points on a tangent plane of the celestial sphere.  With some additional assumptions, this phase space can be nicely expressed in a very compact form.  If we assume that there is an azimuthal symmetry about the boost axis, then we can integrate over the azimuthal angle and produce
\begin{align}
d\Phi^{(N=2)}_\text{coll} &= \frac{E^2}{8\pi^2} \,
dz_1\, z_1 \, \theta_1\,d\theta_1\, \frac{dz_2}{z_2}\, \delta\left(1 -  z_1-z_2\right)\,.
\end{align}
Next, we can express the angle $\theta_1$ of particle $1$ with respect to the boost axis instead as the angle $\theta$ that separates particles $1$ and $2$.  By conservation of transverse momentum in the collinear limit (and illustrated in \Fig{fig:2bodps}), this angle is
\begin{align}
\theta_ 1 = z_2\theta\,.
\end{align}
Then, this two-body collinear phase space is
\begin{align}
d\Phi^{(N=2)}_\text{coll} &= \frac{E^2}{8\pi^2} \,
 z (1-z)\, dz\,\theta\,d\theta\,.
\end{align}
Here $z$ is an energy fraction of one of the final state collinear particles, $z\in[0,1]$.

This is often re-expressed in terms of the energy fraction $z$ and the invariant mass of the two particles, $s$.  In the collinear limit, this invariant mass is
\begin{align}
s = z(1-z)E^2\theta^2\,,
\end{align}
and the corresponding expression for collinear phase space is
\begin{align}
d\Phi^{(N=2)}_\text{coll} &= \frac{dz\, ds}{(4\pi)^2} \,,
\end{align}
which is exceptionally compact.  We will study this phase space from a rather different perspective and set of initial assumptions in \Sec{sec:qvgdisc}.

\subsection{Collinear and Soft Factorization}\label{sec:softcollfact}

QCD is a gauge theory for which there is a spin-1, massless boson, the gluon, that is ultimately responsible for the forces that color charged particles act on one another.  In this way, QCD is analogous to electromagnetism for which the photon is exchanged between electrically charged particles and correspondingly leads to Coulomb's law, the Biot-Savart law, and the full weight of Maxwell's equations.  Phenomenologically, we also know that electromagnetism has many interesting limits, such as acting coherently like a wave at long distances, or, by Fourier transformation, at low energies, or when a charged particle is rapidly accelerated, it emits most of its power in radiation that is approximately collinear to the particle's direction of motion.  These phenomena are also present in QCD, and, especially the latter, is what is ultimately responsible for jets.  

Further, it isn't so much that QCD has an analogous Larmor formula (which it does, by the way), but rather that these soft and collinear limits {\bf factorize} from the dynamics of whatever else is going on in the event of interest.  This factorization or separation of a squared matrix element into a product of factors enables a self-similarity to the emission structure of QCD events.  Radiation collinear to a high-energy quark, for instance, factorizes from the dynamics of the quark, which can then emit more collinear radiation that factorizes from the quark, etc., producing a whole shower or cascade of collinear particles, manifesting itself as a jet.  We will actually start from this scale-invariant assumption in the following section and see how far its consequences go, but here we must first review this soft and collinear factorization and present the explicit forms of the squared matrix elements that we will manipulate and study in these notes.

\subsubsection{Collinear Splitting Factorization}\label{sec:collsplitfact}

In a theory of massless particles, such as QCD at high energies where quark masses can be ignored, propagators in the calculation of Feynman diagrams can diverge in one of two ways.  This can be directly observed by simply writing down the expression for a propagator formed from the sum of two massless momenta $p_1,p_2$:
\begin{align}
\frac{1}{(p_1+p_2)^2} = \frac{1}{2p_1\cdot p_2} = \frac{1}{2E_1E_2(1-\cos\theta_{12})}\,,
\end{align}
where $E_1,E_2$ are the energies of the two momenta and $\theta_{12}$ is their relative angle.  As the focus of this subsection, two massless particles, for example, can become collinear in which they travel in the same direction.  In this case, their invariant mass (the inverse of the propagator) vanishes because two massless particles traveling in the same direction act as an effective meta-massless particle itself.  The fact that the propagator diverges in the collinear limit means that such a configuration is likely to occur; particles {\it want} to be collinear with one another.  We will address the soft limit of the propagator, when, say, $E_2\to 0$, in the next subsection.

Indeed, these divergences can be interpreted through another lens.  Feynman diagram perturbation theory is a degenerate perturbation theory and as a degenerate perturbation theory, we know that divergences arise in intermediate calculations, but cancel when all exactly degenerate states are inclusively summed over.  We will revisit this point again and again, but for now, this interpretation means that collinear divergences in Feynman diagram perturbation theory correspond to states that are indistinguishable from other states in the theory with a different number of collinear particles.  That is, the quantum numbers of exactly collinear particles in QCD correspond to the quantum numbers of individual quarks and gluons, and so by the rules of degenerate perturbation theory, we must sum over states with any possible number of collinear particles.  That is, there is no experiment that you can perform to determine the number of exactly collinear particles produced in any process, because all you can measure are net quantum numbers.  We will see how to do this in our first example next lecture, but the starting point for this analysis is to determine the squared matrix element in the collinear limit.

\begin{figure}[t!]
\begin{center}
\includegraphics[width=14cm]{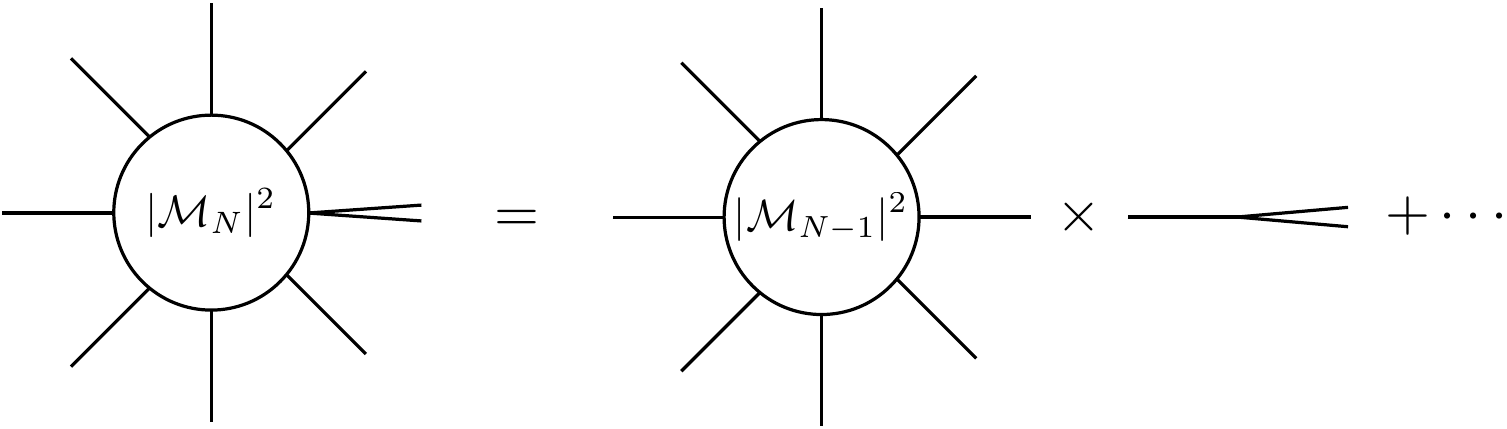}
\caption{\label{fig:collsplit}
Schematic illustration of the collinear limit of an initial $N$ point squared S-matrix element $|{\cal M}_N|^2$ that factorizes into a product of an $N-1$ point squared matrix element $|{\cal M}_{N-1}|^2$ and the collinear splitting function (denoted as the $1\to 2$ splitting at right). 
}
\end{center}
\end{figure}

From the early days of QCD, and from analogy with electromagnetism, it has been known that squared matrix elements factorize in the collinear limit.  A squared matrix element of $N$ final state particles $|{\cal M}(p_1,\dotsc, p_N)|^2$ calculated in perturbation theory takes the following form when two particles $i$ and $j$ become collinear:
\begin{align}\label{eq:collfact}
\left.|{\cal M}(p_1,\dotsc, p_i, p_j,\dotsc, p_N)|^2 \right|_{i\parallel j} =\frac{8\pi \alpha_s}{s_{ij}} P_{(ij)\to ij}(z)\, |{\cal M}(p_1,\dotsc, p_{(ij)},\dotsc, p_N)|^2 +\cdots\,.
\end{align}
By ``collinear'', I mean that the angle between particles $i$ and $j$ is much, much smaller than any other angle between particles $i$ and $j$ and other particles in the event.  At right, $ |{\cal M}(p_1,\dotsc, p_{(ij)},\dotsc, p_N)|^2$ is the squared matrix element in which momenta of individual particles $i$ and $j$ are replaced by their sum $p_{(ij)} = p_i +p_j$.  The factor multiplying this squared matrix element has the propagator $1/s_{ij}$ explicit, as well as the coupling $\alpha_s$ that controls the probability to split at all.  Ellipses hide terms that are less singular in the collinear limit (i.e., do not have the factor of the propagator).  This equation is drawn schematically in \Fig{fig:collsplit}.

The most interesting part of this formula, however, are the {\bf splitting functions}, $P_{(ij)\to ij}$ which describe the probability of the distribution of energy fraction $z$ carried by particle $i$ in the splitting.  These are {\bf universal} in that they are independent of the squared matrix element or process that one considers, and are intrinsic to the theory of QCD.  All possible $1\to 2$ splitting functions are referred to as the {\bf Dokshitzer-Gribov-Lipatov-Altarelli-Parisi (DGLAP) kernels} \cite{Dokshitzer:1977sg,Gribov:1972ri,Gribov:1972rt,Lipatov:1974qm,Altarelli:1977zs}:
\begin{align}
&P_{q\to qg}(z) =C_F\,\frac{1+z^2}{1-z} \,, &P_{q\to gq}(z) =C_F\,\frac{1+(1-z)^2}{z} \,,\\
&P_{g\to q\bar q}(z) = T_R\left[
z^2+(1-z)^2
\right]\,, &P_{g\to gg}(z) = C_A\left(
\frac{z}{1-z}+\frac{1-z}{z}+z(1-z)
\right) \,.
\nonumber
\end{align}
These splitting functions are further controlled by color factors $C_F$, $C_A$, and $T_R$ which quantify how the three colors of QCD are shared between the quarks and gluons involved in the splitting.  $C_F$ and $C_A$ are the quadratic Casimir operators of the fundamental and adjoint representations of SU(3) color, respectively, and $T_R$ is the normalization of the fundamental representation or Gell-Mann matrices of QCD (i.e., the Killing form).  In an SU($N$) gauge theory, the quadratic Casimirs are
\begin{align}
&C_F = \frac{N^2-1}{2N}\,, &C_A = N\,,
\end{align}
and so are $C_F =4/3$ and $C_A = 3$ in QCD.  By convention, we also fix $T_R = 1/2$.

Note, however, that these splitting functions are in general not integrable on $z\in[0,1]$, but typically diverge if $z\to 0,1$.  The additional soft divergences correspond to a gluon produced in the splitting carrying 0 energy, which again is a further degenerate state.  (More on this in a second.)  Individual quarks, on the other hand, have no corresponding soft divergence, for a few reasons.  For example, quarks are fermions and carry spin-1/2 and there is no possible degenerate state in which a spin-1/2 particle becomes soft and ``disappears''.  By angular momentum conservation, you will always know if a quark has become soft.

We will almost exclusively only consider $1\to 2$ collinear splittings in our analyses that follow, but in general, $1\to n$ collinear splitting functions exist and can be calculated.  For this more general splitting, the factorization in the collinear limit takes basically the same form as \Eq{eq:collfact}, with the factors of the propagator and powers of the coupling modified to describe additional particle emission.  Concretely, the $1\to 3$ splitting functions have been worked out \cite{Campbell:1997hg,Catani:1999ss}, and the factorization in this limit takes the form:
\begin{align}
&\left.|{\cal M}(p_1,\dotsc, p_i, p_j,p_k,\dotsc, p_N)|^2 \right|_{i\parallel j \parallel k} \\
&\hspace{3cm}=\left(\frac{8\pi \alpha_s}{s_{ijk}}\right)^2 P_{(ijk)\to ijk}(\{z\},\{s\})\, |{\cal M}(p_1,\dotsc, p_{(ijk)},\dotsc, p_N)|^2 +\cdots\,,\nonumber
\end{align}
for three collinear particles $i,j,k$.  The novel aspect starting at $1\to 3$ splitting is that the splitting function $P_{(ijk)\to ijk}(\{z\},\{s\})$ depends on three energy fractions, $z_i,z_j,z_k$ with $z_i+z_j+z_k = 1$, and, in general, all pairwise invariant masses $s_{ij}, s_{ik}, s_{jk}$ with $s_{ij}+ s_{ik}+ s_{jk} = s_{ijk}$.\footnote{For a complete collinear calculation at this order in perturbation theory, one also needs the one-loop, $1\to 2$ splitting functions \cite{Bern:1998sc,Bern:1999ry,Kosower:1999rx}.  Even higher order splitting functions are now known, as well \cite{Badger:2004uk,Bern:2004cz,Catani:2003vu,Badger:2015cxa,DelDuca:2019ggv,DelDuca:2020vst,Czakon:2022fqi}.}

\subsubsection{Soft Gluon Factorization}

The second possibility for particle configuration in which a propagator diverges is if a soft or low energy gluon is emitted at arbitrary angles to the other, harder, particles in the event.  For jet physics specifically, in which we focus on the collinear, high energy dynamics, soft, wide-angle gluon emission is less directly relevant (at least for these lectures), but we include its discussion for completeness and we will encounter a few scenarios where we will need to understand why and where gluons become soft. As hinted at in introducing this section, the soft limit by Fourier transformation corresponds to the long distance or classical limit.  Soft divergences have a bit of a different interpretation than collinear divergences, as well.  There is no way to determine the number of arbitrarily soft gluons produced because they carry no energy and so never ``light'' up any physical detector.  However, with long wavelength, they are sensitive to the global flow of color charge and act coherently and so conspire in the large number limit as a classical wave.

In the limit that a gluon $k$ carries arbitrarily low energy, also called the {\bf eikonal limit}, the matrix element factorizes as \cite{Low:1958sn,Weinberg:1965nx,Burnett:1967km,Bassetto:1983mvz}
\begin{align}\label{eq:softfact}
\left.|{\cal M}(k,p_1,\dotsc, p_N)|^2 \right|_{k\to 0} =-8\pi\alpha_s \sum_{i,j = 1}^N {\mathbf T}_i\cdot {\mathbf T}_j\frac{s_{ij}}{s_{ik}s_{kj}} |{\cal M}(p_1,\dotsc, p_N)|^2 +\cdots\,.
\end{align}
This has a few more moving parts than for a collinear splitting, because of the global sensitivity of soft gluons.  First, the sum runs over all hard particles in the event $i,j$, including $i < j$, $ i > j$, and $i = j$.  The propagator, or eikonal factor, is the ratio of invariant masses $s_{ij}/s_{ik}s_{kj}$, and accounts for the divergent propagator if the soft gluon $k$ is emitted from either particle $i$ or $j$.

This form of the factorization takes the form of a sum over {\bf dipoles}, for exactly the same reason that an accelerating dipole in electromagnetism emits radiation.  Sensitivity to electric charge in a radiation formula is replaced by sensitivity to the product of color matrices ${\mathbf T}_i\cdot {\mathbf T}_j$ that describes how particles $i$ and $j$ share color; i.e., the color that flows along the dipole formed by particles $i$ and $j$.  This completely inclusive double sum is convenient to represent the sum over dipoles because it nicely manifests color conservation.  We have that
\begin{align}
\sum_{i,j = 1}^N {\mathbf T}_i\cdot {\mathbf T}_j = \left(
\sum_{i=1}^N {\mathbf T}_i
\right)^2=0\,,
\end{align}
but the total color charge of {\it any} event is 0 (summing over all initial and final state particles), because interactions cannot create or annihilate net charge.  To actually evaluate the color matrix dot products can be tricky in general, and typically requires working in an explicit color basis and then simply keeping track of all color matrices in the evaluation of Feynman diagrams.\footnote{A very efficient organizing principle for color management is the topological expansion of 't Hooft \cite{tHooft:1973alw,tHooft:1974pnl}.  Color-ordering and diagrammatic computation of traces of products of color matrices is a standard practice for multi-loop or multi-leg amplitude calculations \cite{cvitanovic2008group,Berends:1987cv,Mangano:1987xk,Mangano:1988kk,Bern:1990ux,Dixon:1996wi}. See Malin Sj\"odahl's lectures at this same school for more details, as well.}  However, for low point amplitudes, with 2 or 3 quarks or gluons, the color factors can be evaluated exactly simply from color conservation and the square of color matrices.  Specifically, the square of the color matrix ${\bf T}_i$ is the corresponding quadratic Casimir of particle $i$:
\begin{align}
{\mathbf T}_i^2 = C_i\,.
\end{align}

It is correspondingly useful to validate consistency of the soft and collinear limit from this factorization, with the expressions from collinear factorization earlier.  Let's say that soft gluon $k$ is collinear with particle $i$.  Then, note that the invariant mass of $k$ with another particle $j \neq i$ can be expressed as
\begin{align}
s_{kj} = zs_{ij}\,,
\end{align}
where $z \ll 1$ is the energy fraction of gluon $k$ with respect to the energy of particle $i$.  Then, the factorization of the squared matrix element becomes
\begin{align}
\left.|{\cal M}(k,p_1,\dotsc, p_N)|^2 \right|_{k\to 0, k\parallel i} &=-\frac{16\pi\alpha_s}{zs_{ik}} |{\cal M}(p_1,\dotsc, p_N)|^2 \sum_{j = 1,\, j\neq i}^N {\mathbf T}_i\cdot {\mathbf T}_j +\cdots\\
&=-\frac{16\pi\alpha_s}{zs_{ik}} |{\cal M}(p_1,\dotsc, p_N)|^2\, {\mathbf T}_i\cdot\sum_{j = 1,\, j\neq i}^N  {\mathbf T}_j +\cdots\nonumber\\
&=\frac{16\pi\alpha_s}{zs_{ik}}{\mathbf T}_i^2 |{\cal M}(p_1,\dotsc, p_N)|^2 +\cdots\nonumber\,,
\end{align}
where we used color conservation on the third line.  Note that we have multiplied by a factor of 2 because the original sum over dipole particles was completely inclusive in both $i$ and $j$.  This indeed agrees with the soft limit of the collinear factorization formulas we presented, with the ellipses hiding terms that are higher order in the soft and/or collinear limits.

In the past decade or so, there has been a flurry of interest in the next-to-leading soft limit behavior of scattering amplitudes in gauge theories, like QCD, and gravity, see, e.g., \Refs{Gross:1968in,Jackiw:1968zza,White:2011yy,Cachazo:2014fwa,Larkoski:2014hta,Casali:2014xpa,Broedel:2014fsa,Bern:2014vva}.  In gauge theories, the next term in the soft gluon expansion is referred to as the Low-Burnett-Kroll theorem \cite{Low:1958sn,Burnett:1967km}, and is sensitive to the angular momentum of the two particles $i$ and $j$ in the dipole off of which the soft gluon is emitted.  However, unlike the leading, eikonal limit, the Low-Burnett-Kroll term only holds for tree-level amplitudes and does not generalize simply for arbitrary loop amplitudes for the simple reason that real emission soft divergences start to mix with soft and collinear virtual divergences starting at one-loop order.  Loop level subleading soft theorems can systematically be constructed \cite{Bern:2014oka,Larkoski:2014bxa}, but they are rather a mess and bookkeeping becomes a serious issue.  At any rate, we will only need to understand the leading soft behavior here.

\subsection{Infrared and Collinear Safety}

In identifying the limits in which collinear and soft factorization of squared matrix elements exist, we uncovered a problem for interpretation of our master formula.  Feynman diagram perturbation theory is degenerate, and this manifests as honest non-integrable divergences of squared matrix elements in the soft and/or collinear limit.  Non-integrability kills hope of interpretation of our master formula as a probability distribution because, if anything, probabilities are integrable.  However, we also know that this cannot be the whole story because degenerate perturbation theory is only problematic if you consider a subset of all possible degenerate states.  Again, summing over all degenerate states eliminates the divergences and renders predictions finite and, correspondingly, integrable.  So, for our master formula to be useful, we need to ensure that all degenerate states are indeed inclusively summed over.

The first step in doing this is to identify what these degenerate states are in our perturbation theory of QCD.  In the previous section, we discussed {\it real} particle emission in the collinear or soft limit.  These real contributions are positive because they correspond to honest absolute squared amplitudes.  So, if their divergences are to cancel, then there must be a negative divergent contribution to account for.  This negative contribution is from, of course, the virtual, or loop, diagrams.  An $N$-point amplitude has a loop expansion of the form
\begin{align}
{\cal M}(p_1,\dotsc,p_N) = {\cal M}^{(0)}(p_1,\dotsc,p_N)+\alpha_s {\cal M}^{(1)}(p_1,\dotsc,p_N)+\cdots\,,
\end{align}
where the superscript $(l)$ denotes the number of loops in the diagram; e.g., $(0)$ is the tree-level diagram and $(1)$ is the one-loop diagram.  In our master formula we need the absolute square of this amplitude, which, through one-loop is
\begin{align}
|{\cal M}(p_1,\dotsc,p_N)|^2 &= \left(
{\cal M}^{(0)}(p_1,\dotsc,p_N)+\alpha_s {\cal M}^{(1)}(p_1,\dotsc,p_N)+\cdots
\right)\\
&\hspace{4cm}\times\left({\cal M}^{(0)}(p_1,\dotsc,p_N)+\alpha_s {\cal M}^{(1)}(p_1,\dotsc,p_N)+\cdots\right)^*\nonumber\\
&=|{\cal M}^{(0)}(p_1,\dotsc,p_N)|^2 + 2\alpha_s\,\text{Re}\left(
{\cal M}^{(0)}(p_1,\dotsc,p_N){\cal M}^{(1)}(p_1,\dotsc,p_N)^*
\right)+\cdots\,.\nonumber
\end{align}
Simply matching orders in $\alpha_s$, the virtual contribution, the real part of the product of the tree-level and conjugate of the one-loop amplitudes, must be negative if there is any hope of divergence cancellation.

Additionally, this loop expansion produces no additional particles, and so all of the terms in this loop expansion are to be evaluated on $N$-body phase space.  By contrast, a collinear splitting from an $N$-point amplitude produces an additional real particle, and so are to be evaluated on $N+1$ body phase space.  Actual infinities are only exactly located at points on phase space where particles are exactly collinear, where there is perfect confusion over how many particles are produced collinearly.  Thus, for the real infinities on $N+1$ body phase space from collinear or soft emission to cancel with the virtual infinities of $N$ point loop amplitudes, we must ensure that any observable that we measure on these events cannot individually identify exactly collinear or exactly 0 energy particles.

This requirement for the success of degenerate Feynman diagram perturbation theory is called {\bf infrared and collinear safety} or IRC safety, and has historical understanding and formalization in the Bloch-Nordseieck theorem of QED and the Kinoshita-Lee-Nauenberg (KLN) theorem more generally in any quantum field theory \cite{Bloch:1937pw,Kinoshita:1962ur,Lee:1964is}.  The first use of IRC safety in QCD was to provide a robust, and correspondingly calculable, definition of jets \cite{Sterman:1977wj}, which is now referred to as the Sterman-Weinberg algorithm.\footnote{See \Refs{Yao:1975fa,Appelquist:1976ay,Tyburski:1976yc,Poggio:1976qr,Sterman:1976jh,Krausz:1977nb} for earlier work on cancellation of infrared divergences in QCD and non-Abelian gauge theories.}  We will review jet algorithms in the following section, but while the Sterman-Weinberg algorithm was extremely important historically, it is not very practical and is therefore not used experimentally, as it defines jets as containing a fraction $\epsilon > 0$ of the total energy within an angular size $\delta > 0$.  Nevertheless, its IRC safety is easy to understand, because exactly collinear particles are always within any angle $\delta > 0$ and exactly 0 energy particles always have less than any fraction $\epsilon > 0$ of the total energy of an event.  So, the Sterman-Weinberg jet definition inclusively sums over degenerate states.

A modern definition of IRC safety for a general observable $\hat {\cal O}(\Phi)$ is typically stated as \cite{Ellis:1996mzs}:
\begin{quote}
The observable $\hat {\cal O}(\Phi)$ is IRC safe if, for any three-momentum of a particle $\vec p_i$ on phase space $\Phi$, the observable's value is invariant under the splitting $\vec p_i \to \vec p_j + \vec p_k$ if the momenta $\vec p_j$ and $\vec p_k$ are parallel or one of them is small.
\end{quote}
This will be our guiding principle in constructing observables for which our master formula can be applied, and for which we can make systematically-improvable predictions.  In particular, a very simple observable structure that is immediately IRC safe is if the observable is only sensitive to linear sums of particle three-momenta.  While I know no proof of this statement, in practice this linear sum rule seems to nearly be necessary and sufficient for IRC safety and even some of the rather complicated observables we will discuss later can all be expressed as a function of sums of momenta, even if we write them for compactness in a rather different way.\footnote{A related statement is proved, however.  Multilinear symmetric functions of particle momenta span the space of IRC safe observables, see \Refs{Sveshnikov:1995vi,Tkachov:1995kk,Cherzor:1997ak,Tkachov:1999py,Komiske:2017aww}, in the sense that arbitrary linear combinations of them can approximate any IRC safe observable with arbitrary accuracy.  The statement that I know of no proof for is if there exist IRC safe observables that are expressed as functions that are irreducibly not linear in particle momenta.}  One can then say the art of jet substructure is in the construction of observables that are both sensitive to the phenomena of interest as well as IRC safe for predictability.

However, it must be said in the same breath that there is nothing ``bad'' or ``undesirable'' about observables that are not IRC safe.  Perhaps the simplest such observable is the total particle multiplicity (just the number of particles produced), as an arbitrarily soft emitted particle will change this quantity by one unit.  IRC unsafe observables are simply those for which our master formula does not apply, but we will nevertheless need an understanding of them.  As these lectures progress, we will stray further and further from the light of IRC safe observables, and find strange situations in which our master formula secretly works for IRC unsafe observables, and for others, like multiplicity, where we simply have to throw up our hands and attack the problem from a completely different direction.  To emphasize again, our master formula and IRC safe observables have a use and are interesting, but the world is much larger so we will need more and powerful tools.  In fact, we will see when IRC safety can guide us no further, in many cases the central limit theorem is there to lead on.

\subsubsection{Master Formula for Jet Distributions}

With phase space, factorization, and IRC safety understood, we can now write down a master Fermi's Golden Rule for jet substructure.  The formula we will return to over and over as a first point of understanding in calculating distributions of observables ${\cal O}$ on jets with two particles in the collinear limit.  For many of the observables we study, this is the first interesting case, or leading-order in the coupling when distributions are non-trivial.  This master formula is then
\begin{align}
\frac{d\sigma^{(0)}}{d{\cal O}} &= \int d\Phi_\text{coll}^{(N=2)}\, \frac{8\pi\alpha_s}{s}\, P_{(ij)\to ij}(z)\,\delta\left({\cal O} - \hat {\cal O}(\Phi_\text{coll}^{(N=2)})\right)\\
&=\frac{\alpha_s }{\pi}\int  \frac{d\theta}{\theta}\,dz\, P_{(ij)\to ij}(z)\,\delta\left({\cal O} - \hat {\cal O}(z,\theta)\right)
\nonumber\,.
\end{align}

\subsection{Aside: Jet Algorithms}

Given a proton collision event at the LHC, like that displayed in \Fig{fig:jetdisplay}, we can see by eye the jets as the high-energy, collimated deposits in the experimental detector.  However, at the extreme data collection rates of 1000 recorded events per second (or more), we can't simply have humans look at every event display and circle the jets for further analysis.  Instead, we need a procedure or algorithm for identifying jets in an event that clusters particles that are close in angle and correlated in energy, and further, is ideally IRC safe so that we can calculate rates for jet production in perturbation theory.  Due to the soft and collinear divergences in QCD, jets in experiment are highly correlated with a single initiating parton that subsequently radiated many particles that were ultimately detected.  Actually, given a jet algorithm and unlabeled particle physics data, the clustering of particles in that data is an example of {\bf unsupervised learning} in machine learning because the algorithm identifies structure without human intervention or explicit ground truth correlation.  The jets that a jet algorithm finds are then proxies for individual quarks and gluons produced in the short distance collision event.

Historically, there have been many jet algorithms defined and used (we already mentioned the Sterman-Weinberg algorithm), but as theoretical and experimental goals sharpened, one class of jet algorithm has come to be standard.  {\bf Pairwise clustering} or {\bf recombination algorithms} proceed by building structure in the event by clustering together pairs of particles that are closest together by a particular measure until some termination criteria is met.  In this way, such algorithms effectively undo the collinear splitting process, and so, at least in the strict collinear limit, can be understood in this way.  However, away from exact collinearity, there is significant freedom in how this clustering is actually done.

The specific algorithms for pairwise clustering were developed in the early 1990s but in some sense was ``perfected'' in 2008 \cite{Catani:1991hj,Ellis:1993tq,Catani:1993hr,Dokshitzer:1997in,Wobisch:1998wt,Cacciari:2008gp}.\footnote{See \Refs{JADE:1986kta,JADE:1988xlj,Brown:1990nm,Brown:1991hx,Bethke:1991wk} for earlier examples of pairwise clustering algorithms.}  They consist of two, logically distinct, parts, the {\bf clustering metric} and the {\bf recombination scheme}.  The clustering metric defines {\it which} particles are combined while the recombination scheme defines {\it how} particles are combined.  We will first present the recombination algorithm, and through it discuss the different metrics and recombination schemes that are used.

\subsubsection{Clustering Metric}

The specific algorithm for pairwise reclustering is as follows:
\begin{enumerate}
\item Calculate the clustering metric 
\begin{align}
d_{ij}^{(n)} = \min\left[
E_i^{2n},E_j^{2n}
\right]\frac{\theta_{ij}^2}{R^2}
\end{align}
between all pair of particles $i,j$ in the event.  Here, $n$ is a parameter that defines the particular algorithm.  The most common choices are $n = 1$ which is the $\bm{k_T}$ {\bf algorithm}, $n = 0$ which is the {\bf Cambridge/Aachen (C/A) algorithm}, and $n = -1$ which is the {\bf anti-}$\bm{k_T}$ {\bf algorithm}.  In general, this clustering metric defines a class of algorithms that are referred to as $k_T$-type.

\item For the pair of particles with the smallest $d_{ij}^{(n)}$, combine their momentum into a meta-particle $(ij)$.  The standard way to do this is through $\bm{E}${\bf -scheme} recombination by which the four-vectors are summed:
\begin{align}
p_{(ij)} = p_i + p_j\,.
\end{align}
We will say more about other recombination schemes shortly.  Then, remove $i$ and $j$ from the list of particles, and add $(ij)$ in their place.

\item Continue this procedure, recalculating the pairwise metric $d_{ij}^{(n)}$ and clustering the smallest pair.  In addition, also calculate the ``beam'' metric
\begin{align}
d_{ib}^{(n)} = E_i^{2n}\,.
\end{align}
If $d_{ib}^{(n)}$ is smaller than all of the $d_{ij}^{(n)} $, then remove all particles from the event list that had been clustered into $i$ and call them jet $i$.  Once all jets are clustered in this way, the algorithm terminates.
\end{enumerate}
Specifically, the algorithm presented here uses energies and angles for simplicity of expressions, but these should be modified to transverse momenta and azimuth-(pseudo)rapidity distances at a hadron collider.  This algorithm produces jets with a characteristic angular size of $R$; that is, all particles within an angle of about $R$ of hard, collinear radiation are put into a single jet.

$k_T$-type pairwise clustering algorithms are IRC safe.  Low energy particles don't affect the clustering sequence and further don't affect global properties of jets, like their total energy or direction of net momentum.  Collinear particles are always clustered in the same jet because of the pairwise angular weighting factor in the clustering metric.  So, as long as you ask IRC safe questions about jets or their substructure, $k_T$-type jets are themselves IRC safe. However, historically, it was not always the case that IRC safe algorithms were used in experiment.  For a few decades, from the early QCD era in the 1970s to really until the late 20-aughts (more about why below), experiments typically clustered jets with iterative, seeded cone algorithms (see, e.g., \Refs{UA1:1983gpz,Huth:1990mi,CDF:1991juq,Abbott:1997fc}).\footnote{My undergraduate research advisor, Steve Ellis, was for many years one of the very few voices in the wilderness pushing back on experimentalists' preferences for seeded cone algorithms, strongly advocating for widespread use of the IRC safe $k_T$ algorithm, or even small modifications of the seeded cone algorithms that rendered them IRC safe \cite{Seymour:1997kj}.  Ellis is also perhaps most famous for a talk he gave in winter 1985 that countered Carlo Rubbia's claim that supersymmetry had been discovered in the UA1 and UA2 experiments \cite{Ellis:1985xw}.  This episode is documented in Gary Taubes's book {\it Nobel Dreams: Power, Deceit and the Ultimate Experiment}.}  A seeded cone algorithm starts with a seed (some particle in the event), and places a cone, or disk of radius $R$, centered about the seed.  Then, the net momentum of the particles in the jet is calculated, and the jet center moved to the location of net momentum.  This procedure continues until the distance that the cone jet center moves is less than some criteria.  Seeded cone algorithms are not IRC safe because arbitrarily soft particles can seed a cone and, by chance, cluster two hard particles that are further than $R$ apart in angle.  

One thing to note about jet algorithms in QCD in general is that there is no natural value for the jet radius $R$.  QCD at high energies is an approximately scale-invariant theory, and correspondingly has no intrinsic angular scales.  Therefore, regardless of the value of $R>0$ you use to cluster jets, you will always find jets!  In practice, however, experimental detectors break scale invariance, and experimentalists weigh several factors to determine an optimal jet radius, including particle resolution, average jet energy from fixed collision energy, maximal pseudorapidity range, etc.  For all-purpose jet finding, the radius that is typically used at both ATLAS and CMS is about $R = 0.4$, but for some applications, a large or ``fat'' jet radius of $R \sim 1$ may be used.  This latter case is especially useful when there is a known angular scale for the jets of interest; namely, those jets that indeed have an intrinsic mass.  Such an example is in searching for energetic top quarks that decay hadronically, which have, because of the large mass of top quarks, $m_t \approx 173$ GeV, a characteristic angular size of about $R\sim 2m_t / E$.  

\begin{figure}[t!]
\begin{center}
\includegraphics[width=7cm]{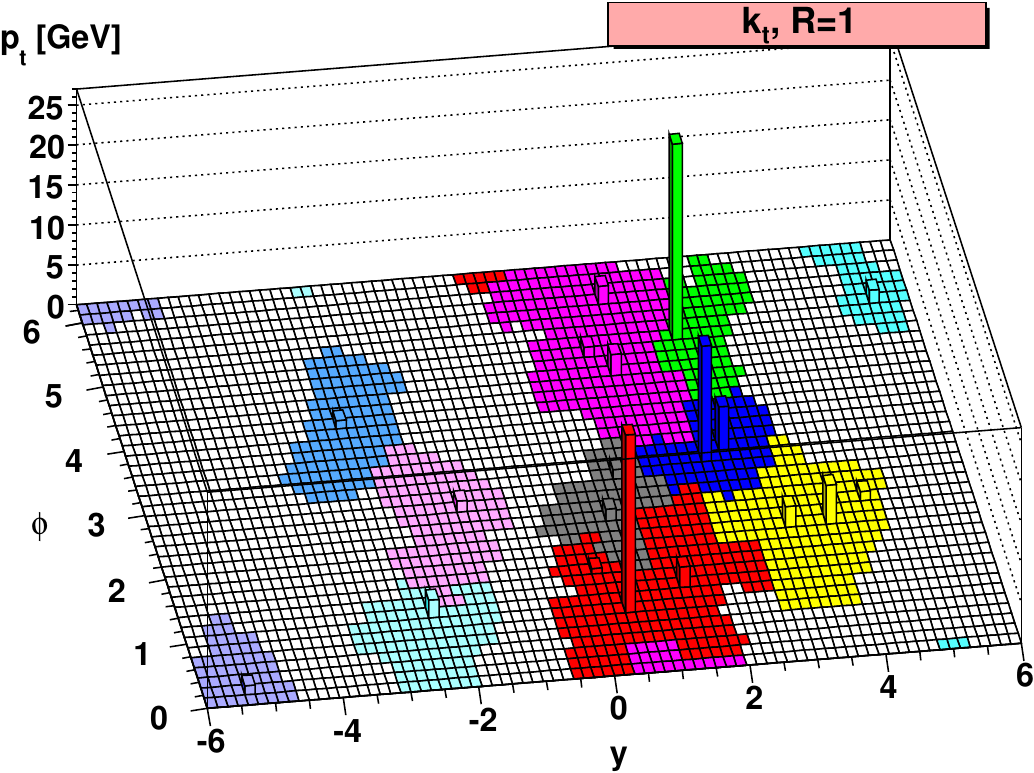}\hspace{1cm}
\includegraphics[width=7cm]{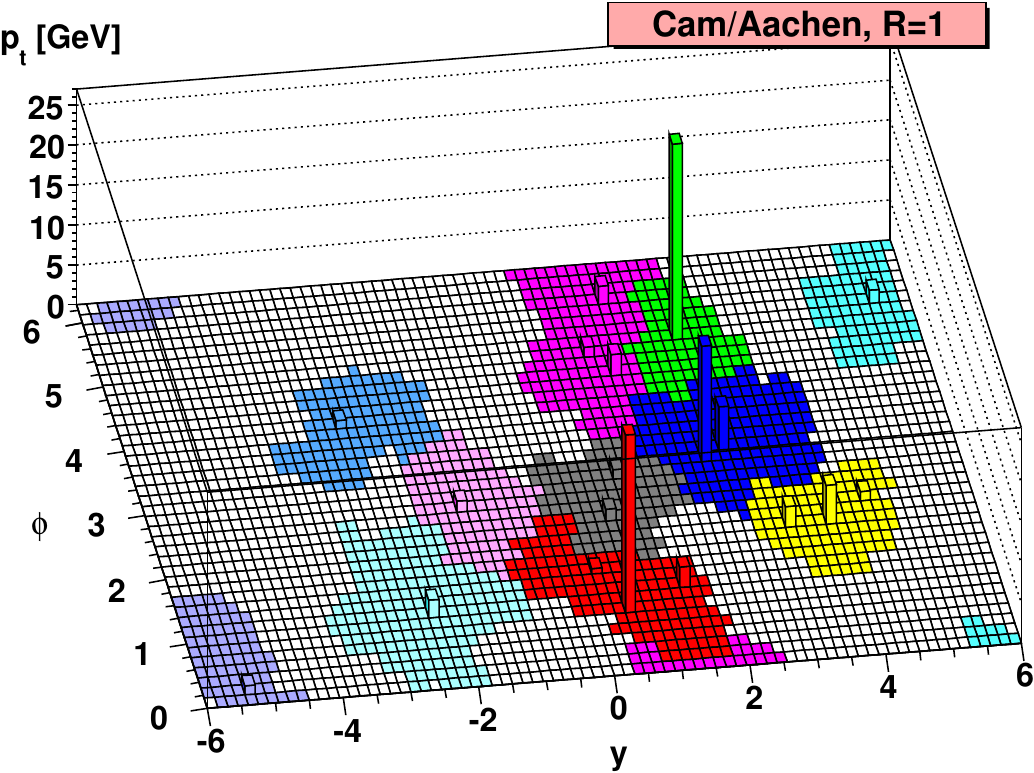}\\
\includegraphics[width=7cm]{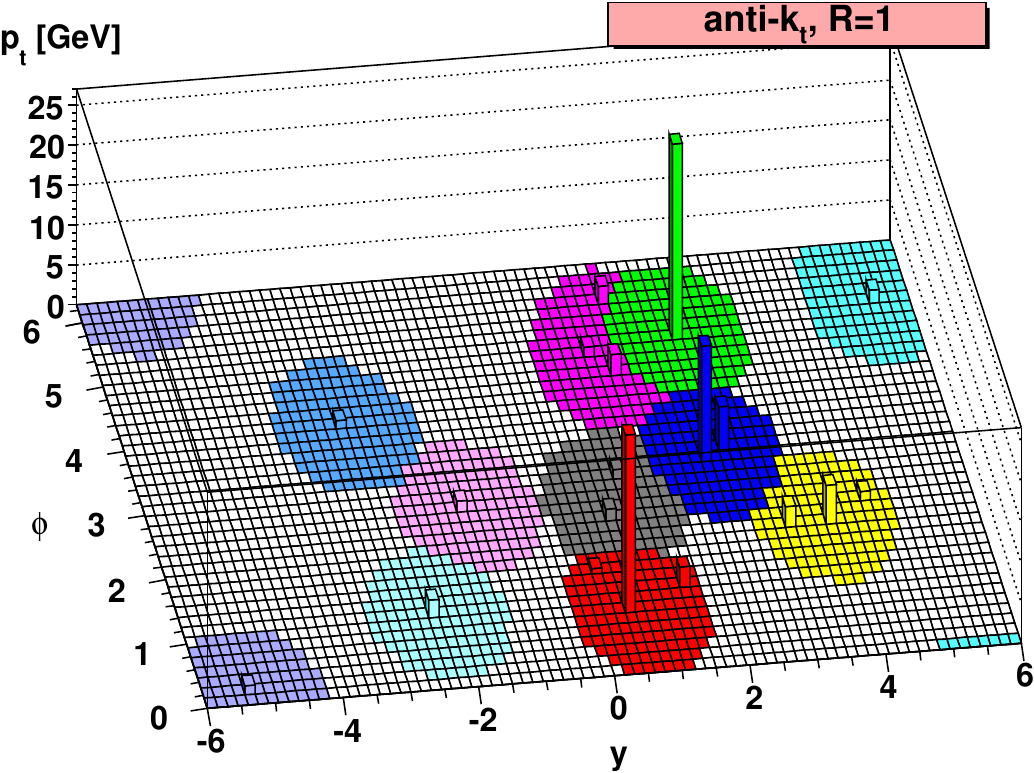}
\caption{\label{fig:jetalgscomps}
Demonstration of jet shape dependence from simulated collision events at the LHC.  The $k_T$ (top left), C/A (top right), and anti-$k_T$ (bottom) algorithms are used to cluster particles in the same event.  Extremely low energy particles, called ``ghosts'', are uniformly distributed over the event and distinct clustered jet regions are colored.  Figure from \InRef{Salam:2010nqg}.
}
\end{center}
\end{figure}

The $k_T$ and C/A algorithms tend to cluster soft, wide angle particles first, and then cluster those structures with hard, collinear radiation.  As such, the shape of $k_T$ or C/A jets is highly irregular, which is undesirable for experimental calibration of such jets (and one reason for significant foot-dragging by experiments to adopt such algorithms).  However, in February 2008, just 7 months before magnets would explode at the LHC delaying its scientific mission by over a year, the anti-$k_T$ algorithm was introduced \cite{Cacciari:2008gp}.  Unlike $k_T$ or C/A, anti-$k_T$ tends to cluster hard, collinear cores first, and then soft radiation about that, and as such produces very regular, nearly perfectly disk shaped jet regions of radius $R$. Representative shapes of jets clustered with these algorithms are illustrated in \Fig{fig:jetalgscomps} from \InRef{Salam:2010nqg}.  Such jets have very consistent areas and are therefore easily calibrated experimentally, and before the LHC physics program started again in 2009, the anti-$k_T$ algorithm was universally adopted as the jet finding algorithm of choice.

In concert with the development of the anti-$k_T$ algorithm, at the same time (and even by some of the same authors!)~another historical problem with pairwise recombination algorithms was solved.  The naive computational complexity of a pairwise recombination algorithm is very expensive.  For an event with $N$ particles, we need to calculate the $\sim N^2$ pairwise metrics $d_{ij}^{(n)}$, and then we have to perform the clustering step $\sim N$ times to put every particle in a jet.  Thus, a pairwise recombination algorithm costs $\sim N^3$, which is seriously slow and was a major sticking point for experiments to adopt it.  This large time cost meant, for example, that jet finding could never be done in early triggering because there simply would not be enough time between bunch collisions for jets to be found.  However, these authors noted that for particles beyond an angular scale of $R$, clustering proceeds independently, until large enough structures are produced.  So, by tessellating the experiment into independent regions according to a fixed minimal distance between particles by the clustering metric (which itself is a fundamental problem in computational geometry \cite{lejeune1850reduction,voronoi1908nouvelles,aurenhammer1991voronoi,fortune1986sweepline,devillers1992fully,devillers1999deletion}), clustering could be dramatically more efficiently calculated, leading to huge gains in compute times, reducing the complexity to a mere $\sim N \log N$ scaling.  This development led to the software program FastJet \cite{Cacciari:2005hq}, which is universally used for jet finding and analysis by theorists and experimentalists alike.

\subsubsection{Recombination Scheme}

Unlike the clustering metric, there has in general been very little study of the recombination algorithm.  At one level, it is ``obvious'' that one should use $E$-scheme recombination in which particle momenta are summed at every step of the clustering.  This indeed has some nice properties, namely, that the jet's net momentum as a sum of momenta of its constituents is sequentially built up by the clustering, and, because momentum is conserved in the collinear splitting process that produced all those particles anyway, the net momentum of a jet can be interpreted as the momentum of the initiating parton.  $E$-scheme recombination was advocated as a standard in a Les Houches Accord \cite{Blazey:2000qt}, is the default setting in FastJet, and is what experiments use almost exclusively, though that need not be the case.

For its simplicity and ubiquity, $E$-scheme recombination does have some issues.  Perhaps the biggest is its sensitivity to {\bf recoil} from contamination radiation that just happens to land in the jet.  Because jet radii must be finite in size, $R > 0$, there is some probability that radiation uncorrelated or only weakly correlated with the jet's dynamics will land in the jet's area.  Such radiation can be produced as dipole radiation off other colored particles in the event, from the {\bf underlying event}, secondary parton collisions in each proton collision at the LHC, or from {\bf pile-up}, secondary proton collisions in each bunch crossing.\footnote{A significant theoretical issue with the measurement of observables on a jet or other restricted region of phase space are {\bf non-global logarithms} that arise from correlated emissions that land inside and outside of the region of interest \cite{Dasgupta:2001sh}.  The resummation of non-global logarithms is accomplished by a non-linear equation, the BMS equation \cite{Banfi:2002hw}, rather unfamiliar to usual linear renormalization group evolution that resums ``global'' logarithms.  Several groups have developed techniques for systematic improvement of the description of non-global logarithms, see, e.g., \Refs{Weigert:2003mm,Hatta:2013iba,Caron-Huot:2015bja,Larkoski:2015zka,Becher:2015hka,AngelesMartinez:2018cfz}}  Even in the extreme case when there is a single hard particle of energy $E$ and a single soft particle from contamination of energy $E_s$, $E$-scheme recombination will result in a jet direction that is displaced from the direction of the hard particle.  Radiation uncorrelated with the jet will be approximately uniformly distributed over the jet's area, and further most of the area is near the edge of the jet, an angle $R$ away.  The angle between the direction of the hard particle and the $E$-scheme jet axis will then be
\begin{align}
\Delta\theta \approx \frac{E_s}{E}\, R\,,
\end{align}
which may be significant, depending on the application.

It is therefore at the very least interesting to consider other recombination schemes, to see if another prescription might assuage the effects of recoil.  The net momentum or $E$-scheme recombination axis of a jet is like the direction of mean momentum flow, and like the mean of a set of numbers, is very sensitive to outliers.  Other averages are less sensitive to outliers, such as the median, so if there were a way to determine the ``median'' jet axis that might solve the problem.  Actually, the median axis can be at least algorithmically defined \cite{Larkoski:2014uqa}, but is a computationally hard problem because jet finding proceeds in the two dimensions of the celestial sphere.  (The median in one dimension is simple to evaluate, however.)  Determining the median jet axis is closely related to optimal transport theory \cite{monge1781memoire,dupinapplications,appell1887memoire,kantorovich1939mathematical,vaserstein1969markov,dobrushin1970prescribing,ollivier2009looking,ollivier2009reduction,kuhn1955hungarian} for jet physics in two dimensions, which has recently seen a burst of interest (see, e.g., \Refs{Komiske:2019fks,Komiske:2020qhg,Cai:2020vzx,Cai:2021hnn,Larkoski:2023qnv}).\footnote{For a modern mathematical treatise on optimal transport theory, see \InRef{villani2009optimal}, by C\'edric Villani.  Villani won the Fields Medal in 2010 for his work on optimal transport theory \cite{otto2000generalization,lott2009ricci}, and has recently been active in politics, having been elected and served in the French National Assembly.  He is also strongly associated with wearing a distinctive spider brooch.}

Is there another way to proceed to define an approximate median jet axis, that is computationally simple yet is insensitive to recoil?  Let's see if we can make sense of recombination schemes in general, in a similar way as the general form of the $k_T$-type metric.  At every step of the clustering, the clustered particle must carry the total energy of its constituents, by collinear safety, but its direction can in principle lie anywhere between them.  This ensures that in the exact collinear limit, the direction and energy of the clustered particle is exactly that of what you could in principle measure.  So, one way to combine clustered momenta $i$ and $j$ into $(ij)$ is as follows:
\begin{align}
E_{(ij)} &= E_i + E_j\,,\\
\hat p_{(ij)} &= \frac{E_i^n \hat p_i + E_j^n \hat p_j}{\left|E_i^n \hat p_i + E_j^n \hat p_j\right|}\,.
\end{align}
The first line, the sum of energies is straightforward, and on the second line, the directions are weighted by powers of the energies of the particles (and then normalized to return the unit vector direction of the clustered particle).  $E$-scheme recombination corresponds to $n = 1$ here, but infrared and collinear safety only limits us to consider $n> 0$.  The $E^2$-scheme has been considered historically \cite{Catani:1993hr,Butterworth:2002xg}, where one takes $n = 2$.  This weights the direction of the clustered particle in such a way that it is much more strongly correlated with the harder particle, but there is still residual sensitivity to the softer particle.\footnote{Mike Seymour used to have a website with his homebrewed software that included an implementation of the $k_T$ algorithm and allowed for changing the recombination scheme, as well EVENT2 \cite{Catani:1996vz}, but the site seems to have dissolved.  I don't know if the site still exists but hosted elsewhere.}

We aren't quite yet at a median, but let's keep going, and while we're at it, might as well crank $n$ up to infinity.  For $n\to\infty$, the direction of the clustered particle lies {\it exactly} on the direction of its harder constituent, and is correspondingly referred to as the {\bf Winner-Take-All (WTA)} recombination scheme \cite{Bertolini:2013iqa,Larkoski:2014uqa,gsalamwta}:
\begin{align}
\hat p_{(ij)}^\text{WTA} = \Theta(E_i-E_j)\,\hat p_i+\Theta(E_j-E_i)\,\hat p_j\,.
\end{align}
This simply ignores the softer of the two particles and therefore is almost even better than a median.  Indeed, the resulting jet axis from WTA recombination is like the jet's {\bf medioid} of momentum flow, the single particle that best represents the net momentum flow.  Because at every stage of the clustering the WTA axis always lies on a particle, the WTA axis of the entire jet necessarily lies on one of the particles in the jet (rather unlike the $E$-scheme).  The WTA recombination axis is therefore insensitive to recoil, solving that issue with the $E$-scheme.  Unfortunately, the WTA axis introduces other issues or confusions that are not with the $E$-scheme, like the fact that at every stage of the clustering, the WTA scheme forbids a generation of non-zero mass, and instead every clustered particle must be treated as massless.  Nevertheless, the WTA recombination scheme has proved very useful for applications of studying structure inside an already clustered jet, some applications of which we will touch on in the following section.

\begin{figure}[t!]
\begin{center}
\includegraphics[width=7.5cm]{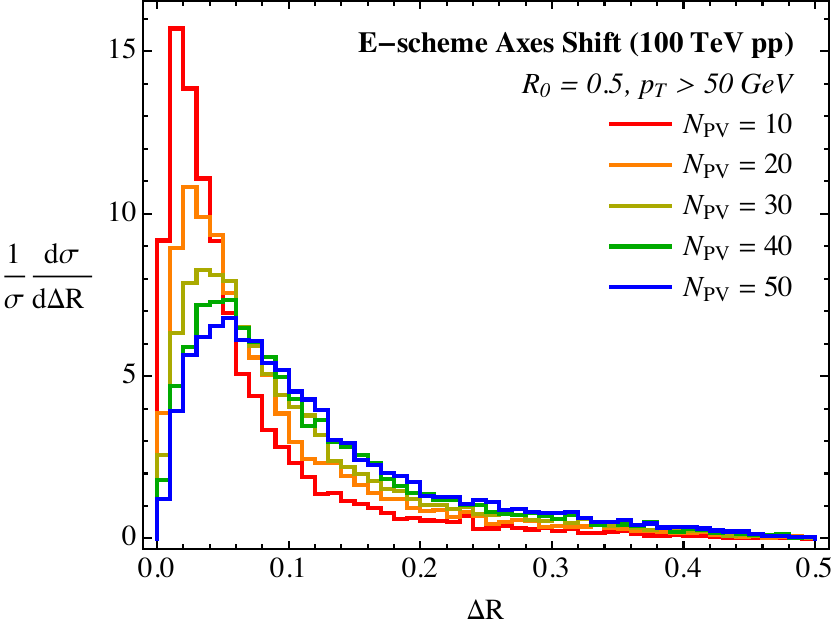}
\hspace{0.5cm}
\includegraphics[width=7.5cm]{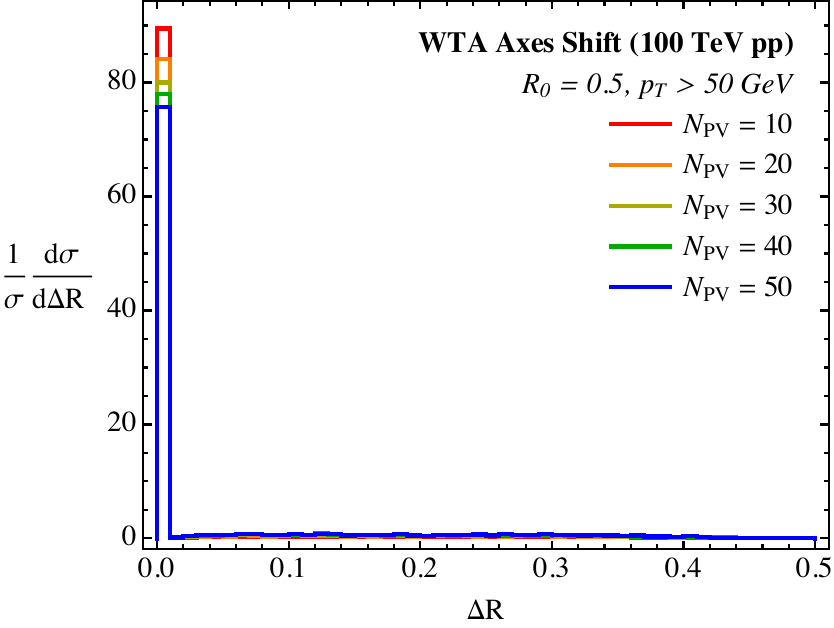}
\caption{\label{fig:axesshift}
Plots of the distribution of angular shift $\Delta R$ in the jet axis as the amount of contamination radiation in a jet is increased.  Jets are clustered with the anti-$k_T$ algorithm with a radius $R = 0.5$ with a minimum transverse momentum of 50 GeV in simulated data from a future 100 TeV hadron collider.  The number of primary vertices $N_\text{PV}$, corresponding to the number of simulated proton collisions per bunch crossing, is a proxy for increased contamination.  Left: Jet axis shift with $E$-scheme recombination.  Right: Jet axis shift with WTA recombination.  From \InRef{Larkoski:2014bia}.
}
\end{center}
\end{figure}

The effect of recoil on the direction of the reconstructed jet axis is displayed in \Fig{fig:axesshift}, reproduced from a study in \InRef{Larkoski:2014bia}.  Here, jets have been simulated at a future 100 TeV hadron collider and clustered with the anti-$k_T$ algorithm with a radius $R = 0.5$.  In this sample of events, contamination radiation has been included through simulation of additional proton collisions per bunch crossing, defined by the number of primary vertices, $N_\text{PV}$.  Then, the angle between the original jet axis, with no additional proton collisions, is compared to the jet axis after including this contamination radiation.  With $E$-scheme recombination, as the number of primary vertices increases, the distribution of angles extends to larger and larger values, with a long tail even out to the jet radius itself.  By contrast, jets constructed with WTA recombination have nearly identical axes regardless of the amount of contamination radiation.

For their enormous practical importance, jet algorithms will play only a minor role in the rest of these lectures.  We will only need the consequence of IRC safety that collinear radiation is always captured in a single jet, regardless of the specific algorithm, so we will always assume that we have already clustered and found jets, and are then looking deep inside of them.

\subsection*{Exercises}

\begin{enumerate}

\item\label{ex:threebodps} Express three-body collinear phase space in terms of three energy fractions $z_1,z_2,z_3$, two angles between pairs of particles $\theta_{12},\theta_{13}$, and one azimuthal angle $\phi$.  You should find \cite{Gehrmann-DeRidder:1997fom,Ritzmann:2014mka}
\begin{align}
d\Phi_\text{coll}^{(N=3)} = \frac{4E^4}{(4\pi)^5}\,d\theta_{12}^2\,d\theta_{13}^2\,d\phi\, dz_1\, dz_2\, dz_3\, z_1\, z_2\, z_3\,\delta(1-z_1-z_2-z_3)\,.
\end{align}

\item Calculate the distribution of the angle between the harder particle and the net momentum axis in a two-particle quark jet.  Assume that the two particles in the jet are within an angle $R$ of each other.  Now, calculate the distribution of the angle between the harder particle and the general $E^n$-scheme axis, as a function of $n$.  See \InRef{Cal:2019gxa} for more details.

\item On flat phase space in the center-of-mass frame with total energy $Q$, determine the single-particle energy distribution, $p(E)$.  Take the large number of particles, $N\to \infty$, limit; what does the distribution reduce to?

\end{enumerate}

\section{First Example: Quark vs.~Gluon Jet Discrimination}\label{sec:qvgdisc}

With our introductions to the necessary tools of machine learning and perturbative quantum field theory, we now turn our focus on using those results in practice, to understand a number of problems of central importance for jet substructure.  We start in this section, and will often return back later, with the problem of discrimination of jets initiated by quarks versus those initiated by gluons.  This problem is essentially as old as QCD itself, through the discovery of the gluon through the observation of three-jet events in $e^+e^-$ collisions \cite{TASSO:1979zyf,Barber:1979yr,PLUTO:1979dxn,JADE:1979rke,Nilles:1980ys}.  Quarks are electrically charged, and so can be probed by deeply-inelastic electron scattering on a proton, but the gluon is electrically neutral and so must be produced directly from the quarks as color-charged particles themselves.  The validity of QCD as {\it the} theory of the strong force requires the existence of the gluon so its unambiguous observation is clearly of paramount importance.\footnote{The gluon, however, is one of the few elementary particle discoveries for which there was no associated Nobel Prize.}

\subsection{Some History and Context}

As stated, this is a binary discrimination problem, we just want to distinguish quark jets (background) from gluon jets (signal) through a detailed study of the internal dynamics of the jets.  Beyond validation of QCD, a powerful quark versus gluon discriminator would have profound impact in nearly the entire particle collider physics program.  Its importance (and, as we will see, its challenge), then, has prompted some to refer to this as the ``white whale of jet substructure'',\footnote{I believe this phrase is credited to my post-doc advisor Jesse Thaler.} but I hope that this ominous description just implies that our quarry is worth the effort and not the associated doom of its literary inspiration.

Regardless of your interests, a quark versus gluon discriminator can help you.  If you are interested in parton distribution function (pdf) extractions, quark pdfs are accurately determined from deeply-inelastic scattering data or Drell-Yan production, but direct sensitivity to the gluon pdf requires final-state jet (or Higgs) production at the LHC \cite{Hou:2019efy,NNPDF:2021njg}.  If you are interested in final state fragmentation, fragmentation functions are accurately measured on jets in $e^+e^-$ collisions, but the vast majority of those jets are initiated by quarks, and gluon jets at high energies are only produced in significant numbers at a hadron collider \cite{deFlorian:2007aj,deFlorian:2007ekg,Albino:2008fy,deFlorian:2014xna,deFlorian:2017lwf,Borsa:2021ran}.  In a related vein, if you are interested in tuning Monte Carlo event generators \cite{Buckley:2009bj,Bellm:2016voq,Mrenna:2016sih,Bothmann:2016nao,Bellm:2019owc,Krishnamoorthy:2021nwv,Bierlich:2023fmh}, $e^+e^-$ collision data strongly constrains quark jets, but without a pure sample of gluon jets, data from hadron colliders is much more challenging to interpret.  If you are interested in physics beyond the Standard Model, cascade decay chains initiated by new, unstable particles can often produce huge numbers of quark jets in a single event, e.g., \Refs{CMS:2020cpy,CMS:2020fia,CMS:2020bfa,ATLAS:2021yyr,ATLAS:2021yqv,CMS:2022vpy,CMS:2022spe,ATLAS:2024tqe}, while background Standard Model processes would dominantly produce gluon jets copiously.  And then, if you are like me and are just interested in the substructure of jets on its own, you just want to determine the method for most efficient extraction of the type of jet simply from the measurable momenta of their constituent particles.

Quark versus gluon discrimination as a problem in its own right has a very long history in particle physics, and in fact some of the first neural networks applied to analyze simulated collider events were employed to this task \cite{Jones:1988ay,Fodor:1989ir,Lonnblad:1990bi,Lonnblad:1990qp,Csabai:1990tg,Jones:1990rz,Pumplin:1991kc,OPAL:1993uun}.  With the rise of jet substructure, early work in the LHC era was focused around designing bespoke observables for quark versus gluon discrimination and combining and slicing high-dimensional observable spaces just to wrap one's head around where on this space different types of jets lived \cite{Gallicchio:2011xq,Gallicchio:2012ez,Larkoski:2013eya}.  More and more techniques were subsequently developed, and the community took up the problem and worked to formalize it in a Les Houches study \cite{Gras:2017jty}.  Since that time, more and more of the discrimination task has been pushed onto deep neural networks, with lower and lower level data as the input.

However, while machine learning discrimination performance is extremely impressive, we want to understand why it is and what physics of quark and gluon jets are responsible for the differences.  This is our task in these lectures and is a vital handrail and guide for machine learning studies because they nearly exclusively are trained on simulated data from an event generator simulation program.  While at their core event generators use fixed-order matrix elements or DGLAP evolution, there are still many tuned parameters and fits and it may be that a machine is merely learning idiosyncratic differences of quark and gluon jets from the way they are simulated in a particular program, and not actually some fundamental physical distinction.  This is especially relevant as mentioned above because a pure sample of gluon jets in data is challenging to produce and it has been noted that while quark jets look more or less the same in different event generators, gluon jets can vary significantly \cite{Gras:2017jty}.  So, part of our goal here will also be to determine those features that are robust, that can inform better, more principled, machine learning, even when the simulated data may be noisy.

\begin{figure}[t!]
\begin{center}
\includegraphics[width=12cm]{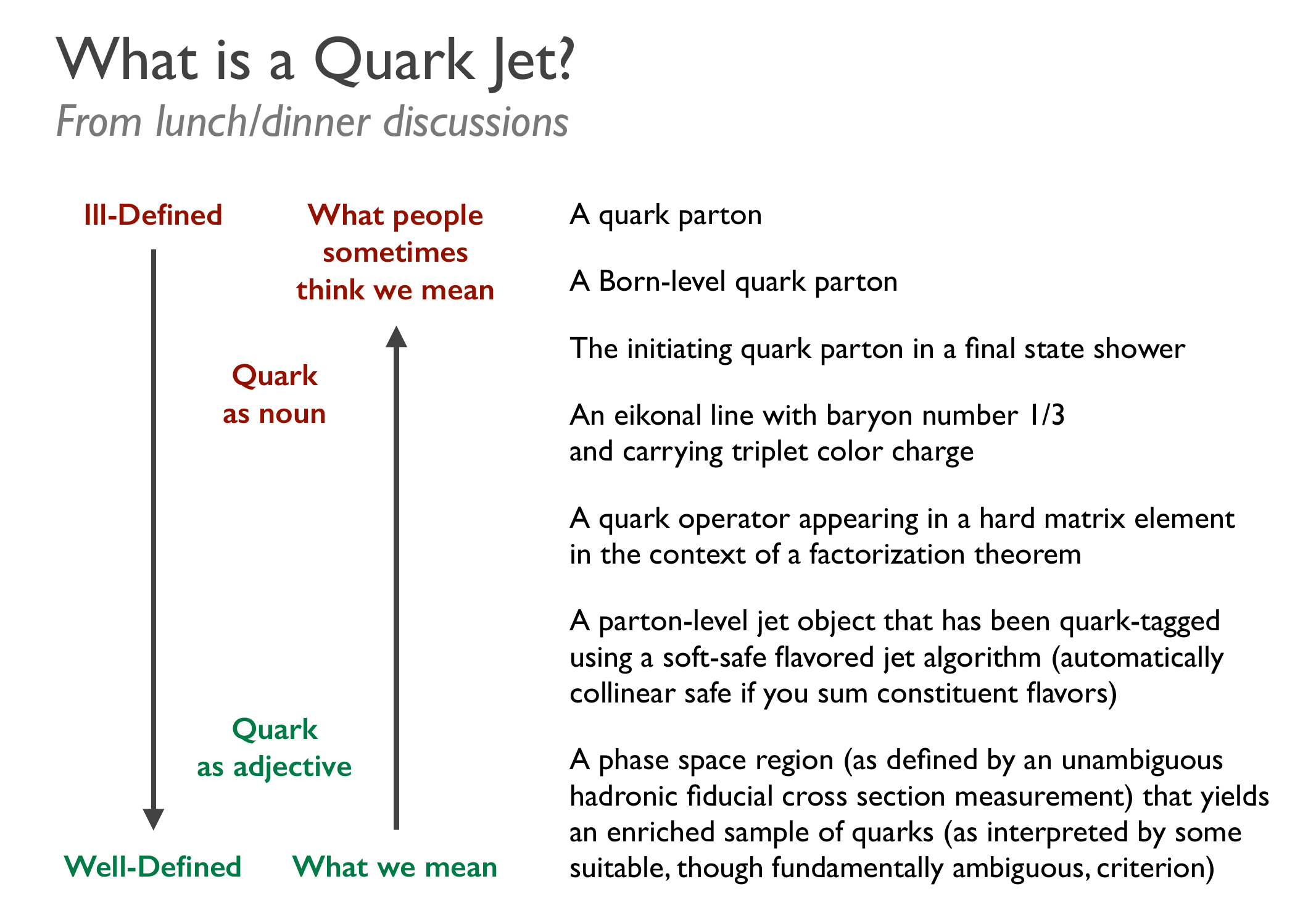}
\caption{\label{fig:lhdefs}
Summary of thoughts and discussions of a more robust and physical definition of a ``quark jet'' from the 2015 Les Houches study.  From \InRef{Gras:2017jty}.
}
\end{center}
\end{figure}

A bigger problem with a quark versus gluon discriminant endeavor is even more basic.  As discussed in the Neyman-Pearson lemma proof, of course the likelihood ratio is the optimal discriminant.  However, the likelihood ratio requires perfect ground truth of what ``signal'' and ``background'' are; we condition on the label ``signal'' and ``background'', and then measure other quantities on these samples.  Jets at the LHC obviously do not come with labels on them (or our job would be very easy!), but jets in simulation do, and this can lead to confusion or at least imprecision.  In an event simulation, you tell the simulator what process to generate, typically defined at leading-order in the strong coupling.  In practice, this ``Monte Carlo truth label'' is then used to define the ground truth and corresponding conditioned distributions of signal and background: if you requested events with a final state gluon, then the resulting event contains a gluon jet.  While the type or flavor of a jet is well-defined at leading order, such a naive label does not generalize to higher orders \cite{Banfi:2006hf}.  This point was identified in the community study of \InRef{Gras:2017jty}, and there, worked to broaden how we refer to jet types to have a more physical foundation.  The outcome of informal discussions during the 2015 Les Houches study on more robust definitions of a ``quark jet'' is presented in \Fig{fig:lhdefs}.  Nevertheless, Monte Carlo truth labeling remains the common practice, and so another aspect of our work here will be to establish when and where such a label is meaningful and accurate.

\subsection{Simplicity of Jets at High Energies}

We will start our analysis of quark versus gluon jet discrimination as simple as possible, but no simpler.  We could use our master formula derived in the previous section to calculate distributions from the known splitting functions, but I want to take a different route here that clearly illustrates the assumptions we employ.\footnote{The approach especially of this lecture is strongly influenced by TASI 2001 lecture notes by my undergraduate research advisor, Matt Strassler \cite{Strassler:2003qg}.}  We will thus ignore the details of QCD, ignore Feynman diagrams, etc., and, like ignorant scientists taking the first great leap into the unknown, simply make some hypotheses and follow them to the edge of logic.  These will be so important that we will refer to them as axioms.

For our first foray into quark versus gluon discrimination, our axioms will be:
\begin{enumerate}
\item At high energies, QCD is an approximately scale-free, conformal field theory.  This means that, to first approximation, the coupling of QCD, $\alpha_s$, is constant, independent of energy.

\item The coupling constant of QCD, $\alpha_s$, is small, $\alpha_s \ll 1$, and so quarks and gluons are good quasi-particles.

\item The only difference between quarks and gluons are their color charges $C_F$ and $C_A$, respectively, the quadratic Casimirs of the fundamental and adjoint representations of SU(3) color, with $C_A > C_F$.
\end{enumerate}

With these axioms, we will ultimately establish baseline discrimination performance for this problem, calculate the ROC curve, and all that, but let's first start much simpler and just attempt to calculate the rate or probability that a quark emits a gluon in this framework, $P_{q\to qg}$.  In the context of Feynman diagrams, what we would want to compute is
\begin{equation}
P_{q\to qg} = \left|
\raisebox{-1.2cm}{\includegraphics[width=5cm]{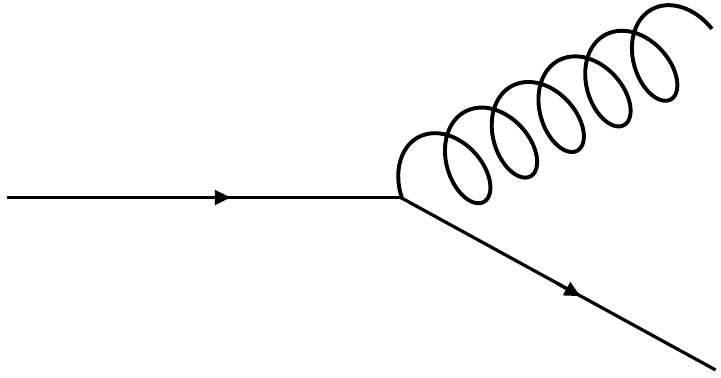}}
\right|^2\,,
\end{equation}
but we don't have Feynman diagrams, so we need to use our axioms alone.

First things first: let's determine what possible quantities this probability can depend on.  Specifically, we want to determine this probability as a function of the properties of the emitted gluon.  So, what possible properties can the gluon have?  If we only consider measuring momentum, then the four vector of the gluon $p_g$ completely specifies it.  Let's choose some convenient coordinates to express the four vector.  In particular, there is no notion of absolute momentum in a Lorentz-invariant theory, so we can express the gluon's four vector in relationship to the final state quark.  Further, the gluon is massless (by the scale invariant assumption), and so its four-vector can be expressed as
\begin{align}
p_g = E_g(1,\sin\theta\cos\phi,\sin\theta\sin\phi,\cos\theta)\,.
\end{align}
Here, $\theta$ is the angle between the quark and the gluon and $\phi$ is the azimuthal angle of the gluon about the quark:

\begin{center}
\includegraphics[width=5cm]{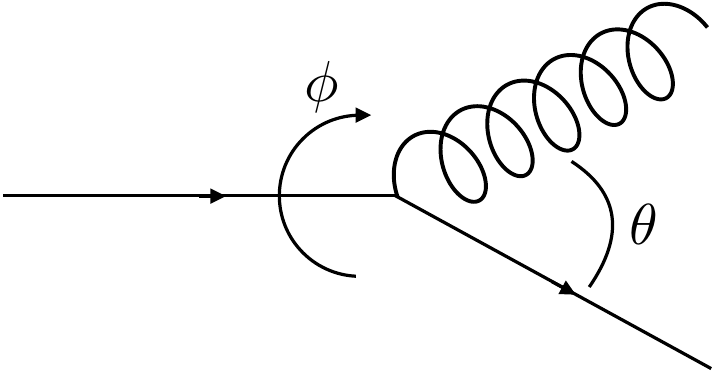}
\end{center}
However, with no preferred spatial orientation, we can freely set the azimuthal angle to 0 with no physical consequence (there is an SO(2) $\simeq$ U(1) symmetry about the initial quark).

We can express the gluon's energy $E_g$ in a convenient way, too.  The initial quark has some fixed energy $E$ set by the experimental set-up, and the gluon energy is some fraction $z$ of that: $E_g = zE$, where $z\in [0,1]$.  With these identifications, the four-vector can be expressed as
\begin{align}
p_g = zE(1,\sin\theta,0,\cos\theta)\,,
\end{align}
and this makes it clear that there are two degrees of freedom of the gluon: the angle $\theta$ and the energy fraction $z$.  We can then write the original probability that we want as
\begin{align}
P_{q\to qg} = p(z,\theta)\, dz\, d\theta\,,
\end{align}
where $p(z,\theta)$ is a probability density to be determined.

Let's see what we can do about this probability density.  ``Scale invariant'' means that the physics is unchanged if you change the scale at which you view the system.  That is, physical quantities are unaffected by a scaling, and a physical quantity is indeed the rate of gluon emission.  So, somehow this probability for gluon emission must be unchanged by a scale transformation.  One possible scale transformation is to modify the splitting angle by some $\lambda > 0$, as $\theta \to \lambda \theta$.  Let's write this as an infinitesimal transformation with $\lambda = e^\epsilon$, for $\epsilon \ll 1$.  Then, under this transformation, the measure becomes
\begin{align}
dz\, d\theta \to dz\, d\theta + \epsilon\, dz\, d\theta +\cdots\,,
\end{align}
suppressing terms higher order in $\epsilon$.  The probability density, by contrast, transforms as
\begin{align}
p(z,\theta) \to p(z,e^\epsilon \theta) =p(z,\theta)  + \epsilon\theta \frac{d}{d\theta}p(z,\theta) +\cdots\,.
\end{align}

If the probability is to be invariant under this transformation, then combining these results we must have
\begin{align}
P_{q\to qg} &\to p(z,e^\epsilon \theta)\, dz\, d(e^\epsilon \theta) =\left(
p(z,\theta)  + \epsilon\theta \frac{d}{d\theta}p(z,\theta)+\cdots
\right)\left(
dz\, d\theta + \epsilon\, dz\, d\theta+\cdots
\right) \\
&=p(z,\theta)\, dz\, d\theta\nonumber\,.
\end{align}
Now, we expand in $\epsilon$, and cancel terms.  The $\epsilon^0$ term explicitly cancels with the untransformed probability on the second line, while at $\epsilon^1$, we have the requirement that
\begin{align}
\theta \frac{dp(z,\theta)}{d\theta} + p(z,\theta) = 0 \,,
\end{align}
tossing away the benign differential elements.  The solution of course is that
\begin{align}
p(z,\theta) \propto \frac{1}{\theta}\,.
\end{align}

The exact same argument goes through with scaling the energy fraction $z$ as well, and one finds the same functional form for the $z$ dependence.  We can, as we did here, independently scale energies or angles and so the only possible solution consistent with scale invariance that allows for independent scalings is
\begin{align}
p(z,\theta)\, dz\, d\theta \propto \frac{dz}{z}\, \frac{d\theta}{\theta}\,.
\end{align}
This result is rather interesting, and one thing to note at this stage is that demanding scale invariance highly constrains the gluon's dynamics.  The only possible way that you can have a valid scale transformation of the energy fraction $z\in[0,1]$ and maintain energy conservation is if the gluon's energy was parametrically smaller than the quark's energy, $z\lll 1$.  Similarly, if there was any constraint on the emission angle of the gluon, like it must lie in a jet of radius $R$, scale invariance enforces that the emission angle is also parametrically small, $\theta \lll R$.  Thus, we see that we are forced into the soft and collinear regime by our axioms.

Now, we turn to our second and third axioms, which set the scale of the probability density.  The coupling $\alpha_s$ determines the strength with which the gluon couples to the quark, and the quadratic Casimir $C_F$ quantifies how the quark and gluon share color.  Thus, the probability density is proportional to both, $p(z,\theta) \propto \alpha_s C_F$.  We'll now cheat a tiny bit and use our knowledge of the soft and collinear limit of the splitting functions from \Sec{sec:collsplitfact}, and get all of the factors of 2 and $\pi$ correct:
\begin{align}
P_{q\to qg} = p(z,\theta) \, dz\, d\theta= \frac{2\alpha_s C_F}{\pi}\, \frac{dz}{z}\,\frac{d\theta}{\theta}
\end{align}
Of course, this is exactly what we had written down earlier for the soft and collinear limit of gluon emission off of a quark, but now we see that it follows from the strong constraint of scale invariance.

\subsubsection{The Lund Plane}

With this result, we can now attempt to interpret it.  The first thing to notice, and you might have for some time already, is that this is not a probability.  By scale invariance, we are forced into the regime where $z,\theta \to 0$, but the probability density is non-integrable there
\begin{align}
\int_0 dz \int_0 d\theta\, p(z,\theta) =\frac{2\alpha_s C_F}{\pi} \int_0\frac{dz}{z}\int_0 \frac{d\theta}{\theta} \to\infty\,.
\end{align}
Lacking integrability, $P_{q\to qg}$ is also not a probability, and so we seem to have lost all of the gains that the axioms had earned us.  However, let's think a bit more critically about what happens in the problematic limits where this purported probability density is non-integrable.  Consider the $z\to 0$ limit, corresponding to the emission of a 0 energy gluon.  Such a gluon is, by definition, not observable because no experiment you can make can detect exactly 0 energy.  Additionally, strictly 0 energy gluons carry, well, no energy, and so we can emit an arbitrary number of them without any effect whatsoever on any observable quantity.  So, the problem with our probability so far is not really that it diverges, it is that we haven't fully accounted for what it describes.  Because no measurement can distinguish 1 emitted  0 energy gluon from 143,872 emitted 0 energy gluons, we need to sum up all possible numbers of emitted 0 energy gluons, by the rules of quantum mechanics.

An analogous argument holds for a 0 angle emission.  No experiment can identify a gluon emitted exactly collinearly with a quark, and so again, we need to sum up all possible numbers of exactly collinear gluon emissions, by the rules of quantum mechanics.  Actually, now is where we see the KLN theorem arise.  We had been studying a system in degenerate perturbation theory (single soft-collinear gluon emission), and if we do that, infinities blow up in our face.  As we will demonstrate shortly, by summing up all possible degenerate states that have exactly the same experimental signatures but different numbers of gluons, we will find finite quantities and true probability distributions, just as KLN promised.

Let's see how to do this.  Back to our emission ``probability'', we can rewrite it as
\begin{align}
P_{q\to qg} =  \frac{2\alpha_s C_F}{\pi}\, \frac{dz}{z}\,\frac{d\theta}{\theta} = \frac{2\alpha_s C_F}{\pi}\, d\left(\log\frac{1}{z}\right)\, d\left(\log\frac{1}{\theta}\right)\,,
\end{align}
which is now flat logarithmically in energy fraction and angle.  This now provides a justification and interpretation of what is going on and how to fix it all up.  Where this probability is small, specifically $P_{q\to qg}< 1$, this is indeed sensible as a probability.  However, where it gets large, it no longer should be interpreted as a probability, but rather as an {\it expectation value} of the number of gluons emitted with energy fractions and splitting angles less than given values (i.e., less resolvable than some values of $z,\theta$).\footnote{This was likely known or at least appreciated long before (perhaps first by Wigner \cite{Wigner:1932eb}), but Dirac \cite{dirac1942bakerian} and Feynman \cite{Feynman:1984ie} noted that predictions from quantum field theory of probabilities that are larger than 1 (as in the case at hand), must be associated with probabilities that are negative, so that their sum produces physical probabilities that lie on $[0,1]$.  Further, such negative probabilities simply must make sense on their own.}  The uniform distribution renders this interpretation immediate, because an integral over a uniform distribution just returns the upper bound.  Therefore, we nicely visualize a large number of gluons emitted according to this uniform distribution.

\begin{figure}[t!]
\begin{center}
\includegraphics[width=6cm]{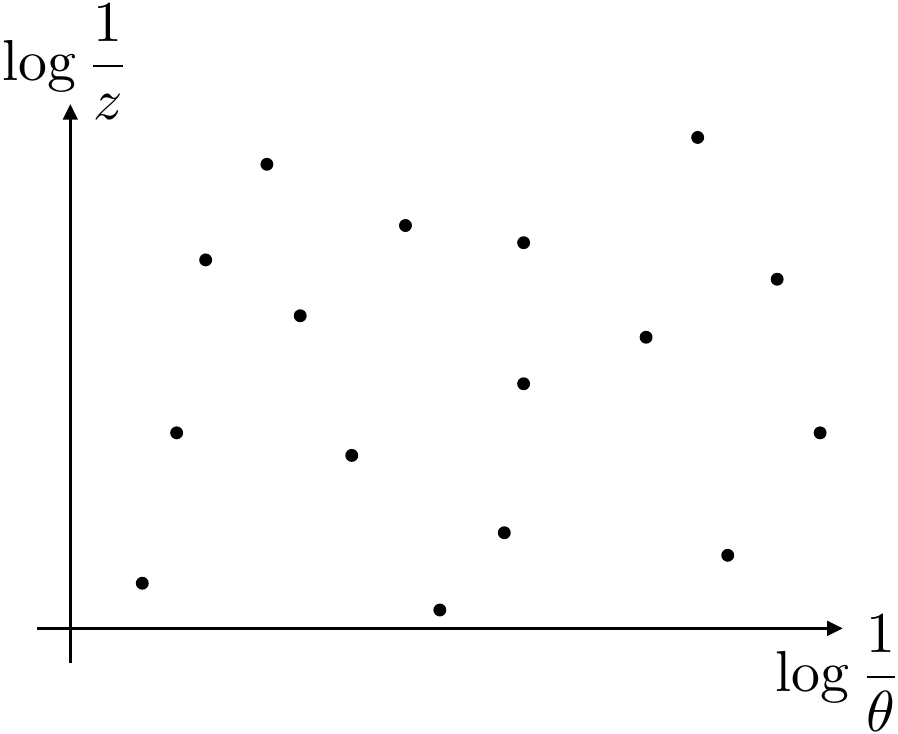}
\hspace{2cm}\raisebox{-0.5cm}{\includegraphics[width=7cm]{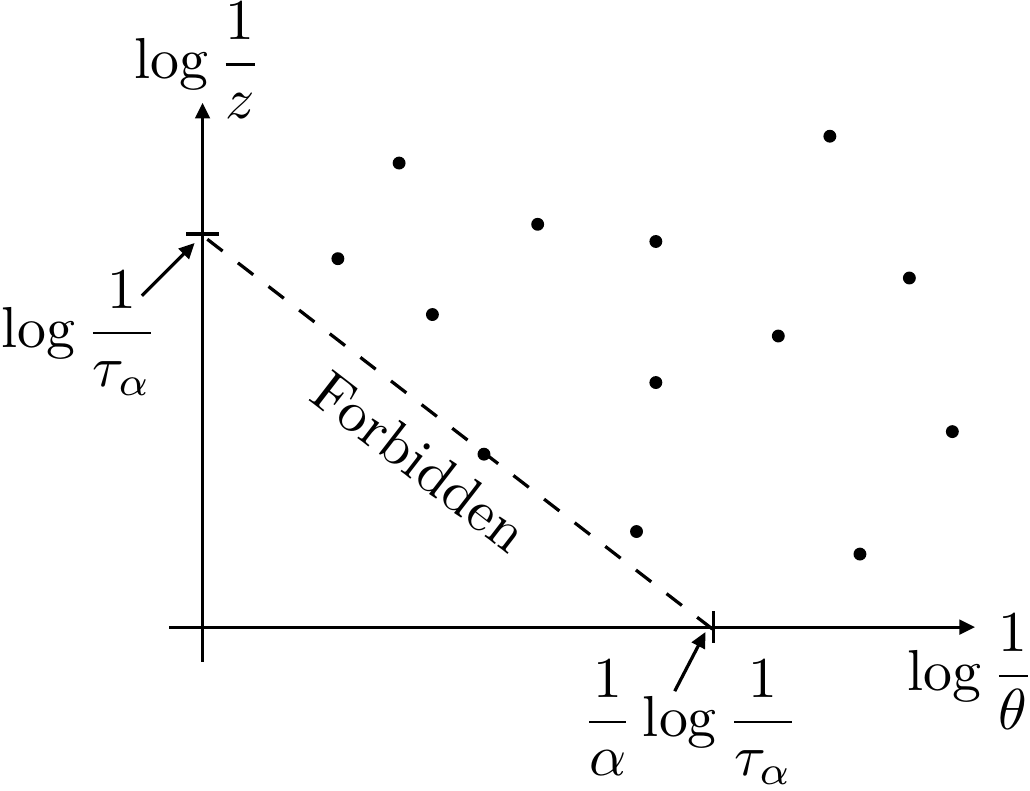}}
\caption{\label{fig:lundpic}
Left: Visualization of emissions uniformly populating the Lund plane.  Right: A measured value of the angularity $\tau_\alpha$ enforces no emissions below the line of constant angularity.}
\end{center}
\end{figure}

The visualization of this is shown at left in \Fig{fig:lundpic} in which phase space of gluon emissions is represented as a semi-infinite plane in $\log\frac{1}{z}\,,\log\frac{1}{\theta}$ and each dot represents an emitted gluon with that value of energy fraction and angle.  This representation is called the {\bf Lund plane} \cite{Andersson:1988gp} (see also \InRef{Bjorken:1991ft} and recent work and measurements of \Refs{Dreyer:2018nbf,Dreyer:2020brq,ATLAS:2020bbn,Lifson:2020gua,ALICE:2021yet,Dreyer:2021hhr,CMS:2023lpp}) after the researchers in Sweden that introduced it.  This uniform emission plane is the starting point for particle production in a modern parton shower Monte Carlo generator, like Pythia, Herwig, or Sherpa \cite{Bierlich:2022pfr,Bahr:2008pv,Bellm:2015jjp,Sherpa:2019gpd}, though they include much, much more physics, like the full splitting functions, the running coupling (which increases emission rate further from the origin), or hadronization (which cuts off perturbative emissions at some scale).\footnote{There has been significant work recently in developing parton showers that are formally accurate to much higher orders than these workhorse generators, e.g., \Refs{Hamilton:2020rcu,Forshaw:2020wrq,Nagy:2020rmk,Herren:2022jej,Preuss:2024vyu}.  A particularly important point of these parton showers, and for their generalization to even higher orders, is an {\bf ordering variable}, which is an observable quantity that monotonically increases across the Lund plane and ensures that emissions are produced in such a way to correctly incorporate all of the logarithmic structure to the accuracy one is working.}  Those pesky divergences are pushed off to infinity, the soft divergence vertically and the collinear divergence horizontally, and the only vestiges of divergences that remain is the fact that this Lund plane, as we have constructed here, is infinite in area.  Once we start asking reasonable, IRC safe questions, about such jets on the Lund plane, however, divergences will be no where in sight.

\subsubsection{Resummation and the Sudakov Form Factor}\label{sec:sudder}

The Lund plane captures the distribution of soft and collinear emissions off of a hard initiator particle and somehow this distribution must encode properties of that initiator particle.  Our goal, for quark versus gluon jet initiator discrimination, is to find out how that information is encoded.  As a first baby step in that direction, we will just calculate one of the oldest jet substructure observables called the {\bf angularities} \cite{Berger:2003iw,Almeida:2008yp,Ellis:2010rwa}.  The angularities measure the net flow of energy and momentum away from the jet axis or initial hard particle and is defined as
\begin{align}
\tau_\alpha = \sum_i z_i \theta_i^\alpha\,.
\end{align}
The sum runs over all particles in the jet, and the parameter $\alpha > 0$ ensures IRC safety.  Note that this is linear in the energy fractions $z_i$, which is a requirement of collinear safety to ensure that the energies of exactly collinear emissions sum correctly.  The angle $\theta_i$ is measured from particle $i$ to the jet axis, and here, because all emissions are soft and collinear, the jet axis coincides with the initial hard particle.  Thus, the only particles that contribute and have a non-zero angle from the jet axis are the emitted gluons.  For the rest of this section, we will determine how the angularities can be used for quark versus gluon discrimination.

Angularities, in one form or another, have been studied for over 50 years in QCD, with the $\alpha = 2$ case corresponding to thrust \cite{Brandt:1964sa,Farhi:1977sg}.  Thrust is one of the most studied, calculated, and experimentally measured jet observables, because, in the collinear limit, thrust is proportional to the jet mass, which is perhaps the most basic jet substructure observable whatsoever \cite{Clavelli:1979md}.  Another ancient angularity is the broadening \cite{Rakow:1981qn,Ellis:1986ig}, corresponding to $\alpha = 1$ exponent.  Broadening has the interesting feature (or bug) that it is sensitive to recoil \cite{Catani:1992jc,Dokshitzer:1998kz}, and correspondingly is sensitive to the precise jet axis definition.\footnote{Spherocity \cite{Georgi:1977sf} is a broadening-like observable that was introduced shortly after thrust that defines the jet or event axis as the direction that minimizes the sum of magnitudes of momentum transverse to it.  As such it is actually insensitive to recoil.  We will discuss such ``broadening axes'' a bit more later with higher-order effects.}  We will discuss this property a bit more later when we discuss the effect of higher orders on discrimination power.  Observables that are related to angularities are energy correlation functions \cite{Parisi:1978eg,Donoghue:1979vi,Banfi:2004yd,Larkoski:2013eya,Moult:2016cvt} which quantify average pairwise radiation patterns.  Energy correlation functions can be identified as moments of energy-energy correlators \cite{Basham:1978bw,Hofman:2008ar,Dixon:2018qgp,Dixon:2019uzg,Henn:2019gkr,Chen:2020adz,Komiske:2022enw,CMS:2024mlf} introduced long ago, but are becoming a very hot topic because their simplicity enables extremely precise theoretical predictions.

But back to angularities for now.  The first thing we need to do to determine the probability distribution of the angularity $\tau_\alpha$ is to evaluate it on our jet of interest.  Emissions on the Lund plane are uniformly distributed in the logarithm of the splitting angle and the energy fraction.  Therefore, in ``real'' space, linear in splitting angle and energy fraction, emissions are exponentially far apart.  This observation makes evaluation of the angularity very simple.  There will be one emission in the jet for which $z_i\theta_i^\alpha$ is the largest, and then all other emissions' contributions to the angularity will be exponentially small.  So, to exponential accuracy, we can consider the value of the angularity set by that one emission alone.  This is illustrated at right in \Fig{fig:lundpic}.

Then, with this one emission, note that a fixed value of the angularity defines a line on the Lund plane:
\begin{align}
\log\frac{1}{\tau_\alpha} = \log\frac{1}{z} + \alpha \log\frac{1}{\theta}\,,
\end{align}
which is dashed at right in \Fig{fig:lundpic}.  If $\tau_\alpha$ is indeed to be the measured value of the angularity on a jet, then there must be no emissions below that dashed line, toward the origin of the Lund plane.  Emissions below that dashed line correspond to larger values of the angularity and are forbidden by our measurement.  However, emissions above the dashed line are completely unconstrained, by our assumption that they do not contribute to the angularity anyway.

Then, to calculate the probability distribution of the angularity, we just need to calculate the probability that there are no emissions below that dashed line whatsoever.  To do this, we will break up the forbidden triangle region into many small pieces:

\begin{center}
\includegraphics[width=3.5cm]{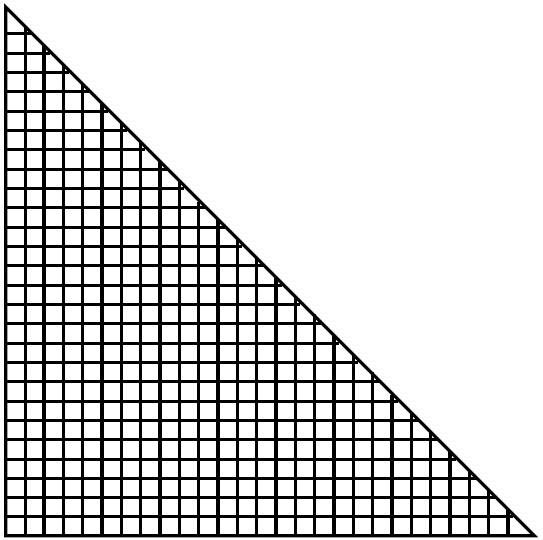}
\end{center}
There can be no emissions in any of the subregions of the triangle, which is an ``and'' statement in probability and so we must multiply all of the individual no emission probabilities together.  If there are $N$ equal area subregions, then the probability for emission in any subregion is equal and is
\begin{equation}
\text{Probability for emission} = \frac{2\alpha_s}{\pi}C_F \frac{\includegraphics[width=0.5cm]{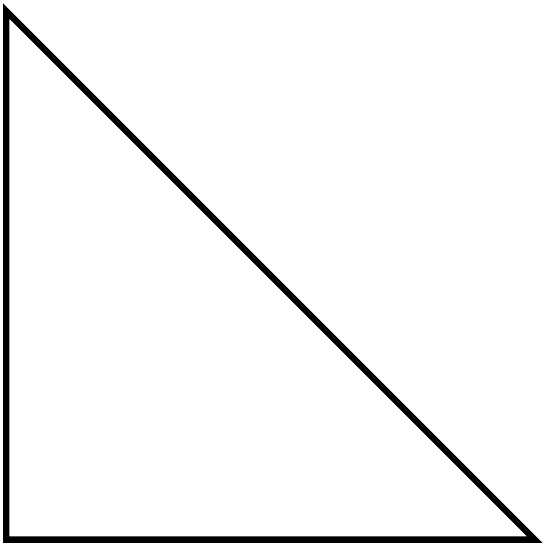}}{N}\,,
\end{equation}
where \includegraphics[width=0.5cm]{triangle} is the area of the forbidden triangle:
\begin{equation}
\includegraphics[width=0.5cm]{triangle} = \frac{1}{2\alpha}\log^2\tau_\alpha\,.
\end{equation}
This follows because emissions are uniformly distributed on the Lund plane and so probabilities are proportional to the area of the region of interest.

The probability of no emission in a small subregion is simply 1 minus the emission probability (either a particle is emitted or not):
\begin{equation}
\text{Probability for no emission} = 1-\frac{2\alpha_s}{\pi}C_F \frac{\includegraphics[width=0.5cm]{triangle}}{N} = 1-\frac{\alpha_s}{\pi}\frac{C_F}{\alpha}\frac{\log^2\tau_\alpha}{N}\,.
\end{equation}
Now, we just need to multiply this no emission probability by itself $N$ times to determine the no emission probability anywhere in the triangle.  Taking the $N\to\infty$ limit to make the subregions infinitesimally small, we have
\begin{equation}
P(\tau_\alpha \text{ less than measured value}) = \lim_{N\to\infty}\left(
1-\frac{\alpha_s}{\pi}\frac{C_F}{\alpha}\frac{\log^2\tau_\alpha}{N}
\right)^N = \exp\left[
-\frac{\alpha_s}{\pi}\frac{C_F}{\alpha}\log^2\tau_\alpha
\right]\,.
\end{equation}
The product transmogrifies into an exponential!  This exponential factor is called the Sudakov form factor \cite{Sudakov:1954sw}, and is simply a manifestation of the scale-invariant Poisson process of particle emission in high-energy QCD.  Note that we have called this the probability that $\tau_\alpha$ is less than the measured value because smaller $\tau_\alpha$ means that the leading emission is farther from the origin of the Lund plane, somewhere, anywhere above that dashed line.

Therefore, this probability is actually a cumulative probability distribution of the angularity of a quark jet, 
\begin{align}
\Sigma_q(\tau_\alpha) = \exp\left[
-\frac{\alpha_s}{\pi}\frac{C_F}{\alpha}\log^2\tau_\alpha
\right]\,.
\end{align}
The differential probability, the probability distribution, is the derivative of this 
\begin{equation}\label{eq:angdist_q}
p(\tau_\alpha) = \frac{d}{d\tau_\alpha}\exp\left[
-\frac{\alpha_s}{\pi}\frac{C_F}{\alpha}\log^2\tau_\alpha
\right] = -\frac{2\alpha_s}{\pi}\frac{C_F}{\alpha}\frac{\log\tau_\alpha}{\tau_\alpha}\exp\left[
-\frac{\alpha_s}{\pi}\frac{C_F}{\alpha}\log^2\tau_\alpha
\right]\,,
\end{equation}
and is normalized on $\tau_\alpha \in[0,1]$.  Note that these expressions are Taylor series in $\alpha_s$ to all orders.  Each emission costs $\alpha_s$, so this exponential Sudakov factor indeed requires explicit summation over all degenerate states, with any and all numbers of emitted gluons, to render the distribution finite and integrable.  In fact, this ``summation'' over degenerate states is referred to as {\bf resummation} and is a necessary theoretical tool for precision control of the soft and collinear regions of phase space.

The particular resummation we have performed here is the lowest, non-trivial accuracy, referred to as {\bf double-logarithmic accuracy}.  It is double logarithmic in the sense that there are two factors of logarithm for every factor of $\alpha_s$ in the exponent.  More generally, observables like angularities that are both IRC safe and additive (expressed as an explicit sum over contributions from each particle) enjoy a resummed cumulative distribution of the form \cite{Catani:1992ua}
\begin{align}
\Sigma(L) =  \exp\left[
L\,g_1(\alpha_s L)+g_2(\alpha_s L) +\alpha_s g_3(\alpha_s L)+\cdots
\right]\,,
\end{align}
where $L = -\log \tau_\alpha$.  The functions in the exponent are ordered in their logarithmic accuracy, with $g_1(\alpha_s L)$ the leading-logarithmic term, $g_2(\alpha_s L)$ the next-to-leading logarithmic term, etc.  Resummation formally treats the region of phase space where $\alpha_s L \sim 1$ as $\alpha_s \to 0$, so that the importance of the terms in the exponent are controlled by the explicit powers of $\alpha_s$.  The double-logarithmic term we calculated is the lowest-order part of the leading logarithmic function, $g_1(\alpha_s L)$.

\subsubsection{Casimir Scaling and a Discrimination Baseline}

\begin{figure}[t!]
\begin{center}
\includegraphics[width = 7.5cm]{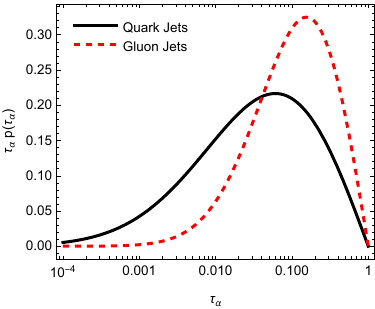}\hspace{0.5cm}
\includegraphics[width = 7.5cm]{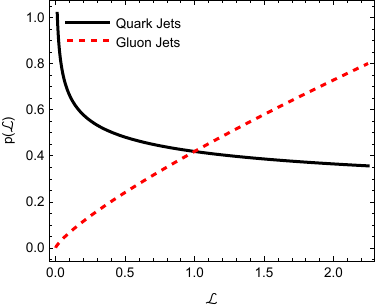}
\caption{\label{fig:angsdist}
Left: Plots of the probability distributions of the angularity $\tau_\alpha$ as measured on quark (black) and gluon (red) jets from the double logarithmic picture of uniform emissions on the Lund plane, \Eq{eq:angdist_q} for quarks (and the corresponding distribution for gluons). Right: Distribution of the likelihood ratio ${\cal L}$, \Eq{eq:likeangs}, as measured on quark (black) and gluon (red) jets from measurement of the angularity $\tau_\alpha$.
}
\end{center}
\end{figure}

We could do this whole rigmarole again for a gluon jet, but by axiom \#3, we know that the only difference would be in the replacement of the color factor $C_F$ with $C_A$ everywhere.  So, let's just do that directly to the cumulative distribution:
\begin{align}
\Sigma_g(\tau_\alpha) = \exp\left[
-\frac{\alpha_s}{\pi}\frac{C_A}{\alpha}\log^2\tau_\alpha
\right]\,.
\end{align}
A plot comparing these double logarithmic distributions for angularities measured on quark and gluon jets is displayed at left in \Fig{fig:angsdist}. This immediately shows us that the quark and gluon cumulative distributions satisfy the scaling relationship
\begin{align}
\Sigma_g = \Sigma_q^{C_A/C_F}\,.
\end{align}
This is referred to as {\bf Casimir scaling} \cite{Polyakov:1980ca,Korchemsky:1987wg,Correa:2012nk,Grozin:2014hna,Grozin:2015kna}, because quark and gluon jets only differ in their color Casimir factors and further, the dependence on the Casimirs is exponential.\footnote{Casimir scaling typically refers to the $C_F\leftrightarrow C_A$ relationship between the cusp anomalous dimensions $\Gamma$ of the quark and gluon form factors.  The one-loop cusp anomalous dimension is the coefficient of the logarithm in the exponent of the Sudakov factor, with $\Gamma_q = \frac{C_F}{C_A}\Gamma_g = \frac{\alpha_s}{\pi}C_F+\cdots$.  Quadratic color Casimir scaling is known to be violated starting at four-loop order, when quartic color invariants first appear \cite{Moch:2018wjh,Henn:2019swt,vonManteuffel:2020vjv}.}

Now we're cooking because if we remember from very early on, this cumulative distribution relationship is precisely the ROC curve.  Then, in the soft and collinear limit, where quarks and gluons only differ by their color charge, the ROC curve of measuring an angularity $\tau_\alpha$ is
\begin{align}
\text{ROC}(x) = x^{C_A/C_F} = x^{9/4}\,.
\end{align}
As a baseline for discrimination power, let's consider a couple of metrics.  First, the ROC curve at $x = 0.5$, where $50\%$ of the quark jets are kept, then
\begin{align}
\text{ROC}(0.5) \approx 0.21\,,
\end{align}
21\% of the gluon jets are also kept.  Correspondingly, the fraction $x$ of quark jets at which 50\% of the gluon jets are kept, where ROC$(x) = 0.5$ is
\begin{align}
1-\text{ROC}^{-1}(0.5) \approx 0.27\,,
\end{align}
so about 27\% of quark jets are kept.  Finally, we can calculate the AUC,
\begin{align}
\text{AUC} = \int_0^1dx\, x^{C_A/C_F} = \frac{C_F}{C_F+C_A}\approx 0.31\,.
\end{align}

We can also calculate the reducibility factors corresponding to maximal attainable sample purity.  In this case, the likelihood ratio is
\begin{align}\label{eq:likeangs}
{\cal L} = \frac{p_g(\tau_\alpha)}{p_q(\tau_\alpha)} = \frac{C_A}{C_F}\exp\left[
-\frac{\alpha_s}{\pi}\frac{(C_A-C_F)}{\alpha}\log^2\tau_\alpha
\right]\,.
\end{align}
The distribution of this observable on quark and gluon jets is plotted at right in \Fig{fig:angsdist}.  As $\tau_\alpha\to 0$, the exponential factor vanishes because $C_A > C_F$, and so the minimum value of the likelihood is ${\cal L}_{\min} = 0$.  That is, a pure sample of quark jets can be isolated by restricting the value of $\tau_\alpha$ to be sufficiently small.  By contrast, as $\tau_\alpha\to 1$, the likelihood ratio approaches its maximum of
\begin{align}
{\cal L}_{\max} = \frac{C_A}{C_F}\,.
\end{align}
This means that, by taking $\tau_\alpha\to 1$, you can create a sample for which gluon jets and quark jets exist in the ratio $C_A : C_F$.  If you only measure the angularity $\tau_\alpha$, you cannot isolate pure gluon jets with no quark jet contamination.

As yet another measure of discrimination power, we can quantify the {\bf significance}, or the ratio of signal to the expected statistical fluctuation in the background.  This metric is relevant for new particle discovery, and was, for example, used to quantify the $5\sigma$ discovery of the Higgs boson in 2012 \cite{ATLAS:2012yve,CMS:2012qbp}.  If we believe quantum mechanics, and collisions are truly independent, then events at the LHC are i.i.d., and the distribution of the number of signal or background events (i.e., the sum of all events satisfying a classification criteria) is governed by the central limit theorem.  Therefore, the expected statistical fluctuation of the background events, the standard deviation of the number of background events, scales like the square-root of the number of background events.  Then, the significance ${\cal S}$ is the ratio of the number of signal events $S$ to the square-root of the number of background events $B$:
\begin{align}
{\cal S} = \frac{S}{\sqrt{B}}\,.
\end{align}
For the discovery of the Higgs, the significance was about 5, with several thousands of background events measured in the bins around $125$ GeV, with an excess observed of only a few hundred events.

The significance can be massaged into a form where its relationship to the ROC can be made apparent.  Let's denote the total number of signal (background) events in our sample as $N_s$ ($N_b$).  Now, given the likelihood ratio discriminant ${\cal L}$, we would like to make a cut on ${\cal L}$ and determine the significance with that additional cut.  The fraction of signal (background) events that pass this cut is $1-\Sigma_s({\cal L})$ ($1-\Sigma_b({\cal L})$), the corresponding cumulative distributions of the likelihood.  Note that we subtract the cumulative distributions from 1 because (by our definition of the likelihood ratio), signal events accumulate at large likelihood, and so the cut value ${\cal L}_\text{cut}$ is the minimal value of the likelihood.  Then, the significance with a cut on the likelihood ratio is
\begin{align}
{\cal S}({\cal L}) = \frac{N_s}{\sqrt{N_b}}\frac{1-\Sigma_s({\cal L})}{\sqrt{1-\Sigma_b({\cal L})}}\,.
\end{align}
The ratio $N_s/\sqrt{N_b}$ is the nominal significance, but the ratio of cumulative distributions can improve it.  The improvement factor in the significance from a cut on the likelihood is then
\begin{align}
\Delta {\cal S} = \frac{1-\Sigma_s({\cal L})}{\sqrt{1-\Sigma_b({\cal L})}} = \frac{1-\text{ROC}(x)}{\sqrt{1-x}}\,,
\end{align}
where on the right, we have related the cumulative distributions of the likelihood ratio to the ROC curve.  For quark versus gluon discrimination, this significance improvement factor is then
\begin{align}
\Delta S = \frac{1-x^{C_A/C_F}}{\sqrt{1-x}} \,.
\end{align}
For $C_A/C_F = 9/4$ in QCD, this is maximized at $x \approx 0.41$, at which $\Delta S \approx 1.13$.  So, there is gluon discovery improvement power of only about 13\% over the nominal expectation.

\subsection{Incorporation Higher Order Effects}

This analysis is but the starting point for quark versus gluon discrimination.  Essentially, no matter what you do or measure on jets, discrimination performance is bounded from below by this simple Casimir scaling result.  By contrast, if you find that your quark versus gluon discriminant performs much worse than this, there is definitely something else going on!  Nevertheless, this discrimination power isn't that great, and suggests that we should look for ways to improve it.  In this section, we will discuss a couple of physics points that arise at higher orders, and in later sections, will revisit the Lund plane and establish the optimal discriminant given this emission phase space.

\subsubsection{Resolving Multiple Emissions}\label{sec:multemit}

The first thing we could consider as an improvement to the discrimination power of the angularities exclusively is to measure additional observables that are correspondingly sensitive to additional emissions on the Lund plane.  The angularities are sensitive to radiation about a single hard particle, or the jet axis, and in the soft and collinear limit, a single emission dominates.  This suggests that if we want sensitivity to additional emissions in the jet, then we should consider radiation about additional axes within the jet.  The observable that accomplishes this is called $N$-subjettiness \cite{Thaler:2010tr,Thaler:2011gf} (see also \Refs{Brandt:1978zm,Stewart:2010tn,Kim:2010uj}), in which one places $N$ axes in the jet along the directions of the $N$ most dominant energy flows and then calculates the effect of additional radiation about those axes:
\begin{equation}
\tau_N^{(\alpha)} = \sum_{i\in J}z_i\, \min\left\{
\theta_{i1}^\alpha\,,\theta_{i2}^\alpha\,,\dotsc\,, \theta_{iN}^\alpha
\right\}\,,
\end{equation}
where again $\alpha>0$ for IRC safety.  Here, $\theta_{iK}$ is the angle between particle $i$ and axis $K$ in the jet.

\begin{figure}[t!]
\begin{center}
\includegraphics[width = 8cm]{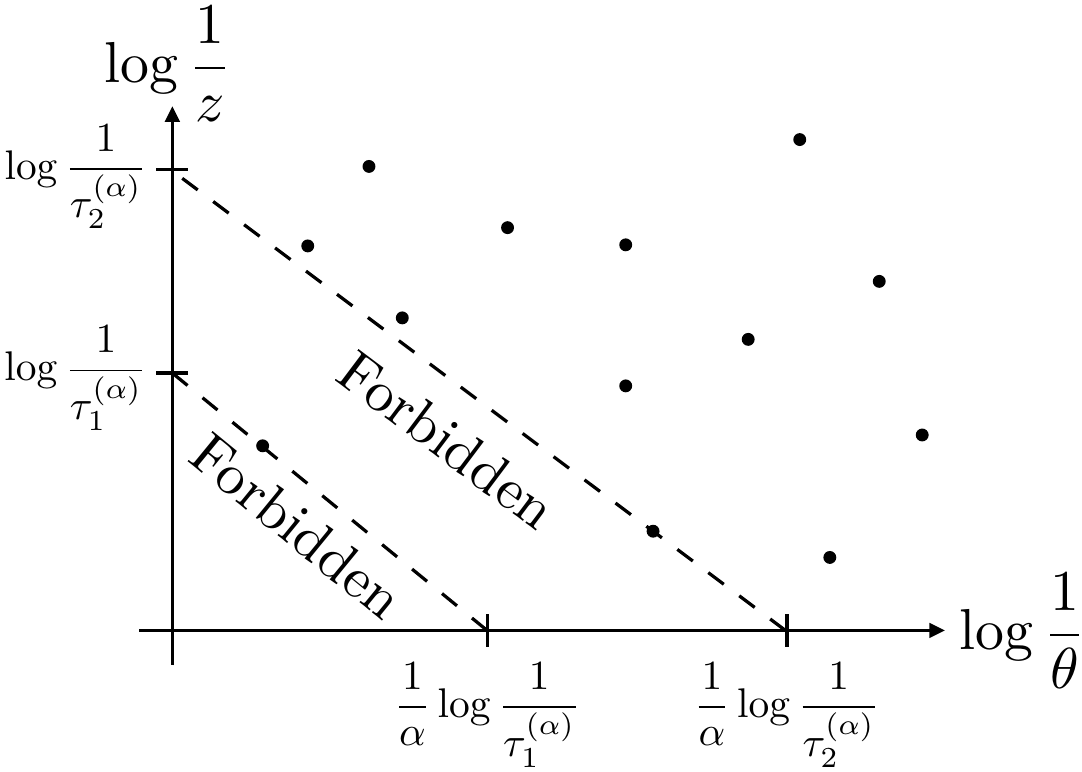}
\caption{\label{fig:twojet}
Representation on the Lund plane of the measurement of 1- and 2-subjettinesses, $\tau_1^{(\alpha)}$ and $\tau_2^{(\alpha)}$.
}
\end{center}
\end{figure}

Let's consider the effect on quark versus gluon discrimination on jets on which we measure both the angularity $\tau_\alpha = \tau_1^{(\alpha)}$, which is equivalently 1-subjettiness, and 2-subjettiness, $\tau_2^{(\alpha)}$.  What is especially convenient in the soft and collinear limit of the Lund plane is that the $N$-subjettiness axes always lie exactly on the first $N$ particles ordered in their contribution to the angularity $\tau_\alpha$.  This makes calculation of the joint distribution of $\tau_1^{(\alpha)}$ and $\tau_2^{(\alpha)}$ very simple.  In \Fig{fig:twojet}, we have illustrated the Lund plane in which both $\tau_1^{(\alpha)}$ and $\tau_2^{(\alpha)}$ are measured.\footnote{We are just considering emissions off of the leading, hard particle.  A complete calculation for 2-subjettiness would also need to include emission off of the particle that set the value of 1-subjettiness.  However, our simpler calculation is sufficient for the conclusions we wish to draw here.}  This figure illustrates everything we need to know to directly write down the joint probability distribution.  First, the total forbidden emission area is set by the value of $\tau_2^{(\alpha)}$, and so the Sudakov factor has no explicit dependence on $\tau_1^{(\alpha)}$.  Next, there are two explicit emissions that set $\tau_1^{(\alpha)}$ and $\tau_2^{(\alpha)}$, and, as emissions are uniform on the Lund plane, the functional form of the differential probability of these emissions is identical.  Therefore, we can immediately write down the joint probability distribution as
\begin{align}
p_q(\tau_1^{(\alpha)},\tau_2^{(\alpha)}) = \left(
\frac{2\alpha_s C_F}{\pi \alpha}
\right)^2\frac{\log \tau_1^{(\alpha)}}{\tau_1^{(\alpha)}}\frac{\log \tau_2^{(\alpha)}}{\tau_2^{(\alpha)}}\,\exp\left[
-\frac{\alpha_s}{\pi}\frac{C_F}{\alpha}\log^2 \tau_2^{(\alpha)}
\right]\,.
\end{align}
As a consistency check of this distribution, we can integrate over $0<\tau_2^{(\alpha)}<\tau_1^{(\alpha)}$ and verify that the correct distribution of $\tau_1^{(\alpha)}$ is produced:
\begin{align}
\int_0^{\tau_1^{(\alpha)}}d\tau_2^{(\alpha)} \, p_q(\tau_1^{(\alpha)},\tau_2^{(\alpha)}) = -
\frac{2\alpha_s C_F}{\pi \alpha}
\frac{\log \tau_1^{(\alpha)}}{\tau_1^{(\alpha)}}\,\exp\left[
-\frac{\alpha_s}{\pi}\frac{C_F}{\alpha}\log^2 \tau_1^{(\alpha)}
\right]\,,
\end{align}
which agrees with what we calculated earlier for the angularities.

As earlier, there's no need for another calculation with gluon jets, we just need to replace $C_F \to C_A$:
\begin{align}
p_g(\tau_1^{(\alpha)},\tau_2^{(\alpha)}) = \left(
\frac{2\alpha_s C_A}{\pi \alpha}
\right)^2\frac{\log \tau_1^{(\alpha)}}{\tau_1^{(\alpha)}}\frac{\log \tau_2^{(\alpha)}}{\tau_2^{(\alpha)}}\,\exp\left[
-\frac{\alpha_s}{\pi}\frac{C_A}{\alpha}\log^2 \tau_2^{(\alpha)}
\right]\,.
\end{align}
Correspondingly, the likelihood ratio is
\begin{align}
{\cal L} = \frac{p_g(\tau_1^{(\alpha)},\tau_2^{(\alpha)})}{p_q(\tau_1^{(\alpha)},\tau_2^{(\alpha)})} = \left(\frac{C_A}{C_F}\right)^2\exp\left[
-\frac{\alpha_s}{\pi}\frac{C_A-C_F}{\alpha}\log^2 \tau_2^{(\alpha)}
\right]\,.
\end{align}
We immediately see two properties of this more differential likelihood ratio.  First, the likelihood is monotonic in $\tau_2^{(\alpha)}$, so $\tau_2^{(\alpha)}$ itself is the optimal discriminant.  Next, note that the reducibility factor for the gluon, the maximum value of the likelihood, has increased with respect to just measuring an angularity.  Now, the maximal purity of a gluon sample is restricting to the region where $\tau_2^{(\alpha)} \to \tau_1^{(\alpha)} \to 1$, and where gluon and quark jets are found in the ratio $C_A^2 : C_F^2$.  This latter point is intriguing: by resolving two emissions, the gluon reducibility factor is squared; could resolving three emissions result in the gluon reducibility factor cubed, from just measuring an angularity?  If so, why don't we just measure all possible emissions on the Lund plane?  We will return to this point in a couple of lectures, but for more details about following this $N$-subjettiness trail, see \Refs{Larkoski:2019nwj,Kasieczka:2020nyd}.

\begin{figure}[t!]
\begin{center}
\includegraphics[width=10cm]{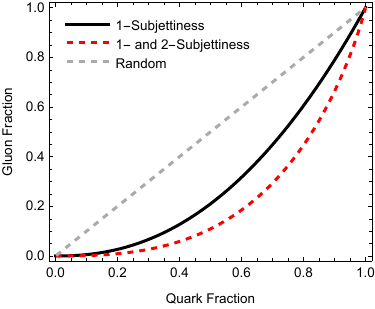}
\caption{\label{fig:1and2roc}
Comparison of the ROC curves for just measuring one angularity (black) to measuring 1- and 2-subjettiness, which resolves two emissions in the jet.  We have also plotted the random AUC (gray), corresponding to the diagonal line.
}
\end{center}
\end{figure}

Calculating the ROC curve isn't as simple in this case as it was for the angularities, but we can nevertheless calculate the distribution of the optimal discriminant, $\tau_2^{(\alpha)} $.  We just need to marginalize over $\tau_1^{(\alpha)}$ to produce
\begin{align}
p_q(\tau_2^{(\alpha)} ) = \int_{\tau_2^{(\alpha)} }^1 d\tau_1^{(\alpha)} \, p_q(\tau_1^{(\alpha)},\tau_2^{(\alpha)}) = -2\left(
\frac{\alpha_s C_F}{\pi\alpha}
\right)^2\frac{\log^3 \tau_2^{(\alpha)}}{\tau_2^{(\alpha)}}\exp\left[
-\frac{\alpha_s}{\pi}\frac{C_F}{\alpha}\log^2 \tau_2^{(\alpha)}
\right]\,.
\end{align}
As always, the gluon distribution immediately follows.  While its functional form is not enlightening, we can calculate the ROC curve in this case, which we have plotted in \Fig{fig:1and2roc}.  This clearly illustrates that by resolving two emissions on the Lund plane, discrimination power is indeed improved throughout all of phase space.  To quantify this, we can calculate the AUC with two resolved emissions, and the result is simple
\begin{align}
\text{AUC} = \int_0^1 d\tau_2^{(\alpha)}\, p_q(\tau_2^{(\alpha)} ) \,\Sigma_g(\tau_2^{(\alpha)} )  = \frac{C_F^2(3C_A+C_F)}{(C_A+C_F)^3} \approx 0.23\,.
\end{align}
Recall that this is smaller by about $0.08$ than the AUC when just a single angularity is measured.

\subsubsection{The Importance of Being Recoil Free}\label{sec:recoilfreeobs}

Though we discussed recombination schemes in our aside on jet algorithms, we have been cavalier about the specific jet axis about which we measure the angularities $\tau_\alpha$ in this section.  At the double-logarithmic level, this ignorance and imprecision is fine, but at higher orders, the specific jet axis about which angularities are measured is important.  Further, at the double logarithmic level, we saw no dependence of discrimination power on the specific value of the angular exponent $\alpha$, but at higher orders this will be relevant.

Let's see how the jet axis and the angular exponent of the angularities $\tau_\alpha$ come into play.  We will just consider two axes as exemplar of the full range of possibilities: the $E$-scheme axis (the direction of mean flow of momentum) and the WTA axis, along the direction of the harder particle.  Further, we will limit our analysis here to first non-trivial order, when the jets have but two particles in them.  With this restriction, we will be able to use our master formula for jet observables to calculate appropriate distributions.

The first thing we need to do, however, is to determine the functional form of the angularities measured about these two different axes.  In general, the angularities on a jet with two particles can be expressed as
\begin{align}
\tau_\alpha = z\theta_1^\alpha + (1-z)\theta_2^\alpha\,,
\end{align}
where $z$ ($1-z$) is the energy fraction of particle 1 (2), and $\theta_1$ ($\theta_2$) is its angle from the appropriate axis.  For the net momentum or $E$-scheme axis, the angle to the axis is set by the angle between particles $\theta$ times the energy fraction of the other particle.  Specifically,
\begin{align}
&\theta_1 = (1-z)\theta\,,&\theta_2 = z\theta\,.
\end{align}
With this, we can then evaluate the angularity about the $E$-scheme axis to be
\begin{align}
\tau_\alpha^{(E\text{-scheme})} = z(1-z)^\alpha \theta^\alpha + z^\alpha(1-z)\theta^\alpha \xrightarrow{z\to 0} \left\{\begin{array}{c}
z\theta^\alpha\,, \qquad \alpha > 1\,,\\
2z\theta\,, \qquad \alpha = 1\,,\\
(z\theta)^\alpha\,, \qquad \alpha < 1\,.
\end{array}
\right.
\end{align}
At right, we have taken the soft, $z\to 0$, limit of the expression, which demonstrates some rather interesting behavior as a function of exponent $\alpha$.

In the soft limit, for $\alpha> 1$, the angularity takes the ``expected'' form, $\tau_\alpha = z\theta^\alpha$, where the dynamics of the soft particle about the hard particle dominate.  Something strange happens when $\alpha = 1$, and for $\alpha < 1$, the term that dominates in the soft limit is the recoil of the hard particle from the jet axis due to momentum conservation with the soft particle.  This is then no longer directly sensitive to the dynamics of the soft particle, and is somewhat rather random and arbitrary, because additional radiation can push the hard particle around further.  Additionally, note that the soft limit of the angularity for $\alpha < 1$ is monotonically related to the angularity with $\alpha = 1$.  For discrimination performance, this means that angularities $\tau_\alpha$ with $\alpha < 1$ will have identical discrimination power to the angularity with $\alpha = 1$.  If there is sensitivity to $\alpha$ in discrimination performance, then that performance saturates at and below $\alpha = 1$.

By contrast, let's consider the angularity measure about the WTA axis.  This expression is exceptionally simple because the WTA axis lies on the harder particle, so only the softer particle contributes:
\begin{align}
\tau_\alpha^{(\text{WTA})} = \min[z,1-z] \,\theta^\alpha  \xrightarrow{z\to 0} z\theta^\alpha\,.
\end{align}
Again, at right, we have taken the soft $z\to 0$ limit, but unlike for the $E$-scheme axis, the functional form of the WTA axis expression is the same for all $\alpha > 0$.  Thus, if there is dependence on $\alpha$ in the discrimination power, we expect that there is no saturation at $\alpha = 1$ for angularities measured about the WTA axis.

Let's now calculate the distribution of the angularities on quark and gluon jets with two particles.  From this analysis, we see that we only need to calculate the distribution with functional form $\tau_\alpha =z \theta^\alpha$, and further, we will demand that the particles are within an angle $R$.  For quark jets, we have
\begin{align}
\frac{d\sigma_q}{d\tau_\alpha} &= \frac{\alpha_s C_F}{\pi}\int_0^R \frac{d\theta}{\theta}\int_0^1 dz\,\frac{1+(1-z)^2}{z}\, \delta(\tau_\alpha - z\theta^\alpha)\\
&=-\frac{2\alpha_s C_F}{\pi \alpha}\frac{1}{\tau_\alpha}\left(
\log\frac{\tau_\alpha}{R^\alpha} + \frac{3}{4}
\right)\,,\nonumber
\end{align}
where we have only kept the leading terms in the $\tau_\alpha\to 0$ limit.  The corresponding calculation for gluon jets is
\begin{align}
\frac{d\sigma_g}{d\tau_\alpha} &= \frac{\alpha_s}{\pi}\int_0^R \frac{d\theta}{\theta}\int_0^{1/2} dz\,\left[
2C_A\left(
\frac{z}{1-z}+\frac{1-z}{z}+z(1-z)
\right)+2n_fT_R\left(
z^2+(1-z)^2
\right)
\right]\, \delta(\tau_\alpha - z\theta^\alpha)\nonumber\\
&=-\frac{2\alpha_s}{\pi \alpha}\frac{1}{\tau_\alpha}\left(
C_A\log\frac{\tau_\alpha}{R^\alpha} + \frac{11}{12}C_A - \frac{n_f T_R}{3}
\right)\,.
\end{align}
Here, we have introduced $n_f$ as the number of active quarks to which the initial gluon can split (the number of quarks whose mass is less than half the initial energy of the gluon).

Both of these expressions will have to be dressed up with their appropriate Sudakov factor to account to arbitrary additional gluon emission, but we can understand where $\alpha$ dependence in discrimination power arises without any more detailed calculations.  Note that in both distributions, the $\alpha$ dependence is an overall factor; however, with the full splitting functions, there is a logarithmic term and a non-logarithmic term, and depending on the value of $\alpha$, these can play off one another in interesting ways.  To see how this works, let's stare at the Sudakov factor (for the quark), and make some estimates.  The quark's Sudakov factor is
\begin{align}
\Sigma_q(\tau_\alpha) = e^{-\frac{\alpha_s}{\pi}\frac{C_F}{\alpha}\log^2\tau_\alpha}\,.
\end{align}
As an exponential, most of its support lies where its argument is order-1, and so we can estimate that the Sudakov restricts the distribution to where
\begin{align}
\frac{\alpha_s}{\pi}\frac{C_F}{\alpha}\log^2\tau_\alpha \sim 1\,.
\end{align}
Just focusing on the relationship between $\alpha$ and $\tau_\alpha$, this implies that
\begin{align}
\log\tau_\alpha \sim -\sqrt{\alpha}\,,
\end{align}
at the value of $\alpha$ where the bulk of the distribution lies.

Using this scaling, let's plug it into the leading order distributions we had just calculated.  We can ignore overall factors common to both quarks and gluons and just focus on the terms in parentheses.  Note that we have the scaling
\begin{align}
\frac{d\sigma_q}{d\tau_\alpha}&\sim C_F\log\frac{\tau_\alpha}{R^\alpha} + \frac{3}{4}C_F \sim -\sqrt{\alpha}+ \frac{3}{4} C_F\,,\\
\frac{d\sigma_g}{d\tau_\alpha}&\sim C_A\log\frac{\tau_\alpha}{R^\alpha} + \frac{11}{12}C_A - \frac{n_f T_R}{3} \sim -\sqrt{\alpha}+ \frac{11}{12}C_A - \frac{n_f T_R}{3}\,.\nonumber
\end{align}
As $\alpha\to \infty$, the logarithm term dominates the distribution and therefore the discrimination power of the angularities reduces to the Casimir scaling that we established earlier.  However, as $\alpha$ decreases, the importance of the logarithmic term decreases, and the new, constant, terms become more important.  The values of the constant terms differ significantly in QCD, namely
\begin{align}
\frac{3}{4} C_F &= 1\,,\\
\frac{11}{12}C_A - \frac{n_f T_R}{3} &= \frac{23}{12}\,,
\end{align}
where we have set $n_f = 5$, and so as their importance increases, so too does discrimination power.  Therefore, as mentioned earlier, we expect that the WTA angularities are better quark versus gluon discriminants at small $\alpha$ than the $E$-scheme angularities, whose performance saturates around $\alpha = 1$.

\begin{figure}[t!]
\begin{center}
\includegraphics[width=8cm]{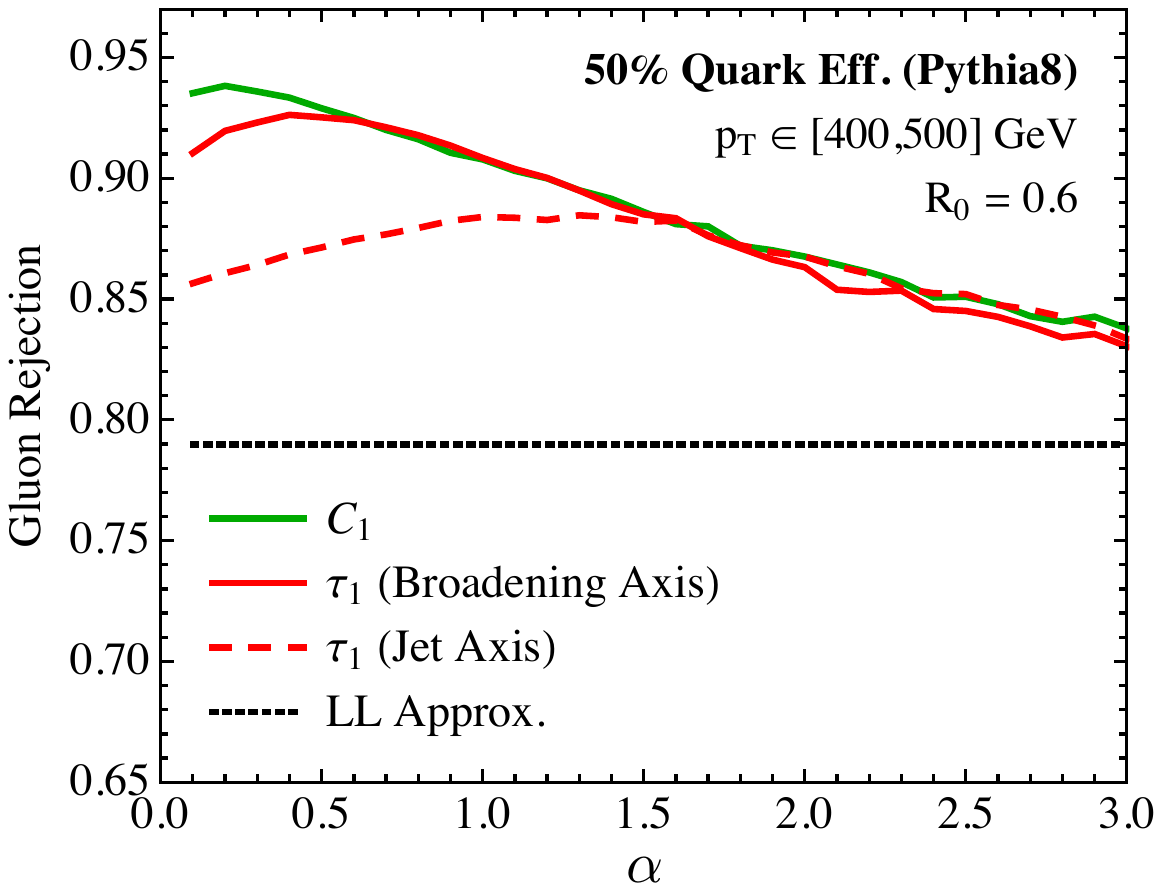}
\caption{\label{fig:recqvg}
Plot of the gluon jet rejection rate ($1-$gluon efficiency) at $50\%$ quark jet efficiency, as a function of angular exponent in the angularities, $\alpha$, in simulated events at the LHC.  The dashed black line corresponds to the Casimir scaling expectation, while the dashed red line is the rejection with angularities measured about the $E$-scheme jet axis, and the solid lines are the rejection rate for angularities measured about a recoil-free axis.  Better discrimination power is higher on this plot.  Slightly modified version of a plot from \InRef{Larkoski:2013eya}.
}
\end{center}
\end{figure}

This observation is illustrated in simulated data in \Fig{fig:recqvg}, reproduced from \InRef{Larkoski:2013eya}.  This plot shows the gluon jet rejection rate ($1-$gluon jet efficiency) at $50\%$ quark jet efficiency, as a function of the angular exponent $\alpha$ of the angularity.  The black dashed line is the Casimir scaling result at double logarithmic accuracy, which is independent of $\alpha$.  The color curves show the rejection rate for angularities measured about different axes: the $E$-scheme axis angularities are dashed red, and recoil-free axis angularities are solid.  The WTA axis was not yet invented at the time of this reference, but the solid red curve (the ``broadening axis'') is in practice nearly identical to the WTA axis, and is indeed the same for a jet with two particles.  As $\alpha$ decreases, we indeed see better discrimination power (higher rejection rate), but the $E$-scheme angularities stop improving below about $\alpha \sim 1.5$, close to the $\alpha = 1$ point we had predicted.  By contrast, the recoil-free axis angularities continue to improve discrimination power down to very small values of $\alpha$, where other physical effects start to dominate.\footnote{These effects are likely dominantly non-perturbative in origin.  Recall that in the Sudakov form factor, the coupling $\alpha_s$ and the angular exponent $\alpha$ appear in the ratio $\alpha_s/\alpha$.  A perturbative analysis assumes that $\alpha_s/\alpha \ll 1$, so that truncation of the expansion is justified; however, at sufficiently small $\alpha$, we will have $\alpha_s/\alpha \sim 1$, and perturbation theory breaks down.}

\subsection*{Exercises}

\begin{enumerate}

\item Consider the measurement of two angularities, $\tau_\alpha$ and $\tau_\beta$, with, say $\alpha> \beta$.  
\begin{enumerate}
\item Calculate the Sudakov form factor for two angularities measured on quark jets, the joint probability distribution $p_q(\tau_\alpha,\tau_\beta)$.  Further, ensure that the joint probability distribution marginalizes to the correct single probability distributions.  That is,
\begin{equation}
\int_{\tau_\text{lo}}^{\tau_\text{hi}}d\tau_\beta\, p_q(\tau_\alpha,\tau_\beta) = p_q(\tau_\alpha)\,,
\end{equation}
for particular bounds $\tau_\text{lo}< \tau_\beta < \tau_\text{hi}$ (that you should determine).  For a hint to this problem, see Ref.~\cite{Larkoski:2013paa}.

\item Determine the likelihood ratio for quark versus gluon discrimination from this joint probability distribution,
\begin{align}
{\cal L} = \frac{p_g(\tau_\alpha,\tau_\beta)}{p_q(\tau_\alpha,\tau_\beta)}\,.
\end{align}
What is the gluon reducibility factor with this likelihood ratio?

\item Can you determine the distribution of the likelihood ratio on quark and gluon jets?  What is the ROC curve and how does it compare to the case when only a single angularity is measured?

\end{enumerate}

\item\label{ex:aleph} (This is an extension of Exercise 9.3 in Ref.~\cite{Larkoski:2019jnv}.)  The ALEPH experiment at the Large Electron-Positron Collider (LEP) measured the number of jets produced in $e^+e^-\to$ hadrons collision events.  The experiment counted $n$ jets, if, for every pair $i,j$ of jets the following inequality is satisfied:
\begin{equation}
2\min[E_i^2,E_j^2](1-\cos\theta_{ij}) > y_{\text{cut}} E_{\text{cm}}^2\,,
\end{equation}
for $y_\text{cut}<1$, $E_i$ is the energy of jet $i$, $\theta_{ij}$ is the angle between jets $i$ and $j$ and $E_\text{cm}$ is the center-of-mass collision energy.  In the soft and collinear limits, determine the probability $p_n$ for observing $n$ jets, as a function of $y_\text{cut}$.

Note that the minimum number of jets is 2 ($e^+e^- \to q\bar q$) and gluons can be emitted from either the quark or the anti-quark.  Compare your result to figure 7 of Ref.~\cite{ALEPH:2003obs}.  What value of $\alpha_s$ fits the data the best?  This fit is imperfect because we're omitting a lot of important physics, but it will be qualitatively close.

\item Consider the number of jets as defined by the procedure introduced in Exercise \ref{ex:aleph} above.  Consider that procedure measured on $e^+e^-\to q\bar q+X$ and $e^+e^-\to gg+X$ events, where $X$ is any other hadronic activity.  Using the discrete probability distribution $p_n$ for the quark and gluon final states, determine the likelihood ratio, ROC curve and AUC for this number of jets observable for discrimination of the quark from gluon final states, as a function of the parameter $y_\text{cut}$.  Does the AUC for this observable ever correspond to better discrimination to that for $\tau_\alpha$ as derived in this section?

\end{enumerate}

\section{Second Example: $H\to b\bar b$ Identification and IRC Safe Binary Discrimination}\label{sec:hvsg}

By the mid-2000s, just a year or two before the LHC was scheduled to turn on, the ATLAS and CMS experiments had clearly identified the path they would take to achieve the central goal of the LHC program: discovery of the Higgs boson.  Much of that story is well known because, of course, the Higgs was discovered in 2012 \cite{ATLAS:2012yve,CMS:2012qbp} through observation in the ``golden channels'' of $H\to \gamma\gamma$ and $H\to 4$ leptons, which have clean experimental signatures and very small Standard Model backgrounds.\footnote{The story of the discovery of the Higgs is documented in the fantastic film ``Particle Fever'', starring and produced by physicist David E.~Kaplan, himself an author on a very early top tagging algorithm \cite{Kaplan:2008ie}.  I happened to be at CERN on 4 July, 2012, but had caught a cold on the trans-Atlantic flight and was therefore unable to stay up all night saving a precious early spot in line for entrance into the auditorium. I was thus resigned to watching the discovery announcement from my computer in my hostel room.}  However, Higgs bosons only decay to photons or four leptons (through off-shell $Z$ bosons) about 0.5\% of the time combined, so just by discovering the Higgs in these modes, one can clearly not claim victory in knowing that what you saw was actually {\it the} Higgs.  Most of the time, the vast majority of the time, nearly 90\% of all decays of the Higgs produce hadrons in relatively low-energy jets (a scale around half the Higgs mass), but hadrons at these low energies are produced in huge quantities at the LHC.

Initial analyses were done by the experiments to quantify the status of searching for the Higgs in these hadronic decay modes, but the conclusion was that the backgrounds are simply too large.  In particular, the dominant decay mode of the Higgs, $H\to b\bar b$ which occurs about 50\% of the time itself, was not considered viable to observe at the LHC in the ATLAS and CMS Technical Design Reports \cite{ATLAS:1999vwa,CMS:2007sch}.  Then, again in February 2008 (perhaps the historically most important month for jet substructure!),\footnote{In addition to the anti-$k_T$ algorithm and the BDRS mass drop-filtering algorithm for $H\to b\bar b$ identification, February 2008 also saw the posting of a method for removing pile-up radiation from jets \cite{Cacciari:2008gn}, and then in early March 2008, a re-evaluation of energy correlators from a strong coupling/string theory perspective \cite{Hofman:2008ar}.} a group composed of theorists and experimentalists introduced a new technique that rendered the bottom decay mode of the Higgs observable \cite{Butterworth:2008iy}, and subsequently kick-started the modern jet substructure endeavor.\footnote{See also \Refs{Seymour:1993mx,Butterworth:2002tt,Butterworth:2007ke} for historical foundations.  As an undergraduate, I worked on a project starting in fall 2006 on a closely related problem, but it wasn't published and I then moved on to graduate school.  See my abstract for a presentation at the 2007 University of Washington Undergraduate Research Symposium on page 150 of \url{https://uw-s3-cdn.s3.us-west-2.amazonaws.com/wp-content/uploads/sites/73/2014/06/24075140/2007SympProceedings.pdf}}  The key observations of \InRef{Butterworth:2008iy} were two-fold: first, Standard Model backgrounds decrease much more rapidly than Higgs production at higher energies, and second, that the inside of a ``Higgs jet'' should contain two hard prongs (the two bottom quarks), but multiple prong substructure was rare in QCD (essentially because of approximate scale invariance).  Exploiting these features enabled the authors to show that discovery significances of 5 or more were possible.  It took another decade, but direct evidence for $H\to b\bar b$ decays has now been observed \cite{ATLAS:2018kot,CMS:2018nsn}.

The $H\to b\bar b$ decay has remained an interesting and widely studied benchmark case, especially for machine learning discrimination studies (see, e.g., \Refs{Lin:2018cin,Datta:2019ndh,Moreno:2019neq,Chakraborty:2019imr,CMS:2020poo,Chung:2020ysf,Tannenwald:2020mhq,Guo:2020vvt,Abbas:2020khd,Jang:2021eph,Khosa:2021cyk}).  The dominant background to this decay at high energies relevant for jet substructure is the splitting of an off-shell gluon to bottom quarks, $g\to b\bar b$.  No other hadronic backgrounds really need to be considered because bottom quark tagging in experiment is now nearly perfect \cite{ATLAS:2022qxm,CMS:2023tlv}.  So, we can formulate the identification of $H\to b\bar b$ decays at high energies as a binary discrimination problem.  However, unlike for our quark versus gluon discrimination in the previous section, the jets of interest now are definitely not scale invariant because the Higgs is massive.  Instead, what we will do is work within fixed-order perturbation theory, identifying signal and background distributions on phase space, their likelihood ratio, and the corresponding discrimination power.  Because of the explicit scale of the Higgs mass and the requirement of bottom quarks in the problem, leading-order distributions calculated in fixed-order perturbation theory will actually be true probabilities because there is no degeneracy of states that needs to be accounted for.

First, however, we will formulate the binary discrimination problem in general in fixed-order perturbation theory, without assuming a particular problem or goal until later.  Thus, $H\to b\bar b$ versus $g\to b\bar b$ discrimination will be the exemplar of this formalism, but can be applied more broadly.  As another interesting case, significant work has also been devoted to developing techniques for discrimination of hadronic top quark decays from jets initiated by light QCD partons.  In 2019, a review and comparison of a large number of machine learning approaches to this problem were studied and summarized in \InRef{Kasieczka:2019dbj}.  One conclusion of this study is that top quark discrimination and identification algorithms are now extremely powerful, but another conclusion is that all of these machine learning approaches (which at least naively differ significantly in their implementation, internal architecture, etc.) have very similar performance.  Further, combining different algorithms seems to not improve performance very much, and so suggests that there is a maximal upper bound to possible discrimination.  Machine learning cannot answer this question, because the models are all trained on simulated data, but from a principled, order-by-order analysis of the problem, we can hope to do so.\footnote{I thank Gregor Kasieczka for inspiring my interest in this problem.}  Top quark discrimination is much more challenging than that for Higgs bosons (at leading order, top quarks decay to three final state particles), but this framework can be extended to that case.

While we will be able to understand $H\to b\bar b$ decays versus $g\to b\bar b$ splittings in detail, there are other, related searches and discrimination problems for which the goal or signal is less well-defined.  In searches for new physics or new particles, the mass of the particle is unknown, and so one needs a more general framework or reduced set of assumptions to construct the likelihood ratio.  In this lecture, we will also review and apply power counting methods to optimal observable construction, given a parametric and hierarchical picture of what signal and background are expected to be, with no strict assumption on a particular value of the mass.  Through this, we will also see our first glimpses beyond the cosy confines of perturbation theory, and demonstrate that seemingly innocuous questions have shocking answers.  But first, to binary discrimination in fixed-order perturbation theory.

\subsection{Fixed-Order Expansion of the Likelihood Ratio}

We will study the general problem of binary discrimination in fixed-order perturbation theory, building up a description of the likelihood ratio, ROC curve, etc., in powers of the coupling $\alpha_s$.  The analysis presented here and in the specific example of $H\to b\bar b$ identification later closely follows that from \InRef{Larkoski:2023xam}, where more details can be found, including a complete calculation at next-to-leading order.  Our starting point will be understanding the signal and background distributions on phase space $\Phi$.  We can generically write the signal $s$ and background $b$ distributions as a series in $\alpha_s$ as
\begin{align}
\tilde p_s(\Phi) &= p_s^{(0)}(\Phi^{(0)})+\frac{\alpha_s}{2\pi}\,p_s^{(1)}(\Phi^{(1)})+{\cal O}(\alpha_s^2)\,,\\
\tilde p_b(\Phi) &= p_b^{(0)}(\Phi^{(0)})+\frac{\alpha_s}{2\pi}\,p_b^{(1)}(\Phi^{(1)})+{\cal O}(\alpha_s^2)\,,
\end{align}
just working through next-to-leading order explicitly.  Here, the over-tilde represents that (so far) these are unnormalized distributions, which we will work to fix up.  The superscript $(n)$ denotes the order in $\alpha_s$ of that contribution, with $(0)$ being leading order, $(1)$ next-to-leading order, etc.  If there is any hope that these are probability distributions, the leading order distributions must be integrable themselves.  For simplicity, we assume that they are also already normalized, where
\begin{align}
1 = \int d\Phi^{(0)}\, p(\Phi^{(0)})\,.
\end{align}
This will make future expressions significantly more compact.

We will make the additional assumption that the signal and background final states are indistinguishable.  That is, at each order in perturbation theory and on each corresponding phase space $\Phi^{(n)}$, signal and background events have the same particle content.  Thus, as quantum mechanical processes, we should coherently sum the signal and background states together, and the take their combined absolute square.  This then suggests a related discrimination problem of unambiguously observing the quantum interference of the signal and background states.  This problem can again be formulated as a binary discrimination problem, distinguishing the quantum pure state density matrix of signal from the classically mixed density matrix of background \cite{Larkoski:2022lmv}.  While several avenues for quantum interference at colliders have been proposed, from interference of Standard Model processes with new physics to interference between intermediate on-shell gluon spin states, e.g., \Refs{Gottfried:1964nx,Webber:1986mc,Collins:1987cp,Knowles:1987cu,Buckley:2007th,Buckley:2008pp,Alves:2008up,Boudjema:2009fz,Murayama:2009jz,Chen:2020adz,Karlberg:2021kwr,Hamilton:2021dyz,Larkoski:2022qlf}, we won't study this fascinating direction further here.

While the leading-order distribution is normalized, the full distribution $\tilde p(\Phi)$ is not, so let's work to calculate the normalization factor ${\cal N}$ that accomplishes this.  We require that
\begin{align}
1 &= {\cal N}\int d\Phi\, \tilde p(\Phi) \\
&=\left( 1+\frac{\alpha_s}{2\pi}{\cal N}^{(1)}+\cdots\right)\left(1+\frac{\alpha_s}{2\pi}\int d\Phi^{(1)}p^{(1)}(\Phi^{(1)})+\cdots\right)\nonumber\\
&=1 + \frac{\alpha_s}{2\pi}\left(
\int d\Phi^{(1)}\, p^{(1)}(\Phi^{(1)})+{\cal N}^{(1)}
\right)+\cdots\nonumber\,.
\end{align}
On the second line, we have introduced the next-to-leading order normalization factor ${\cal N}^{(1)}$ and ellipses hide terms at higher orders in $\alpha_s$.  Through next-to-leading order, then, the normalization factor has the expansion
\begin{align}
{\cal N} = 1 - \frac{\alpha_s}{2\pi}\int d\Phi^{(1)}\, p^{(1)}(\Phi^{(1)}) + \cdots\,.
\end{align}
This can then be used to construct unit normalized distributions on phase space (no over-tildes) as
\begin{align}
p(\Phi) &= {\cal N}\tilde p(\Phi) \\
&= p^{(0)}(\Phi^{(0)})+\frac{\alpha_s}{2\pi}\,p^{(1)}(\Phi^{(1)}) - \frac{\alpha_s}{2\pi}\,p^{(0)}(\Phi^{(0)})\int d\Phi^{(1)}\, p^{(1)}(\Phi^{(1)}) +\cdots
\nonumber\,.
\end{align}

With normalized probability distributions, we can then evaluate the likelihood ratio on phase space to establish the optimal observable possible, given that we only measure particle momenta.  The form of the likelihood ratio we will consider here is the inverse of that we introduced with the Neyman-Pearson Lemma (signal distribution is in the denominator now), which we will find is convenient when applied to Higgs decays.  As a function of phase space variables, the likelihood ratio is
\begin{align}
\hat {\cal L}(\Phi) &= \frac{p_b(\Phi)}{p_s(\Phi)}\\
&=\frac{p_b^{(0)}(\Phi^{(0)})+\frac{\alpha_s}{2\pi}\,p_b^{(1)}(\Phi^{(1)}) - \frac{\alpha_s}{2\pi}\, p_b^{(0)}(\Phi^{(0)}\int d\Phi^{(1)}\, p_b^{(1)}(\Phi^{(1)})) +\cdots
}{p_s^{(0)}(\Phi^{(0)})+\frac{\alpha_s}{2\pi}\,p_s^{(1)}(\Phi^{(1)}) - \frac{\alpha_s}{2\pi}\, p_s^{(0)}(\Phi^{(0)})\int d\Phi^{(1)}\, p_s^{(1)}(\Phi^{(1)}) +\cdots
}
\nonumber\\
&= \frac{p_b^{(0)}(\Phi^{(0)})}{p_s^{(0)}(\Phi^{(0)})}\left[1+\frac{\alpha_s}{2\pi} \left(
\int d\Phi^{(1)}\left( p_s^{(1)}(\Phi^{(1)})- p_b^{(1)}(\Phi^{(1)})\right)+\frac{p_b^{(1)}(\Phi^{(1)})}{p_b^{(0)}(\Phi^{(0)})}-\frac{p_s^{(1)}(\Phi^{(1)})}{p_s^{(0)}(\Phi^{(0)})}
\right)+\cdots\right]\nonumber\,,
\end{align}
expanding through next-to-leading order on the final line.  We can now identify terms in the Taylor expansion of the likelihood as an observable itself:
\begin{align}
\hat {\cal L}(\Phi) = \hat {\cal L}^{(0)}(\Phi)+\frac{\alpha_s}{2\pi}\hat {\cal L}^{(1)}(\Phi)+\cdots\,,
\end{align}
where
\begin{align}
\hat {\cal L}^{(0)}(\Phi) &=\frac{p_b^{(0)}(\Phi^{(0)})}{p_s^{(0)}(\Phi^{(0)})}\,,\\
\hat {\cal L}^{(1)}(\Phi) &=\frac{p_b^{(0)}(\Phi^{(0)})}{p_s^{(0)}(\Phi^{(0)})}\left[
\int d\Phi^{(1)}\left( p_s^{(1)}(\Phi^{(1)})- p_b^{(1)}(\Phi^{(1)})\right)+\frac{p_b^{(1)}(\Phi^{(1)})}{p_b^{(0)}(\Phi^{(0)})}-\frac{p_s^{(1)}(\Phi^{(1)})}{p_s^{(0)}(\Phi^{(0)})}
\right]\,,
\end{align}
so the likelihood ratio as the optimal observable itself is modified order-by-order.

If this likelihood ratio is to be well-defined and IRC safe, we now see immediately some requirements it must satisfy.  First and foremost, the leading order distribution ratio
\begin{align}
\hat {\cal L}^{(0)}(\Phi) &=\frac{p_b^{(0)}(\Phi^{(0)})}{p_s^{(0)}(\Phi^{(0)})}\,,
\end{align}
must be well-defined and finite on leading-order phase space $\Phi^{(0)}$ if this is to make any sense whatsoever.  Starting at next-to-leading order, there are terms in the likelihood ratio that mix leading and next-to-leading order distributions which correspondingly live on different phase spaces, $\Phi^{(0)}$ and $\Phi^{(1)}$.  If these are to be consistent and well-defined, we must map next-to-leading order phase space $\Phi^{(1)}$ onto leading order phase space $\Phi^{(0)}$ in an IRC safe way.  We know how to do this: we can just use a sequential recombination jet algorithm to cluster the particles at next-to-leading order into the correct number of particles that live on leading order phase space.  In the strict soft and collinear limits, IRC safety requires that this map is unambiguous and universal, but away from those limits, algorithm dependence will be introduced.  We will not study effects of different algorithms here because they will require a complete and detailed calculation at next-to-leading order.

Regarding the functional form of the next-to-leading order contribution to the likelihood ratio in the square brackets, there are two distinct parts.  One is the integral
\begin{align}
\int d\Phi^{(1)}\left( p_s^{(1)}(\Phi^{(1)})- p_b^{(1)}(\Phi^{(1)})\right)\,,
\end{align}
which is simply a number that ensures proper normalization.  As such, we won't consider it further here.  The more interesting piece is the difference of the background to signal distribution ratios:
\begin{align}
\frac{p_b^{(1)}(\Phi^{(1)})}{p_b^{(0)}(\Phi^{(0)})}-\frac{p_s^{(1)}(\Phi^{(1)})}{p_s^{(0)}(\Phi^{(0)})}\,,
\end{align}
which has explicit dependence on next-to-leading order phase space and so will be doing the heavy lifting of discrimination at this order.  

Let's dig a bit deeper into these ratios of distributions on phase space.  We can immediately write down explicit expressions for these ratios from our understanding of soft and collinear factorization from \Sec{sec:softcollfact}.  From our results on soft and collinear factorization, we can express the ratio of next-to-leading to leading order distributions in the following way: 
\begin{align}
\frac{p^{(1)}(\Phi^{(1)})}{p^{(0)}(\Phi^{(0)})} = \sum_{\text{NLO partons }i,j} \frac{(4\pi)^2}{s_{ij}}\,P_{(ij)\to ij}(z)  - \sum_{\text{LO partons }i,j}(4\pi)^2{\mathbf T}_i\cdot{\mathbf T}_j\,\frac{s_{ij}}{s_{ik}s_{kj}}+\text{non-singular}\,.
\end{align}
There are a few things going on, so let's explain.  The first sum encodes all hard, collinear emissions, as described by the invariant mass of particles $i$ and $j$, $s_{ij}$, and corresponding splitting functions $P_{(ij)\to ij}(z)$.  (We should also explicitly subtract the soft and collinear limit to avoid double counting with soft emissions, but we will see in a second that such terms cancel anyway.)  The second sum encodes soft gluon emission $k$ from every dipole $i,j$ at leading order.  Finally, furthest at right, terms in the ratio of distributions that are non-singular on phase space have not been explicitly written.

Let's now apply this factorization to understanding the difference between the background and signal ratio distributions.  First, there is a significant cancelation we can take account of.  Earlier, we had assumed that the signal and background states were indistinguishable and as such have the same particle content.  As such, at next-to-leading order, the collinear emission contributions of the signal and background are identical, and therefore cancel in the difference.  Collinear emissions, then, apparently do not modify the likelihood ratio, at least through next-to-leading order.  Soft emissions are a different case because of sensitivity to the color flow along and connection amongst dipoles at leading order.  At any rate, we can compactly represent the difference of the ratio as
\begin{align}\label{eq:likenlocolor}
\frac{p_b^{(1)}(\Phi^{(1)})}{p_b^{(0)}(\Phi^{(0)})}-\frac{p_s^{(1)}(\Phi^{(1)})}{p_s^{(0)}(\Phi^{(0)})} = -(4\pi)^2\sum_{\text{LO partons }i,j}\left({\mathbf T}^{(b)}_i\cdot{\mathbf T}^{(b)}_j-{\mathbf T}^{(s)}_i\cdot{\mathbf T}^{(s)}_j\right)\,\frac{s_{ij}}{s_{ik}s_{kj}}+\text{non-singular}\,.
\end{align}
Here, the superscripts on the color matrices represent whether they are from signal or background.  We see then, completely generally, that the likelihood ratio at next-to-leading order is explicitly sensitive to differences in the flow of color from signal and background.

We could continue, and calculate the signal and background distributions of the likelihood ratio order-by-order in $\alpha_s$, and then further combine those results into the ROC curve, but we won't do that here.  Our observation in \Eq{eq:likenlocolor} that differences in color flow are the dominant contribution to the likelihood ratio at next-to-leading order will be sufficient to deeply understand $H\to b\bar b$ versus $g\to b\bar b$ discrimination.

\subsection{$H\to b\bar b$ versus $g\to b\bar b$ and Sensitivity to Color Representation}

For the problem at hand, we will make a number of simplifying assumptions so that our analysis is contained.  The decay width the of Higgs boson is very small compared to its mass, $\Gamma_H\approx 3$ MeV compared to $m_H\approx 125$ GeV, so we can work in the narrow width approximation in which there is no quantum interference between signal and background. We will work in the highly-boosted limit, in which the energy of the Higgs boson $E$ is much larger than its mass, $m_H$.  As such, the decay products of the Higgs boson will be highly collimated, and we can work in the collinear limit to describe the splitting of the gluon to bottom quarks.  Additionally, in this highly boosted limit, the rest of the event is effectively an ``infinite'' angle away from the Higgs, and so can be considered as a single particle off of which the Higgs recoils.  For an actual calculation, we will also need to use a jet algorithm and impose a jet radius about the Higgs and gluon, but we won't worry about that detail here.  We will just assume that the jet radius is sufficiently large such that all decay products of the Higgs are comfortably contained within the jet.

Before doing any analysis or performing any calculation, you should always ask yourself what result you expect.\footnote{This advice is attributed to John Wheeler and was so important that he referred to it as ``Wheeler's First Moral Principle'' \cite{taylor1992spacetime}.}  Of course this can't necessarily be quantitative (for that is the point of a calculation!), but we can understand some limiting behavior and our results must respect that.  Let's consider discrimination between $H\to b\bar b$ and $g\to b\bar b$ jets in the truly infinite boost limit, for $E/m_H\to \infty$.  In this limit, the decay products of the Higgs boson are exactly collinear.  Further, the Higgs boson is a color singlet, and so there is no net color observable in this Higgs.  As such, there is no possible color dipole that can emit soft gluons a finite angle from the Higgs.  So, a Higgs jet in the infinite boost limit just consists of a single, collimated core of energy.  By contrast, the background gluon is a color octet, and so even in the infinite boost limit, its $b\bar b$ splitting products form a collinear color octet state.  This then has a net color charge and therefore can emit soft gluons at finite angles.  So, if you observe even a {\it single} soft gluon at any finite angle from the infinitely-boosted jet core, then that jet was necessarily initiated by a gluon.  Thus, we expect that in the infinite boost limit, there is perfect discrimination of color-singlet Higgs bosons from color-octet gluons.

In an actual, realistic experiment like ATLAS or CMS at the LHC this thought experiment is a bit unrealistic because there are many other sources of radiation that might happen to deposit in the jet of interest, such as errant radiation off of other produced jets, underlying event radiation, initial state radiation, etc., so no perfect discrimination is ever expected.  Nevertheless, we do expect that discrimination power improves as jet energy increases for the reasons explained above, which is something we can then test in simulated data with our machine learning algorithm, if so desired.

\subsubsection{Discrimination at Leading Order}

Let's now put all of this esoterica to use and actually calculate likelihoods and discrimination power for this problem.  Working in the collinear limit, we know that relevant two-body phase space is two dimensional, and can be expressed as the invariant mass $s$ of the particles and the energy fraction $z$ of one of them.  Because we are searching for a Higgs boson, we know its mass and so fix $s = m_H^2$ in this approximation, because if the mass of the jet is not $m_H$, there's no way that that jet is a Higgs.  Then, there is but a single variable on leading order phase space to consider, the energy fraction $z$.

Because the Higgs boson decays on-shell and is a scalar, spin-0 particle, the probability distribution of the energy fraction of one of its decay products is flat in $z$:
\begin{align}
p_H^{(0)}(z)  =1\,,
\end{align}
which is normalized on $z\in[0,1]$.  We are being slightly cavalier here, because there should be lower and upper bounds on the energy fraction as imposed by the constraint on the invariant mass.  A truly 0 energy particle has no invariant mass with any other particle, however, in the high energy limit, this restriction is suppressed by $m_H^2/E^2$, which we can safely ignore for simplicity.  Correspondingly, for the background process $g\to b\bar b$, we can use the corresponding splitting function $P_{g\to q\bar q}(z)$, appropriately normalized on $z\in[0,1]$:
\begin{align}
p_g^{(0)}(z) = \frac{3}{2}\left[
z^2+(1-z)^2
\right]\,.
\end{align}
Though the bottom quark is massive, $m_b \approx 4.2$ GeV, this is much less than the mass of the Higgs boson, $m_H \approx 125$ GeV, and so is safely ignorable for our purposes here.

Then, with these distributions, the leading-order likelihood ratio is
\begin{align}
\hat {\cal L}^{(0)}(z) = \frac{p_g^{(0)}(z)}{p_H^{(0)}(z)} = \frac{3}{2}\left[
z^2+(1-z)^2
\right]\in \left[\frac{3}{4},\frac{3}{2}\right]\,,
\end{align}
and we clearly see why we put the signal Higgs in the denominator.  This likelihood ratio has a very small dynamic range, just a factor of 2 from minimum to maximum, so purifying a sample of Higgs bosons only through measuring the energy fraction of one of the decay products is not possible.  Nevertheless, for good exercise, let's continue on, and calculate the ROC curve.  To do this, we need the cumulative distributions of the likelihood on signal and background:
\begin{align}
\Sigma_H^{(0)}({\cal L}) &= \int_0^1 dz\, \Theta\left({\cal L} -\frac{3}{2}\left[
z^2+(1-z)^2
\right] \right) = \sqrt{\frac{4}{3}{\cal L} - 1}\,,\\
\Sigma_g^{(0)}({\cal L}) &= \int_0^1 dz\, \frac{3}{2}\left[
z^2+(1-z)^2
\right] \Theta\left({\cal L} -\frac{3}{2}\left[
z^2+(1-z)^2
\right] \right) = \left(
\frac{1}{2}+\frac{{\cal L}}{3}
\right)\sqrt{\frac{4}{3}{\cal L} - 1}\,.
\end{align}

The leading order ROC curve is then
\begin{align}
\text{ROC}^{(0)}(x) = \Sigma_g^{(0)}\left(
{\Sigma_H^{(0)}}^{-1}(x)
\right) = \frac{3}{4}x+\frac{x^3}{4}\,.
\end{align}
The AUC is correspondingly very large
\begin{align}
\text{AUC}^{(0)} = \int_0^1 dx\, \text{ROC}^{(0)}(x) = \frac{7}{16} = 0.4375\,,
\end{align}
close to completely random AUC = 0.5, and so at leading order there is very little discrimination power.  What gives?  We had argued that discrimination power should get arbitrarily good at high energies, but that argument required radiation off of the leading-order dipoles.  Radiation only first arises at next-to-leading order, so let's go to higher order and see what we find!

\subsubsection{Discrimination at Next-to-Leading Order and Beyond}

At next-to-leading order, our signal jets now consist of $H\to b\bar b g$ final states and background jets consist of $g\to b\bar b g$ final states.  As just mentioned, the additional radiated gluon will be doing the heavy lifting of discrimination at this order, which we will show shortly.  Before that, however, I want to comment on something that we mentioned in the general expression for the likelihood ratio at next-to-leading order.  We had found that we needed an IRC safe map or projection ${\cal P}$ from next-to-leading to leading order phase space: ${\cal P}:\Phi^{(1)}\to \Phi^{(0)}$.  Not only must this map be IRC safe, but further, the particle content at next-to-leading order must be consistently mapped to particles at leading order.  For the case at hand, leading order jets are defined by having both a bottom $b$ and anti-bottom $\bar b$ quark, and so our projection ${\cal P}$ must respect this.  To do this requires enforcing that the $b$ and $\bar b$ are resolved at next-to-leading order (i.e., with sufficient energy and a sufficiently large angle from each other).  As such, the projection ${\cal P}$ always clusters the gluon with one of the bottoms, and never the bottoms together.

This prescription is fine and practical at next-to-leading order, but attempting its naive application at higher orders, one finds that flavor identification of the clustered subjets as bottom quarks is not IRC safe itself \cite{Banfi:2006hf}.  Soft quark emission at next-to-next-to-leading order can spoil simple flavor sums of clustered subjets, and so a more complicated procedure must be introduced to ensure calculability.  Constructing and calculating with IR (and possibly C) safe flavor algorithms is a rather hot topic now \cite{Caletti:2022hnc,Caletti:2022glq,Czakon:2022wam,Gauld:2022lem,Caola:2023wpj}, but we can ignore these subtleties at next-to-leading order, and just assume that whatever we do can be absorbed into a more principled approach at higher orders.

\subsubsection{Likelihood Ratio in the Soft Limit}

With those caveats out of the way, let's determine the likelihood ratio for $H\to b\bar b$ versus $g\to b\bar b$ at next-to-leading order.  Let's focus on the component of the likelihood that is doing the heavy lifting for us, namely, the difference of ratios of distributions:
\begin{align}
\frac{p_g^{(1)}(\Phi^{(1)})}{p_g^{(0)}(\Phi^{(0)})}-\frac{p_H^{(1)}(\Phi^{(1)})}{p_H^{(0)}(\Phi^{(0)})} = -(4\pi)^2\sum_{\text{LO partons }i,j}\left({\mathbf T}^{(g)}_i\cdot{\mathbf T}^{(g)}_j-{\mathbf T}^{(H)}_i\cdot{\mathbf T}^{(H)}_j\right)\,\frac{s_{ij}}{s_{ik}s_{kj}}+\text{non-singular}\,.
\end{align}
To evaluate this, we need to calculate the various color matrix products, but even before that, we need to establish what the sum over leading order particles is.  At leading order there are, of course the bottom quarks, $b$ and $\bar b$, but additionally, there is the ``particle'' that represents the rest of the event, off of which the Higgs or gluon recoils.  We will call this particle $\bar n$ and, in the collinear limit, the direction of $\bar n$ is exactly opposite to that of the jet.  Further, because the Higgs is a color singlet, the color of $\bar n$ in its event is 0, ${\mathbf T}_{\bar n}^{(H)} = 0$.  Then, expanding out the sum over leading order partons, this contribution to the likelihood ratio is
\begin{align}
\frac{p_g^{(1)}(\Phi^{(1)})}{p_g^{(0)}(\Phi^{(0)})}-\frac{p_H^{(1)}(\Phi^{(1)})}{p_H^{(0)}(\Phi^{(0)})} &= -(4\pi)^2\left({\mathbf T}^{(g)}_b\cdot{\mathbf T}^{(g)}_{\bar b}-{\mathbf T}^{(H)}_b\cdot{\mathbf T}^{(H)}_{\bar b}\right)\,\frac{2s_{b\bar b}}{s_{bk}s_{k\bar b}}\\
&\hspace{1cm}-(4\pi)^2{\mathbf T}^{(g)}_b\cdot{\mathbf T}^{(g)}_{\bar n}\,\frac{2s_{b\bar n}}{s_{bk}s_{k\bar n}}-(4\pi)^2{\mathbf T}^{(g)}_{\bar b}\cdot{\mathbf T}^{(g)}_{\bar n}\,\frac{2s_{\bar b\bar n}}{s_{\bar bk}s_{k\bar n}}+\text{non-singular}\,.\nonumber
\end{align}

\begin{figure}[t!]
\begin{center}
\includegraphics[height=3cm]{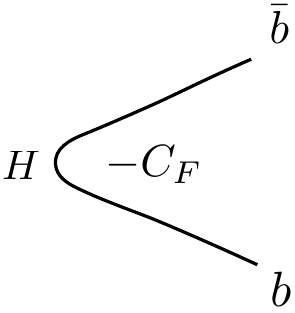}
\hspace{2cm}
\includegraphics[height=3cm]{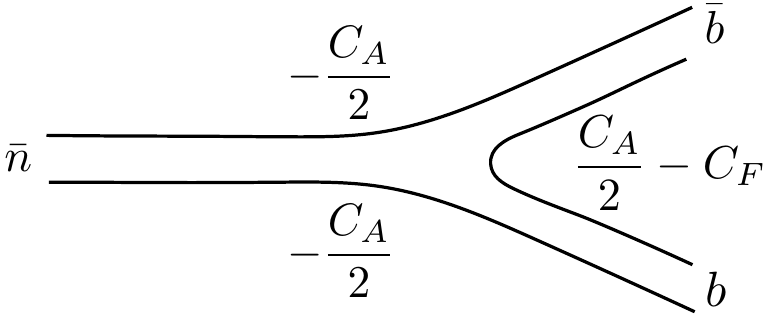}
\caption{\label{fig:colorflow}
Illustrations of color flow between bottom quarks in Higgs decay (left) and in gluon splitting (right).  Because the gluon is a color octet, there are dipole connections to the rest of the event, represented by the $\bar n$ direction.  The lines indicate color matrix products between the particles at the endpoints.  Figure adapted from \InRef{Buckley:2020kdp}.
}
\end{center}
\end{figure}

Let's calculate some color matrix products!  These cases are simple enough that we just need to use color conservation.  Because the Higgs is a color singlet, the sum of its color matrices vanishes:
\begin{align}
\left({\mathbf T}^{(H)}_b+{\mathbf T}^{(H)}_{\bar b}\right)^2 = 0 = 2 C_F+2{\mathbf T}^{(H)}_b\cdot{\mathbf T}^{(H)}_{\bar b}\,,
\end{align}
because the color of a bottom quark is $C_F = {\mathbf T}_b^2$.  The gluon is just a bit more complicated because there is an additional color direction.  Nevertheless, color conservation requires
\begin{align}
\left(
{\mathbf T}^{(g)}_{b} + {\mathbf T}^{(g)}_{\bar b} + {\mathbf T}^{(g)}_{\bar n}
\right)^2 = 0 = 2C_F + C_A+2{\mathbf T}^{(g)}_b\cdot{\mathbf T}^{(g)}_{\bar b}+2{\mathbf T}^{(g)}_b\cdot{\mathbf T}^{(g)}_{\bar n}+2{\mathbf T}^{(g)}_{\bar b}\cdot{\mathbf T}^{(g)}_{\bar n}\,,
\end{align}
where we have used that the color of the recoiling event must be $C_A = {\mathbf T}^2_{\bar n}$.  Additionally, we have the constraints that the color products of the bottom and anti-bottom with $\bar n$ are equal ${\mathbf T}^{(g)}_b\cdot{\mathbf T}^{(g)}_{\bar n} = {\mathbf T}^{(g)}_{\bar b}\cdot{\mathbf T}^{(g)}_{\bar n}$, and that the color of the gluon is
\begin{align}
\left({\mathbf T}^{(g)}_b+{\mathbf T}^{(g)}_{\bar b}\right)^2 = C_A = 2 C_F+2{\mathbf T}^{(g)}_b\cdot{\mathbf T}^{(g)}_{\bar b}\,.
\end{align}
These results can then be simplified to evaluate all necessary color matrix products.  These are summarized in \Fig{fig:colorflow}, and then the expression for the difference of the ratio of distributions becomes
\begin{align}
\frac{p_g^{(1)}(\Phi^{(1)})}{p_g^{(0)}(\Phi^{(0)})}-\frac{p_H^{(1)}(\Phi^{(1)})}{p_H^{(0)}(\Phi^{(0)})} &= -(4\pi)^2C_A\left(\frac{s_{b\bar b}}{s_{bk}s_{k\bar b}}-\frac{s_{b\bar n}}{s_{bk}s_{k\bar n}}-\frac{s_{\bar b\bar n}}{s_{\bar bk}s_{k\bar n}}\right)+\text{non-singular}\,.
\end{align}
Thus, the likelihood is explicitly sensitive to color of the gluon, $C_A$.

It is useful to further manipulate this expression to isolate the geometric dependence in the kinematic factor in parentheses.  Recall that, in the collinear limit, the invariant mass of particles 1 and 2 is
\begin{align}
s_{12} = z_1z_2 E^2\theta_{12}^2\,,
\end{align}
where $z_1$ is the energy fraction of particle $1$, $E$ is the total jet energy, and $\theta_{12}$ is their pairwise angle.  Additionally, we have to evaluate invariant masses with the rest of the event, e.g., $s_{b\bar n}$.  By assumption in the high-boost limit, the $b$ and $\bar n$ are back-to-back, and so this invariant mass is
\begin{align}
s_{b\bar n} = 4E_{b}E_{\bar n}\,.
\end{align}
With these results, the contribution to the likelihood can be written as
\begin{align}
\frac{p_g^{(1)}(\Phi^{(1)})}{p_g^{(0)}(\Phi^{(0)})}-\frac{p_H^{(1)}(\Phi^{(1)})}{p_H^{(0)}(\Phi^{(0)})} &= \frac{(4\pi)^2C_A}{z_k^2E^2}\,\frac{\theta_{bk}^2+\theta_{\bar bk}^2-\theta_{b\bar b}^2}{\theta_{bk}^2\theta_{\bar bk}^2}+\text{non-singular}\,.
\end{align}
In the collinear limit, we can then use the law of cosines to exchange $\theta_{b\bar b}^2$ for an opening angle $\phi$ in the triangle with vertices at the $b$, $\bar b$, and gluon $k$.  We have
\begin{align}
\theta_{b\bar b}^2 = \theta_{bk}^2+\theta_{\bar bk}^2 - 2\theta_{bk}\theta_{\bar b k}\cos\phi\,,
\end{align}
so that the difference can be nicely and compactly expressed as
\begin{align}
\frac{p_g^{(1)}(\Phi^{(1)})}{p_g^{(0)}(\Phi^{(0)})}-\frac{p_H^{(1)}(\Phi^{(1)})}{p_H^{(0)}(\Phi^{(0)})} &= \frac{2(4\pi)^2C_A}{z_k^2E^2}\,\frac{\cos\phi}{\theta_{bk}\theta_{\bar bk}}+\text{non-singular}\,.
\end{align}

\subsubsection{Discrimination Power at Next-to-Leading Order}

Let's now put this all together and write the expression for the likelihood ratio for $H\to b\bar b$ versus $g\to b\bar b$ discrimination through next-to-leading order:
\begin{align}
\hat {\cal L}(\Phi) =  \frac{3}{2}\left(
z_b^2+z_{\bar b}^2
\right)\left(
1 + (4\pi)^2\frac{\alpha_s}{2\pi}\frac{2C_A}{z_k^2E^2}\,\frac{\cos\phi}{\theta_{bk}\theta_{\bar bk}}+\text{non-singular}+{\cal O}(\alpha_s^2)
\right)\,.
\end{align}
For the leading order likelihood, we have expressed the gluon splitting function $P_{g\to b\bar b}(z)$ as a function of the energy fractions of the bottom and anti-bottom, $z_b$ and $z_{\bar b}$, respectively.  At next-to-leading order, note that this is explicitly dimensionful, with the factor of squared jet energy in the denominator.  However, this is canceled by the corresponding factor of squared energy from three-body collinear phase space of \Eq{eq:collpsN}, where we also include the $\delta$-function $\delta(s_{b\bar b k} - m_H^2)$ to constrain the jet mass appropriately.  Further, when expressed in appropriate coordinates, three-body collinear phase space is flat in the squared angles of the emitted gluon to both bottom quarks and the squared energy of the soft gluon, $d\Phi^{(1)} \propto d\theta_{bk}^2\, d\theta_{\bar b k}^2\, dz_k\, z_k$ (see Exercise \ref{ex:threebodps} of \Sec{sec:qftfundy}), and so there is no collinear divergence of the likelihood ratio at next-to-leading order.  There is only a single soft divergence which is logarithmic, scaling as $dz_k / z_k$ on phase space.

\begin{figure}[t!]
\begin{center}
\includegraphics[width=4cm]{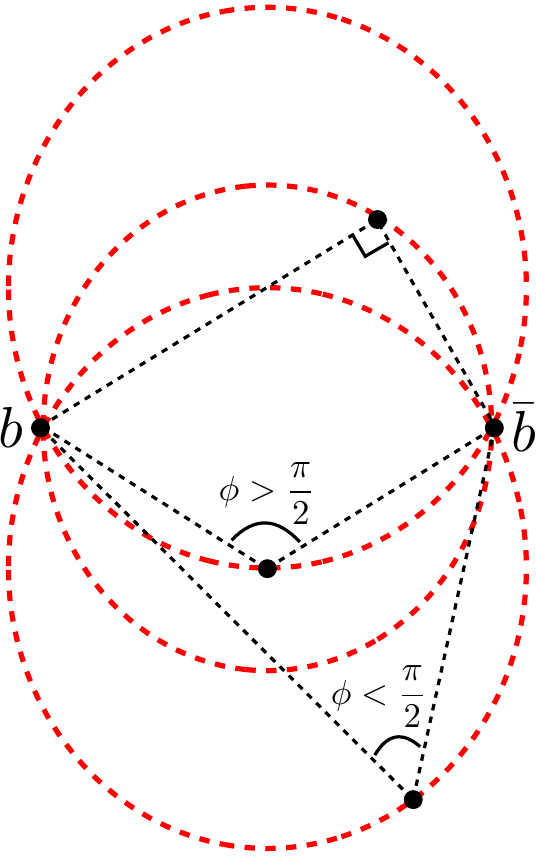}
\caption{\label{fig:azicolor}
Regions selected by the structure of the next-to-leading order contribution to the likelihood ratio, with respect to the $b\bar b$ dipole.  Between the $b\bar b$ dipole, radiation is suppressed ($\phi > \pi/2$ and so $\cos\phi < 0$) while outside the dipole, radiation is enhanced ($\phi < \pi/2$ and so $\cos\phi > 0$).}
\end{center}
\end{figure}

The dependence on the azimuthal angle $\phi$ in the term at next-to-leading order can be geometrically understood by drawing a picture of the $b\bar b$ dipole structure on the celestial sphere, as illustrated in \Fig{fig:azicolor}.  On this figure, we have drawn characteristic field value contours of a dipole, which are also contours of the cosine of the opening angle $\phi$ of the sides of the triangle formed from the emitted gluon and the $b$ and $\bar b$.  If the gluon lies ``inside'' the region of the dipole, then the angle $\phi > \pi/2$ and so $\cos\phi < 0$, which decreases the value to the likelihood here.  The Higgs is most likely to radiate between the dipole by its color singlet structure, and indeed, small values of the likelihood are more Higgs-like.  By contrast, if the emitted gluon lies outside the dipole region, then $\phi < \pi/2$ and so $\cos\phi > 0$, which increases the value of the likelihood.  By their color octet structure, gluons are more likely to emit at wide angles, and again, large values of the likelihood are more gluon-like.  Various historical jet substructure observables have been constructed to be sensitive to this specific radiation pattern as well, exhibiting much the same dependence as discussed here; see, e.g., \Refs{Thaler:2008ju,Almeida:2008yp,Gallicchio:2010sw,Hook:2011cq,Buckley:2020kdp}.

We can estimate the range of the likelihood now with some understanding about the energy fraction of the emitted gluon, $z_k$.  First focus on the Higgs.  The Higgs is a color singlet, and so will only radiate in and around the dipole region, so that the angle between the gluon and either the $b$ or $\bar b$ is of the order of the angle between the $b\bar b$ itself: $\theta_{bk}\sim \theta_{\bar b k} \sim \theta_{b\bar b}$.  We can estimate this angle then, through the mass constraint on the Higgs:
\begin{align}
m_H^2 = z_bz_{\bar b}E^2\theta_{b\bar b}^2 \approx \frac{1}{4}E^2\theta_{b\bar b}^2\,,
\end{align}
or that
\begin{align}
\theta_{b\bar b} \approx \frac{2m_H}{E}\,.
\end{align}
Next, for the gluon emission to still be perturbative and resolved, its relative transverse momentum to either $b$ or $\bar b$ must be perturbative, above about $1$ GeV, or so.  Let's call this transverse momentum cut $k_\perp$.  This relative transverse momentum is then
\begin{align}
k_\perp \lesssim z_k E \theta_{b\bar b} \approx 2z_k m_H\,,
\end{align}
or that 
\begin{align}
z_k \gtrsim \frac{k_\perp}{2m_H}\,.
\end{align}
Now, using this result, we can estimate the lower bound of the likelihood ratio as measured on Higgs decays:
\begin{align}
\hat {\cal L}(\Phi) \gtrsim \frac{3}{4}\left(
1-\frac{\alpha_s}{2\pi}2C_F\frac{2m_H}{k_\perp}+\cdots
\right)\,,
\end{align}
where we have used that $\cos\phi < 0$ for the region inside the dipole, and dropped constant factors for this estimate.  Perturbation theory assumes an expansion about $\alpha_s = 0$ and so formally here $\alpha_s$ is arbitrarily close to 0.  However, in practice, we set $\alpha_s$ to its non-zero, measured value, which in this case would be measured at the Higgs mass because this is an on-shell decay for which $\alpha_s(m_H)\approx 0.11$.  Thus, the likelihood estimate can potentially become negative depending on the precise values for the cutoff scale $k_\perp$ and the numerical prefactors.  Such behavior would obviously be unphysical, and is just telling us that we need even higher-order corrections to correctly describe the region of small likelihood.  Nevertheless, we see that starting at next-to-leading order, there is really no non-zero lower bound on the likelihood.

What about the upper bound?  The upper bound is set by the dynamics of the emitted gluon itself and the corresponding estimate for the energy fraction must be modified.  Now, because the gluon is a color octet, the soft radiated gluon can be emitted at any angle up to the jet radius $R$ (so we actually see the radiation in the jet).  Then, from the perturbative transverse momentum cut the minimal energy fraction $z_k$ is
\begin{align}
k_\perp \lesssim z_k E R\,,
\end{align}
or that
\begin{align}
z_k \gtrsim \frac{k_\perp}{ER}\,.
\end{align}
The upper bound of the likelihood ratio is therefore
\begin{align}
\hat {\cal L}(\Phi) \lesssim \frac{3}{2}\left(
1+\frac{\alpha_s}{2\pi}2C_A\frac{ER}{k_\perp} + \cdots
\right)\,.
\end{align}
As energy $E$ increases, this grows without bound, for fixed jet radius $R$, regardless of the value of the cutoff scale $k_\perp$.  Combining these results, we therefore observe what we had anticipated from physical considerations at the beginning: at high energies, the difference in color representation of the Higgs and gluon can enable arbitrarily good discrimination, but this is only first observed at next-to-leading order.

\subsection{Two-Prong Identification Without Knowledge of Mass}\label{sec:d2sec}

The fixed-order analysis of the likelihood ratio is clearly very powerful when there is both a known constraint on the mass of the signal jets as well as a fixed and known particle content, both properties of which are explicit in $H\to b\bar b$ identification.  However, there are numerous cases, especially for new particle searches, where it is known that a signature of new physics would be massive jets, but what that mass is or what Standard Model particles the new physics decays to is unknown a priori.  What can we do in that case?

Without explicit knowledge of what the new physics is, we can't just take squared matrix elements from Feynman diagrams as the probability distributions on phase space because we simply do not know them; we need another tactic.  What we can do instead is to parametrize phase space as relevant for the particular signal you wish to be sensitive to by some set of IRC safe observables.  Then, given that set of observables, we can determine the parametric or scaling relationships amongst them according to our expectation of signal and background jets.  This technique is familiar from analyses of effective field theories and establishes a robust {\bf power counting} that provides systematic expansion parameters about idealized limits.  In this section, we will apply this power counting to the problem of discrimination of scale-invariant jets (jets with a single hard core of radiation) from jets with an intrinsic but unknown mass (jets with two hard cores of radiation).  These techniques were developed and fleshed out in \InRef{Larkoski:2014gra}.\footnote{Earlier related work on power counting for multi-differential measurements are \Refs{Walsh:2011fz,Bauer:2011uc,Larkoski:2014tva,Procura:2014cba}.}

Our first task is to determine what set of IRC safe observables we want to use to define the relevant phase space.  These observables should be sufficiently expressive that they take very different values on one- or two-prong jets.  We have already encountered one possible set of observables, the $N$-subjettinesses \cite{Brandt:1978zm,Stewart:2010tn,Thaler:2010tr,Thaler:2011gf,Kim:2010uj} $\tau_1$ and $\tau_2$ (suppressing angular exponent dependence).  For a jet with a single particle, $\tau_1 = \tau_2 = 0$, because there is no radiation displaced from the jet axis, while for a jet with two particles, $\tau_2 = 0$, but $\tau_1 > 0$.  So these observables seem to fit the bill.  However, we will instead work with the two- and three-point energy correlation functions \cite{Banfi:2004yd,Larkoski:2013eya,Moult:2016cvt}, $e_2^{(\beta)}$ and $e_3^{(\beta)}$, defined as
\begin{align}
&e_2^{(\beta)} = \sum_{i<j} z_i z_j\theta_{ij}^\beta\,,
&e_3^{(\beta)} = \sum_{i<j<k} z_i z_jz_k\theta_{ij}^\beta\theta_{ik}^\beta\theta_{jk}^\beta\,.
\end{align}
These observables make no reference to a jet axis and are only sensitive to products of pairwise angles between particles in the jet.  As such, energy correlation functions are also recoil-free and exhibit all of the nice properties we established for quark versus gluon discrimination earlier.  At any rate, analogous to $\tau_1$ and $\tau_2$, both $e_2^{(\beta)}$ and $e_3^{(\beta)}$ vanish on one-particle jets, while $e_3^{(\beta)} = 0$ but $e_2^{(\beta)} > 0$ on two-particle jets.  So these observables will also work to parametrize the desired phase space.

As an amusing historical anecdote, the optimal observable that we will derive for this discrimination problem, called $D_2^{(\beta)}$, was (one of the many) used by experimentalists to test the most exciting result at the LHC since the discovery of the Higgs.  In December 2015, both ATLAS and CMS observed an excess in the di-photon invariant mass distribution at about 750 GeV \cite{cmsdigamma,atlasdigamma,ATLAS:2016gzy,CMS:2016xbb}.  This is the same search, though at a different mass, that was instrumental in discovering the Higgs.  That both experiments observed an excess of about 3$\sigma$ significance at the same mass was extremely enticing, and theorists went crazy over it, producing over 500 (!!!) papers with every possible new physics explanation presented in the following months.\footnote{An amusing ``theory'' of jumping on the bandwagon of any excesses in particle physics data was proposed in \InRef{Backovic:2016xno}, that also analyzed the 750 GeV excess.  For a snapshot of that time, the community response, problems with statistical interpretation, problems with citation count culture, etc., see Jester's blog post and the hundreds of comments at {\it R\'esonaances} here: \url{https://resonaances.blogspot.com/2016/06/game-of-thrones-750-gev-edition.html}.}  However, as more and more data were collected, and the interesting events analyzed in new ways like testing the consistency with the hypothesis that such a new particle should produce two-prong jets, the excess shrunk and had completely disappeared by summer 2016.  Such is the life of a model builder.

\subsubsection{Power Counting One- Versus Two-Prong Discrimination}

\begin{figure}[t!]
\begin{center}
\includegraphics[width=6cm]{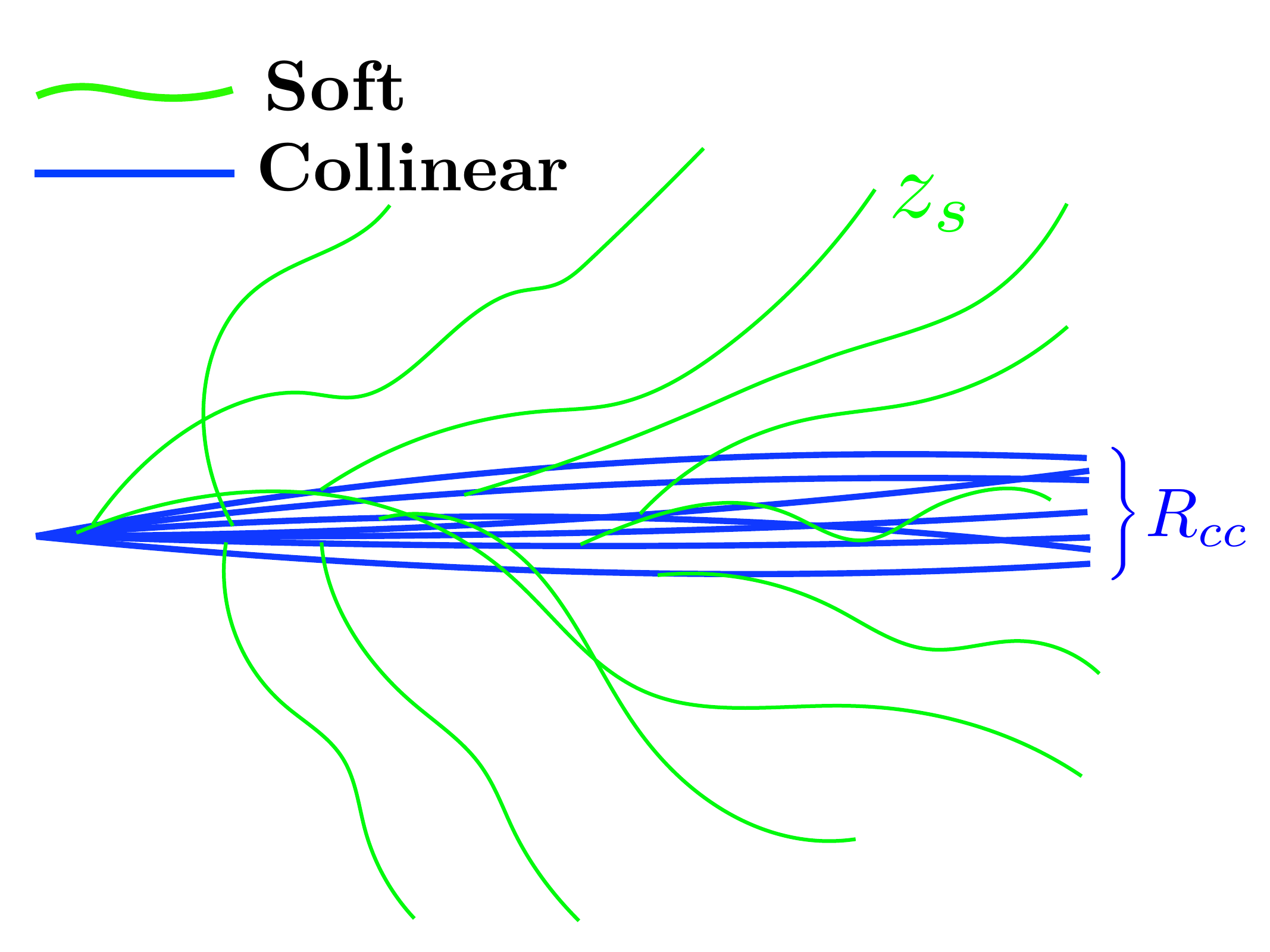}
\hspace{2cm}\includegraphics[width=6cm]{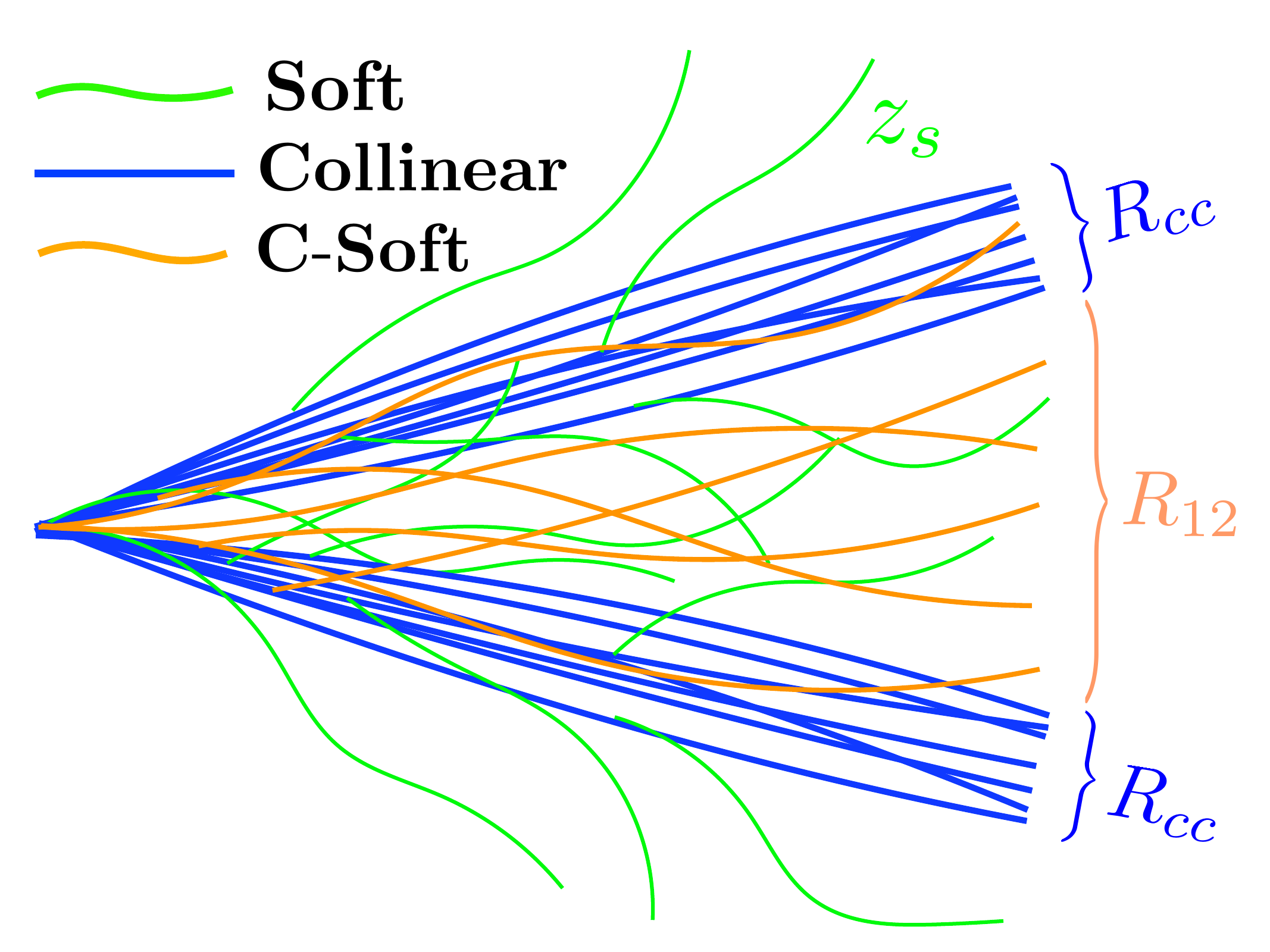}
\caption{\label{fig:onevstwprong}
Illustrations of the dominant modes in one-prong (left) and two-prong (right) jets. Hard, collinear radiation is illustrated in blue, soft, wide-angle radiation in green, and for two-prong jets, soft-collinear radiation emitted from the dipole of the two prongs is orange.  Figure reproduced from Ref.~\cite{Larkoski:2014gra}.
}
\end{center}
\end{figure}

Let's see how this power counting approach, which focuses on the parametric scaling of emissions in the jets, proceeds.  The first thing we need to do is to identify the relevant emission structure, or {\bf modes}, in the jets of interest that we wish to discriminate.  For the problem at hand, of one- versus two-prong jet discrimination, the relevant modes are illustrated in \Fig{fig:onevstwprong}.  Here, by ``one-prong'' jet, we mean, in a parametric limit, that there is a single hard core of collinear radiation $c$ with characteristic angular size $R_{cc}\ll 1$ surrounded by wide-angle, soft radiation $s$ of characteristic energy fraction $z_s\ll 1$.  For a two-prong jet, each prong is high energy and again has characteristic angular size $R_{cc}$, and is surrounded by soft wide-angle radiation of energy fraction $z_s$.  Additionally, there is a new characteristic scale corresponding to the opening angle between the two prongs, $R_{12}$.  Radiation that fills this region about the hard prong dipole is boosted soft radiation, the collinear-soft mode $cs$, that has characteristic angular size $R_{12}$ and energy fraction $z_{cs}\ll 1$.  For two-prong jets to be clearly two-prongy, the angular scales must be hierarchically separated, $R_{cc} \ll R_{12}\ll 1$.  These modes and their corresponding energy and angular scales are presented in the table below.
\begin{align}
\begin{array}{c||c|c}
\text{Modes} & \text{Energy} & \text{Angle}\\
\hline\hline
\text{Soft } s & z_s\ll 1 & 1\\
\text{Collinear } c & 1 & R_{cc} \ll R_{12}\ll 1,\\
\text{Collinear-Soft } cs & z_{cs}\ll 1 & R_{12} \ll 1
\end{array}\nonumber
\end{align}

With these identified modes, we then use them, and only them, to evaluate the two energy correlation functions.  For example, for one-prong jets, we can express its parametric contributions from soft and collinear modes by taking every possible combination of soft and collinear particles in the expression for the observables.  We then plug in the energy fraction and characteristic angle into the formulas for the observables.  Namely, we write
\begin{align}
e_2^{(\beta)} &\sim \sum_{s} z_s^2 + \sum_{s} z_s + \sum_{c}R_{cc}^\beta\,,\\
e_3^{(\beta)} &\sim \sum_{s} z_s^3+\sum_{s} z_s^2 + \sum_{s\,,c} z_s R_{cc}^\beta+\sum_c R_{cc}^{3\beta}\,.
\end{align}
For $e_2^{(\beta)}$, for example, the first term correlates two soft particles, the second term correlates a soft and collinear particle, and the third term correlates two collinear particles.

Next, we can drop terms that are explicitly subleading, e.g., $z_s^2 \ll z_s$.  Further, because we only care about scaling relations, we will drop the explicit summation over particles in the modes, and leave that implicit.  Then, on one-prong jets, the energy correlation functions have the dominant contributions from the soft and collinear modes of:
\begin{align}
&e_2^{(\beta)} \sim z_s +R_{cc}^\beta\,, &e_3^{(\beta)} \sim z_s^2 + z_s R_{cc}^\beta + R_{cc}^{3\beta}
\end{align}
We can now determine the relationship between these observables on one-prong jets.  If $z_s \gtrsim R_{cc}^\beta$, then $e_3^{(\beta)} \sim \left(e_2^{(\beta)}\right)^2$, because then $R_{cc}^{3\beta}$ is subleading.  The other possibility is if $R_{cc}^\beta \gg z_s$ for which we find the relationship $e_3^{(\beta)} \sim \left(e_2^{(\beta)}\right)^3$.  Therefore, one-prong jets live in the region of phase space defined by the dual measurement of the energy correlation functions
\begin{align}
\text{One-Prong Region: } \left(e_2^{(\beta)}\right)^3 \lesssim e_3^{(\beta)} \lesssim \left(e_2^{(\beta)}\right)^2
\end{align}

For two-prong jets, the hard splitting itself fixes the measurement of $e_2^{(\beta)} \sim R_{12}^\beta$, so we only need to determine leading-power scaling for $e_3^{(\beta)}$.  Also, for $e_3^{(\beta)}$ for two-prong jets, it is obvious that at leading-power we must include the correlation of the two hard cores, and then the third particle can be from anywhere else.  This results in the dominant contributions to the energy correlation functions of
\begin{align}
&e_2^{(\beta)} \sim R_{12}^\beta\,, &e_3^{(\beta)} \sim z_sR_{12}^\beta+z_{cs}R_{12}^{3\beta} + R_{cc}^\beta R_{12}^{2\beta}
\end{align}
Now, there are two possibilities to consider.  First, for color-singlets, like Higgs, $Z$, or $W$ bosons in the Standard Model, or, for example, electroweakinos in supersymmetry, wide-angle soft radiation exclusively comes from initial state radiation or other sources external to the physics of decay.  If we assume these are negligible, by, say, going to sufficiently high boost or small jet radius, then the color-singlet two-prong region is
\begin{align}
\text{Color-Singlet Two-Prong Region: }  e_3^{(\beta)} \ll \left(e_2^{(\beta)}\right)^3\,.
\end{align}
That is, at sufficiently high boosts and small jet radii, we have $z_s \to 0$ and so wide-angle soft radiation does not contribute.

This color-singlet resonance is typically the most relevant case (and which we will focus more on below), but it is interesting to consider the case when the two-prong jet is not a color-singlet, or when there is significant contamination radiation in the jet (at high pile-up, say).  In this case, the wide-angle soft radiation may be relevant, if $ R_{12}^{2\beta}\lesssim z_s \ll R_{12}^\beta$, where the upper bound is required to define ``two-prong''.  Because we make no more measurements and assume a relatively large jet radius, $R \sim 1$, this soft radiation actually dominates $e_3^{(\beta)}$ so that the scaling is
\begin{align}
\text{Non-Color-Singlet Two-Prong Region: }  e_3^{(\beta)} \ll \left(e_2^{(\beta)}\right)^2\,,
\end{align}
which is not parametrically different from the one-prong case.  So, for identification of two-prong non-color-singlet jets, other measurements must be made to tease out individual contributions from soft, collinear, and collinear-soft radiation.

\begin{figure}[t!]
\begin{center}
\includegraphics[width=6cm]{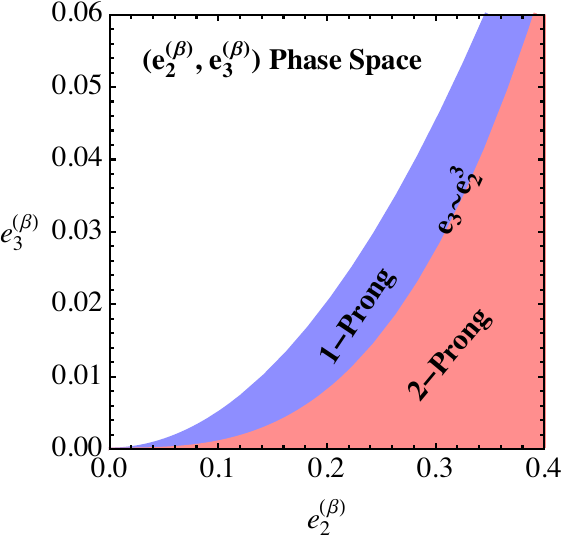}
\hspace{2cm}\includegraphics[width=6cm]{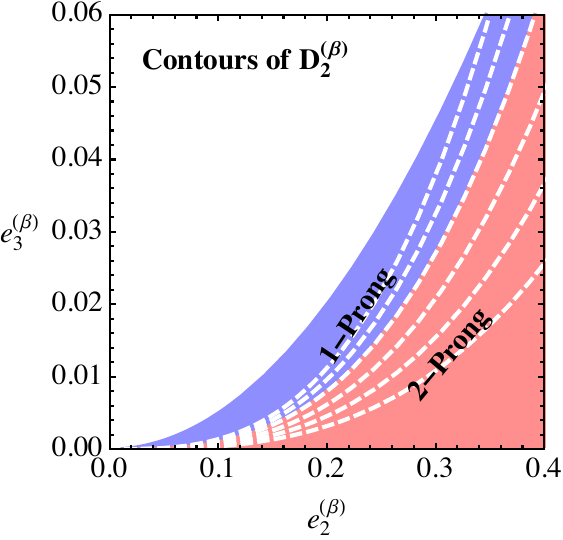}
\caption{\label{fig:d2space}
Plots of the phase space defined by simultaneous measurements of the energy correlation functions $e_2^{(\beta)}$ and $e_2^{(\beta)}$.  Left: Illustration of the regions dominated by one-prong (purple) and two-prong jets (red).  Right: Contours of the observable $D_2^{(\beta)}$ on this phase space. From Ref.~\cite{Larkoski:2014gra}.
}
\end{center}
\end{figure}

Then, for one-prong versus color-singlet two-prong jets, the parametric separatrix is the curve $e_3^{(\beta)} \sim \left(e_2^{(\beta)}\right)^3$, therefore the parametrically optimal discriminant is the ratio of this scaling, an observable called $D_2^{(\beta)}$:
\begin{align}
D_2^{(\beta)} \equiv \frac{e_3^{(\beta)}}{\left(e_2^{(\beta)}\right)^3}\,.
\end{align}
In \Fig{fig:d2space}, we show plots of this phase space, with the regions dominated by one- and two-prong jets in distinct colors.  At right, we also show the contours of constant values of $D_2^{(\beta)}$, which is the only observable whose contours do not cross the parametric scaling boundary $e_3^{(\beta)} \sim \left(e_2^{(\beta)}\right)^3$.  Any other observable as a function of $e_2^{(\beta)}$ and $e_3^{(\beta)}$ will mix these regions and so a cut on another observable will necessarily have reduced discrimination power from a cut on $D_2^{(\beta)}$.

\subsubsection{Boost Invariance of One- Versus Two-Prong Discrimination}

There is another way to observe the optimality of $D_2^{(\beta)}$ as a discriminant for this problem.  While we are working in the high-boost limit, so everything is collimated, our power counting made no specific requirements on angular scales, for example, but only needed parametric relationships.  Therefore, the optimal discrimination observable for one- versus two-prong jets must necessarily be invariant to rescalings of angles in the jets that maintains the parametric relationships.  However, the only physical way to rescale angles in the jets is by a Lorentz boost along the jet's direction.  That is, the optimal discrimination observable separates jets into one- and two-prong classes in a Lorentz invariant way.

Let's see how this works with the functional forms of the energy correlation functions.  First, we note that energy fractions $z_i$ are invariant to boosts along the jet direction in the collimated limit because individual particle energy $E_i$ and the total jet energy $E$ scale in the same way:
\begin{align}
z_i = \frac{E_i}{E} \to \frac{\gamma E_i}{\gamma E} = z_i\,,
\end{align}
where $\gamma$ is the boost factor.  By contrast, the angle between two particles $i,j$, $\theta_{ij}$, scales inversely with the boost factor:\footnote{This simply follows from noting that the invariant mass of two massless momenta, $2p_1\cdot p_2 = E_1 E_2\theta_{12}^2$ in the collinear limit, is Lorentz invariant.}
\begin{align}
\theta_{ij} \to \gamma^{-1} \theta_{ij}\,.
\end{align}
With this property, we then see that under a Lorentz boost, the energy correlation functions scale as
\begin{align}
e_2^{(\beta)} &= \sum_{i<j}z_i z_j\theta_{ij}^\beta \to \sum_{i<j}z_i z_j \gamma^{-1}\theta_{ij}^\beta = \gamma^{-1}e_2^{(\beta)}\,,\\
e_3^{(\beta)} &= \sum_{i<j}z_i z_jz_k\theta_{ij}^\beta\theta_{ik}^\beta\theta_{jk}^\beta \to \sum_{i<j}z_i z_j z_k\gamma^{-1}\theta_{ij}^\beta\gamma^{-1}\theta_{ik}^\beta\gamma^{-1}\theta_{jk}^\beta = \gamma^{-3}e_3^{(\beta)}\,.\nonumber
\end{align}
Therefore, under a Lorentz boost along the jet direction, the ratio
\begin{align}
D_2^{(\beta)} =\frac{e_3^{(\beta)}}{\left(e_2^{(\beta)}\right)^3} \to \frac{\gamma^{-3}e_3^{(\beta)}}{\left(\gamma^{-1}e_2^{(\beta)}\right)^3} = D_2^{(\beta)}\,,
\end{align}
is invariant.  Another way to say this, in relationship to the phase space illustrated at right in \Fig{fig:d2space}, is that jets move along contours of constant $D_2^{(\beta)}$ under Lorentz boosts, never crossing the parametric separatrix.

\subsubsection{Perturbative Calculation Beyond Perturbation Theory}

With $D_2^{(\beta)}$ established as the likelihood ratio for optimal discrimination of one- versus two-prong jets with phase space defined by simultaneous measurement of $e_2^{(\beta)}$ and $e_3^{(\beta)}$, we would now typically move on to calculating its probability distribution on signal and background jets.  The natural way to do this, within our master formula, is through the double-differential master formula
\begin{align}
\frac{d^2\sigma}{de_2^{(\beta)}\,de_3^{(\beta)}} = \int d\Phi\, |{\cal M}|^2\,\delta\left(e_2^{(\beta)} - \hat e_2^{(\beta)}(\Phi)\right)\delta\left(e_3^{(\beta)} - \hat e_3^{(\beta)}(\Phi)\right)\,.
\end{align}
Both $e_2^{(\beta)}$ and $e_3^{(\beta)}$ are IRC safe, so we can calculate this distribution order-by-order in perturbation theory to our heart's content.  Then, our next step would be to marginalize this double differential distribution, slicing out contours of constant $D_2^{(\beta)}$, where
\begin{align}\label{eq:ddifd2}
\frac{d\sigma}{dD_2^{(\beta)}} = \int de_2^{(\beta)}\, de_3^{(\beta)}\, \frac{d^2\sigma}{de_2^{(\beta)}\,de_3^{(\beta)}}\, \delta\left(D_2^{(\beta)} - \frac{e_3^{(\beta)}}{(e_2^{(\beta)})^3}\right)\,.
\end{align}
So far, this seems kosher, just following the rules of probability.  

However, try as you might, this integral \Eq{eq:ddifd2} in fixed-order perturbation theory does not exist \cite{Soyez:2012hv}.  We can see why simply from an analysis of the contours of $D_2^{(\beta)}$ from \Fig{fig:d2space} and the definition of IRC safety.  First, in the calculation of the fixed-order distribution of $D_2^{(\beta)}$, there will be a virtual contribution which lives exclusively at $D_2^{(\beta)} = 0$.  This is negative and divergent, but if $D_2^{(\beta)}$ were IRC safe, then the real contribution would also be divergent at $D_2^{(\beta)} = 0$ with exactly the opposite strength, and then, when summed, divergences would cancel and all would be right with the world.  However, the real contribution to $D_2^{(\beta)}$ actually diverges at {\it all} possible values, because all contours of $D_2^{(\beta)}$ pass through the divergent region of phase space, where $e_2^{(\beta)}, e_3^{(\beta)}\to 0$.  There is simply no consistent prescription to eliminate divergences over all of phase space that should be concentrated exclusively at $D_2^{(\beta)} = 0$.  Thus, $D_2^{(\beta)} $ is not IRC safe, and so its distribution cannot be calculated order-by-order in perturbation theory.  

The problem with this observable is that it is formed from the ratio of two IRC safe observables, and there is a $0/0$ ambiguity if the entire phase space is allowed to contribute.  A simple solution, then, is to make a cut on the denominator observable, here $e_2^{(\beta)}$, that removes the ambiguous region of phase space from consideration.  However, by the approximate scale invariance of QCD at high energies, we know that the singular regions of phase space are always associated with an exponential Sudakov suppression, when all-orders effects are included.  This would then seem sufficient for producing a finite distribution for $D_2^{(\beta)}$, with the caveat that we marginalized over the resummed double differential cross section:
\begin{align}
\frac{d\sigma}{dD_2^{(\beta)}} = \int de_2^{(\beta)}\, de_3^{(\beta)}\, \frac{d^2\sigma^{(\text{resum})}}{de_2^{(\beta)}\,de_3^{(\beta)}}\, \delta\left(D_2^{(\beta)} - \frac{e_3^{(\beta)}}{(e_2^{(\beta)})^3}\right)\,.
\end{align}
Thus, a prediction for the inclusive distribution of $D_2^{(\beta)}$ cannot exist independently of resummation, and the Sudakov factor is vital for finiteness.  This class of observables that are not IRC safe, but whose distribution can be rendered finite by an exponential Sudakov factor are therefore referred to as {\bf Sudakov safe} \cite{Larkoski:2013paa,Larkoski:2015lea,Larkoski:2015kga}. 

We could perform this resummed calculation to study this behavior in $D_2^{(\beta)}$, but the phase space restrictions on the Lund plane imposed by measurement of $e_2^{(\beta)}$ and $e_3^{(\beta)}$ is rather complicated and the non-linear phase space constraints make bookkeeping a bit annoying.  Because we already calculated it, let's instead consider calculating the $N$-subjettiness ratio, $\tau_{2,1}^{(\alpha)} = \tau_2^{(\alpha)}/\tau_1^{(\alpha)}$, which exhibits the same phenomena.  The complete calculation for $D_2^{(\beta)}$ is presented in \InRef{Larkoski:2015kga}, with related calculations in \Refs{Dasgupta:2015lxh,Larkoski:2019nwj}.  For the ratio observable $\tau_{2,1}^{(\alpha)}$, we had already calculated the joint resummed probability distribution of $\tau_1^{(\alpha)}$ and $\tau_2^{(\alpha)}$ in \Sec{sec:multemit}:
\begin{align}
p_q(\tau_1^{(\alpha)},\tau_2^{(\alpha)}) = \left(
\frac{2\alpha_s C_F}{\pi \alpha}
\right)^2\frac{\log \tau_1^{(\alpha)}}{\tau_1^{(\alpha)}}\frac{\log \tau_2^{(\alpha)}}{\tau_2^{(\alpha)}}\,\exp\left[
-\frac{\alpha_s}{\pi}\frac{C_F}{\alpha}\log^2 \tau_2^{(\alpha)}
\right]\,.
\end{align}
The distribution of their ratio is then found from the marginalization integral:
\begin{align}
p_q(\tau_{2,1}^{(\alpha)}) &= \int d\tau_1^{(\alpha)}\, d\tau_2^{(\alpha)}\,p_q(\tau_1^{(\alpha)},\tau_2^{(\alpha)})\,\delta\left(
\tau_{2,1}^{(\alpha)} - \frac{\tau_2^{(\alpha)}}{\tau_1^{(\alpha)}}
\right)\\
&=\sqrt{\frac{\alpha_s C_F}{\alpha}}\frac{1}{\tau_{2,1}^{(\alpha)}}\left(
1+\text{erf}\left(
\sqrt{\frac{\alpha_s C_F}{\pi\alpha}}\log\tau_{2,1}^{(\alpha)}
\right)
\right)
\nonumber\\
&=\sqrt{\frac{\alpha_s C_F}{\alpha}}\frac{1}{\tau_{2,1}^{(\alpha)}} + \frac{2\alpha_s C_F}{\pi\alpha}\frac{\log \tau_{2,1}^{(\alpha)}}{\tau_{2,1}^{(\alpha)}} - \frac{2}{3}\left(
\frac{\alpha_s C_F}{\pi\alpha}
\right)^2\frac{\log^3 \tau_{2,1}^{(\alpha)}}{\tau_{2,1}^{(\alpha)}} + {\cal O}(\alpha_s^3)\,.\nonumber
\end{align}
On the second line, we have written the result in closed form, in terms of an error function $\text{erf}(x)$, and then on the third line, Taylor expanded the result in powers of $\alpha_s$.  This makes the IRC unsafety manifest.  First, in fixed-order perturbation theory, $\tau_{2,1}^{(\alpha)}$ would need two emissions in the jet to be non-zero, and so its non-trivial distribution would start at ${\cal O}(\alpha_s^2)$.  However, the Taylor expansion of this, inclusive, distribution has contributions at orders lower than this, and further, has contributions at fractional powers of $\alpha_s$!  By the structure and rules of Feynman diagrams, there's no way to predict especially the $\sqrt{\alpha_s}$ term, and therefore this distribution cannot be computed in fixed-order perturbation theory, and so $\tau_{2,1}^{(\alpha)}$ (and also $D_2^{(\beta)}$) is IRC unsafe, as argued.

Why the distribution starts at $\sqrt{\alpha_s}$ is rather simple to see from the structure of the Sudakov factor.  The Sudakov factor has the form of a exponential distribution in $\alpha_s \log^2 \tau_2^{(\alpha)}$.  Therefore, when integrated over, $\log\tau_2^{(\alpha)}$ has most of its support displaced from $\tau_2^{(\alpha)} = 0$ by an amount proportional to $1/\sqrt{\alpha_s}$, in the region where the value of the exponent is order-1.

By contrast, let's now consider bounding $\tau_1^{(\alpha)}$ from below by any non-zero $\tau_1^{(\alpha)}>\epsilon > 0$.  We then find
\begin{align}
p_q(\tau_{2,1}^{(\alpha)}|\tau_1^{(\alpha)}>\epsilon ) &= \int d\tau_1^{(\alpha)}\, d\tau_2^{(\alpha)}\,p_q(\tau_1^{(\alpha)},\tau_2^{(\alpha)})\,\delta\left(
\tau_{2,1}^{(\alpha)} - \frac{\tau_2^{(\alpha)}}{\tau_1^{(\alpha)}}
\right)\Theta(\tau_1^{(\alpha)}-\epsilon )\\
&=-\frac{2}{3}\left(
\frac{\alpha_s C_F}{\pi\alpha}
\right)^2\frac{\log^2\epsilon\left( 2\log \epsilon + 3\log \tau_{2,1}^{(\alpha)} \right)}{\tau_{2,1}^{(\alpha)}}+{\cal O}(\alpha_s^3)\,.
\nonumber
\end{align}
As promised, this distribution starts at the right order in perturbation theory and has a Taylor expansion in $\alpha_s$, and so indeed represents the distribution of an IRC safe observable.  The cost of this, though, is logarithmic sensitivity to the value of the cutoff, $\epsilon$, order-by-order, which can be large if $\epsilon$ is small.

Many examples of Sudakov safe observables are now known, with two broad classes of non-IRC safety established.  The first historically (and that we discussed) are those observables that are functions of IRC safe observables and whose contours on a continuously connected set of non-zero measure of the observable pass through the singular or degenerate regions of phase space.  Such observables generically have an expansion in $\sqrt{\alpha_s}$, for the reasons we discussed.  A second class, and in some sense much more interesting than the first, are observables whose expansion in $\alpha_s$ has a non-trivial contribution at $\alpha_s^0$, e.g., \Refs{Larkoski:2014wba,Larkoski:2014bia,Larkoski:2015lea,Dreyer:2018tjj,Cal:2020flh,Mehtar-Tani:2019rrk,Caletti:2022glq,Alipour-Fard:2023prj}.  The leading-order piece of such observables therefore corresponds to an ultraviolet fixed point of renormalization group evolution of QCD.  By asymptotic freedom, $\alpha_s$ vanishes at high energies, and therefore only the leading contribution in such observables survives, resulting in a probe of universality classes of theories as defined by their non-coupling parameters (e.g., rank of gauge group, gauge representation of matter, number of quark flavors, etc.).

\subsection{Linearizing the Function Basis}

Let's put our machine learning hats back on and think about implementing the approaches to the problem of general resonant particle jets versus QCD jets through optimization of parameters of a neural network through gradient descent of an objective function.  In this section, we were able to produce compact, functional forms for the likelihoods (and with a little more work, for the ROC curve) as functions of the phase space variables we worked with.  What made this especially simple was that the dimensionality of the phase spaces were very small (at most, maybe 4 or 5) and so our brain could rather easily parse the physics content of the expressions.  By contrast, if you want a machine to learn the likelihood from (simulated) data, you want that data to be representative of real physics and as such jets contain many particles with a phase space that is high dimensionality (like, say, 100).  Not only can our feeble brains not imagine 100 dimensional spaces well, if you asked the machine for the functional form of the likelihood it constructed, it would return some rather nasty function that was some linear combination of compositions of compositions of the non-linear activation functions.  Just by staring at the explicit functional form of the machine's likelihood ratio, you would never know if there was some simpler approximation represented by only a few variables without significantly more work.

However, as we have emphasized here, as long as we ensure that the input data spans the data space of interest, we are free to represent the data in any way that we want, and the universal approximation theorem guarantees that the machine relaxes to the same functional form of the likelihood ratio.  With a sufficiently large and general basis of functions, the likelihood can be expressed as a linear combination of additive, IRC safe observables that generalize energy correlation functions, by the Stone-Weierstrauss theorem \cite{weierstrass1885analytische,stone1937applications,stone1948generalized}.  On a fixed manifold, this is intimately familiar from harmonic analysis in electromagnetism or quantum mechanics where, using the superposition principle, we find the complete, orthonormal basis of functions on the space of interest (spherical harmonics, Legendre polynomials, etc.).  In jet physics, particle multiplicity fluctuates event-by-event and so on an ensemble of data, jets populate phase spaces of different dimensionality.  To resolve this issue, one can define an overcomplete polynomial basis of IRC safe functions, referred to as energy flow polynomials \cite{Komiske:2017aww}, to accomplish this linearization.  Then, one can simply use linear regression to find the coefficients of the expansion for the given task to approximate the likelihood ratio to arbitrary accuracy.  What could be simpler!  What's more, our brains can make sense of a power spectrum of coefficients of such a linear combination and correspondingly ascribe meaning to it, and can meaningfully filter this representation to compress the data and pull out its most important contributions.  (Our brains do this all the time with Fourier series and transforms whenever we listen to music, for example.)  In the soft and/or collinear limit, many of the polynomials become redundant or linearly dependent, and power counting methods can be used to establish a (less over-)complete basis \cite{Komiske:2019asc,Cal:2022fnm}.

While we won't discuss this linear basis for likelihood estimation more, in many ways this can be the most direct way to actually figure out what the machine is learning \cite{Komiske:2018cqr,Faucett:2020vbu,Romero:2021qlf} because the learned function through linear regression is unambiguous through (pseudo)inversion of a linear equation, and the corresponding calculation is straightforward (if tedious when phase space is high-dimensional) because all observables are themselves IRC safe.

\subsection*{Exercises}

\begin{enumerate}

\item We discussed in detail the fixed-order analysis of the IRC safe discrimination problem $H\to b\bar b$ versus $g\to b\bar b$.  After $H\to b\bar b$ and $H\to WW^*$ ($W^* = $ an off-shell $W$ boson), the third most common decay mode of the Higgs is $H\to gg$, which happens about 10\% of the time.  However, unlike the dominant hadronic decay mode $H\to b\bar b$, there is little hope to observe the $H\to gg$ channel.  With our understanding of the IRC safe likelihood ratio, can we help improve the situation at all?  Consider the binary discrimination problem $H\to gg$ versus $g\to gg$ in highly-boosted, collimated jets.  What are the leading and next-to-leading order likelihood ratios now?  How is this different than $H\to b\bar b$ identification?  What challenges with $H\to gg$ identification do you envision and are they surmountable?

\item  Let's consider one- versus two-prong discrimination again, but let's now define phase space by simultaneous measurement of $\tau_1^{(\alpha)}$ and $\tau_2^{(\alpha)}$.  Perform the corresponding power counting analysis like we did with the energy correlation functions.  What is the expression for the likelihood ratio on this phase space now?  Where do signal and background events live?  Have we seen this observable before?

\item As stated in the beginning of this section, one of the goals of jet substructure (at least historically, e.g., \Refs{Kaplan:2008ie,Thaler:2008ju,Ellis:2009su,Ellis:2009me,Almeida:2010pa,Thaler:2010tr,Jankowiak:2011qa}) was the identification of boosted, hadronically-decaying top quarks.  At leading order, top quarks decay to three massless partons through the weak interaction, and in general, all three partons carry an order-1 energy fraction and are all a similar angle from one another.  Additionally, the top quark has a fixed mass, $m_t \approx 173$ GeV.  Therefore, background jets initiated by light QCD partons must also have this mass and as such must exhibit a strong two-prong structure.  Therefore, top versus QCD discrimination is in general a two- versus three-prong jet discrimination problem, on jets with a fixed mass.

\begin{enumerate}

\item Perform the power counting analysis for this problem on the phase space defined by simultaneous measurement of $\tau_2^{(\alpha)}$ and 3-subjettiness, $\tau_3^{(\alpha)}$.  What is the resulting likelihood observable you find?  Where on this phase space do signal and background events live?

\item Perform the power counting analysis for this problem on the phase space defined by the simultaneous measurement of the two- and three-point correlation functions, $e_2^{(\beta)}$ and $e_3^{(\beta)}$, and the four-point correlation function $e_4^{(\beta)}$, where
\begin{align}
e_4^{(\beta)} = \sum_{i<j<k<l}z_iz_jz_kz_l\theta_{ij}^\beta\theta_{ik}^\beta\theta_{il}^\beta\theta_{jk}^\beta\theta_{jl}^\beta\theta_{kl}^\beta\,.
\end{align}
What is the likelihood in terms of these observables?  Where on phase space do signal and background events live?  See \InRef{Larkoski:2014zma} for more details.

\end{enumerate}

\end{enumerate}

\section{Third Example: Quark vs.~Gluon Jet Discrimination Redux}\label{sec:qvgagain}

In the past two lectures, we had established a couple of properties of the likelihood ratio for different discrimination problems in QCD that were not evident a priori.  First, in our analysis of quark versus gluon discrimination, even from the simple Lund plane picture of scale invariant soft and collinear emissions, we found that sensitivity to multiple emissions improved discrimination power.  Second, in our analysis of one- versus two-prong discrimination useful in new physics searches, we found that the likelihood ratio was not IRC safe, even though it was constructed as a function of IRC safe observables.  Putting these things together, perhaps a fixed-order or even a perturbative analysis is insufficient for actually establishing the form of the likelihood ratio for general binary discrimination problems in QCD.  In this lecture, we will revisit quark versus gluon discrimination, with a reconsideration of the Lund plane and work to identify the likelihood ratio for emissions on any subset of the plane with finite area.  This work will suggest that total hadronic multiplicity is actually {\it the} likelihood ratio for quark versus gluon discrimination, which is not an IRC safe observable.  So, we will have to stretch and strengthen our tools for going outside the friendly confines of Feynman diagrams and our master formula.  This will prepare us for the final lecture in which we cast off any pretense of a perturbative analysis whatsoever.

\subsection{Likelihood on the Lund Plane and Perturbative Multiplicity}

\begin{figure}[t!]
\begin{center}
\includegraphics[width = 8cm]{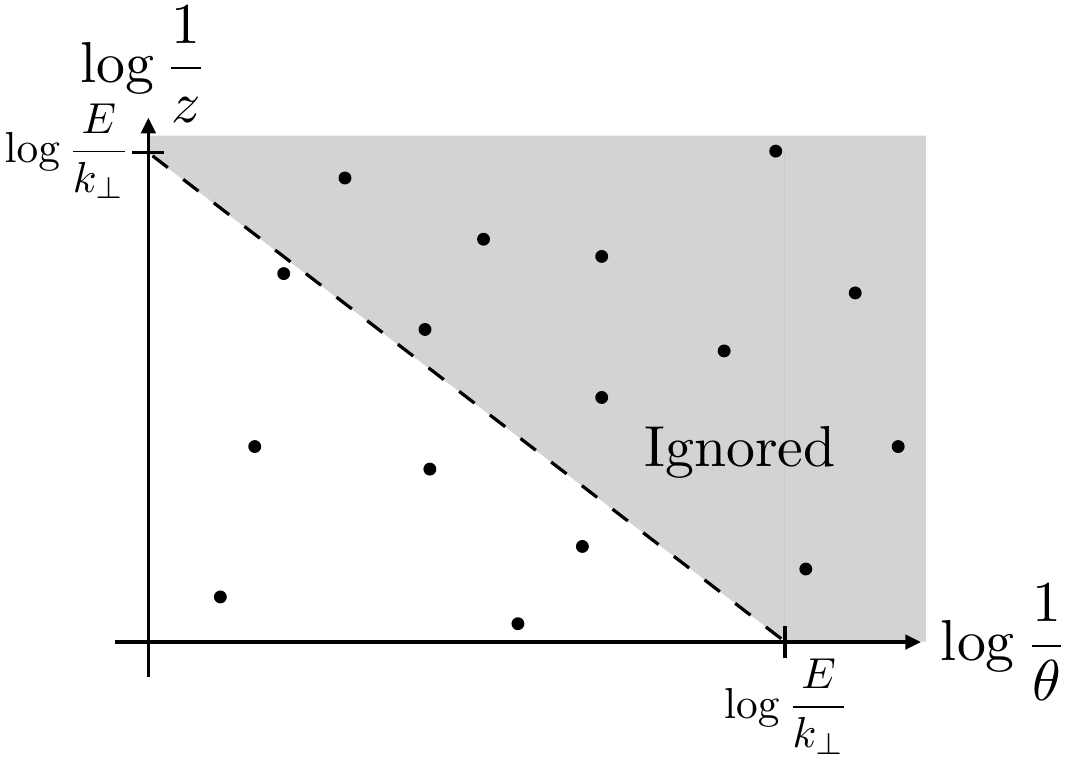}
\caption{\label{fig:lund_multreg}
Region of interest on the (primary) Lund plane of emissions.  Here, emissions are ignored if their relative transverse momentum $k_\perp$ to the jet axis is below a cut value.}
\end{center}
\end{figure}

Let's go back to the Lund plane picture of scale-invariant emissions in a jet and see if, simply from this and the difference in color factors $C_F$ versus $C_A$, we can identify the likelihood observable for quark versus gluon discrimination.  To ensure that intermediate steps of this argument are always finite, let's consider some finite area region of the Lund plane as our region of interest; we only consider emissions in that region as contributing to the likelihood, and all emissions outside of that region are ignored.  An example of this is illustrated in \Fig{fig:lund_multreg}.  Recall that in this simple picture, the density of emissions on the Lund plane is uniform, with
\begin{align}\label{eq:lunddens}
&\rho_q \propto \frac{2\alpha_s}{\pi}\, C_F\,,
&\rho_g \propto \frac{2\alpha_s}{\pi}\, C_A\,,
\end{align}
the emission densities of quark and gluon jets, respectively.  Because the emission density is uniform, the likelihood observable is independent of the area of interest on the Lund plane and correspondingly must be independent of the particular energy fractions $z_i$ and splitting angles $\theta_i$ of the emissions.  Therefore, the only remaining piece of information about the jets that can possibly be used for discrimination on the Lund plane is the number or {\bf multiplicity} of emissions in the region of interest.  The emission multiplicity is therefore the optimal quark versus gluon discriminant within the Lund plane approximation \cite{benchat,Frye:2017yrw}.  For now, we will only consider emissions in the so-called ``primary'' Lund plane, corresponding to direct emission from the hard, initiating particle, ignoring secondary (or later) emissions from already emitted particles.

Let's make this argument that the multiplicity of emissions on the Lund plane is the likelihood ratio more precise.  Given emissions on the Lund plane, the joint probability distribution of anything we could possibly measure on it is
\begin{align}
p(\{z_i,\theta_i\}_{i=1}^n,n) = p(\{z_i,\theta_i\}_{i=1}^n|n)\,p(n) = p(n)\prod_{i=1}^n p(z_i,\theta_i)\,,
\end{align}
where $n$ is the multiplicity of emissions and $\{z_i,\theta_i\}$ are the energy fraction and angle of emission $i$ with respect to the jet axis.  In the middle, we have simply used the definition of conditional probability and at right, we note that all emissions in the Lund plane are independent, so the energy fraction and angles of distinct emissions are all uncorrelated.  Given a finite area of interest on the Lund plane ${\cal A}$, the probability distributions of the energy fractions and angles are
\begin{align}
p(z,\theta) = \frac{\int_{\cal A}\frac{dz'}{z'}\,\frac{d\theta'}{\theta'}\,\delta(z-z')\delta(\theta-\theta')}{\int_{\cal A}\frac{dz'}{z'}\,\frac{d\theta'}{\theta'}}\,.
\end{align}
Because this is a normalized distribution, there is no sensitivity to the color Casimir of the jet and this distribution is not only identical for all emissions, but insensitive to the initiating particle of the jet, as well.  Now, with this construction, we simply take the ratio of this multi-differential distribution on gluon and quark jets to produce the likelihood:
\begin{align}
{\cal L} = \frac{p_g(\{z_i,\theta_i\}_{i=1}^n,n)}{p_q(\{z_i,\theta_i\}_{i=1}^n,n)} = \frac{p_g(n)\prod_{i=1}^n p(z_i,\theta_i)}{p_q(n)\prod_{i=1}^n p(z_i,\theta_i)} = \frac{p_g(n)}{p_q(n)}\,,
\end{align}
exclusively dependent on the multiplicity $n$, with no dependence on the kinematics of the emissions in the jet whatsoever.

Now, we see the necessity of the restriction to a finite area on the Lund plane.  Assuming true scale invariance, of course any jet will emit an infinite number of particles, extending to arbitrarily low scales.  QCD, of course, is not really a conformal or scale-invariant field theory, but is only approximately at relatively large energies, energies well above the scale at which the coupling becomes large and where hadron masses are, at around 1 GeV.  So, for our perturbative analysis to make sense and to indeed be a decent approximation, we should restrict emissions on the Lund plane to lie in the perturbative region.  Emissions become non-perturbative, sensitive to hadronic as opposed to partonic physics, when the relative transverse momentum of the particle to the jet axis falls below about 1 GeV.  In general, however, we will introduce a transverse momentum cut $k_\perp$ and demand that the region of interest on the Lund plane are those emissions with relative transverse momentum above this cut.  This region of interest is shown in \Fig{fig:lund_multreg} for jets with energy $E$.  Note that the cut is a diagonal line through the relationship of the Lund plane coordinates and the transverse momentum of
\begin{align}
k_\perp = zE \theta\,.
\end{align}

\subsubsection{Derivation of Poissonian Distribution}

Calculating the distribution of emission multiplicity is simple using the facts we already know about the Lund plane.  Emissions in this region of interest are independent of one another, with constant probability proportional to the density from \Eq{eq:lunddens}.  We could just skip to the answer, but, as was the case with the Sudakov factor, the derivation of the distribution of multiplicity is illuminating, so we will proceed slowly.  Just like we did for the Sudakov factor in \Sec{sec:sudder}, we will split up this region into $N$ equal area pieces, and then take the $N\to\infty$ limit.  Let's assume that we find $k$ emissions in the region of \Fig{fig:lund_multreg} on a quark jet.  That is, $k$ of the $N$ total pieces have emissions, while $N-k$ have no emissions, and the area of each small piece is
\begin{align}
A_N = \frac{\log^2\frac{E}{k_\perp}}{2N}\,.
\end{align} 
That is, the emission probability into any one piece is the area times the density:
\begin{align}
p_\text{emit} = A_N \rho_q = \frac{\alpha_sC_F}{\pi}\frac{\log^2\frac{E}{k_\perp}}{N}\,.
\end{align}

Then, the probability for there to be $k$ emissions in the region of interest $p_q(k)$ is
\begin{align}
p_q(k)\propto p_\text{emit}^k(1-p_\text{emit})^{N-k} = \left(
\frac{\alpha_sC_F}{\pi}\frac{\log^2\frac{E}{k_\perp}}{N}
\right)^k\left(
1-\frac{\alpha_sC_F}{\pi}\frac{\log^2\frac{E}{k_\perp}}{N}
\right)^{N-k}\,.
\end{align}
With all $N$ pieces of the region of interest identical, these $k$ emissions can of course be located anywhere, so we need a binomial factor to sum over all possible emission combinatorics:
\begin{align}
p_q(k)={N \choose k} \left(
\frac{\alpha_sC_F}{\pi}\frac{\log^2\frac{E}{k_\perp}}{N}
\right)^k\left(
1-\frac{\alpha_sC_F}{\pi}\frac{\log^2\frac{E}{k_\perp}}{N}
\right)^{N-k}\,.
\end{align}
Now, we can take the $N\to\infty$ limit of this expression, with fixed value for $k$, so that $k/N \to 0$.  In this limit, we see that the no emission factor exponentiates, and the binomial factor reduces to
\begin{align}
\lim_{N\to \infty}  {N \choose k} \frac{1}{N^k} =\frac{N^k}{k!}\frac{1}{N^k} = \frac{1}{k!}\,.
\end{align}
Then, putting these pieces together, the probability that there are $k$ emissions in a quark jet's perturbative region of interest is Poissonian distributed with
\begin{align}
p_q(k)=\frac{1}{k!} \left(
\frac{\alpha_sC_F}{\pi}\log^2\frac{E}{k_\perp}
\right)^k\exp\left[-\frac{\alpha_sC_F}{\pi}\log^2\frac{E}{k_\perp}
\right]\,.
\end{align}
For both quark and gluon jets, this can be expressed compactly as
\begin{align}
&p_q(k) = \frac{\lambda_q^k \,e^{-\lambda_q}}{k!}\,, &p_g(k) = \frac{\lambda_g^k \,e^{-\lambda_g}}{k!}\,,
\end{align}
where $\lambda_q$ and $\lambda_g$ are the mean multiplicities in the perturbative region where,
\begin{align}\label{eq:dlogpoissonmean}
&\lambda_q = \frac{\alpha_sC_F}{\pi} \log^2\frac{k_\perp}{E}\,, &\lambda_g = \frac{\alpha_sC_A}{\pi} \log^2\frac{k_\perp}{E}\,.
\end{align}

These Poissonian distributions are very interesting and of a completely different form than that of the Sudakov distributions we had derived for angularities, for example.  In particular, as Poissonian distributions, their means are equal to their variances, and so both exhibit Casimir scaling:
\begin{align}\label{eq:casscalpois}
&\frac{\langle n_q\rangle}{\langle n_g\rangle} = \frac{C_F}{C_A}\,,
&\frac{\sigma^2_q}{\sigma^2_g} = \frac{C_F}{C_A}\,,
\end{align}
where $\langle n_q\rangle$ is the mean quark jet multiplicity on the Lund plane and $\sigma_q^2$ is its variance.  By contrast, no such simple Casimir scaling relationship exists for moments of Sudakov distributions (which you can verify!), but rather the cumulative distributions of the angularities are related by Casimir scaling.  Further, this comparison between the Sudakov distribution of angularities and these Lund plane multiplicity distributions is a bit unfair, because we made no such perturbative cut on emissions for the angularities.  If one does this, to have an apples-to-apples comparison of discrimination power, the story for angularities changes significantly, which you will study in the exercises.

\begin{figure}[t!]
\begin{center}
\includegraphics[width=10cm]{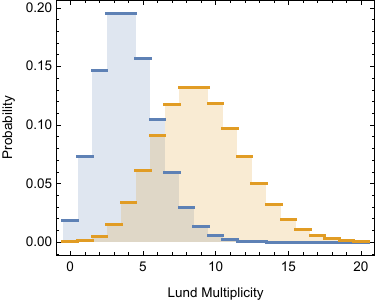}
\caption{\label{fig:lundmultdist}
Plots of the Poisson distributions of the Lund plane multiplicity for quark jets (blue) and gluon jets (orange).  For this plot, the mean values $\lambda_q = 4$ and $\lambda_g = 9$ were used, which have the correct ratio as expected from Casimir scaling.
}
\end{center}
\end{figure}

A useful measure of discrimination power that we will employ in other cases when cumulative distributions are a bit unwieldy is the ratio of the difference of means to the square root of the sum of variances:
\begin{align}\label{eq:discdist}
\eta \equiv \frac{|\mu_q-\mu_g|}{\sqrt{\sigma_q^2+\sigma_g^2}}\,.
\end{align}
We have illustrated example Poisson distributions in \Fig{fig:lundmultdist}, where the means (and variances) satisfy Casimir scaling in QCD, \Eq{eq:casscalpois}.  The sum of the variances sets a ruler or natural distance scale between the distributions, and so this measure is the distance between distributions in these natural units.  If this distance is small, then the distributions have large overlap and hence weak discrimination power, and conversely, if the distance is large then the overlap is small, and discrimination power is strong.  For Poissonian Lund plane multiplicity, we have
\begin{align}
\eta = \frac{|\langle n_q\rangle-\langle n_g\rangle|}{\sqrt{\sigma_q^2+\sigma_g^2}} = \frac{\lambda_g - \lambda_q}{\sqrt{\lambda_q+\lambda_g}} = \sqrt{\frac{\alpha_s}{\pi}}\frac{C_A-C_F}{\sqrt{C_F+C_A}}\,\log\frac{E}{k_\perp}\,.
\end{align}
The scale of the cutoff $k_\perp$ is fixed above the mass scale of the hadrons, $k_\perp \sim 1$ GeV, while the energy $E$ of the jet is something we can control.  As we increase the energy of the jet, this distance between the distributions grows without bound, and so discrimination power can become arbitrarily good at high energies.

This observation is rather similar to testing the fairness of a flipped coin as the number of flips grows.  As the energy or mean multiplicity increases, the distributions become more and more narrow as compared to their mean values.  Further, gluon jets will have approximately twice as many emissions as quark jets on the Lund plane.  Therefore, if you observe twice as many emissions as you expect for a quark jet, the probability that the jet was actually a quark jet vanishes extremely  quickly with mean multiplicity.  

As another interesting measure of discrimination power, let's look at the reducibility factors, and the maximal purity of quark and gluon jets with this Lund plane multiplicity.  The likelihood ratio of the Poisson distributions is
\begin{align}
{\cal L} = \frac{p_g(k)}{p_q(k)} = \left(
\frac{\lambda_g}{\lambda_q}
\right)^k e^{-(\lambda_g - \lambda_q)}=\left(
\frac{C_A}{C_F}
\right)^k e^{-\frac{\alpha_s}{\pi}(C_A-C_F)\log^2\frac{E}{k_\perp}}\,.
\end{align}
The maximal purity of a quark jet sample with this observable is the minimal value of the likelihood, where $ k = 0$:
\begin{align}
{\cal L}_{\min} = e^{-\frac{\alpha_s}{\pi}(C_A-C_F)\log^2\frac{E}{k_\perp}}\,.
\end{align}
Interestingly, and unlike the case for angularities, a pure sample of quark jets is not possible with the Lund plane multiplicity.  Nevertheless, this minimal value is exponentially suppressed in the large jet energy limit, so the gluon contamination to a pure quark sample is practically very small.  By contrast, the maximum gluon jet purity is determined by the maximal value of the likelihood ratio, but because $C_A > C_F$, this is divergent:
\begin{align}
{\cal L}_{\max} = \lim_{k\to\infty}\left(
\frac{C_A}{C_F}
\right)^k e^{-\frac{\alpha_s}{\pi}(C_A-C_F)\log^2\frac{E}{k_\perp}} \to\infty\,.
\end{align}
Thus, a pure sample of gluons is possible with the Lund plane multiplicity, again, unlike the situation with measurement of an angularity.

\subsubsection{Large Multiplicity Approximation}

From these expressions for the Poisson distributions of the Lund plane multiplicity, we could calculate ROC curves, AUCs, etc., but the discrete nature of the Poisson distribution renders these expressions complicated and unilluminating.  So, here we will continue and make an additional approximation in which the average number of emissions is very large, $\lambda_q,\lambda_g \gg 1$, for which the distributions can be very well approximated by a continuous distribution.  Again, we could just write down the answer where the Poissonians transform into Gaussians, but the derivation is illuminating in itself, so we provide the details here.

Just focusing on the multiplicity distribution for quark jets, we can express this distribution as
\begin{align}
p_q(k) = \frac{\lambda_q^k \,e^{-\lambda_q}}{k!}= e^{-\lambda_q+k\log\lambda_q-\log k!}\,,
\end{align}
putting everything in the exponent.  This is an exact relationship, but this exponential form will make approximation in the $\lambda_q\to \infty$ limit straightforward.  As an exponential, this distribution will be dominated by the region about where its argument takes its maximal value.  Expanding about this maximum is called the {\bf saddlepoint approximation}, which here is effectively a version of the central limit theorem.  Note that if $k\ll \lambda_q$, then the first term in the exponent dominates, and the distribution is exponentially suppressed.  Therefore, the multiplicity $k$ itself must be large, $k\gg 1$, and so we can expand the factorial with Stirling's approximation.  The leading contributions from Stirling's approximation for the asymptotic behavior of the factorial are
\begin{align}
\lim_{k\to\infty}\log k! = k\log k - k+{\cal O}(\log k)\,.
\end{align}

Using this approximation, the exponent can be expressed as
\begin{align}
-\lambda_q+k\log\lambda_q-\log k! \approx -\lambda_q+k\log\lambda_q-k\log k + k\,.
\end{align}
Its maximum, or stationary point, is where its derivative vanishes:
\begin{align}
\frac{d}{dk}\left(
-\lambda_q+k\log\lambda_q-k\log k + k
\right) = \log\lambda_q-\log k=0\,,
\end{align}
or that the extremum is where $k_0 = \lambda_q$.  We can then approximate the exponent as a quadratic function near its maximum by taking a second derivative, where
\begin{align}
\left.\frac{d^2}{dk^2}\left(
-\lambda_q+k\log\lambda_q-k\log k + k
\right) \right|_{k = \lambda_q} = -\frac{1}{\lambda_q}\,.
\end{align}
Then, the Taylor expansion through quadratic order of the exponent is
\begin{align}
-\lambda_q+k\log\lambda_q-\log k! \approx -\frac{(k-\lambda_q)^2}{2\lambda_q}+\cdots\,.
\end{align}
Then the Poisson probability distribution in this large mean and continuous limit becomes
\begin{align}
p_q(x) \propto e^{-\frac{(x-\lambda_q)^2}{2\lambda_q}}\,,
\end{align}
which is a Gaussian.  We only write proportional here, because we need to fix up the normalization, but that is easy:
\begin{align}
&p_q(x) = \frac{1}{\sqrt{2\pi \lambda_q}}\, e^{-\frac{(x-\lambda_q)^2}{2\lambda_q}}\,, &p_g(x) = \frac{1}{\sqrt{2\pi \lambda_g}}\, e^{-\frac{(x-\lambda_g)^2}{2\lambda_g}}\,,
\end{align}
writing both the quark and gluon large multiplicity (or, correspondingly, high energy) limiting distributions.

The likelihood ratio in this continuous emission limit takes the form
\begin{align}
{\cal L} = \frac{p_g(x)}{p_q(x)} = \sqrt{\frac{\lambda_q}{\lambda_g}}\exp\left[
-\frac{\lambda_g - \lambda_q}{2\lambda_g\lambda_q}(x^2 - \lambda_g\lambda_q)
\right]\,,
\end{align}
which is (still!)~monotonic in the multiplicity $x$, and so $x$ itself is the optimal discrimination observable.  We can correspondingly directly calculate the AUC as our familiar ordered integral, where
\begin{align}
\text{AUC} &= \int dx \, dy\, p_q(x) \, p_g(y)\,\Theta(x-y) \\
&= \int dx\, dy\,  \frac{1}{\sqrt{2\pi \lambda_q}}\, e^{-\frac{(x-\lambda_q)^2}{2\lambda_q}}\, \frac{1}{\sqrt{2\pi \lambda_g}}\, e^{-\frac{(y-\lambda_g)^2}{2\lambda_g}}\,\Theta(x-y)\nonumber\\
&=\frac{1}{2}\left(
1 - \text{erf}\left(
\frac{\lambda_g-\lambda_q}{\sqrt{2}\sqrt{\lambda_q+\lambda_g}}
\right)
\right)
\nonumber\,.
\end{align}
Fascinatingly, the argument of the error function is exactly the discrimination distance measure we defined in \Eq{eq:discdist} (which justifies it a posteriori).  As the means of the distributions get large, $\lambda_q,\lambda_g\to \infty$, the argument of the error function also diverges, and the AUC vanishes, because $\text{erf}(\infty) = 1$.  Thus, again we see that perfect discrimination power is possible in this limit.

\subsubsection{Lund Plane Multiplicity with Running Coupling}

This analysis of the Lund plane multiplicity can be extended to higher orders, including the effects of physics that start only formally beyond the simple double logarithmic limit in which we have worked.  Many more details and calculations are presented in the literature, e.g., \Refs{Frye:2017yrw,Dreyer:2021hhr,Medves:2022ccw,Medves:2022uii}, but here we will just discuss one effect, the running, or scale-dependence, of the coupling $\alpha_s$.  QCD exhibits asymptotic freedom in which its coupling $\alpha_s$ decreases in value as the relevant energy scale increases.  As a consequence, because emission rates are proportional to the value of the coupling, fewer particles are emitted at higher energies as compared to lower energies.  This has the effect of increasing the density of emissions on the Lund plane as you move away from the origin, toward low scales off at infinity.  Let's calculate the effect of this variation of emission density.

First, we need to establish the scale dependence of the coupling itself.  This is governed by the $\beta$-function of QCD, where the dependence of $\alpha_s$ on the energy scale $\mu$ is
\begin{align}
\mu\frac{\partial \alpha_s}{\partial \mu} = \beta(\alpha_s) = -\frac{\beta_0}{2\pi} \alpha_s^2+\cdots\,,
\end{align}
only keeping terms to leading-logarithmic or one-loop accuracy in the $\beta$-function.  The coefficient $\beta_0$ is \cite{Gross:1973id,Politzer:1973fx} (see also \InRef{tHooft:1998qmr,watershed::1734884})
\begin{align}
\beta_0 = \frac{11}{3}C_A - \frac{4}{3}n_f T_R\,,
\end{align}
which depends on the number of active quark flavors, $n_f$.  Actually, we have already seen this before, in \Sec{sec:recoilfreeobs}, as the contribution to the distribution of angularities from hard, collinear emissions in a gluon jet.  The solution to the $\beta$-function equation with one-loop running is
\begin{align}
\alpha_s(\mu) = \frac{2\pi}{\beta_0\,\log\frac{\mu}{\Lambda_\text{QCD}}}\,,
\end{align}
where $\Lambda_\text{QCD}$ is the scale at which the value of the (perturbative) coupling diverges.  This scale can be determined by matching the value of $\alpha_s$ at a fixed scale, typically taken to be the mass of the $Z$ boson, and accounting for the variation of the number of active quarks as the scale $\mu$ increases (especially the effect of the bottom quark must be accounted for).  Experimental numbers can be taken from the PDG \cite{Workman:2022ynf}, but we won't worry about that here, and just use this functional form for our analysis.

Even though the value of the coupling varies with scale, and therefore location on the Lund plane, at this approximation, emissions are still independent and so the resulting multiplicity distribution is still Poissonian.  Therefore, all we need to calculate is the mean value of the multiplicity with this running coupling.  This calculation is now a non-trivial integral, and not simply proportional to the area of the region of interest, but is nevertheless not complicated to evaluate.  For quark jets, for example, the mean multiplicity with running coupling is then
\begin{align}\label{eq:lundquarkmultaverun}
\lambda_q &= \frac{2 C_F}{\pi}\int \frac{dz}{z}\,\frac{d\theta}{\theta}\,\alpha_s(z\theta E) \, \Theta\left(z\theta - \frac{k_\perp}{E}\right) \\
&= \frac{4C_F}{\beta_0}\log\frac{E}{\Lambda_\text{QCD}}\left(
\log\frac{\log\frac{E}{\Lambda_\text{QCD}}}{\log\frac{k_\perp}{\Lambda_\text{QCD}}}-1
\right)+\frac{4C_F}{\beta_0}\log\frac{k_\perp}{\Lambda_\text{QCD}}\,,
\nonumber
\end{align}
where the scale at which the coupling is evaluated is the relative transverse momentum of the emission, $z\theta E$.  The mean Lund plane multiplicity on gluon jets is simply related by replacing $C_F \to C_A$, as usual, so the distributions still exhibit Casimir scaling of their moments.  As energy increases, this mean multiplicity increases more slowly than the double logarithmic result of \Eq{eq:dlogpoissonmean}.  The leading contribution in the $E\to\infty$ limit, with fixed $\Lambda_\text{QCD}$ and $k_\perp$, scales like
\begin{align}
\lim_{E\to \infty}\lambda_q &\to \frac{4C_F}{\beta_0}\log\frac{E}{\Lambda_\text{QCD}}\,
\log\log\frac{E}{\Lambda_\text{QCD}}\,,
\end{align}
which has an interesting logarithm of a logarithm structure.  Nevertheless, this still diverges as $E\to\infty$, and so perfect discrimination is still possible in this limit.

\subsection{Baby Steps Beyond Perturbation Theory: Discrimination with Hadronic Multiplicity}

This analysis that simply the number of emissions in the primary Lund plane is the likelihood observable, at least to this approximation, suggests that the particle multiplicity in jets in general is either the likelihood observable, or is, at least, a very good discriminant (if not technically optimal).  However, once we open up to an observable that simply counts the number of particles in the jet, we lose IRC safety.  The argument that multiplicity is not IRC safe is very simple; consider an initial jet with $n$ particles.  Now, emit an exactly 0 energy gluon, so that the total number of particles in the jet is now $n+1$, but there is no way to observe a 0 energy gluon.  Thus, the value of the multiplicity changes from emission of 0 energy gluons, and so it cannot be IRC safe.  The solution we had arrived at earlier was to cut off the Lund plane or require a minimal resolution for any emission that contributed to the multiplicity.  However, we now want to remove that restriction, counting {\it all} particles in the jet, which suffers from IRC unsafety.

It needs to be emphasized that a lack of IRC safety does not place multiplicity (or any IRC unsafe observable) into second-class, or that somehow IRC safe observables are ``better'' than their non-IRC safe counterparts.  We had already seen in the case of Sudakov safe observables, which are not IRC safe, a way to calculate them outside of fixed-order perturbation theory.  Similarly, for multiplicity, we will have to broaden our scope and possible tools for understanding it, and can't rely on our Fermi's Golden Rule master formula with Feynman diagrams as the be-all, end-all of a theoretical analysis.  This illustrates the flexibility necessary for working with (more) realistic jets: we have to let Nature and the rules of statistics determine optimal observables for discrimination, and not be biased by our theoretical prejudices.  

For the remainder of this lecture, we will study the total multiplicity of hadrons inside jets and introduce techniques and approximations that have been identified as useful starting point approximations when Feynman rules are of limited use.  We will see that scale dependence of the mean hadronic multiplicity can be calculated perturbatively, but for discrimination, we need the full functional form of the distribution, not just its mean.  The key observation that allows us to make progress is that of KNO scaling, where the functional form of the multiplicity distribution seems to be universal, and just rescaled by the mean multiplicity.  This then connects hadronic multiplicity distributions at various energy scales with no additional parameters.  To end this section, we compare the discrimination power ROC curves for quark versus gluon discrimination for the various observables we have considered throughout these lectures.\footnote{Consideration of and bias to IRC safe observables is absolutely justified for precision studies, in which a systematically-improvable perturbative calculation is compared to experimental data.  In such a case, it is necessary to quantify theoretical uncertainties.  For our study of optimal discrimination observables and likelihoods outside of perturbation theory, our goal is markedly distinct: we want to establish robust correlations and scaling relationships between observables that can inform a more principled machine learning study.}

\subsubsection{Scale Dependence of Mean Multiplicity}

\begin{figure}[t!]
\begin{center}
\includegraphics[width=8cm]{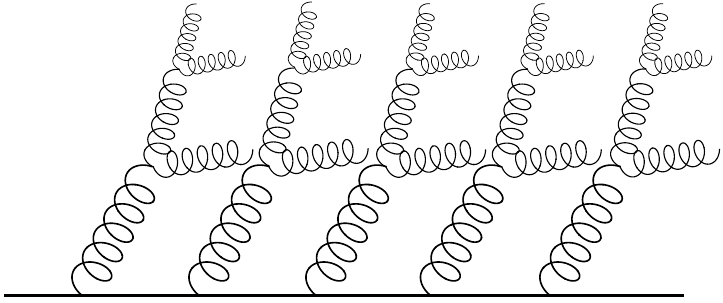}
\caption{\label{fig:emitstructmult}
Schematic picture of dominant perturbative emission structure.  Gluons are emitted off of the hard, initiating particle (solid line) with a rate proportional to its quadratic color Casimir.  Then, those secondary gluons emit further gluons that emit further gluons in a quasi-scale invariant process.
}
\end{center}
\end{figure}

To calculate the mean hadronic multiplicity, we need to establish a number of assumptions, and then we can get to calculating.  First things first: let's establish a dominant picture for the total partonic multiplicity.  This is illustrated in \Fig{fig:emitstructmult}.  The picture we have in our mind is as follows.  First, gluons are emitted off of the initiating, hard particle in proportion to its quadratic color Casimir, just like in our picture of the Lund plane.  This is the dominant contribution for primary emissions because these gluons exhibit both soft and collinear divergences.  Now, with gluons emitted, the dominant source of secondary, tertiary, or higher emissions will be more gluons from these primary gluons for the two reasons that gluons have a larger color factor than quarks and again,  the squared matrix element for emitted gluons exhibits both soft and collinear singularities.  The number of emissions off of the initiating particle is effectively fixed, and so the total multiplicity of partons is dominantly gluons and dominantly from gluons that emit gluons that emit gluons that emit gluons$\dotsc$.

We don't want the partonic multiplicity, we want the hadronic multiplicity, so we need to connect this picture with that of measured hadrons.  We will make the assumption, which has some theoretical justification in a phenomenon called {\bf preconfinement} \cite{Amati:1979fg,Bassetto:1979vy,Field:1982dg,Webber:1983if}, that after this showering process, once the partons have a relative scale of about 1 GeV from one another, they arrange themselves into color singlet clusters that independently fragment or ``hadronize'' into a fixed number of hadrons.  This clustering and fragmentation process therefore multiplies the partonic multiplicity by a constant, fixed factor which we cannot predict, but can nevertheless extract from experimental data.  That is, if we can derive a homogeneous differential equation for the scale dependence of mean multiplicity, this overall constant factor will not appear, and we will only need a description of the generation of perturbative, partonic multiplicity.

To derive such a differential equation, we need two remaining assumptions.  Because we want to determine the scale $\mu$ dependence of the mean hadronic multiplicity $\langle n\rangle$, the differential equation will be in the variable $\mu$, which correspondingly orders the emission structure.  So, we will need to determine a reasonable (and defensible!)~ordering prescription to the generation of emissions.  To determine this, we look to results in electromagnetism first.  In the study of high-energy cosmic ray showers, it was observed that ionization radiation from electron-position pairs was emitted at subsequently smaller angles as the shower progressed, a phenomenon now referred to as the {\bf Chudakov effect} \cite{chudakov,doi:10.1080/14786441008521131}.  This is a consequence of quantum coherence of emission radiation in electromagnetism, and suppresses radiation emitted at wider angles than that of the initial dipole.  An exactly analogous phenomenon exists in QCD through the quantum coherence of color, and soft gluon emission is suppressed at angles wider than that of their emitting dipole.  The Chudakov effect or color coherence then implies that emissions are ordered in the scale set by their relative angle, $\mu = \theta E$, for a jet with energy $E$.

This property of color coherence or angular ordering can be observed by azimuthal averaging of the eikonal emission factor of a gluon $k$ off of a dipole.  Recall that the eikonal emission factor from a $q\bar q$ dipole is, for example,
\begin{align}
S_{q\bar q}(k)\propto \frac{s_{q\bar q}}{s_{qk}s_{k\bar q}}\,.
\end{align}
This expression is independent of the energy of the $q$ or $\bar q$ in the dipole, and the energy of the gluon $k$ is just an overall factor.  For azimuthal averaging, we can ignore that energy dependence.  In the collinear limit, then, the angular dependence in the eikonal factor is
\begin{align}
S_{q\bar q}(k)\propto \frac{\theta_{q\bar q}^2}{\theta_{qk}^2\theta_{k\bar q}^2} = \frac{\theta_{q\bar q}^2+\theta_{k\bar q}^2-\theta_{qk}^2}{2\theta_{qk}^2\theta_{k\bar q}^2}+\frac{\theta_{q\bar q}^2+\theta_{qk}^2-\theta_{k\bar q}^2}{2\theta_{qk}^2\theta_{k\bar q}^2}\,.
\end{align}
At right, we have separated the eikonal factor into two terms, each of which has a single collinear divergence.  For example, the expression
\begin{align}
\frac{\theta_{q\bar q}^2+\theta_{k\bar q}^2-\theta_{qk}^2}{2\theta_{qk}^2\theta_{k\bar q}^2} 
\end{align}
has a collinear divergence when $k\parallel q$, but no divergence if $k\parallel \bar q$.  Thus, this representation of the eikonal factor separates it into two distinct sectors, where in each we can identify a unique emitter, at least in the well-defined collinear limit.

What makes this sectoring interesting is that angular ordering is manifest in each sector separately.  To see this, let's use the law of cosines to re-express the non-collinear angle $\theta_{k\bar q}^2$ as an azimuthal angle, and write
\begin{align}
\frac{\theta_{q\bar q}^2+\theta_{k\bar q}^2-\theta_{qk}^2}{2\theta_{qk}^2\theta_{k\bar q}^2}  &= \frac{\theta_{q\bar q}^2 - \theta_{q\bar q}\theta_{qk}\cos\phi}{\theta_{qk}^2\left(
\theta_{q\bar q}^2+\theta_{qk}^2-2\theta_{q\bar q}\theta_{qk}\cos\phi
\right)}\\
&=\frac{\theta_{q\bar q}^2(\theta_{q\bar q}^2+\theta_{qk}^2) - 2\theta_{q\bar q}^2\theta_{qk}^2\cos^2\phi}{\theta_{qk}^2\left(
(\theta_{q\bar q}^2+\theta_{qk}^2)^2-4\theta_{q\bar q}^2\theta_{qk}^2\cos^2\phi
\right)} + \text{odd in } \phi
\nonumber\,.
\end{align}
In the second line, we have expressed the denominator as a manifestly even function on $\phi\in[0,\pi]$, and in the numerator have ignored terms odd in $\phi$ on that same domain.  Then, azimuthal averaging this is easy:
\begin{align}
\int_0^{2\pi}\frac{d\phi}{2\pi}\, \frac{\theta_{q\bar q}^2+\theta_{k\bar q}^2-\theta_{qk}^2}{2\theta_{qk}^2\theta_{k\bar q}^2} &= \int_0^\pi \frac{d\phi}{\pi}\,\frac{\theta_{q\bar q}^2(\theta_{q\bar q}^2+\theta_{qk}^2) - 2\theta_{q\bar q}^2\theta_{qk}^2\cos^2\phi}{\theta_{qk}^2\left(
(\theta_{q\bar q}^2+\theta_{qk}^2)^2-4\theta_{q\bar q}^2\theta_{qk}^2\cos^2\phi
\right)}\\
&=\frac{1}{\theta_{qk}^2}\,\Theta(\theta_{q\bar q}^2 - \theta_{qk}^2)\,.\nonumber
\end{align}
That is, after azimuthal averaging, the sectored eikonal factor consists exclusively of the collinearly divergent angular factor, and further, is only non-zero if the emission angle is smaller than the opening angle of the dipole.  

The final piece of the puzzle is the relevant scale for particle production.  For a particle to be produced, that is, to actually {\it observe} that a particle has been emitted, that particle must have a sufficiently large relative transverse momentum to its emitter.  Note that the relative transverse momentum of a particle with energy fraction $z$ and splitting angle $\theta$ is $k_\perp = z\theta E = z\mu$, where $\mu$ is the angular ordering scale established earlier.  Finally, these considerations motivate the following homogeneous evolution equation for the mean multiplicity:
\begin{align}
 \langle n(\mu+\delta \mu)\rangle =  \langle n( \mu)\rangle+\frac{\delta \mu}{\mu}\, \frac{2\alpha_sC_A}{\pi}\int_0^1 \frac{dz}{z}\,  \langle n(z\mu)\rangle\,.
\end{align}
Here, $\delta\mu$ is a small change in the evolution scale $\mu$.  As $\mu$ increases, two things can happen: either nothing (first term on the right), or a soft gluon can be emitted at a relative transverse momentum of $z\mu$, and we need to sum over all possible emission energies, consistent with the scale $\mu$.  By taking $\delta \mu \to 0$, this can equivalently be expressed as an integro-differential equation:
\begin{align}
\mu\frac{\partial}{\partial \mu} \langle n(\mu)\rangle = \frac{2\alpha_sC_A}{\pi}\int_0^1 \frac{dz}{z}\,   \langle n(z\mu)\rangle\,.
\end{align}

To solve this, we make a power-law ansatz, where
\begin{align}
\langle n(\mu)\rangle \propto \mu^\gamma\,,
\end{align}
where $\gamma$ is called the {\bf anomalous dimension}.  This equation then becomes
\begin{align}
\gamma =  \frac{2\alpha_s C_A}{\pi}\int_0^1 dz\, z^{-1+\gamma} =  \frac{2\alpha_s C_A}{\pi}\,\frac{1}{\gamma}\,.
\end{align}
The anomalous dimension is therefore
\begin{align}
\gamma = \sqrt{\frac{2\alpha_s C_A}{\pi}}\,.
\end{align}
This dependence on $\sqrt{\alpha_s}$ can be explained as the fractal dimension \cite{hausdorff1918dimension,doi:10.1126/science.156.3775.636} of the shape of the multiparticle Lund planes produced as the shower progresses \cite{Gustafson:1991ru}.  The solution to this evolution equation with fixed coupling in a truly scale-invariant theory is then
\begin{align}\label{eq:scaleinvmeanmult}
\langle n(\mu)\rangle \propto \mu^{\sqrt{\frac{2\alpha_s C_A}{\pi}}}\,.
\end{align}
That is, the mean hadronic multiplicity in a jet, in the extreme high energy limit that QCD is scale-invariant, scales with a fractional power of the energy scale $\mu$ of the jet.  Note that this grows faster with $\mu$ than any power of $\log\mu$ and so the total multiplicity increases much faster with energy than the primary Lund multiplicity.

This description of the mean multiplicity can be improved through incorporation of higher-order effects. With this result and one-loop running of the coupling, note that the mean multiplicity satisfies the differential equation
\begin{align}
\mu \frac{\partial}{\partial \mu}\langle n(\mu)\rangle = \sqrt{\frac{2\alpha_s C_A}{\pi}}\, \langle n(\mu)\rangle = \sqrt{\frac{4C_A}{\beta_0 \log\frac{\mu}{\Lambda_\text{QCD}}}}\, \langle n(\mu)\rangle\,.
\end{align}
The solution is
\begin{align}\label{eq:hadmultave}
\langle n(\mu)\rangle \propto e^{\sqrt{\frac{16C_A}{\beta_0}\log\frac{\mu}{\Lambda_\text{QCD}}}} =e^{\sqrt{\frac{32\pi C_A}{\beta_0^2}\frac{1}{ \alpha_s(\mu)}}} \,,
\end{align}
where at right we have replaced the explicit scale dependence with the value of the running coupling, $\alpha_s(\mu)$.  More details about this calculation and how higher-order effects can be included can be found in the original literature \cite{Bassetto:1979nt,Furmanski:1979jx,Mueller:1981ex,Ermolaev:1981cm,Catani:1991pm,Dremin:2000ep,Medves:2022ccw}.

\begin{figure}[t!]
\begin{center}
\includegraphics[width=10cm]{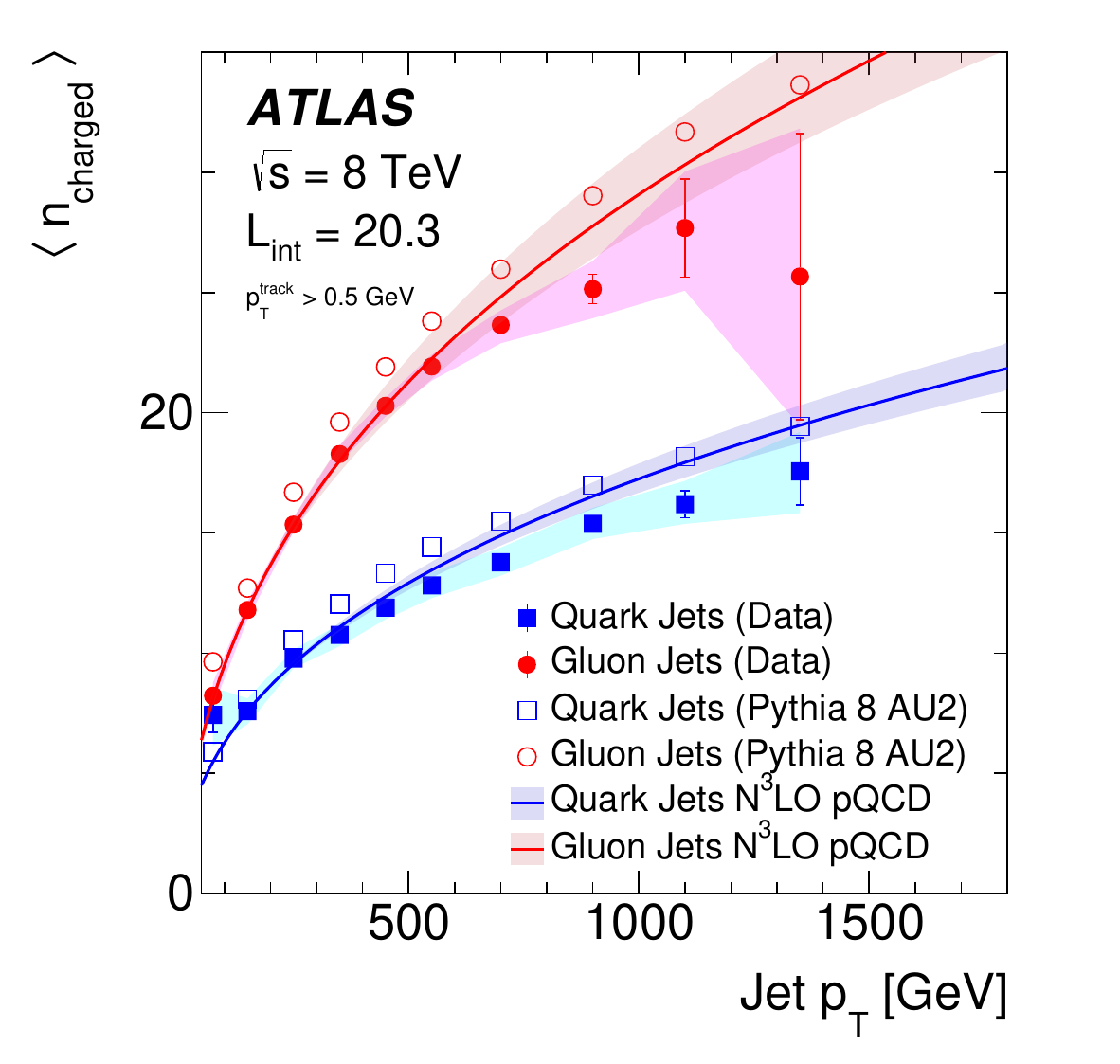}
\caption{\label{fig:atlas_qgmult}
Plot of the mean charged particle multiplicity on quark (blue) and gluon (red) jets as measured at the ATLAS experiment, as a function of jet energy.  Data (solid dots) are compared to simulated data (empty dots), and to the theoretical calculation (solid curves), an extension of \Eq{eq:hadmultave} including three higher orders in the perturbative expansion of the anomalous dimension.  From \InRef{ATLAS:2016vxz}.
}
\end{center}
\end{figure}

From our stated assumptions, we expect that the mean multiplicity is proportional to the appropriate quadratic color Casimir of the initiating particle; $\langle n_q\rangle \propto C_F$ for quark jets and $\langle n_g\rangle \propto C_A$ for gluon jets.  However, this intuition comes from a perturbative analysis and understanding, and so the only limit in which a perturbative description dominates is in the extreme high-energy limit, $\mu\to \infty$.  The approach to true Casimir scaling of the mean multiplicities as a function of the jet scale $\mu$ is obviously very important for discrimination power of multiplicity, and is experimentally observed to be extremely slow.  Recall that in QCD, $C_A/C_F = 2.25$, but the observed ratio of average multiplicity of gluon to quark jets at energies of $200$ GeV is about 1.5, while even at an energy of $1.2$ TeV is still only about 1.7 \cite{ATLAS:2016vxz}.  This data from ATLAS is plotted in \Fig{fig:atlas_qgmult} and compared with both simulated data and the calculated energy dependence of the mean multiplicity, as in \Eq{eq:hadmultave}, though to much higher perturbative accuracy.  We will see in \Sec{sec:threetechs} that there is strong sensitivity of discrimination power to the assumed relationship between quark and gluon jet multiplicity.

\subsubsection{Higher Moments of Multiplicity and KNO Scaling}

One can continue this analysis of the moments of the hadronic multiplicity distribution, and derive differential equations for their scale dependence.  These results can then be re-interpreted as the scale dependence of parameters in the complete multiplicity distribution itself, and so one can observe and predict how the distribution is warped and molded by physics at different energy scales.  Here, instead of this, we will listen to the experimentalists' results for the measurement of these higher moments in collision events at various energies.  Even by the late 1960s, it was observed in data that the second moment of the hadronic multiplicity was proportional to the square of the mean:
\begin{align}
\langle n^2\rangle \propto \langle n\rangle^2\,.
\end{align}
Even higher moments were measured, and the pattern continued, with the $k$th moment of the multiplicity distribution proportional to the $k$th power of the mean:
\begin{align}
\langle n^k\rangle \propto \langle n\rangle^k\,.
\end{align}
In the early 1970s, this was formalized through a simple scaling relationship of the hadronic multiplicity distribution, $p(n)$, where
\begin{align}
p(n) = \frac{1}{\langle n\rangle} \, \psi\left(
\frac{n}{\langle n\rangle}
\right)\,,
\end{align}
where $\psi(x)$ is some universal, normalized, distribution with unit mean.  This relationship is called {\bf KNO scaling} \cite{Polyakov:1970lyy,Koba:1972ng}.\footnote{It was actually first established by Polyakov, but was named after Koba, Nielsen, and Olesen.  This is merely one of an infinity of examples of Stigler's law of eponymy \cite{https://doi.org/10.1111/j.2164-0947.1980.tb02775.x} in particle physics.}

It is important to note that KNO scaling is a property of the total hadronic multiplicity, and is not apparent in different definitions of (sub)multiplicities.  For instance, the Poisson distribution of the primary Lund plane multiplicity that we studied earlier this lecture clearly does not satisfy KNO scaling.  The mean and variance of the Poisson distribution are equal, $\mu = \sigma^2$, but KNO scaling would predict that the variance is proportional to the mean squared, $\sigma^2 \propto \mu^2$.  The original justification for KNO scaling followed from Bjorken and Feynman scaling \cite{Bjorken:1968dy,Feynman:1969ej,Wilson:1970zzb}, early identified properties of the strong interactions (before the development of QCD) that are a consequence of its scale invariance.\footnote{My Ph.D.~advisor Michael Peskin made the defensible claim that James ``BJ'' Bjorken was the first person to understand conformal field theories.}

The minimal assumption from which KNO scaling is a consequence is that the multiplicity distribution is invariant to scaling of jet energies.  By ``scale-invariant'' we mean that the multiplicity distribution at a given energy scale $E$, $p(n|E)$, transforms simply when energy is scaled by $\lambda > 0$:
\begin{align}
p(n|E) = \lambda^{-\gamma}\, p(\lambda^\gamma n|\lambda E)\,.
\end{align}
That is, as the energy is scaled by $\lambda$, the multiplicity $n$ is correspondingly scaled by $\lambda^\gamma$.\footnote{Of course, this relationship isn't perfect because the multiplicity $n$ is integral valued and so arbitrary scale transformations are not allowed, but we can assume that the support of $p(n|E)$ is over a sufficiently large domain that the continuous and smooth approximation is valid.}  Thus, the relationship between the multiplicity and energy must take the form
\begin{align}
p(n|E) = \frac{1}{k E^\gamma}\,\psi\left(\frac{n}{k E^\gamma}\right)\,,
\end{align}
where $\psi(x)$ is some energy-independent function and $k$ is a constant.  Note that the mean multiplicity from this scale-invariant assumption is
\begin{align}
\langle n\rangle = \int dn \, n\, p(n|E) = \int dn\, \frac{n}{k E^\gamma}\,\psi\left(\frac{n}{k E^\gamma}\right) = k E^\gamma\int dx\, x\, \psi\left(x\right)=k E^\gamma\,.
\end{align}
In the penultimate expression, we can set the mean value of $\psi(x)$ to 1 with appropriate normalization because it is energy-independent.  As demonstrated in \Eq{eq:scaleinvmeanmult}, the mean multiplicity does indeed satisfy this fractional power dependence on energy, in the limit that QCD is scale-invariant.  Therefore, we indeed have that
\begin{align}
p(n) = \frac{1}{\langle n\rangle}\,\psi\left(\frac{n}{\langle n\rangle}\right)\,,
\end{align}
which is KNO scaling.  The particular form of the KNO function $\psi(x)$, as well as violations of KNO scaling, can be predicted in perturbative QCD, see, e.g.,  \Refs{Konishi:1979cb,Bassetto:1979nt,Dokshitzer:1982ia,Malaza:1984vv,Munehisa:1986sz,dokshitzer1991basics,Cuypers:1991hm,Gustafson:1992uh,Dokshitzer:1993dc,Gustafson:1993dd,Khoze:1996dn,Dremin:2000ep}.  While we will present a specific parametrization of the KNO function shortly, in the analysis of these notes, however, we will mostly only need the existence of KNO scaling.

\begin{figure}[t!]
\begin{center}
\includegraphics[width=7.7cm]{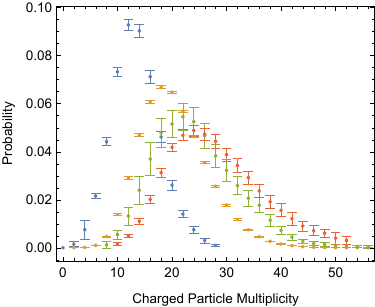}\hspace{0.5cm}
\includegraphics[width=7.7cm]{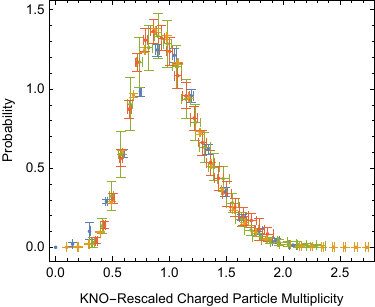}
\caption{\label{fig:knomult} Charged particle multiplicity distributions (left) in $e^+e^-\to$ hadrons collisions at $\sqrt{s} = 34, 91, 161, 206$ GeV, and the same distributions rescaled by their means (right).  Data from \Refs{TASSO:1983cre,L3:2004cdh,OPAL:1997asf}.}
\end{center}
\end{figure}

KNO scaling can be directly observed by comparing multiplicity data from experiments at different center-of-mass collision energies, for example.  At left in \Fig{fig:knomult} are the total charged hadronic multiplicity distributions in $e^+e^-$ collision events at four different center-of-mass collision energies \cite{TASSO:1983cre,L3:2004cdh,OPAL:1997asf}.  Naturally, as the collision energy increases, so too does the multiplicity, as follows from the result in \Eq{eq:hadmultave}.  However, if each of these distributions is rescaled by its respective mean, as prescribed by KNO scaling, the rescaled multiplicity distributions snap into very nearly the exact same functional distribution, as illustrated at right in \Fig{fig:knomult}.  One small thing to note is that these experimental data only correspond to observed charged particles because individual charged particles can be observed by their ionization track in the detector.  To good approximation, however, the relationship between charged and all hadronic particles is just an overall constant factor, which follows from approximate isospin conservation.  We will say much more about this in the next lecture.

\subsubsection{Discrimination from KNO Scaling of Multiplicity}\label{sec:knodiscsec}

With the assumption of KNO scaling of the hadronic multiplicity distribution, let's now work to calculate its discrimination power.  In this section, we will work to establish a bound on the discrimination power as defined through the AUC, and in the next section, use a convenient parametrization from which we can calculate the full ROC curve.  For quark versus gluon discrimination with hadronic multiplicity, the AUC can be calculated from
\begin{align}
\text{AUC} = \int dn_q\, dn_g\, p_q(n_q)\, p_g(n_g)\,\Theta(n_q - n_g)\,,
\end{align}
where $p_q(n)$ and $p_g(n)$ are the quark and gluon multiplicity distributions, respectively.  Let's now assume KNO scaling for the multiplicity distributions, with
\begin{align}
&p_q(n) = \frac{1}{\langle n_q\rangle}\,\psi\left(\frac{n}{\langle n_q\rangle}\right)\,,
&p_g(n) = \frac{1}{\langle n_g\rangle}\,\psi\left(\frac{n}{\langle n_q\rangle}\right)\,.
\end{align}
Not only do we assume KNO scaling of the distributions, but that the universal distribution $\psi(x)$ is identical for quarks and gluons.  This seems like a reasonable assumption, and consistent with all of the results we have derived thus far, but is potentially interesting to test and understand its implications in more detail.  In \Sec{sec:knoqvg}, we will test this hypothesis from a couple different experimental analyses.
 
With these KNO scaled distributions, the AUC can be written as 
\begin{align}\label{eq:knoauc}
\text{AUC} &= \int dn_q\, dn_g\,  \frac{1}{\langle n_q\rangle}\,\psi\left(\frac{n_q}{\langle n_q\rangle}\right)\,  \frac{1}{\langle n_g\rangle}\,\psi\left(\frac{n_g}{\langle n_g\rangle}\right)\,\Theta(n_q - n_g)\\
&=\int dx\, dy\, \psi(x)\,\psi(y)\,\Theta\left(
\frac{\langle n_q\rangle}{\langle n_g\rangle}\,x -  y
\right)\nonumber\\
&=\int dx\, \psi(x)\,\Psi\left(\frac{\langle n_q\rangle}{\langle n_g\rangle}\,x\right)
\nonumber\,.
\end{align}
Here, we have introduced the cumulative KNO function
\begin{align}
\Psi(x) = \int_0^x dx'\, \psi(x')\,.
\end{align} 
With Casimir scaling, the ratio of mean multiplicities is set by the ratio of quadratic Casimirs,
\begin{align}
\frac{\langle n_q\rangle}{\langle n_g\rangle} = \frac{C_F}{C_A}\,,
\end{align}
but even relaxing Casimir scaling, we still expect that $\langle n_q\rangle < \langle n_g\rangle$, and so the AUC is necessarily less than 1/2, indicating useful discrimination power.  If  $\langle n_q\rangle = \langle n_g\rangle$, the AUC would be 1/2 (the multiplicity distributions would be identical), so let's see if we can provide an estimate for the AUC in terms of an expansion about the point $\langle n_q\rangle = \langle n_g\rangle$.

Let's introduce the deviation from unit ratio ${\cal R}_{qg}$, where
\begin{align}
\frac{\langle n_q\rangle}{\langle n_g\rangle} = 1-{\cal R}_{qg}\,,
\end{align}
and $0 < {\cal R}_{qg} < 1$ because $0<\langle n_q\rangle < \langle n_g\rangle$.  Just writing out the first few orders in the expansion explicitly, the AUC can be expressed as
\begin{align}
\text{AUC} &=\int dx\, \psi(x)\,\Psi\left((1-{\cal R}_{qg})\,x\right)\\
&=\int dx\,\psi(x)\,\Psi(x) - {\cal R}_{qg}\int dx\, x\, \psi(x)^2 + \frac{{\cal R}_{qg}^2}{2}\int dx\, x^2\, \psi(x)\, \psi'(x)+\cdots\nonumber\\
&=\frac{1}{2} - \left(
{\cal R}_{qg}+\frac{{\cal R}_{qg}^2}{2}
\right)\int dx\, x\, \psi(x)^2 + \cdots
\nonumber\,.
\end{align}
In going from the second to third lines, we used integration by parts, and assumed that the KNO distribution vanished at the endpoint $x = 0$.  The remaining integral of the KNO function expanded to this order, 
\begin{align}
\int dx\, x\, \psi(x)^2\,,
\end{align}
has no upper bound (which can easily be seen by taking the extreme distribution $\psi(x) = \delta(x-1)$), but does have a non-trivial lower bound.
 
To establish this lower bound, let's collect properties and assumptions of the KNO function.  We have and assume:
\begin{enumerate}
\item The KNO function is non-negative on $x\in[0,\infty)$ and is normalized:
\begin{align}
1 = \int_0^\infty dx\, \psi(x)\,.
\end{align}

\item The mean value of the KNO function is 1:
\begin{align}
1 = \int_0^\infty dx\, x\, \psi(x)\,.
\end{align}

\item We assume that the probability for a multiplicity of 0 is 0:
\begin{align}
\psi(0) = 0\,.
\end{align}

\item Because the KNO function is observable, all of its moments exist:
\begin{align}
\int_0^\infty dx\, x^n \,\psi(x) < \infty\,, \qquad\forall\, n>0\,.
\end{align}
\end{enumerate}

To establish the lower bound,\footnote{This is the best argument for a maximum lower bound I could construct.  If you have a better approach with fewer assumptions, I would be greatly interested!} we can use the Cauchy-Schwarz inequality, for which we have
\begin{align}
\left(
\int dx\, e^{-tx}\,x\,\psi(x)
\right)^2 \leq \left(\int dx\, x\, \psi(x)^2\right)\left( \int dx\,x\, e^{-2tx}\right)\,,
\end{align}
where we have introduced the exponential as a helper function and $t>0$ is some parameter.  Rearranging and evaluating the exponential integral, we have a bound on the integral of interest:
\begin{align}
4t^2\left(
\frac{d}{dt}\int dx\, e^{-tx}\,\psi(x)
\right)^2 \leq \int dx\, x\, \psi(x)^2\,,
\end{align}
In this way, the integral in the parentheses is the KNO function's Laplace transform:
\begin{align}
{\cal L}[\psi](t) = \int dx\, e^{-tx}\,\psi(x)\,,
\end{align}
and so the inequality is
\begin{align}
4t^2\left(
\frac{d}{dt}{\cal L}[\psi](t)
\right)^2 \leq \int dx\, x\, \psi(x)^2\,.
\end{align}

We would like to establish the greatest lower bound, and so let's maximize with respect to $t$.  Differentiating the left side and setting it to 0, we find
\begin{align}
\frac{d}{dt}\left[
4t^2\left(
\frac{d}{dt}{\cal L}[\psi](t)
\right)^2
\right] = 0\,,
\end{align}
or that
\begin{align}\label{eq:extremelaplace}
\left(\frac{d}{dt} + t\, \frac{d^2}{dt^2}\right){\cal L}[\psi](t) = 0\,.
\end{align}
Because we assume that all moments of the KNO distribution $\psi(x)$ exist, the Taylor expansion of its Laplace transform is meaningful and converges.  Through $t^2$, the Taylor expansion of the Laplace transform is
\begin{align}
{\cal L}[\psi](t) = 1 - t + \frac{\langle x^2\rangle t^2}{2}+\cdots = e^{-t}\left(
1+\frac{\sigma^2t^2}{2}+\cdots
\right)\,.
\end{align}
On the right, we have just extracted the overall exponential suppression, which modifies the coefficients of the expansion from moments about 0 to central moments, about the mean.  By the third and fourth assumptions, the central moment expansion of the Laplace transform rapidly converges because the KNO function is necessarily rather narrowly distributed about its mean, $\sigma^2 \lesssim 1$.  So, we can safely ignore higher-order contributions in the central moment expansion.  Terminating the raw moment expansion at low orders is much less justified.

Plugging this expansion through the variance term into the extremization formula, \Eq{eq:extremelaplace}, we find that the value of $t$ that maximizes the lower bound is $t = 1$, independent of the value of the variance, $\sigma^2$.  Then, we can evaluate the lower bound at $t = 1$:
\begin{align}
\left.4t^2\left(
\frac{d}{dt}{\cal L}[\psi](t)
\right)^2\right|_{t = 1} &= \left.4t^2\left(
\frac{d}{dt}e^{-t}\left(
1+\frac{\sigma^2t^2}{2}+\cdots
\right)
\right)^2\right|_{t = 1}\\
&=\frac{4}{e^2}\left(
1-\frac{\sigma^2}{2}+\cdots
\right)^2\,.\nonumber
\end{align}
This is generically non-zero.  Further, with our assumptions, the variance of the KNO distribution is bounded from above by $\sigma^2\lesssim 1$, and so we can plug in $\sigma^2 = 1$ to relax the bound a bit more:
\begin{align}
\left.\frac{4}{e^2}\left(
1-\frac{\sigma^2}{2}+\cdots
\right)^2\right|_{\sigma^2 = 1} \approx \frac{1}{e^2}\approx0.135\,. 
\end{align}

Now, returning way back to the estimate of the AUC and putting these results together, we have
\begin{align}
\text{AUC} =\frac{1}{2} - \left(
{\cal R}_{qg}+\frac{{\cal R}_{qg}^2}{2}
\right)\int dx\, x\, \psi(x)^2 + \cdots \lesssim \frac{1}{2}-\frac{1}{e^2}\left(
{\cal R}_{qg}+\frac{{\cal R}_{qg}^2}{2}
\right)\,.
\end{align}
For relatively small ${\cal R}_{qg} =1 - \langle n_q\rangle/\langle n_g\rangle$, ${\cal R}_{qg}\lesssim 0.7$ (so the term in parentheses with explicit ${\cal R}_{qg}$ is less than 1), this should be a rather robust upper bound on the AUC.

\subsubsection{An Explicit Parametrized KNO Function}

Just using abstract KNO scaling and the few other assumptions we made about the properties of the KNO function, we can only provide so much detail as to the discrimination power of the hadronic multiplicity distribution.  Essentially since the time that KNO scaling was identified, groups have presented numerous parametrized expressions for the KNO distribution (e.g., \Refs{Furmanski:1979jx,Bassetto:1979nt,Konishi:1979ft,Hayot:1982mi,Cai:1983tj,Liu:1982ut,Dokshitzer:1982xr,Pancheri:1982ai,Chou:1983xg,Carruthers:1983my,Bassetto:1987fq,Liu:2022bru,Liu:2023eve}), and here we will study one such representation.  By our assumption that all moments of the KNO distribution exist, the KNO distribution must vanish as $x\to\infty$ more rapidly than any power of $x$.  This is easily accomplished by assuming that the KNO distribution has an exponential tail:
\begin{align}
\psi(x )\to e^{-a x}\,,
\end{align}
for $a > 0$, as $x\to\infty$.  Further, we also assume that the KNO distribution vanishes at $x = 0$, which is easily accomplished with a non-negative power of $x$ near $x = 0$:
\begin{align}
\psi(x) \to x^b\,,
\end{align}
for $b > 0$, as $x\to 0$.  Then, along with the normalization and mean restrictions, this suggests the one-parameter form
\begin{align}\label{eq:knonegbin}
\psi(x|k) = \frac{k^k}{\Gamma(k)}\, x^{k-1}e^{-kx}\,,
\end{align}
where $k > 1$.  This distribution is normalized and has unit mean:
\begin{align}
1 = \int_0^\infty dx\, \psi(x|k) = \int_0^\infty dx\,x\, \psi(x|k)
\end{align}
Its variance is
\begin{align}
\sigma^2 = \int_0^\infty dx\,(x - 1)^2\, \psi(x|k) = \frac{1}{k}\,,
\end{align}
which is indeed less than 1, for $ k > 1$.  For the data in Fig.~\ref{fig:knomult}, the KNO function is fit extremely well to this function with $k = 10$.\footnote{This particular parametrized functional form for the KNO function is also often the same that is used to parametrize non-perturbative contributions to additive IRC safe observables and is called a {\bf shape function} \cite{Korchemsky:1999kt,Korchemsky:2000kp,Bosch:2004th,Hoang:2007vb,Ligeti:2008ac,Stewart:2014nna}.  In that context, the non-perturbative shape function is convolved with the perturbative differential cross section to produce a distribution that can describe data over all of phase space.  Further, the low moments of the non-perturbative corrections for many observables are known to take a universal form \cite{Akhoury:1995sp,Dokshitzer:1995zt,Lee:2006fn,Lee:2006nr}.  There may be a rather direct connection between KNO scaling and shape functions as perturbative calculations are accurate where resolved particle multiplicity is small (just a few hard emissions), while non-perturbative physics dominates when multiplicity is large (every pion is resolved), but I am unaware of such an analysis in the literature.}

This functional form of the KNO distribution is the continuous generalization of the negative binomial distribution.  The negative binomial distribution can be constructed from a mixture of a Gamma distribution and a Poisson distribution in the following way.  The mean of the Poisson distribution $\lambda$ can be considered a random variable itself, distributed according to a Gamma distribution.  The Gamma distribution is the maximal entropy probability distribution with fixed, positive mean and fixed expectation value of the logarithm of the random variable.  Thus, the interpretation of the negative binomial distribution for multiplicity is that the particle production is independent and uniform over a volume of phase space that fluctuates with maximal entropy.  Planck was apparently the first to identify the negative binomial as the distribution of bosons among $k$ clusters with equal probability \cite{Carruthers:1983my,planck1923sitzunsber,decomps1963distribution}.

With the negative binomial assumption for the KNO distribution, we can then calculate the corresponding AUC for quark versus gluon discrimination.  From the general analysis in \Eq{eq:knoauc}, the AUC can be expressed as
\begin{align}\label{eq:aucmultfull}
\text{AUC} &= \int dx\, dy\, \psi(x|k)\,\psi(y|k)\,\Theta\left(
\langle n_q\rangle\,x -  \langle n_g\rangle\,y
\right)\\
&= \frac{k^{2k}}{\Gamma(k)^2}\int dx\, dy\, x^{k-1} y^{k-1}e^{-k(x+y)}\,\Theta\left(
\langle n_q\rangle\,x -  \langle n_g\rangle\,y
\right)\nonumber\\
&=1 - \frac{\Gamma(2k)}{k\,\Gamma(k)^2}\left(\frac{\langle n_g\rangle}{\langle n_q\rangle}\right)^k\left(1+\frac{\langle n_g\rangle}{\langle n_q\rangle}\right)^{-2k}\,_2F_1\left(
2k, 1,1+k,\frac{\langle n_g\rangle}{\langle n_g\rangle+\langle n_q\rangle}
\right)
\nonumber\,.
\end{align}
Here, $_2F_1(a,b,c,z)$ is the hypergeometric function.  This particular form isn't so illuminating, but we can express $\langle n_q\rangle = (1-{\cal R}_{qg})\langle n_g\rangle$ and Taylor expand in ${\cal R}_{qg}$ to produce
\begin{align}
\text{AUC} = \frac{1}{2}-\frac{4^{-k}\Gamma(2k)}{\Gamma(k)^2}\left( {\cal R}_{qg}+\frac{{\cal R}_{qg}^2}{2}\right) +\cdots\,.
\end{align}
The form of this expansion of course follows the general results we had established with the bounds that we derived in the previous section.  In particular, we can validate the bound, by saturating the Gamma function coefficient at $k = 1$, the minimum value allowed by our assumptions.  At this point, note that
\begin{align}
\left.\frac{4^{-k}\Gamma(2k)}{\Gamma(k)^2}\right|_{k = 1} = \frac{1}{4} > \frac{1}{e^2}\,,
\end{align}
and so the AUC of the negative binomial distribution is indeed less than our bound.  As $k$ increases, the coefficient scales like $\sqrt{k}$, which can be verified through Stirling's formula,
\begin{align}
\frac{4^{-k}\Gamma(2k)}{\Gamma(k)^2} =\frac{\sqrt{k}}{2\sqrt{\pi}}+{\cal O}(k^{-1/2})\,.
\end{align}

With this explicit parametrization of the multiplicity distribution, we can also determine the reducibility factors for samples of maximal quark and gluon purity.  The likelihood ratio of the quark and gluon multiplicity distributions is
\begin{align}
{\cal L} = \frac{p_q(n)}{p_g(n)} = \frac{\langle n_g\rangle \, \psi\left(
\frac{n}{\langle n_q\rangle}|k
\right)}{\langle n_q\rangle\, \psi\left(
\frac{n}{\langle n_g\rangle}|k
\right)} = \left(
\frac{\langle n_g\rangle}{\langle n_q\rangle}
\right)^k\, e^{-kn \frac{\langle n_g\rangle-\langle n_q\rangle}{\langle n_q\rangle\langle n_g\rangle}}\,.
\end{align}
From this, we immediately see that a pure gluon sample can be attained with a cut at large multiplicity,
\begin{align}
\lim_{n\to \infty}{\cal L} = 0
\end{align}
A pure quark sample is technically not possible, as the maximal value of the likelihood is limited by the parameter $k$:
\begin{align}
\lim_{n\to 0}{\cal L} =  \left(
\frac{\langle n_g\rangle}{\langle n_q\rangle}
\right)^k = e^{-k\log\frac{\langle n_q\rangle}{\langle n_g\rangle}}\,.
\end{align}
In practice, however, we demonstrated that $k$ is large, $k\sim 10$, and even if the ratio of means is much weaker than Casimir scaling, say only $\langle n_g\rangle/\langle n_q\rangle \sim 1.5$, the maximum value of the likelihood still exceeds 50.  We note that this pattern of reducibility factors is similar to what we observed with Lund plane multiplicity as well, that a pure gluon sample was attainable, but a quark sample will always be contaminated by an exponentially small amount of gluon jets.

\subsection{Comparison of Three Techniques for Quark vs.~Gluon Discrimination}\label{sec:threetechs}

\begin{figure}[t!]
\begin{center}
\includegraphics[width=10cm]{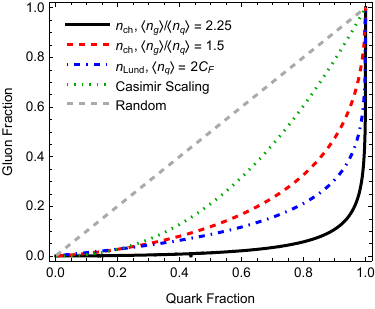}
\caption{\label{fig:roccompsqvg}
Comparison of ROC curves of the classes of quark versus gluon discriminators considered in these lectures: Casimir scaled additive IRC safe observables (green dotted), primary Lund plane multiplicity with $\langle n_q\rangle = 2C_F$ (blue dot-dashed), total (charged) hadronic multiplicity with $\langle n_g\rangle /\langle n_q\rangle = 1.5$ (red dashed), and total (charged) hadronic multiplicity with $\langle n_g\rangle /\langle n_q\rangle = C_A/C_F = 2.25$ (black).  For the total charged multiplicity curves, we use the negative binomial distribution with $k = 10$.  Dashed gray is the random classifier as reference.}
\end{center}
\end{figure}

With the collected results for quark versus gluon discrimination presented throughout these lectures, it is useful to compare them all, as quantified by their ROC curves.  The comparison of Casimir scaling in additive IRC safe observables (like angularities), Lund plane multiplicity, and total charged particle multiplicity is displayed in \Fig{fig:roccompsqvg}.  The ROC curve for IRC safe angularities has no parameters (it is fixed by the ratio $C_A/C_F$), while the multiplicity distributions have parameters that fix the mean multiplicities.  For the primary Lund plane multiplicity, we take $\langle n_q\rangle = 2C_F$, which, from \Eq{eq:lundquarkmultaverun} for the mean multiplicity with one-loop running of the coupling, corresponds to jets with an energy of about $E \sim 200$ GeV.  For the total multiplicity curves, we fix the form of the KNO distribution to be the negative binomial with $k = 10$, but choose two different values for the ratio of quark and gluon expectation values, $\langle n_g\rangle / \langle n_q\rangle = 1.5$ and $=C_A/C_F = 2.25$.

These plots make it clear that observables like the angularities that are dominantly sensitive to a single emission in the jet are not optimal for this discrimination problem.  Multiplicity observables are much more powerful discriminants, and their discrimination power improves as the ratio of the means of gluon to quark jets increases.  Sensitivity to total multiplicity with full infinite energy Casimir scaling of means (solid black in \Fig{fig:roccompsqvg}) produces extremely good discrimination, with an AUC of $\text{AUC}\approx 0.04$ (from \Eq{eq:aucmultfull}).

\subsubsection{Mutual Irreducibility and Quarks and Gluons from Mixed Samples}

As discussed all the way back in the beginning of \Sec{sec:qvgdisc}, practically defining the ground truth labels of ``quark'' and ``gluon'' jets has migrated to the rather lazy and theoretically imprecise practice of whatever jet is asked for in selecting the process of interest in an event simulator.  Through our analysis, and especially through the identification of the multiplicity as such a powerful discriminant, another technique suggests itself.  Namely, quark and gluon jet labels are not set a priori, but rather are an output from modeling mixed samples as some linear combination of two ``topics'', which we can, a posteriori, call ``quark'' and ``gluon'', if so desired.  This technique of jet {\bf topic modeling} \cite{blei2012probabilistic,Metodiev:2018ftz,Komiske:2018vkc} enables extraction of individual quark and gluon jet distributions from unlabeled data, as long as those data satisfy a few reasonable requirements.

First, the number of linearly-independent samples must be equal to the number of topics that one desires to model.  For quark and gluon jets, this means that we need two samples of jets and further, the relative quark and gluon fractions in those samples must be different (but whose precise values do not need to be known).  For example, one might choose the leading jet from, say, $pp\to Z+$ jet and $pp\to $ dijets samples.  Second, the likelihood ratio observable measured on those samples of jets must be mutually irreducible, which means that pure samples of the two topics can be isolated somewhere on phase space.  As we have discussed, mutual irreducibility means that the likelihood ratio ranges over its entire domain, ${\cal L}\in[0,\infty)$.  If these criteria are satisfied, then the quark distribution of the likelihood ratio observable can be defined from the distributions on samples 1 and 2 as:
\begin{align}\label{eq:topicinverse}
p_q({\cal L}) = \frac{f_g^{(2)}p_1({\cal L}) - f_g^{(1)}p_2({\cal L})}{f_q^{(1)}-f_q^{(2)}}\,.
\end{align}
Here, $p_i({\cal L})$ is the distribution of observable ${\cal L}$ on sample $i$, and $f_q^{(i)}$ ($f_g^{(i)}$) is the fraction of quark (gluon) jets in sample $i$, such that $f_q^{(i)} + f_g^{(i)} = 1$.  The corresponding gluon distribution is defined analogously, and the jet fractions $f_q^{(i)},f_g^{(i)}$ can be extracted through a convex optimization problem \cite{arora2012learning,katz2019decontamination}.

Mutual irreducibility is a necessary property for this inverse problem to be well-defined, and, unfortunately, we technically have not identified a quark versus gluon discriminant that is truly mutual irreducible.\footnote{The existence of a positive-definite probability distribution solution to \Eq{eq:topicinverse} only requires knowing the values of the reducibility factors or asserting them by fiat.  However, if the minimum and maximum values of the likelihood are not 0 and $\infty$, respectively, then the definition of the topics can be sensitive to the particular approximations one employs, and may change with more information or a better approximation.  In topic modeling in natural language processing, perfectly pure samples are referred to as {\bf anchor words} that appear exclusively in a writing sample about a given topic.  I thank Eric Metodiev for clarification about this point.}  However, total hadronic multiplicity only has exponentially small contamination to defining a pure quark sample, so it is expected that it should work rather well for this purpose, which is indeed what is observed in simulated data \cite{Metodiev:2018ftz}.  A limitation of this technique is that quark and gluon topic labels are only defined statistically, and not on an event-by-event basis.  So, if we only use multiplicity to extract quark and gluon distributions, only quark and gluon distributions of the multiplicity are returned.  There may be some methods to clean up the statistical extraction to be closer to an event-by-event level, by, say, providing the topic modeling with more differential distributions on phase space, but that would also suffer from the curse of dimensionality as the resolved dimensionality of phase space increases.

\subsubsection[Are Quark and Gluon Particle Multiplicities Related by KNO Scaling?]{Are Quark and Gluon Particle Multiplicities Related by KNO Scaling?\footnote{Hinchliffe's rule \cite{Peon:1988kx} suggests the answer ``No'', but we will see that it is a bit more subtle.}}\label{sec:knoqvg}

In our analysis of quark versus gluon discrimination with hadronic multiplicity, we made the assumption that the fundamental KNO function for both, $\psi(x)$, was the same, and all that differed in their functional form were the average multiplicities, $\langle n_q\rangle < \langle n_g\rangle$.  This seems like a reasonable enough assumption, but it also seems reasonable that particle production in quark and gluon jets may be different enough that their KNO functions would, and perhaps should, be different.  At any rate, this can in principle be tested, and we can see if this assumption is born out in experimental data.

Unfortunately, there isn't much extant experimental data on which (charged) particle multiplicity distributions on quark and gluon jets are extracted separately.  There were a few studies of quark versus gluon jet differences measured at LEP, the electron-positron collider that occupied the current LHC ring until 2001.  In these studies, $e^+e^-\to 3$ jets events were identified, and then the two quark jets and one gluon jet were identified by relative energies, angular distributions, and other correlations that are expected at least at leading-order perturbatively.  Early first studies \cite{OPAL:1991ssr,OPAL:1993uun,ALEPH:1994hlg,OPAL:1995ab,DELPHI:1995nzf} measured quark and gluon jet multiplicity distributions at very low jet energies and in some cases determined the parameter $k$ of the negative binomial distribution of the KNO function.  While mean values of multiplicities differed, the best fits to negative binomial KNO functions were consistent on quark and gluon jets, but the quoted uncertainties were extremely large.

\begin{figure}[t!]
\begin{center}
\includegraphics[height=6cm]{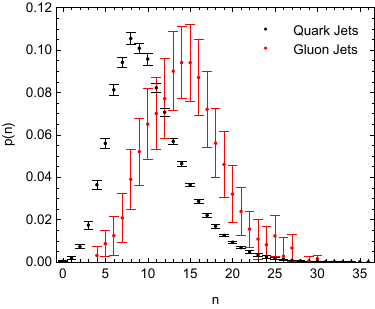}\hspace{0.5cm}
\includegraphics[height=6cm]{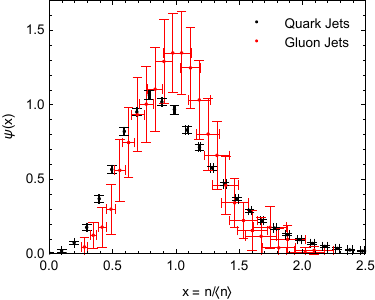}
\caption{\label{fig:opalmults}
Plots of the charged particle multiplicity on quark (black) and gluon (red) jets as extracted from $e^+e^-\to 3$ jets events at the OPAL detector at LEP.  The energy scale of the jets is about 40-45 GeV.  Left: Measured multiplicity distributions.  Right: KNO-scaled distributions, in which the distributions have been rescaled by their respective means.  Data from \InRef{OPAL:1997dkk}.
}
\end{center}
\end{figure}

More definite and precise results were presented in \InRef{OPAL:1997dkk}, using similar techniques for quark and gluon jet extraction, but with significantly smaller uncertainties and at higher energies.  In \Fig{fig:opalmults}, we have plotted the charged particle multiplicity distributions from this study at left, and then the KNO distributions as determined by rescaling by their respective means.  While gluon jets naturally have significantly larger uncertainties, two things are rather clear from these data.  First, as expected, the mean multiplicity of gluon jets is larger than that of quark jets, and second, though uncertainties are still large, the mean-rescaled KNO distribution for gluon jets is narrower than that for quark jets.  In the context of the negative binomial distribution, the value of the $k$ parameter for gluon jets is a bit larger than that for quark jets, $k_g > k_q$.

\begin{figure}[t!]
\begin{center}
\includegraphics[height=6.5cm]{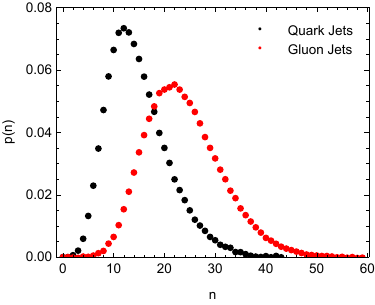}\hspace{0.5cm}
\includegraphics[height=6.5cm]{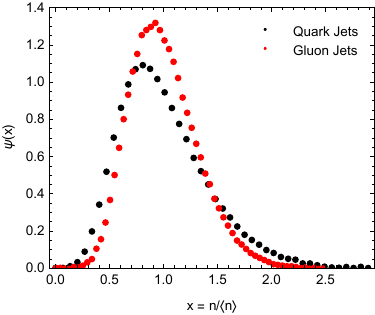}
\caption{\label{fig:atlasmults}
Plots of the charged particle multiplicity (track multiplicity) on quark (black) and gluon (red) jets as extracted from mixed samples with machine learning from jets with transverse momenta of 500-600 GeV in ATLAS events.  Uncertainties are at the level of the size of the data points.  Left: Measured multiplicity distributions.  Right: KNO-scaled distributions, in which the distributions have been rescaled by their respective means.  Data extracted from \InRef{ATLAS:2023dyu}.
}
\end{center}
\end{figure}

This observation is supported by a very recent study on quark versus gluon jets by ATLAS \cite{ATLAS:2023dyu}.  In this study, ATLAS selected jets from distinct samples that are known to have different compositions of quarks and gluons, and was able to extract purified, individual quark and gluon distributions of various observables.  One of those observables was the charged track multiplicity, which is displayed in \Fig{fig:atlasmults}.  At left, we show the plot of the multiplicity on quark and gluon jets, and again, as expected, gluons have a larger mean multiplicity than quarks.  When re-scaled by their respective means, we see a similar relationship of the KNO functions for quark and gluon jets as was observed in the data at LEP, but with significantly smaller uncertainties: the gluon KNO function is narrower than that of quarks.  Further, the rescaled quark and gluon distributions are very similar between LEP and LHC jets, and each can be fit well by essentially the same value of $k$ in the negative binomial, with $k_q \approx 6$ and $k_g\approx 10$.  Further studies will be needed to clearly establish this relationship, and its potential jet energy dependence, as well as provide a theoretical understanding of its origin, but this is a promising direction for improving discrimination of quark and gluon jets.

\subsection*{Exercises}

\begin{enumerate}

\item Redo the Lund plane analysis and calculation of the distribution of an angularity $\tau_\alpha$ when there is an additional perturbative cut on emissions, $zE\theta > k_\perp$.  Is it still true that quark versus gluon discrimination power is independent of angular exponent $\alpha$?  If not, what is the value of $\alpha$ that, say, minimizes the corresponding AUC?

\item Consider the simultaneous measurement of primary Lund plane multiplicity and angularity $\tau_1$.  We choose $\tau_1$ so that its contour is parallel to the perturbative cutoff and demand that $\tau_1 > k_\perp/E$.  Calculate the joint probability distribution of the angularity $\tau_1$ and the number of emissions in the Lund plane above the scale $k_\perp$ on quark and gluon jets, $p_q(\tau_1,k)$.  Show that the likelihood ratio is still monotonic in the multiplicity, and so also measuring the angularity adds no more discrimination information.\\
{\it Hint:} Don't forget to write the joint distribution with conditional probabilities: $p(\tau_1,k) = p(k|\tau_1)p(\tau_1)$.  Where on the Lund plane are the emissions that need to be considered in calculating $p(k|\tau_1)$?

\item From \Eq{eq:knoauc}, we showed that the AUC for KNO scaling multiplicity takes the form
\begin{align}
\text{AUC}({\cal C}) = \int dx\, \psi(x)\, \Psi({\cal C} x)\,,
\end{align}
where $0 < {\cal C} < 1$ is the ratio of quark to gluon mean multiplicities.  Note that this satisfies the differential equation
\begin{align}\label{eq:aucdiffeq}
\frac{d\text{AUC}({\cal C})}{d{\cal C}} =  \int dx\,x\, \psi(x)\, \psi({\cal C} x)\,,
\end{align}
with $\text{AUC}({\cal C} = 1) = 1/2$.  Using methods like that in \Sec{sec:knodiscsec}, provide a robust bound on the integral on the right of \Eq{eq:aucdiffeq}, as a function of ${\cal C}$.  Solve the corresponding bounded differential equation for the AUC.  How does the resulting bound compare to the explicit value of the AUC with the negative binomial model, \Eq{eq:aucmultfull}?

\end{enumerate}

\section{Fourth Example: Quark Flavor Tagging and IRC Unsafe Binary Discrimination}\label{sec:uvsd}

For our final example and lecture, we will pivot to another, distinct, binary jet discrimination problem: classifying jets initiated by different quark flavors.  Like general quark versus gluon jet discrimination, this problem has a lot of motivation from every possible direction in collider physics, from pdf extractions, to extractions of Yukawa couplings, to flavor-sensitive new physics.  For concreteness, we will restrict our focus to the discrimination of jets initiated by quarks in the first generation, up versus down quark jet discrimination.  This problem is of central importance to the goals of the Electron-Ion Collider (EIC) for high-precision extraction of up and down quark pdfs \cite{AbdulKhalek:2021gbh}.\footnote{As of the writing of these notes, the EIC is also the only future collider located in the United States that will definitely be constructed and take data.}

Another, rather orthogonal, motivation for this problem has come from a rather new corner of collider physics: quantum entanglement.  The idea of quantitatively demonstrating that correlations between measurements are due to honest quantum entanglement, and not hidden classical correlations \cite{einstein1935can,bohm1952suggested}, of course dates back to John Bell's thought experiments in the 1960s \cite{Bell:1964kc,bell1966problem}.  Measurements at a collider experiment are classical, all observables commute with one another because only energy is measured, and so this would seem to render any such ability to observe violation of Bell's inequalities impossible.  However, the exclusive left-handed nature of the weak interactions can provide a direct and undiluted connection between particle flavor and spin.  This has motivated several groups to study flavor correlations in top--anti-top pair production events at the LHC (see, e.g., \Refs{Afik:2020onf,Fabbrichesi:2021npl,Takubo:2021sdk,Barr:2021zcp,Severi:2021cnj,Afik:2022kwm}), and then correspondingly reinterpreting them through entanglement of the spins of the top quarks at the point of production.  These techniques have extremely recently been used by ATLAS and CMS to demonstrably observe quantum entanglement in these systems \cite{ATLAS:2023fsd,Collaboration:2024dvz}.\footnote{Really, these experimental results have only demonstrated that the top quark pairs are produced in an entangled state assuming quantum mechanics.  These results do not unambiguously establish non-classical correlations with no such assumption of quantum mechanics, because there are still potential loopholes.  Further, Bell's unambiguous criteria for non-classical entanglement of a pair of particles is expressed as an inequality, but for entanglement of three particles, a boolean test is possible \cite{greenberger1990bell}.  If such a test of entanglement could be extended to collider experimental detectors, there would be no denying the quantum mechanics in your face \cite{mermin1990quantum,mermin1990s,Coleman:2020put}.  I thank Dorival Gon\c calves for clarifications of this point.}

These techniques have mostly focused on the case of leptonic top quark decay, for which the charged lepton from subsequent $W$ boson decay has perfect correlation of its momentum with the spin of the initial top quark.  This is experimentally very clean, because charged leptons are easily identified and their momenta measured to high precision.  However, the top quark pair production decays to visible leptons (electrons or muons) only about 5\% of the time, and so one takes a huge statistical hit in restricting to this decay channel.  The vast majority of the time, the top quark pairs decay hadronically or semi-leptonically, and so one needs techniques to efficiently analyze these processes to make the most of the extant data.  Through the subsequent decay of the $W$ boson, the down-type quark carries perfect correlation of its momentum to the spin of the top, and so an efficient up versus down flavor discriminant could significantly open up the phase space for studying this fundamental of quantum phenomena.  Surprisingly rather little work has focused on hadronic top decays (see, e.g., recent \Refs{Dong:2023xiw,Han:2023fci,Maltoni:2024csn}), with perhaps the largest step forward from \InRef{Tweedie:2014yda} that identified the optimal direction of the spin correlation vector from only measuring the total momentum of the jets from top quark decay.  If we can construct a powerful up versus down discriminant, perhaps this story will change \cite{Dong:2024xsg}.

\subsection{How Do We Define ``Flavor''?}

Our goal is to identify the likelihood ratio observable for jets initiated by up versus down quarks.  As we did with the quark versus gluon or $H\to b\bar b$ versus $g\to b\bar b$ problems, we need to first establish what it is we mean when we say ``up quark'' or ``down quark''.  Naturally, by quantum mechanics, the only observable quantities that can uniquely define a particle are eigenvalues of Hermitian operators.  There are but a limited few Hermitian operators whose eigenvalues are actually (potentially) observable in a collider, so let's simply enumerate them for up and down quarks, and then comment on their consequences:
\begin{center}
\begin{tabular}{c|cc}
Observable & Up Quark & Down Quark\\
\hline
Energy-Momentum & $m_u \approx 2 $ MeV & $m_d \approx 4.5$ MeV\\
Spin & 1/2 & 1/2\\
Color Charge & $C_F$ & $C_F$\\
Electric Charge & $2/3e$ & $-1/3 e$ 
\end{tabular}
\end{center}

There simply isn't much to uniquely define the quarks.\footnote{I am glossing over another potential source for discrimination.  The other light quark, the strange quark, has the same electric charge as the down quark.  As such, mesons formed from strange quarks with first-generation quarks, the kaons, differ in their electric charge.  So, observing charged kaons in a jet is strongly correlated with up-type jets. Strange jet tagging has seen some interest recently, e.g.,~\Refs{Duarte-Campderros:2018ouv,Nakai:2020kuu,Albert:2022mpk,Bedeschi:2022rnj,Erdmann:2020ovh}, especially applied to (exclusive) Higgs decays.}  Both up and down quarks have exceedingly small masses (values from the PDG \cite{Workman:2022ynf}) and such a mass is completely irrelevant both because hadron masses are so much greater than individual quark masses, and because any reasonable jet energy is tens, hundreds, or thousands of GeV.  Up and down quarks are both fermions, and so have identical spin quantum numbers, so there are no games to play with polarization states, for example, to distinguish them.  All quarks carry QCD color in the fundamental representation, with quadratic Casimir $C_F$, and so the radiation pattern of gluons off of these quarks will be indistinguishable.  However, there is one (!!!)~quantum number that is distinct and will have observable consequences: electric charge. 

Non-zero electric charge is relatively easy to measure in a collider, because high-energy charged particles will emit ionization radiation that is observable and their trajectory will be bent when passing through a magnetic field, with curvature proportional to the electric charge.  This would suggest that the net electric charge of a jet,
\begin{align}
Q_\text{jet} = \sum_i Q_i\,,
\end{align}
where $Q_i$ is the electric charge in units of the fundamental charge $e$, would be a good discrimination observable for this task.  There are, however, at least two problems with interpretation of this raw jet charge.  First, every charged particle in the jet contributes with equal strength, and yet a jet can contain a significant amount of very low energy contamination radiation.  This contamination can significantly increase the variance of the distribution of such an observable, decreasing discrimination power.  Second, there is a disconnect with the electric charge of bare quarks and that of actual observable hadrons.  The electric charge of quarks is some fraction of the fundamental electric charge, but the electric charge of hadrons is necessarily an integral multiple of $e$.  So on any jet of interest, it is simply impossible for its observed electric charge to be 1/3 $e$, but perhaps in an ensemble, the mean of the jet charge distribution carries information about the quark charge.

The second problem we will address with our formal analysis in the next section, but for the first problem, there is a relatively easy fix.  We just weight the contribution to the jet charge by some power of the particles' energy fraction,
\begin{align}
{\cal Q}_\kappa \equiv \sum_{i} z_i^\kappa\, Q_i\,.
\end{align}
This definition of ``the'' jet charge was introduced long ago \cite{Field:1977fa}\footnote{A quantitative study of jets and the first Monte Carlo event generator was also presented in this paper, written by Feynman and his post-doc Rick Field.  Field is emeritus professor at the University of Florida and the brother of actress Sally Field.} and experimentally studied in great detail \cite{Fermilab-Serpukhov-Moscow-Michigan:1979zgc,Erickson:1979wa,Berge:1980dx,Aachen-Bonn-CERN-Munich-Oxford:1981lfk,Aachen-Bonn-CERN-Munich-Oxford:1982riw,EuropeanMuon:1984xji,Amsterdam-Bologna-Padua-Pisa-Saclay-Turin:1981hcw,TASSO:1990gda,DELPHI:1991mqi,ALEPH:1991fba,OPAL:1992jsm,OPAL:1994xvz,SLD:1994wsf,DELPHI:1996hzb,ALEPH:1997agc,CDF:1999jfn,L3:1999dig,OPAL:2000rnf,DELPHI:2001kqh,D0:2006kee,CDF:2013tpw,CDF:2013tpw}, and whose interest for LHC and jet physics was initiated in \Refs{Krohn:2012fg,Waalewijn:2012sv} and subsequently measured on jets at the LHC in \Refs{ATLAS:2013mkl,CMS:2014rsx,ATLAS:2015rlw,CMS:2017yer,CMS:2020plq}.  The parameter $\kappa > 0$ so that arbitrarily soft particles do not contribute to the jet charge.  This then renders the jet charge infrared safe, but it is still not collinear safe. Consider, for example, the contribution to the jet charge from a gluon that splits to an exactly collinear $u\bar u$ pair.  The gluon is electrically neutral, and so does not initially contribute to the jet charge.  However, after the splitting the $u\bar u$ pair modifies the jet charge by
\begin{align}
\Delta{\cal Q}_\kappa = \frac{2}{3}(z_u^\kappa - z_{\bar u}^\kappa)\,.
\end{align}
This has probability 1 to be non-zero, and so the jet charge is irreducibly IRC unsafe.  There are no tricks with multi-differential distributions or Sudakov factors that can save us here.  To theoretically understand this observable, we need a totally different approach than Feynman diagrams, Lund planes, etc.

\subsubsection{Non-Perturbative Assumptions and Consequences}\label{sec:nonpertassumpts}

We are at any rate scientists, and starting from nothing has never scared us before.  We just need to identify a good and powerful set of assumptions from which to work and which can be tested in experiment.  As mentioned previously, the results we derive here will not follow any perturbative analysis in the coupling $\alpha_s$.  Instead, we will use justifiable non-perturbative assumptions that follow from basic observations about the structure of hadronic jets at high energies and their particle content.  The assumptions that we exploit were introduced in \InRef{Kang:2023ptt} and are as follows:
\begin{enumerate}
\item Particles (hadrons) in the jet are produced though identical, independent processes.
\item The multiplicity of particles in the jet $n$ is large.
\item The only particles in the jet are the pions: $\pi^+$, $\pi^-$, and $\pi^0$.
\item We assume exact SU(2) isospin symmetry and so quantities measured on any of the pions have identical distributions.
\end{enumerate}
Exclusively from these assumptions, we will be able to quantitatively produce the functional form of the likelihood ratio for up versus down quark discrimination.  Let's see how this is done.

We will start with the goal of determining the probability distribution of the jet charge on jets initiated by quark $q$ itself, $p_q({\cal Q}_\kappa)$.  Conditioning on the hadronic multiplicity $n$ in the jet, this distribution can be expressed as
\begin{align}
p_q({\cal Q}_\kappa) = \int dn\, p_q({\cal Q}_\kappa|n)\,p_q(n)\,.
\end{align}
We can, of course, say something about the KNO scaling of the multiplicity distribution $p(n)$, but we won't worry about that yet. Let's instead focus on the conditioned distribution, $p_q({\cal Q}_\kappa|n)$.  The form of the jet charge is a sum over i.i.d.~random variables (by the first assumption), and we are considering the large multiplicity $n\to\infty$ limit (by the second assumption).  Therefore, the central limit theorem predicts the functional form of this distribution as a Gaussian:
\begin{align}
p_q({\cal Q}_\kappa|n) = \frac{1}{\sqrt{2\pi\sigma^2}}\,e^{-\frac{({\cal Q}_\kappa - \mu_q)^2}{2\sigma^2}}\,,
\end{align}
where we must determine the mean $\mu_q$ and variance $\sigma^2$.

The mean is just the mean of the jet charge on our ensemble of jets with $n$ constituents, where
\begin{align}
\mu_q = \left\langle\sum_{i=1}^n z_i^\kappa Q_i\right\rangle = n\langle z^\kappa\rangle \langle Q\rangle = Q_q \langle z^\kappa\rangle\,.
\end{align}
At right, we note that the mean value of the individual electric charges in the jets initiated by a quark $q$ is $Q_q / n$, where $Q_q$ is the electric charge of the quark $q$.  $\langle z^\kappa\rangle$ is the expectation value of the single particle energy fraction raised to power $\kappa$.  Right now, we have no constraint on what this should be, but we will fix that soon.  The variance of the central limit Gaussian is, correspondingly,\footnote{We are being slightly cavalier here.  This expression for the variance assumes that the mean jet charge is 0, which is not true in general.  However, the variance is a measure of the spread of the distribution about the mean, but is independent of the particular value of the mean, so we are free to set the mean to 0 to evaluate the variance.} 
\begin{align}
\sigma^2 = \left\langle\sum_{i=1}^n z_i^{2\kappa} Q_i^2\right\rangle = \frac{2}{3} n\langle z^{2\kappa}\rangle\,.
\end{align}
The argument of the sum in the middle follows from our proof of the central limit theorem in \Sec{sec:clt}, and what is especially important here is that each term is proportional to the square of particle $i$'s electric charge.  From assumption 3 and 4, because we only measure pions and assume exact isospin symmetry, $Q_i^2 = 0$ or $1$ on pions, and 2/3 of the pions are charged.  Hence, at right, the mean number of charged pions in each jet in this ensemble is $2/3n$.  This variance also needs no subscript decoration because it is universal and identical for all jets, by the assumption of isospin conservation.

We still have to evaluate these tricky moments of the energy fraction distribution, $\langle z^\kappa\rangle$ and $\langle z^{2\kappa}\rangle$.  These can be calculated through an additional conditional distribution of energy fraction, $p(z|n)$, where, for example,
\begin{align}
\langle z^\kappa\rangle = \int_0^1 dz\, z^\kappa\, p(z|n)\,,
\end{align}
where the energy fraction $z\in[0,1]$.  This conditional distribution is of course normalized,
\begin{align}
1 = \int_0^1 dz\, p(z|n) 
\end{align}
Because all jets in the ensemble have $n$ total particles, the mean of this distribution is 
\begin{align}
\langle z\rangle = \int_0^1dz\, z\, p(z|n) = \frac{1}{n}\,,
\end{align}
as the sum total energy fraction must be 1.  However, no other moments of this distribution are known, so we need another approach.

\subsubsection{The Moment Expansion}

Because the Gaussian distribution of jet charge only needs moments of this energy fraction distribution, it is useful to work with a representation of the distribution expressed exclusively in terms of its moments.  The question of consistency of the set of moments of a non-negative, integrable distribution is an ancient problem in probability, called the {\bf moment problem}, with known solutions \cite{stieltjes1894recherches,hamburger1920erweiterung,hausdorff1921summationsmethoden}.  Formally, any probability distribution $p(x)$ can always be expanded in a series of $\delta$-functions and its derivatives as
\begin{align}
p(x) = \sum_{k = 0}^\infty \frac{(-1)^k \langle x^k\rangle}{k!}\,\delta^{(k)}(x)\,,
\end{align}
where $\langle x^k\rangle$ is the $k$th moment.\footnote{However, it is possible that the moments do not uniquely determine the probability distribution.  For example, the distribution
\begin{align}
p(x) = \frac{e^{-x^{1/4}}}{24}\left(1+\epsilon \sin(x^{1/4})\right)\,,
\end{align}  
for $x\in[0,\infty)$ and $|\epsilon| < 1$, has moments
\begin{align}
\int_0^\infty dx\, x^n\, p(x) = \frac{(4n+3)!}{6}\,,
\end{align}
independent of $\epsilon$.  The moment problem is unique if the higher moments at worst diverge sufficiently slowly \cite{carleman1922ueber,krein1945problem}, with a uniqueness bound of roughly $\langle x^n\rangle \lesssim (2n)!$.}  For the energy fraction distribution at hand, $p(z|n)$, we know the mean, so it is much more useful to represent this expansion in terms of central moments $\mu_{z,k}$, where
\begin{align}
p(z|n) = \delta\left(
z - \frac{1}{n}
\right) + \sum_{k = 2}^\infty \frac{(-1)^k \mu_{z,k}}{k!}\,\delta^{(k)}\left(
z - \frac{1}{n}
\right)\,.
\end{align}
Note that there is no term proportional to the first derivative of the $\delta$-function here.  Expanding about the mean is expanding about a stable minima, in exactly the same way as expanding the description of a physical pendulum about its static equilibrium.  Oscillations about that minimum are first described by the moment of inertia, which is a quadratic moment of the distribution of the mass of the pendulum.  The central moments are defined as
\begin{align}
\mu_{z,k} = \int_0^1 dz\, \left(
z - \frac{1}{n}
\right)^k\, p(z|n)\,,
\end{align}
and $\mu_{z,2} = \sigma_z^2$ is the variance.\footnote{The solution to the moment problem involves constructing the Hankel matrix of all moments and demanding that all possible determinants of minors of the matrix are positive.  This constraint has seen use in a number of recent papers as explanation for positivity of Wilson coefficients in an effective field theory, for example \Refs{Green:2019tpt,Chen:2019qvr,Arkani-Hamed:2018ign,Sen:2019lec,Arkani-Hamed:2020blm,Bellazzini:2020cot,Huang:2020nqy,Tolley:2020gtv,Bern:2021ppb,Kang:2023zdx}.}

While you are welcome to incorporate higher moments in the description of the likelihood, we will only explicitly consider through the variance, with
\begin{align}
p(z|n) = \delta\left(
z-\frac{1}{n}
\right) + \frac{\sigma_z^2}{2}\,\delta''\left(
z-\frac{1}{n}
\right)
+\cdots\,,
\end{align}
suppressing higher moments.\footnote{In general, we expect that the moment expansion is only asymptotic because $\delta$-functions and its derivatives have support at a single point.  However, in practice, unless integrating the distribution $p(z|n)$ against a particularly pathological weight, the moment expansion converges quite rapidly.  For example, the mean value of $\log x$ on the uniform distribution on $x\in [0,1]$ is $\langle \log x\rangle = -1$.  Using the central moment expansion for the uniform distribution, the first three non-zero terms produce
\begin{align}
\langle \log x\rangle =\int_0^1dx\, \log x = -\log 2 -\frac{1}{6} -\frac{1}{20}+\cdots \approx -0.9098\,.
\end{align}}  This will be sufficient to draw a number of conclusions.  Importantly, we note that the variance is strictly non-negative, $\sigma_z^2 \geq 0$.  Using this moment expansion, the $\kappa$ moment of the energy fraction distribution is then easy to calculate, where
\begin{align}
\langle z^\kappa\rangle = n^{-\kappa}\left(
1-\frac{\kappa(1-\kappa)}{2}\,\sigma_z^2n^2+\cdots
\right)
\end{align}
Using this result, we can then express the mean and variance of the multiplicity-conditioned jet charge distribution as
\begin{align}
\mu_q &= Q_q n^{-\kappa}\left(
1-\frac{\kappa(1-\kappa)}{2}\,\sigma_z^2n^2+\cdots
\right)\,,\\
\label{eq:chargewidthexp}
\sigma^2 &= \frac{2}{3}n^{1-2\kappa}\left(
1-\kappa(1-2\kappa)\,\sigma_z^2n^2+\cdots
\right)\,.
\end{align}

In general, the moment expansion of $p(z|n)$ will be an expansion in central moments times a power of multiplicity as $\mu_{z,k} n^k$.  The mean of $p(z|n)$ is $\langle z\rangle = 1/n$ and it has support on $z\in[0,1]$ and so general results imply that the largest possible value that the central moment can be is $\mu_{z,k} \sim 1/n$ \cite{doi:10.1080/00029890.2000.12005203}.  In the large $n$ limit, terms in such an expansion would grow unbounded with $n$ and therefore be rather useless.  However, this $1/n$ scaling of the central moment $\mu_{z,k}$ corresponds to the distribution
\begin{align}\label{eq:worstdist}
p_\text{worst}(z|n) = \left(1-\frac{1}{n}\right)\delta(z) + \frac{1}{n}\,\delta(1-z)\,,
\end{align}
which is highly degenerate and unphysical, with a single particle carrying the entire energy of the jet.  A realistic distribution $p(z|n)$ will have most of its support around the mean value, with a tail extending to large values, with strictly 0 probability for $z=1$.  An example of such a distribution would be
\begin{align}\label{eq:simpzdist}
p(z|n) = (n-1)(1-z)^{n-2}\,,
\end{align}
which is normalized and has mean $1/n$.  For such a distribution, its $k$th central moment is
\begin{align}
\mu_{z,k} = \int_0^1 dz\, \left(z-\frac{1}{n}\right)^k\, (n-1)(1-z)^{n-2} \propto n^{-k}\,,
\end{align}
so that in the large multiplicity $n$ limit, $\mu_{z,k} n^k$ is order-1 and the moment expansion is sensible.  Thus, we have replaced a perturbative expansion in the coupling $\alpha_s$ with a perturbative expansion in central moments.

\begin{figure}[t!]
\begin{center}
\includegraphics[width=8cm]{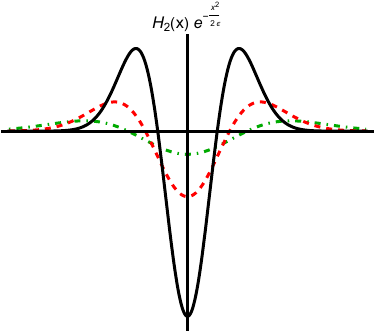}
\caption{\label{fig:hermite}
Plot of the second Hermite polynomial, as defined in \Eq{eq:hermite_rod}, times the Gaussian with variance $\epsilon$.  $\epsilon$ is decreased by a factor of 2 from green--dot-dashed to red-dashed, and then from red-dashed to black-solid.
}
\end{center}
\end{figure}

It's worth saying a bit more about the moment expansion to provide a bit more justification as to where it comes from and why it actually exists.  To do this, let's assume we are considering the probability distribution $p(x)$ of a random variable $x\in(-\infty,\infty)$ with mean $\langle x\rangle =0$.  We can, of course, expand the probability distribution in Hermite polynomials $H_n(x)$ with an overall Gaussian centered at $x = 0$, where
\begin{align}
p(x) = \frac{1}{\sqrt{2\pi \epsilon}}\,e^{-\frac{x^2}{2\epsilon}}\sum_{i = 0}^\infty c_i\, H_i(x)\,.
\end{align}
Here, $\epsilon$ is the variance of the Gaussian, $c_i$ are some coefficients, and the Hermite polynomials are defined through a (non-standard) Rodrigues-like formula as
\begin{align}\label{eq:hermite_rod}
H_n(x) =  e^{\frac{x^2}{2\epsilon}}\,\frac{d^n}{dx^n}e^{-\frac{x^2}{2\epsilon}}\,.
\end{align}
The functional form of the second Hermite polynomial with this definition times the Gaussian kernel is plotted in \Fig{fig:hermite}, as the variance $\epsilon$ decreases.  

With this expansion, note that the Gaussian becomes the $\delta$-function in the limit that $\epsilon\to 0$.  With the Rodrigues formula, we can determine the coefficients of the expansion $c_i$ through their relationship to moments.  Let's calculate the contribution to moment $\langle x^m\rangle$ from the $n$th order Hermite polynomial.  We have
\begin{align}
\langle x^m\rangle \supset \frac{c_n}{\sqrt{2\pi \epsilon}}\int_{-\infty}^\infty dx\, x^m\,e^{-\frac{x^2}{2\epsilon}} H_n(x) = \frac{c_n}{\sqrt{2\pi \epsilon}}\int_{-\infty}^\infty dx\, x^m\,\frac{d^n}{dx^n}e^{-\frac{x^2}{2\epsilon}}\,.
\end{align}
Now, we can integrate-by-parts to move all the derivatives from the Gaussian onto the power of $x$.  Note that if $n > m$, these derivatives eventually kill $x^m$, so we only need to consider $m \geq n$.  We then have
\begin{align}
\langle x^m\rangle \supset (-1)^{n}\frac{m!}{(m-n)!}\frac{c_n}{\sqrt{2\pi \epsilon}}\int_{-\infty}^\infty dx\, x^{m-n}\,e^{-\frac{x^2}{2\epsilon}}\,.
\end{align}
We could just evaluate the integral that remains, but it is more enlightening to rescale the integration variable to remove dependence on $\epsilon$.  With $x\to\sqrt{\epsilon}x$, we then have
\begin{align}
\langle x^m\rangle \supset (-1)^{n}\frac{m!}{(m-n)!}\epsilon^{\frac{m-n}{2}}\frac{c_n}{\sqrt{2\pi}}\int_{-\infty}^\infty dx\, x^{m-n}\,e^{-\frac{x^2}{2}}\,.
\end{align}
We already noted that $m\geq n$, and with our $\delta$-function limit in mind, we can now take the $\epsilon \to 0$ limit.  If $m > n$, then all such contributions are suppressed by positive powers of $\epsilon$, so only a single term remains: $n = m$.  Thus,
\begin{align}
\langle x^m\rangle = \lim_{\epsilon \to 0}\sum_{n=0}^m (-1)^{n}\frac{m!}{(m-n)!}\epsilon^{\frac{m-n}{2}}\frac{c_n}{\sqrt{2\pi}}\int_{-\infty}^\infty dx\, x^{m-n}\,e^{-\frac{x^2}{2}} =m! (-1)^{m} c_m\,.
\end{align}

Therefore, the Hermite polynomial expansion of the probability distribution can be expressed in the limit that:
\begin{align}
p(x) = \lim_{\epsilon\to 0}\frac{1}{\sqrt{2\pi \epsilon}}\,e^{-\frac{x^2}{2\epsilon}}\sum_{i = 0}^\infty c_i\, H_i(x) = \delta(x)+\sum_{k=2}^\infty \frac{(-1)^k \langle x^k\rangle}{k!}\,\delta^{(k)}(x)\,.
\end{align}
So, while the derivative of $\delta$-function expansion for a probability distribution may seem awkward and ill-defined, it is simply the narrow Gaussian limit of the familiar expansion into orthogonal Hermite poynomials.

\subsubsection{Measurements of Discrimination Power and the Likelihood Ratio}

With these results for the Gaussian distributions of jet charge conditioned on multiplicity, we can then analyze and quantify its discrimination power.  We will start with the distance between the mean values $\mu_u$ and $\mu_d$ with respect to the square-root of the sum of the variances of the jet charge distributions.  As we demonstrated earlier, this quantity is the fundamental discrimination power distance and appears, for example, in the argument of error function in the calculation of the AUC.  Using the means and variances we calculated above in the moment expansion of the energy fraction distribution $p(z|n)$, we have
\begin{align}
\frac{|\mu_u - \mu_d|}{\sqrt{\sigma^2 + \sigma^2}} &= \frac{n^{-\kappa}\left(
1-\frac{\kappa(1-\kappa)}{2}\,\sigma_z^2n^2+\cdots
\right)}{\sqrt{\frac{4}{3}n^{1-2\kappa}\left(
1-\kappa(1-2\kappa)\,\sigma_z^2n^2+\cdots
\right)}}\\
&=\frac{\sqrt{3}}{2\sqrt{n}}\left(
1 - \frac{\kappa^2}{2}\sigma_z^2 n^2 +\cdots
\right)\,.
\nonumber
\end{align}

This expression makes two concrete predictions that can be validated in (simulated) data.  First, there is overall dependence on the inverse of the multiplicity, and so the distance between the jet charge distributions increases as multiplicity $n$ decreases.  There is some evidence in previous machine learning studies of the jet charge that discrimination improves as multiplicity decreases \cite{Fraser:2018ieu}, but this can be directly demonstrated in simulated data by binning jets by multiplicity \cite{Kang:2023ptt}.  Second, $\kappa$ dependence appears as the coefficient proportional to the variance of the energy fraction distribution, and the discrimination power increases as $\kappa$ decreases.  This property of the jet charge has been directly observed in several prior studies \cite{Krohn:2012fg,Fraser:2018ieu,Chen:2019uar,Chen:2019uar,Kang:2021ryr,Lee:2022kdn}, but only recently was this simple explanation derived.  We note that both of these observations have limits, simply from the assumptions we made.  Specifically, if multiplicity $n$ is too small, then the central limit theorem no longer applies, and the Gaussian form of the jet charge distribution breaks down.  Similarly, if energy weighting exponent $\kappa$ is too small, then jet charge loses its infrared safety and additionally necessarily takes only integer values, because in this limit the jet charge reduces to a sum over individual particles' integer charges.

Let's now move to evaluating the likelihood ratio for up and down quark jets on which the jet charge ${\cal Q}_\kappa$ and the hadronic multiplicity $n$ is measured.  From both our assumption of isospin conservation and the quantum numbers of up and down quarks, the multiplicity distributions of hadrons in up and down quark jets will be identical:
\begin{align}
p_u(n) = p_d(n) \equiv p(n)\,.
\end{align}
Then, the likelihood ratio as the ratio of the joint probability distributions of the jet charge and multiplicity reduces to the ratio of the jet charge distributions conditioned on the multiplicity:
\begin{align}
{\cal L} = \frac{p_u({\cal Q}_\kappa,n)}{p_d({\cal Q}_\kappa,n)} = \frac{p_u({\cal Q}_\kappa|n)p(n)}{p_d({\cal Q}_\kappa|n)p(n)} = \frac{p_u({\cal Q}_\kappa|n)}{p_d({\cal Q}_\kappa|n)} \,.
\end{align}
These conditional distributions are Gaussians by the central limit theorem, so it is useful to take the logarithm, which is a monotonic function and so is still the optimal discriminant.  The logarithm of the likelihood ratio using the results derived above is then
\begin{align}\label{eq:loglikecharge}
\log {\cal L} = \frac{3}{2} n^{-1+\kappa}{\cal Q}_\kappa - \frac{1}{4n} +\frac{\kappa}{4n}\left(
\kappa+3 n^\kappa {\cal Q}_\kappa(1-3\kappa)
\right) \sigma_z^2 n^2 +\cdots\,,
\end{align}
where we have expanded through the variance contribution of the energy fraction distribution.

\begin{figure}[t!]
\begin{center}
\includegraphics[width=7.5cm]{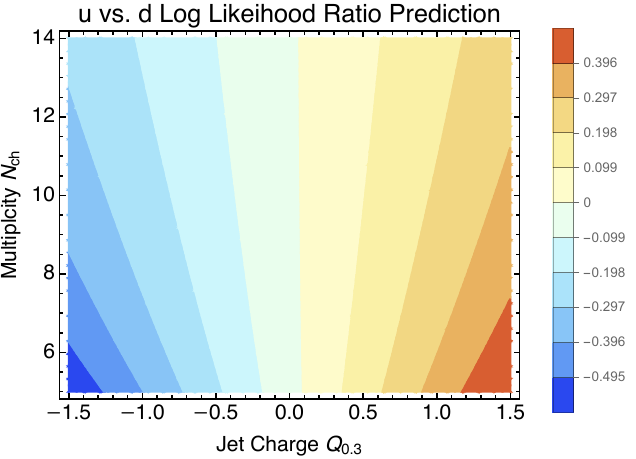}
\hspace{1cm}\includegraphics[width=7.5cm]{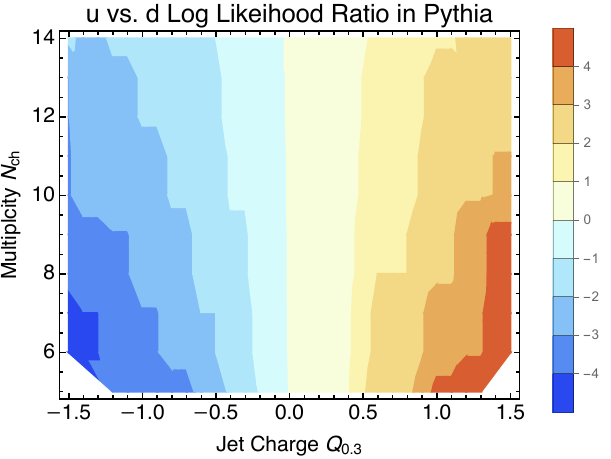}
\caption{\label{fig:qmultcont}
Contours of the logarithm of the likelihood ratio on the plane of the charged particle multiplicity $N_\text{ch}$ and the jet charge ${\cal Q}_{\kappa = 0.3}$.  At left are the contours from the leading terms of \Eq{eq:loglikecharge}, and at right are the contours from simulated data.  Figure from Ref.~\cite{Kang:2023ptt}.}
\end{center}
\end{figure}

The rather remarkable thing about this expression is that the likelihood ratio is not monotonic in the jet charge ${\cal Q}_\kappa$; it is a non-trivial function of both jet charge and multiplicity.  Thus, studies that use the jet charge alone for discrimination of jets initiated by up versus down quarks are necessarily suboptimal, and can be improved by additionally measuring particle multiplicity.  This prediction, which again follows from the eminently reasonable, basic assumptions of things like the central limit theorem and isospin conservation, can be tested in simulated data.  This comparison is shown in \Fig{fig:qmultcont}.  At left is plotted the contours of the log likelihood on the plane of the jet charge ${\cal Q}_{\kappa = 0.3}$ versus charged particle multiplicity, $N_\text{ch}$.  By isospin, the charged particle multiplicity is just $2/3$ of the total multiplicity.  Here, we have just included the first two terms of \Eq{eq:loglikecharge}, ignoring the variance $\sigma_z^2$ term for simplicity.  Indeed, these contours fan out as multiplicity increases, demonstrating the non-trivial interrelation of these observables.  At right on this figure is plotted the log likelihood as extracted from simulated data of jets initiated by up and down quarks.  The data set is finite, and so there are some residual binning effects, and further, the scale of the contours is off by a factor (that is beyond the domain of our assumptions), but the shapes of the predicted and simulated contours are strikingly similar.

\subsubsection{Up and Down Quark Reducibility Factors}

As we have in other situations, the expression for the likelihood itself provides a lower bound estimate on the AUC through the value of its minimum and maximum, the reducibility factors.  Naively, we would think that the jet charge in our approximation is Gaussian distributed and therefore can range over all values, ${\cal Q}_\kappa\in(-\infty,\infty)$, and so the likelihood ranges over all allowed values, ${\cal L}\in[0,\infty)$.  However, the jet charge ${\cal Q}_\kappa$ is conditioned on the multiplicity of particles in the jet $n$, and this fixes finite maximum and minimum values of it.  This correspondingly will affect the bounds on the likelihood, as we will derive now.\footnote{We are working within the constraints and assumptions of the central limit theorem, which is only an accurate description of the jet charge distribution in the large $n$ limit, and relatively close to its mean.  The bounds of the likelihood are determined by the behavior far in the tails, where the Gaussian approximation is not valid.  Nevertheless, we will use these assumptions to derive first quantitative results here as a benchmark for future improved studies.  I thank Eric Metodiev for suggesting this analysis.}

First, note that the expression for the log likelihood of the jet charge and multiplicity, \Eq{eq:loglikecharge}, is monotonic in the jet charge, for fixed multiplicity.  That is, at fixed multiplicity, the bounds on the likelihood are set by the bounds on the jet charge itself.  For fixed multiplicity $n$, recall that the jet charge is
\begin{align}
{\cal Q}_\kappa = \sum_{i=1}^n z_i^\kappa Q_i\,.
\end{align}
With our assumption that jets only consist of pions, the maximum value of the jet charge is accomplished in a jet full of $\pi^+$:
\begin{align}
{\cal Q}_\kappa \leq \sum_{i=1}^n z_i^\kappa\,.
\end{align}
Assuming that $\kappa \leq 1$, the sum over energy fractions is maximized if all are equal, $z_i = 1/n$, and so the jet charge is bounded from above as\footnote{If $\kappa > 1$, the sum over energy fractions is maximized when a single particle takes all of the energy, so the corresponding bound on the jet charge is ${\cal Q}_{\kappa > 1} \leq 1$.}
\begin{align}
{\cal Q}_\kappa \leq n^{1-\kappa}\,.
\end{align}
The minimum value of the jet charge is just the opposite of this, if the jet is filled with $\pi^-$.  Thus, the bounds on the value of the jet charge at fixed multiplicity are
\begin{align}
-n^{1-\kappa}\leq {\cal Q}_\kappa \leq n^{1-\kappa}\,.
\end{align}

Let's use this to determine the bounds on the log likelihood.  With the maximum value of the jet charge, the log likelihood is bounded from above as
\begin{align}
\log {\cal L}\leq  \frac{3}{2}  - \frac{1}{4n} +\frac{\kappa}{4n}\left(
\kappa+3 n(1-3\kappa)
\right) \sigma_z^2 n^2 +\cdots\,.
\end{align}
To maximize this expression over multiplicity $n$ now, we see that we need to take the $n\to\infty$ limit in which this reduces to
\begin{align}
\log {\cal L}\leq  \frac{3}{2}  +\frac{3}{4}\kappa (1-3\kappa)
\sigma_z^2 n^2 +\cdots\,,
\end{align}
where we assume that $\sigma_z^2 n^2\sim 1$, as discussed earlier.  For the minimum value of the log likelihood, we replace ${\cal Q}_\kappa\to -n^{1-\kappa}$ and have
\begin{align}
\log {\cal L}\geq  -\frac{3}{2}  - \frac{1}{4n} +\frac{\kappa}{4n}\left(
\kappa-3 n(1-3\kappa)
\right) \sigma_z^2 n^2 +\cdots\,.
\end{align}
This is minimized for $n\to 0$, for which the likelihood vanishes.  This is obviously very far from the large multiplicity assumption, but nevertheless, let's take $n=0$ here for the expression of the minimum value of the likelihood, but a more honest consideration and consequences of the small multiplicity limit would be fascinating and important for this problem. 

Then, the bounds on the log likelihood for up versus down quark jet discrimination with the jet charge and multiplicity are
\begin{align}
- \infty \leq \log{\cal L}\leq  \frac{3}{2}  +\frac{3}{4}\kappa (1-3\kappa)
\sigma_z^2 n^2 +\cdots\,,
\end{align}
again, assuming $\kappa \leq 1$.  These bounds can then be used to establish a robust lower bound on the AUC, from the expression of \Eq{eq:aucmin}, which we reproduce here:
\begin{align}
\text{AUC} \geq \frac{1-2{\cal L}_{\min}+{\cal L}_{\max}{\cal L}_{\min}}{2({\cal L}_{\max}-{\cal L}_{\min})}\,.
\end{align}
With sufficiently small $\kappa$, the subleading contributions to the bounds of the likelihood are negligible, and so we will estimate the bounds of the likelihood with its leading terms:
\begin{align}
0\leq {\cal L} \lesssim e^{3/2} \approx 4.48\,.
\end{align}
While a pure sample of down quark jets is possible because the minimum likelihood is 0, a pure sample of up quark jets is not because the upper bound is not infinite.  The corresponding bound on the AUC is then
\begin{align}
\text{AUC}\gtrsim \frac{1}{2e^{3/2}}\approx 0.112\,.
\end{align}
This will be an interesting baseline for comparison with quantitative calculations of the AUC from the constructed distributions of the jet charge and multiplicity.

\subsection{Effects of Multiplicity Distribution}

Now that we have the likelihood ratio for up versus down quark discrimination, let's get our hands dirty and actually calculate some explicit discrimination metrics on the various distributions.  Here, we will present calculations of the AUCs for both the measurement of the jet charge alone, and when we measure the likelihood ratio of jet charge and multiplicity, as established above.  In these calculations, we will further be able to include effects of non-trivial multiplicity distributions and demonstrate what effect a finite width in the KNO distribution for the multiplicity has on discrimination.  This will produce additional dependence on the energy weight exponent $\kappa$ and, combined with earlier results with a finite width of the energy fraction distribution $p(z|n)$, provides a potential explanation of observed $\kappa$ dependence in discrimination power.

\subsubsection{Calculating the Jet Charge AUC}

Let's start by just calculating the AUC for up versus down jet discrimination with the jet charge alone.  The integrals we need to calculate are
\begin{align}
\text{AUC}_{{\cal Q}_\kappa} &= \int d{\cal Q}_u\, d{\cal Q}_d\, p_u({\cal Q}_u)\, p_d({\cal Q}_d)\,\Theta({\cal Q}_d-{\cal Q}_u)\,,
\end{align}
but there are additional integrals implicit.  The jet charge distribution, inclusive over particle multiplicity, is
\begin{align}
p({\cal Q}_\kappa) = \int dn \, p({\cal Q}_\kappa|n)\, p(n)\,,
\end{align}
and the conditional distribution takes the familiar Gaussian form.  Then, the AUC in all of its glory is
\begin{align}
\text{AUC}_{{\cal Q}_\kappa} &= \int d{\cal Q}_u\, d{\cal Q}_d\, p_u({\cal Q}_u)\, p_d({\cal Q}_d)\,\Theta({\cal Q}_d-{\cal Q}_u)\\
&=\int d{\cal Q}_u\, d{\cal Q}_d\, dn_u\, dn_d\, p_u({\cal Q}_u|n_u)p(n_u)\, p_d({\cal Q}_d|n_d)p(n_d)\,\Theta({\cal Q}_d-{\cal Q}_u) \nonumber\\
&=\int d{\cal Q}_u\, d{\cal Q}_d\, dn_u\, dn_d\, \frac{1}{\sqrt{2\pi\frac{2}{3}n_u^{1-2\kappa}}}\frac{1}{\sqrt{2\pi\frac{2}{3}n_d^{1-2\kappa}}}
\exp\left[
-\frac{\left({\cal Q}_u - \frac{2}{3} n_u^{-\kappa}\right)^2}{\frac{4}{3}n_u^{1-2\kappa}}-\frac{\left({\cal Q}_d + \frac{1}{3} n_d^{-\kappa}\right)^2}{\frac{4}{3}n_d^{1-2\kappa}}
\right]\nonumber\\
&\hspace{2cm}\times p(n_u)\,p(n_d)\,\Theta({\cal Q}_d-{\cal Q}_u) \nonumber\,.
\end{align}
With our goal of establishing the leading dependence on the shape of the multiplicity distribution $p(n)$, we just use the leading terms for the mean and variance of the Gaussians.  I encourage you to evaluate the AUC with more general expressions, however!

We can explicitly integrate over the jet charges ${\cal Q}_u,{\cal Q}_d$ which is made easier with the change of variables
\begin{align}
&x = n_u^\kappa {\cal Q}_u\,,
&y = n_d^\kappa {\cal Q}_d\,.
\end{align}
This eliminates much of the $\kappa$ dependence, but cannot remove it completely:
\begin{align}\label{eq:aucchargeud}
\text{AUC}_{{\cal Q}_\kappa} &=\int dx\, dy\, dn_u\, dn_d\, \frac{1}{\sqrt{2\pi\frac{2}{3}n_u}}\frac{1}{\sqrt{2\pi\frac{2}{3}n_d}}
\exp\left[
-\frac{\left(x - \frac{2}{3} \right)^2}{\frac{4}{3}n_u}-\frac{\left(y + \frac{1}{3} \right)^2}{\frac{4}{3}n_d}
\right]\\
&\hspace{2cm}\times p(n_u)\,p(n_d)\,\Theta(n_d^{-\kappa}y-n_u^{-\kappa}x) \nonumber\\
&=\frac{1}{2}-\frac{1}{2}\int dn_u\, dn_d\, p(n_u)\, p(n_d)\,\text{erf}\left(
\frac{n_u^\kappa+2n_d^\kappa}{2\sqrt{3}\sqrt{n_u^{2\kappa}n_d+n_d^{2\kappa}n_u}}
\right)\nonumber\,.
\end{align}
For generic random variables $n_u,n_d$, the argument of the error function in the remaining integral over multiplicities is not monotonic in $\kappa$, so to establish the $\kappa$ dependence of discrimination power requires actually integrating over the multiplicity distributions.  Nevertheless, at this point, we can directly observe that discrimination power of the jet charge is necessarily dependent on the energy scaling exponent $\kappa$, which is consistent with our earlier analysis.

We will integrate over the multiplicity distributions shortly, but at this stage it is also useful to stare at some expressions for particularly simple, limiting choices of $\kappa$.  While $\kappa = 0$ has some problems with interpretation as discussed earlier, its AUC is rather simple:
\begin{align}\label{eq:jetchargeauc0}
\text{AUC}_{{\cal Q}_0} =\frac{1}{2}-\frac{1}{2}\int dn_u\, dn_d\, p(n_u)\, p(n_d)\,\text{erf}\left(
\frac{\sqrt{3}}{2\sqrt{n_d+n_u}}
\right)\,.
\end{align}
In the limit that $\kappa = 0$, all particles contribute equally to the jet charge, independent of energy, and note that here, the argument of the error function is a democratic sum of multiplicities, $n_u+n_d$.  In the opposite limit, $\kappa\to \infty$, in which the jet charge is fixed by the hardest particle in the jet alone, its AUC is
\begin{align}
&\text{AUC}_{{\cal Q}_\infty} \\
&\hspace{0.5cm}= \frac{1}{2}-\frac{1}{2}\int dn_u\, dn_d\, p(n_u)\, p(n_d)\,\left[
\Theta(n_u - n_d)\,\text{erf}\left(
\frac{1}{2\sqrt{3}\sqrt{n_d}}
\right)+\Theta(n_d - n_u)\,\text{erf}\left(
\frac{1}{\sqrt{3}\sqrt{n_u}}
\right)
\right]\,.\nonumber
\end{align}
Only keeping the leading expression for the mean and variance of the jet charge distributions effectively means that the energy fraction distribution is just a $\delta$-function:
\begin{align}
p(z|n) = \delta\left(
z - \frac{1}{n}
\right)\,.
\end{align}
Therefore, if the multiplicity of the up quark jet is larger than the down quark jet, $n_u > n_d$, then the hardest particle in the down quark jet is more energetic than that in the up quark jet, and we see that the argument of the error function only depends on the multiplicity $n_d$ in that case.

\subsubsection{Calculating the Jet Charge and Multiplicity Likelihood AUC}

Let's now move to the calculation of the AUC for jets on which the likelihood ratio of jet charge and multiplicity is measured.  By the Neyman-Pearson lemma, we ``know'' that this must be smaller than measuring the jet charge alone, but seeing how it works in action is rather illuminating and informative.  Our first task is to calculate the distribution of the likelihood itself, where
\begin{align}
p({\cal L}) &= \int d{\cal Q}_\kappa\, dn\, p({\cal Q}_\kappa|n)\, p(n)\, \delta\left(
{\cal L} - \frac{3}{2} n^{-1+\kappa}{\cal Q}_\kappa + \frac{1}{4n} +\cdots
\right)\,.
\end{align}
As discussed earlier, this is actually the logarithm of the likelihood ratio (which eliminates extreme exponential suppression), and we will work, as before, to leading-order in the expansion of the likelihood in the moments of the energy fraction distribution.  The AUC of the likelihood can then be expressed as
\begin{align}
\text{AUC}_{\cal L} &=\int d{\cal L}_u\, d{\cal L}_d\, p_u({\cal L}_u)\, p_d({\cal L}_d)\,\Theta({\cal L}_d-{\cal L}_u)\\
&= \int d{\cal Q}_u\, d{\cal Q}_d\, dn_u\, dn_d\, \frac{1}{\sqrt{2\pi\frac{2}{3}n_u^{1-2\kappa}}}\frac{1}{\sqrt{2\pi\frac{2}{3}n_d^{1-2\kappa}}}
\exp\left[
-\frac{\left({\cal Q}_u - \frac{2}{3} n_u^{-\kappa}\right)^2}{\frac{4}{3}n_u^{1-2\kappa}}-\frac{\left({\cal Q}_d + \frac{1}{3} n_d^{-\kappa}\right)^2}{\frac{4}{3}n_d^{1-2\kappa}}
\right]\nonumber\\
&\hspace{2cm}\times p(n_u)\, p(n_d)\,\Theta\left(
\frac{3}{2} n_d^{-1+\kappa}{\cal Q}_d - \frac{1}{4n_d}-\frac{3}{2} n_u^{-1+\kappa}{\cal Q}_u + \frac{1}{4n_u}
\right)\nonumber\,.
\end{align}

While these integrals still look like a huge mess, we can re-introduce the scaled charge variables
\begin{align}
&x = n_u^\kappa {\cal Q}_u\,,
&y = n_d^\kappa {\cal Q}_d\,,
\end{align}
which, in this case of the likelihood, completely eliminates all explicit $\kappa$ dependence on discrimination:
\begin{align}\label{eq:auclikecharge}
\text{AUC}_{\cal L} &= \int dx\, dy\, dn_u\, dn_d\, \frac{1}{\sqrt{2\pi\frac{2}{3}n_u}}\frac{1}{\sqrt{2\pi\frac{2}{3}n_d}}
\exp\left[
-\frac{\left(x - \frac{2}{3} \right)^2}{\frac{4}{3}n_u}-\frac{\left(y + \frac{1}{3} \right)^2}{\frac{4}{3}n_d}
\right]\\
&\hspace{2cm}\times p(n_u)\, p(n_d)\,\Theta\left(
\frac{6y-1}{n_d} - \frac{6x-1}{n_u}
\right)\nonumber\\
&=\frac{1}{2}-\frac{1}{2}\int dn_u\, dn_d\, p(n_u)\, p(n_d)\, \text{erf}\left(
\frac{\sqrt{3}}{4}\sqrt{\frac{1}{n_u}+\frac{1}{n_d}}
\right)\nonumber\,.
\end{align}
This is a rather amazing result.  It is true that if you only measure the jet charge, then discrimination power is a function of $\kappa$, as, as argued earlier, one should choose reasonably small $\kappa$ for the best performance.  However, if you measure jet charge and multiplicity of the jets, then the resulting likelihood ratio has discrimination power performance that is independent of $\kappa$!  It has been observed that the variance in discrimination power among jet charges with different $\kappa$ is significantly smaller when the jet charge is but one input to a neural network that also includes all particle momenta (and also, therefore, total multiplicity), e.g., \InRef{Fraser:2018ieu}.

\subsubsection{Optimality of the Likelihood}

Let's now compare these results from the measurement of the jet charge to the measurement of the likelihood ratio to explicitly demonstrate the latter's dominance as a discriminant for up versus down jet classification.  The first comparison we will make is of the integrand of the AUC for $\kappa = 0$ jet charge and the likelihood.  We first make the following observation: that the geometric mean of two numbers $n_u,n_d$ is always bounded by their arithmetic mean:
\begin{align}
\sqrt{n_u n_d}\leq \frac{n_u + n_d}{2}\,.
\end{align}
This is equivalent to 
\begin{align}
\sqrt{\frac{n_un_d}{n_u+n_d}}\leq \frac{\sqrt{n_u+n_d}}{2}\,,
\end{align}
or, on inverting this relationship and sprinkling in a couple of numerical factors,
\begin{align}
\frac{\sqrt{3}}{2\sqrt{n_u+n_d}}\leq \frac{\sqrt{3}}{4}\sqrt{\frac{1}{n_u}+\frac{1}{n_d}}\,.
\end{align}
This final expression is a direct comparison of the arguments of the error functions in the integrands of the AUCs for ${\cal Q}_0$ and the likelihood ratio, \Eqs{eq:jetchargeauc0}{eq:auclikecharge}.  The error function monotonically increases as a function of its argument, so this demonstrates directly that $\text{AUC}_{{\cal Q}_0} > \text{AUC}_{\cal L}$.

As mentioned earlier, this simple comparison of integrands doesn't work for generic values of $\kappa$ and instead the remaining integrals over the multiplicities $n_u,n_d$ must be evaluated to establish optimal discrimination performance with the likelihood ratio.  To do these integrals, we need an expression for the multiplicity distribution $p(n)$ and through our detailed study of KNO scaling, we have such an expression.  However, the negative binomial, for example, would produce nasty integrals whose dependence on the relevant parameters would be significantly muddied, so let's again do something extremely simple, but nevertheless robust.  As it worked for us before, let's again use the moment expansion, but this time for the multiplicity distribution about its mean, where
\begin{align}
p(n) = \delta(n-\langle n\rangle)+\frac{\sigma_n^2}{2}\delta''(n-\langle n \rangle)+\cdots\,,
\end{align}
and, as before, we will only explicitly work through the variance.

Now, with this simple expression for the multiplicity distribution, we can evaluate the AUCs for jet charge with arbitrary $\kappa$ and the likelihood.  We find
\begin{align}
\text{AUC}_{{\cal Q}_\kappa} =\frac{1}{2}- \frac{1}{2}\text{erf}\left(
\frac{\sqrt{3}}{2\sqrt{2}}\frac{1}{\sqrt{\langle n\rangle}}
\right)+\frac{\sigma_n^2}{2\langle n\rangle^2}\frac{9-84\langle n\rangle+4\kappa^2+48\langle n\rangle(\kappa - 1)^2}{64\sqrt{6\pi}\langle n\rangle^{3/2}}\, e^{-\frac{3}{8\langle n\rangle}}+\cdots\,,
\end{align}
for the jet charge and for the likelihood, we find
\begin{align}
\text{AUC}_{\cal L} =\frac{1}{2}- \frac{1}{2}\text{erf}\left(
\frac{\sqrt{3}}{2\sqrt{2}}\frac{1}{\sqrt{\langle n\rangle}}
\right)+\frac{\sigma_n^2}{2\langle n\rangle^2}\frac{9-84\langle n\rangle}{64\sqrt{6\pi}\langle n\rangle^{3/2}}\, e^{-\frac{3}{8\langle n\rangle}}+\cdots\,.
\end{align}
Here, we have scaled the variance $\sigma_n^2$ by the squared mean, as, from KNO scaling, we expect that $\sigma_n^2/\langle n\rangle^2 \sim 1$.  The first two terms of both are equal, so we just need to compare the third terms, proportional to the variance of the multiplicity distribution.  The $\kappa$ dependent terms in the jet charge AUC are manifestly positive, and so indeed the AUC for the likelihood is smaller and therefore is a better discriminant than the jet charge alone.  

The AUC for the jet charge suggests a rather interesting optimal choice for $\kappa$, due to the finite width of the multiplicity distribution.  For any reasonable average multiplicity $\langle n\rangle$, the $(\kappa - 1)^2$ term will overwhelmingly dominate the AUC unless $\kappa = 1$, and so it is clear that this is the optimal value to minimize the AUC.  Note that this result is not at odds with what we found above, that discrimination power increases as $\kappa \to 0$.  This previous result assumed a finite width of the particle energy fraction distribution with fixed multiplicity, while this current result assumes a fixed and equal energy fraction for each particle but finite width of the multiplicity distribution.  Actually, combining these results may lead to better insight for other results that suggest that small, but non-zero values of $\kappa$ are actually optimal for jet charge \cite{Krohn:2012fg,Fraser:2018ieu}.  That is, these competing effects of finite energy fraction width that prefers $\kappa\to 0$ and finite multiplicity width that prefers $\kappa \to 1$ are both optimized for intermediate $\kappa$, $\kappa \sim 0.3$ -- $0.5$, or so.

\subsection{Quark versus Gluon Jet Discrimination One Final Time}

As we argued in the previous section, if all you measure are particle momenta, then the hadronic multiplicity itself is an extremely good quark versus gluon discrimination observable, and in the double logarithmic limit, is in fact the likelihood ratio for this problem.  With jet charge, however, there is potentially more useful information for this problem because gluons are electrically neutral while quarks are electrically charged, and so, at least on average the jet charge distributions from quark or gluon jets will be displaced from one another.  In this section, we will explore how multiplicity and the jet charge can be used in concert to further improve our quark versus gluon jet discriminant.

\begin{figure}[t!]
\begin{center}
\includegraphics[width=8cm]{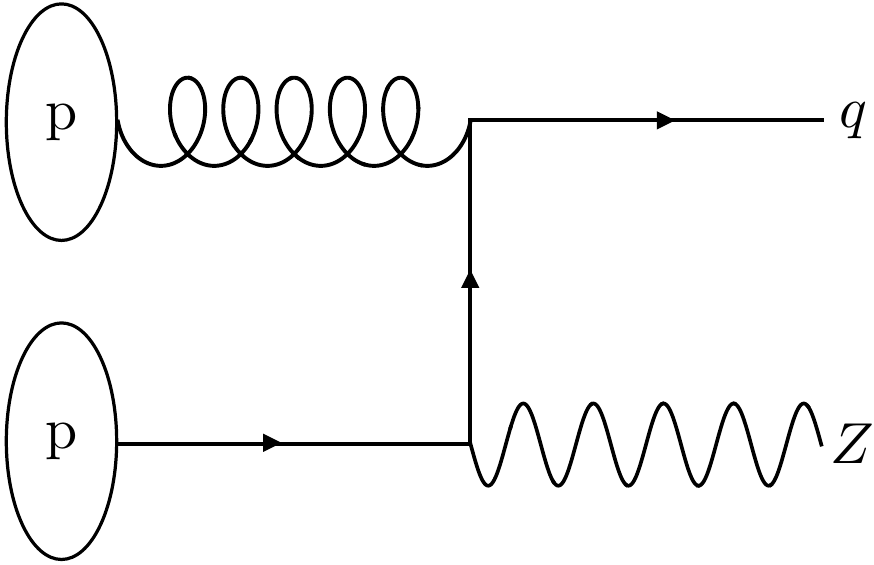}
\caption{\label{fig:ppzjet}
Feynman diagram of $pp\to Z+$ jet events in which a quark parton $q$ is produced at leading order.
}
\end{center}
\end{figure}

Before diving into calculations, there are a few potential confusions to address.  First, in a fully inclusive quark jet sample, where all flavors of quarks initiate jets with equal probability, the average jet charge would be 0.  In your quark jet sample, there would be just as many jets initiated by up quarks as anti-up quarks, and these obviously have opposite electric charge.  However, some simple selection criteria can significantly bias the quark jet flavor; for example, by studying the jet in $pp\to Z+$ jet events.  The leading-order Feynman diagram for this process is illustrated in \Fig{fig:ppzjet}.  Here, the flavor of the quark jet $q$ is exactly the flavor of the parton pulled out of the proton in collision.  The proton is composed of up and down valence quarks, and so dominantly these quark flavors are represented in the jet.  If we just naively assume that the fraction of up and down quarks that initiate the jet in these events is described by the relative quark content in the proton, then $2/3$ of the time the jet is up flavor, while $1/3$ of the time it is down.  The average electric charge of the jet is then
\begin{align}
\langle Q_j\rangle =\frac{2}{3}Q_u + \frac{1}{3}Q_d =  \frac{2}{3}\cdot \frac{2}{3} + \frac{1}{3}\left(
-\frac{1}{3}
\right) = \frac{1}{3}\,.
\end{align}
This is non-zero, and so would be distinct from the charge of gluon jets, at least on average.

For our analysis in this section, we will simply assume that some selection criteria have been made such that gluon jets have 0 average electric charge, while quark jets have some non-zero average charge, $\langle {\cal Q}\rangle_q$.  Other than this and the assumption of KNO scaling of the multiplicity, we will again work within the assumptions that we laid out in \Sec{sec:nonpertassumpts}.

\subsubsection{Jet Charge Distributions on Quarks and Gluons and Discrimination Measures}

Our first step to analyzing this problem is to determine the joint probability distributions of jet charge and multiplicity on quark and gluon jets.  This can be expressed as the product of the jet charge distribution conditioned on multiplicity and the multiplicity itself as
\begin{align}
p_q({\cal Q}_\kappa,n) = p_q({\cal Q}_\kappa|n)p_q(n) = \frac{1}{\sqrt{2\pi\sigma_q^2}}\,\exp\left[
-\frac{({\cal Q}_\kappa-\langle {\cal Q}\rangle_q)^2}{2\sigma_q^2}
\right]\,p_q(n)\,,
\end{align}
for quarks, and
\begin{align}
p_g({\cal Q}_\kappa,n) = p_g({\cal Q}_\kappa|n)p_g(n) = \frac{1}{\sqrt{2\pi\sigma_g^2}}\,\exp\left[
-\frac{{\cal Q}_\kappa^2}{2\sigma_g^2}
\right]\,p_g(n)\,,
\end{align}
for gluons.  Here, we have assumed that the widths of the jet charge distributions are not equal for quark and gluon jets, but they can nevertheless be expressed in terms of moments of the corresponding energy fraction distributions, where
\begin{align}
&\sigma_q^2 = \frac{2}{3}n\langle z^{2\kappa}\rangle_q\,,
&\sigma_g^2 = \frac{2}{3}n\langle z^{2\kappa}\rangle_g\,.
\end{align}
Here, the energy fraction moment is defined with respect to the corresponding energy fraction distribution from a jet of flavor $f$ as
\begin{align}
\langle z^{2\kappa}\rangle_f = \int_0^1 dz\,z^{2\kappa}\, p_f(z|n)\,.
\end{align}
The average value of the quark's jet charge can be expressed similarly, where
\begin{align}
\langle {\cal Q}\rangle_q = Q_q\langle z^\kappa\rangle_q\,.
\end{align}

As before, we use the moment expansions for the energy fraction distributions through the variance,
\begin{align}
p_q(z|n) = \delta\left(
z-\frac{1}{n}\right)+\frac{\sigma_{q,z}^2}{2}\delta''\left(
z-\frac{1}{n}\right)+\cdots\,,\\
p_g(z|n) = \delta\left(
z-\frac{1}{n}\right)+\frac{\sigma_{g,z}^2}{2}\delta''\left(
z-\frac{1}{n}\right)+\cdots\,.
\end{align}
We assume that the widths of these distributions are different on quark and gluon jets, $\sigma_{q,z}^2 \neq \sigma_{g,z}^2$, and further, there is evidence that the energy distribution on gluon jets is narrower, $\sigma_{q,z}^2 > \sigma_{g,z}^2$, which can be extracted from fragmentation functions and mean multiplicity measurements of quark and gluon jets from $e^+e^-$ collisions \cite{DELPHI:1995nzf,DELPHI:1997oih,OPAL:2004prv}.  That is, gluons fragment into hadrons that carry more similar energies than in comparable quark jets.\footnote{Here, we assume that the average jet charge on quark jets is non-zero.  However, one could instead consider the case in which both quark and gluon jets have zero mean jet charge, and only the widths of the jet charge distributions differ.  This would produce a different expression for the likelihood ratio, but would still produce improved discrimination performance over multiplicity alone.}

\begin{figure}[t!]
\begin{center}
\includegraphics[width=7.5cm]{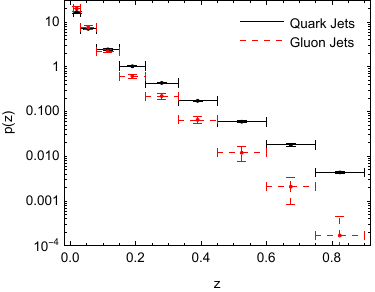}
\hspace{0.5cm}
\includegraphics[width=7.5cm]{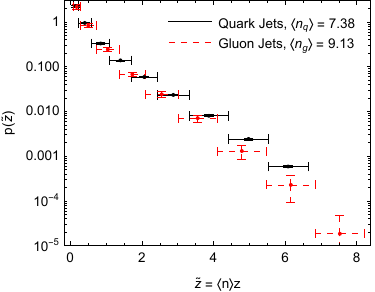}
\caption{\label{fig:fragscale}
Plots of the energy fraction distribution $p(z)$ from $e^+e^-\to 3$ jets events on quark (black) and gluon (red) jets with energies of about 25 GeV.  At left are the normalized energy fraction distributions and at right are those distributions rescaled by the corresponding charged particle multiplicities. Data from \InRef{OPAL:2004prv} and mean multiplicities from \InRef{DELPHI:1995nzf}.
}
\end{center}
\end{figure}

Evidence for this is presented in \Fig{fig:fragscale}.  At left, we plot the energy fraction distributions $p(z)$ as measured on quark and gluon jets in $e^+e^-\to 3$ jets events \cite{OPAL:2004prv}, where the jet energy is about 25 GeV.  The gluon jet distribution is clearly narrower than that of quark jets, but we also expect that the particle multiplicity of gluon jets is larger than that of quarks.  Because of this fact, the average energy fraction $\langle z\rangle$ inclusive in the number of particles $n$ in the jet is expected to scale inversely with the mean multiplicity:
\begin{align}
\langle z\rangle = \int dz\, dn\, z\, p(z|n)\, p(n) = \int dn\, \frac{p(n)}{n} = \frac{1}{\langle n\rangle} + \cdots\,,
\end{align}
where at right, we have expanded the multiplicity distribution $p(n)$ to lowest order in the moment expansion, 
\begin{align}
p(n) = \delta(n-\langle n\rangle)+\cdots\,.
\end{align}
For sensitivity to the variance of the exclusive energy fraction distribution, $p(z|n)$, we rescale the inclusive distribution by the mean multiplicity, $\tilde z = \langle n\rangle z$, at right in \Fig{fig:fragscale}.  While the differences between quarks and gluons are significantly reduced with this rescaling, the gluon jet distribution is still slightly narrower than that of quark jets, suggesting that $\sigma_{q,z}^2 > \sigma_{g,z}^2$.  However, much clearer evidence would be provided by simply binning in particle multiplicity, but to my knowledge, such data do not exist.

With these results, the mean values of the conditional jet charge distribution are
\begin{align}
&\mu_q = Q_q n^{-\kappa}\left(
1-\frac{\kappa(1-\kappa)}{2}\,\sigma_{q,z}^2n^2+\cdots
\right)\,, &\mu_g = 0\,,
\end{align}
and the variances are
\begin{align}
&\sigma_q^2 = \frac{2}{3}n^{1-2\kappa}\left(
1-\kappa(1-2\kappa)\,\sigma_{q,z}^2n^2+\cdots
\right)\,,\\ &\sigma_g^2 = \frac{2}{3}n^{1-2\kappa}\left(
1-\kappa(1-2\kappa)\,\sigma_{g,z}^2n^2+\cdots
\right)\,.\nonumber
\end{align}
The distance between means as measured by the sum of the variances of the conditioned jet charge distributions is then
\begin{align}
\frac{|\mu_q-\mu_g|}{\sqrt{\sigma_q^2+\sigma_g^2}} = \frac{\sqrt{3}}{2}\frac{|Q_q|}{\sqrt{n}}\left[
1-\frac{\kappa}{4}\left(
1 - (1-2\kappa)\frac{\sigma_{g,z}^2}{\sigma_{q,z}^2}
\right)\sigma_{q,z}^2n^2+\cdots
\right]\,.
\end{align}
This distance as a measure of the discrimination power of jet charge and multiplicity is rather interesting.  The overall scaling with the multiplicity is similar to that for up versus down discrimination, and shows that as multiplicity decreases, this distance between distributions increases, and so too does discrimination power.

The $\kappa$ dependence is much less obvious, but let's take a derivative of the $\kappa$-dependent piece and see what we find.  We have
\begin{align}
\frac{d}{d\kappa}\left[
-\kappa\left(
1 - (1-2\kappa)\frac{\sigma_{g,z}^2}{\sigma_{q,z}^2}
\right)
\right] = -1+(1-4\kappa)\frac{\sigma_{g,z}^2}{\sigma_{q,z}^2}\,.
\end{align}
Because we assume that $\sigma_{g,z}^2 < \sigma_{q,z}^2$, this is necessarily negative for all $\kappa > 0$, and so discrimination power is still optimized for small $\kappa$, $\kappa \to 0$.

\subsubsection{The Likelihood Ratio and Its AUC}

While it would be interesting to study the dependence on the quark versus gluon jet AUC as a function of the shape of the energy fraction distributions, we will restrict our study here to the leading scaling dependence.  That is, we will just take
\begin{align}
\langle z^{2\kappa}\rangle_f = n^{-2\kappa}+\cdots\,,
\end{align}
and so the functional form of the distributions we work with are
\begin{align}
p_q({\cal Q}_\kappa,n) &= \frac{1}{\sqrt{2\pi\frac{2}{3}n^{1-2\kappa}}}\,\exp\left[
-\frac{({\cal Q}_\kappa-Q_q n^{-\kappa})^2}{\frac{4}{3}n^{1-2\kappa}}
\right]\,p_q(n)\,,\\
p_g({\cal Q}_\kappa,n) &= \frac{1}{\sqrt{2\pi\frac{2}{3}n^{1-2\kappa}}}\,\exp\left[
-\frac{{\cal Q}_\kappa^2}{\frac{4}{3}n^{1-2\kappa}}
\right]\,p_g(n)\,.
\end{align}
With these distributions, we then find that the logarithm of the likelihood ratio is
\begin{align}
\log{\cal L} =\log \frac{p_g({\cal Q}_\kappa,n)}{p_q({\cal Q}_\kappa,n)} =
-\frac{3}{2}Q_q n^{-1+\kappa}{\cal Q}_\kappa + \frac{3}{4}\frac{Q_q^2}{n}+\cdots
+\log\frac{p_g(n)}{p_q(n)}\,.
\end{align}
As observed with up versus down quark jet discrimination, this likelihood ratio is monotonic in neither the jet charge nor the multiplicity, and so is necessarily a better discriminant than either.  The ellipses hide dependence on the explicit moments of the energy fraction distributions, $p_q(z|n)$ and $p_g(z|n)$.

With these restrictions, the AUC for quark versus gluon discrimination with the likelihood ratio of jet charge and multiplicity is
\begin{align}\label{eq:aucjetchargemultqvg}
\text{AUC}_{{\cal Q}_\kappa,n} &= \int dn_q\, dn_g\, d{\cal Q}_q\, d{\cal Q}_g\, p_q({\cal Q}_q,n_q)\, p_q({\cal Q}_g,n_g)\,\Theta\left(
\log \frac{p_g({\cal Q}_q,n_q)}{p_q({\cal Q}_q,n_q)}-\log \frac{p_g({\cal Q}_g,n_g)}{p_q({\cal Q}_g,n_g)}
\right)\\
&= \int dn_q\, dn_g\, d{\cal Q}_q\, d{\cal Q}_g\,\frac{1}{2\pi\frac{2}{3}\sqrt{n_q^{1-2\kappa}n_g^{1-2\kappa}}}\,\exp\left[
-\frac{({\cal Q}_q-Q_q n_q^{-\kappa})^2}{\frac{4}{3}n_q^{1-2\kappa}}
-\frac{{\cal Q}_g^2}{\frac{4}{3}n_g^{1-2\kappa}}
\right]\,p_q(n_q)\,p_g(n_g)\nonumber\\
&\hspace{1cm}\times\Theta\left(
-\frac{3}{2}Q_q n_q^{-1+\kappa}{\cal Q}_q+ \frac{3}{4}\frac{Q_q^2}{n_q}+\frac{3}{2}Q_q n_g^{-1+\kappa}{\cal Q}_g- \frac{3}{4}\frac{Q_q^2}{n_g}+\log\frac{p_g(n_q)}{p_q(n_q)}-\log\frac{p_g(n_g)}{p_q(n_g)}
\right)\nonumber\\
&=\int dn_q\, dn_g\,p_q(n_q)\,p_g(n_g)\, \left[\frac{1}{2}+\frac{1}{2}\text{erf}\left(
\frac{4n_gn_q \Delta_{\min}-3(n_q+n_g)Q_q^2}{4\sqrt{3}\sqrt{n_qn_g(n_q+n_g)}|Q_q|}
\right)\right]
\nonumber\,.
\end{align}
For compactness, we have introduced
\begin{align}
\Delta_{\min} \equiv\log\frac{p_g(n_q)}{p_q(n_q)}- \log\frac{p_g(n_g)}{p_q(n_g)}\,.
\end{align}

We would like to compare this to the situation in which we only measure the multiplicity on the jets, which would have a corresponding AUC of 
\begin{align}
\text{AUC}_n = \int dn_q\, dn_g\,p_q(n_q)\,p_g(n_g)\,\Theta(n_q-n_g)\,.
\end{align}
Note that to the order of approximations we work, the quark and gluon jet charge distributions have the same width, and only their mean differs by the net charge $Q_q$.  So, if the charge $Q_q = 0$, the quark and gluon jet charge distributions would be identical, and no discrimination information could be gained from measuring the jet charge.  Thus, for comparison to just measuring the multiplicity, it makes sense to expand the expression for the AUC of \Eq{eq:aucjetchargemultqvg} in the value of the net charge $Q_q$ itself, assuming that $|Q_q|\ll 1$, which is expected as discussed earlier.  For the most direct comparison of these two expressions for the AUCs, we will need a singular expansion of the error function, in terms of Heaviside $\Theta$-functions and its derivatives (which are Dirac $\delta$-functions and its derivatives).  However, from this lecture, we are now professionals at this, and this shouldn't scare us one bit.

Our goal is then to expand the error function as
\begin{align}
\frac{1}{2}+\frac{1}{2}\text{erf}\left(
\frac{4n_gn_q \Delta_{\min}-3(n_q+n_g)Q_q^2}{4\sqrt{3}\sqrt{n_qn_g(n_q+n_g)}|Q_q|}
\right) = \Theta(n_q-n_g) + c_1|Q_q|^{a_1} \delta(n_q-n_g)+\cdots\,,
\end{align}
where terms with higher powers of $Q_q$ and more derivatives of the $\delta$-function are encoded in the ellipses.  We will just work to calculate the coefficient $c_1$ and power $a_1$ here.  We first need to validate the lowest-order term.  To do this, we can simplify the argument of the error function in the $Q_q\to 0$ limit as
\begin{align}
\lim_{Q_q\to 0}\left[
\frac{1}{2}+\frac{1}{2}\text{erf}\left(
\frac{4n_gn_q \Delta_{\min}-3(n_q+n_g)Q_q^2}{4\sqrt{3}\sqrt{n_qn_g(n_q+n_g)}|Q_q|}
\right) 
\right] = \frac{1}{2}+\frac{1}{2}\text{erf}\left(
\frac{n_q-n_g}{|Q_q|}
\right) \,.
\end{align}
Note that $\Delta_{\min}$ is monotonic in $n_q-n_g$ and all other factors of multiplicity in the error function are positive, and so can safely be ignored to this order because they do not affect the ordering of the argument.  Now, in the limit that $Q_q\to 0$, the error function rapidly switches from positive to negative values about $n_q-n_g = 0$, and, with the vertical displacement and scaling, exactly corresponds to the step function:
\begin{align}
\lim_{Q_q\to 0}\left[
\frac{1}{2}+\frac{1}{2}\text{erf}\left(
\frac{4n_gn_q \Delta_{\min}-3(n_q+n_g)Q_q^2}{4\sqrt{3}\sqrt{n_qn_g(n_q+n_g)}|Q_q|}
\right) 
\right] &=\lim_{Q_q\to 0}\left[ \frac{1}{2}+\frac{1}{2}\text{erf}\left(
\frac{n_q-n_g}{|Q_q|}
\right)\right] \\
&= \Theta(n_q-n_g)\,.
\nonumber
\end{align}

For the next term, proportional to the $\delta$-function, we take the derivative with respect to $|Q_q|$ in the error function to produce
\begin{align}
&\frac{d}{d|Q_q|}\left[
\frac{1}{2}+\frac{1}{2}\text{erf}\left(
\frac{4n_gn_q \Delta_{\min}-3(n_q+n_g)Q_q^2}{4\sqrt{3}\sqrt{n_qn_g(n_q+n_g)}|Q_q|}
\right) 
\right] \\
&
\hspace{1cm}= -\frac{4n_gn_q \Delta_{\min}+3(n_q+n_g)Q_q^2}{4\sqrt{\pi}\sqrt{3}\sqrt{n_qn_g(n_q+n_g)}Q_q^2}\,\exp\left[-\frac{(4n_gn_q \Delta_{\min}-3(n_q+n_g)Q_q^2)^2}{48n_qn_g(n_q+n_g)Q_q^2}\right]\nonumber\,.
\end{align}
We will first approximate this derivative as
\begin{align}
-\frac{4n_gn_q \Delta_{\min}+3(n_q+n_g)Q_q^2}{4\sqrt{\pi}\sqrt{3}\sqrt{n_qn_g(n_q+n_g)}Q_q^2}\,\exp\left[-\frac{(4n_gn_q \Delta_{\min}-3(n_q+n_g)Q_q^2)^2}{48n_qn_g(n_q+n_g)Q_q^2}\right] =a_1 \tilde c_1 |Q_q|^{a_1-1}\,\delta(4n_gn_q \Delta_{\min})\,,
\end{align}
for some coefficient $\tilde c_1$ and we have taken the derivative of the coefficient of the $\delta$-function term, as well.  To determine this, we simply integrate both sides over all values of $x \equiv 4n_gn_q \Delta_{\min}$ and find
\begin{align}
a_1 \tilde c_1 |Q_q|^{a_1-1} &= -\int_{-\infty}^\infty dx\,\frac{x+3(n_q+n_g)Q_q^2}{4\sqrt{\pi}\sqrt{3}\sqrt{n_qn_g(n_q+n_g)}Q_q^2}\,\exp\left[-\frac{(x-3(n_q+n_g)Q_q^2)^2}{48n_qn_g(n_q+n_g)Q_q^2}\right] \\
&=-6(n_q+n_g)|Q_q|
\nonumber\,.
\end{align}
Matching powers of $|Q_q|$, we see that $a_1 = 2$ and we can expand the error function in powers of the average quark jet charge as
\begin{align}
\frac{1}{2}+\frac{1}{2}\text{erf}\left(
\frac{4n_gn_q \Delta_{\min}-3(n_q+n_g)Q_q^2}{4\sqrt{3}\sqrt{n_qn_g(n_q+n_g)}|Q_q|}
\right)  &= \Theta(n_q-n_g)-3(n_q+n_g)Q_q^2\,\delta\left(
4n_gn_q \Delta_{\min}
\right)+\cdots\nonumber\\
&\hspace{-3cm}=\Theta(n_q-n_g)-\frac{3}{4}\frac{n_q+n_g}{n_qn_g}Q_q^2\,\delta\left(
\log\frac{p_g(n_q)}{p_q(n_q)}- \log\frac{p_g(n_g)}{p_q(n_g)}
\right)+\cdots\,.
\end{align}

With this result, this explicitly demonstrates that the AUC when jet charge and multiplicity are measured is smaller than just that for multiplicity:
\begin{align}\label{eq:aucdiffnvsnq}
\text{AUC}_{{\cal Q}_\kappa,n} - \text{AUC}_{n} &=-\frac{3}{4}Q_q^2 \int dn_q\, dn_g\, p_q(n_q)\, p_g(n_g)\,\frac{n_q+n_g}{n_qn_g}\,\delta\left(
\log\frac{p_g(n_q)}{p_q(n_q)}- \log\frac{p_g(n_g)}{p_q(n_g)}
\right)+\cdots \nonumber\\
&< 0\,.
\end{align}
This result nicely ties together our entire analysis on quark versus gluon jet discrimination.  The most striking differences between quarks and gluons are their QCD color charges (which control the number of particles produced in the jets) and their electric charges (which control the average net electric charge of the jets).  The observables that are most sensitive to these differences in quantum numbers are the total hadronic multiplicity and the jet charge, and their simultaneous measurement on jets improves discrimination performance over using either individually.  We can, of course, continue to refine and improve our picture and definition of what ``quark'' or ``gluon'' jets are to squeeze more discrimination power out, but direct sensitivity to color and electric charges is both extremely robust (ultimately sensitive to particle quantum numbers) and, as we have shown, easy to analyze (data space is only two dimensional).

While we will stop here in these lectures, I encourage you to expand the expressions for the AUCs to higher orders in the moment expansions, or to use the explicit negative binomial representation of the KNO function for the multiplicity distribution to quantitatively evaluate the improvement in discrimination performance.

\subsection*{Exercises}

\begin{enumerate}

\item In our analysis of up versus down quark jet discrimination, we left the value of explicit moments of the energy fraction distribution $p(z|n)$ implicit, to stress the generality of our results.  However, it is enlightening to do some calculations within specific models to gain intuition, which we explore here.

\begin{enumerate}

\item Use the expression for the ``simplest'' reasonable energy fraction distribution from \Eq{eq:simpzdist}, and redo the analyses of this lecture.  Determine exact results from this model, with no moment expansions.

\item In the large multiplicity $n\to \infty$ limit, \Eq{eq:simpzdist} exponentiates which suggests another form of the expansion of this energy fraction distribution as
\begin{align}
p(z|n) = n \, e^{-nz}\left[
1+\sum_{i=2}^\infty c_i\, L_i(nz)
\right]\,.
\end{align}
Here, $L_i(x)$ is a Laguerre polynomial defined as
\begin{align}
L_i(x) = \frac{e^x}{i!}\frac{d^i}{dx^i}\left(
e^{-x}x^i
\right)\,,
\end{align}
and normalized as
\begin{align}
\delta_{ij} = \int_0^\infty dx\, e^{-x}\, L_i(x)\,L_j(x)\,.
\end{align}
Redo the analyses of this lecture with this Laguerre expansion through the first few orders.  What is the value of $c_2$ if the variance saturates the bound set by the distribution of \Eq{eq:worstdist}?

\end{enumerate}

\item Evaluate the AUC for jet charge, \Eq{eq:aucchargeud}, and the likelihood ratio, \Eq{eq:auclikecharge}, with the negative binomial KNO function for the multiplicity distribution, \Eq{eq:knonegbin}.  How do the values of the AUCs depend on the negative binomial parameter $k$?

\item We mentioned that gluon jets are expected to have smaller dispersion in the energy fraction distribution of their constituent particles than quark jets with fixed particle multiplicity.  One interesting observable to consider for quark versus gluon jet discrimination is pTD \cite{CMS:2012rth,CMS:2013wea}, which is exactly this dispersion:
\begin{align}
\text{pTD} = \sum_i z_i^2\,.
\end{align}
We can, more generally, consider observables that are just the sum of energy fractions raised to the power $\kappa$ \cite{Larkoski:2014pca}:
\begin{align}
\lambda^\kappa = \sum_i z_i^\kappa\,.
\end{align}
Such an observable is just like the jet charge, but without sensitivity to particle electric charge.  pTD is $\lambda^2$ while $\lambda^1 = 1$ by energy conservation, and $\lambda^0 = n$, the hadronic multiplicity of the jet, but for other values of $\kappa$, this is a distinct observable.

Use this observable for quark versus gluon jet discrimination.  What value of $\kappa$ provides the best discrimination power?  Does simultaneous measurement of $\lambda^\kappa$ and multiplicity $n$ improve discrimination performance beyond that achieved by only measuring multiplicity?

\end{enumerate}

\section{Summary and Conclusions}\label{sec:summ}

Binary discrimination problems in collider particle physics and especially jet substructure are ubiquitous and a fundamental component of the data inverse problem: given events from collisions at the LHC, what are the physics responsible?  This problem has taken on new breadth and depth in this machine learning era, but beyond and beneath all the noise of the parameters and layers and epochs and architecture of some instantiation of a machine learning algorithm are guiding principles, like the Neyman-Pearson lemma or the central limit theorem, that can be used and exploited to establish a principled prediction of discrimination performance.  We have used these results to study numerous classification problems in these lectures, reviewing established results in the literature and deriving some novel results.

While we have especially taken time to analyze and re-analyze the problem of discrimination of jets initiated by either quarks or gluons, we also studied dominant Higgs boson decays, optimizing general resonance searches, and up versus down quark jet discrimination.  Each of these problems has seen significant, detailed study in the literature.  Other than some passing references, however, perhaps the most glaring omission of discrimination problems is that of identification of boosted hadronic decays of top quarks versus massive jets initiated by light QCD partons.  Its omission was neither an oversight nor a criticism of its relevance (the HEP ML Living Review alone lists at least 30 references that just study top quark identification), but rather due to its significantly more complicated decay phase space.  At leading order, the top quark decays to three partons, and even further enforcing subsequent on-shell decay of the $W$-boson, phase space for top decay is three dimensional.  Further, sensitivity to the flow of color in the decay products of the top only starts with another soft emission, which increases phase space by another three dimensions. We had enough difficulty with two dimensional phase spaces, but perhaps there are simplifications or tricks like those introduced in these lectures that render the problem tractable on a normal sized sheet of paper.

I look forward to observing further progress in both establishing improved performance and deeper understanding of discrimination problems from students of this school or from readers of these lectures.

\section*{Acknowledgements}

I thank Fran\c cois Arleo and St\'ephane Munier for the invitation to lecture in Saint-Jacut-de-la-Mer and to all of the students for their comments, questions, and curiosity throughout the school.

\bibliographystyle{jhep}
\bibliography{refs}

\end{document}